\documentclass[a4paper,oneside]{scrartcl}

\usepackage[T1]{fontenc}
\usepackage[utf8]{inputenc}
\usepackage{hyperref}
\usepackage{amsmath}
\usepackage{amssymb}
\usepackage{slashed}
\usepackage[acronym,nohypertypes={acronym}]{glossaries}
\usepackage{makeidx}
\usepackage[nottoc]{tocbibind}
\usepackage{graphicx}
\usepackage[numbers,sort&compress]{natbib} 
\usepackage[height=25cm,width=16cm]{geometry}
\usepackage[update]{epstopdf}

\makeindex 

\newacronym{qcd}{QCD}{quantum chromodynamics}
\newacronym{qed}{QED}{quantum electrodynamics}
\newacronym{ym}{YM}{Yang--Mills}
\newacronym{dse}{DSE}{Dyson--Schwinger equation}
\newacronym[longplural=Slavnov--Taylor identities]{sti}{STI}{Slavnov--Taylor identity}
\newacronym[longplural=Ward--Takahashi identities]{wti}{WTI}{Ward--Takahashi identity}

\newacronym{1pi}{1PI}{one-particle irreducible}
\newacronym{3pi}{3PI}{three-particle irreducible}

\newacronym{ir}{IR}{infrared}
\newacronym{uv}{UV}{ultraviolet}

\newacronym{rg}{RG}{renormalization group}
\newacronym{frg}{FRG}{functional renormalization group}
\newacronym{ndim}{NDIM}{negative dimension integration method}

\newacronym{st}{ST}{static triangle} 
\newacronym{mag}{MAG}{maximally Abelian gauge}
\newacronym{fmr}{FMR}{fundamental modular region}
\newacronym{rgz}{RGZ}{refined Gribov-Zwanziger}
\newacronym{gz}{GZ}{Gribov-Zwanziger}
\newacronym{ms}{MS}{minimal subtraction}
\newacronym{msbar}{$\overline{\text{MS}}$}{modified minimal subtraction}
\newacronym{mom}{MOM}{momentum subtraction}
\newacronym{njl}{NJL}{Nambu--Jona-Lasinio}
\newacronym{brst}{BRST}{Becchi-Rouet-Stora-Tyutin}
\newacronym{ire}{IR}{infrared exponent}
\newacronym{pt-bfm}{PT-BFM}{pinch technique-background field method}

\begingroup\expandafter\expandafter\expandafter\endgroup
\expandafter\ifx\csname pdfsuppresswarningpagegroup\endcsname\relax
\else
\fi

\title{Nonperturbative properties of Yang-Mills theories}

\author{Markus Q. Huber}

\allowdisplaybreaks 

\newcommand{\de}{\delta}
\newcommand{\eps}{\epsilon}
\newcommand{\ka}{\kappa}

\newcommand{\eref}[1]{Eq.~(\ref{#1})}
\newcommand{\nnnl}{\nonumber\\}	
\newcommand{\fref}[1]{Fig.~\ref{#1}}
\newcommand{\tref}[1]{Tab.~\ref{#1}}
\newcommand{\LL}{{\Lambda^2}}
\newcommand{\LQ}{{\Lambda_\mathrm{QCD}^2}}
\newcommand{\GeV}{\text{GeV}}
\newcommand{\MeV}{\text{MeV}}
\newcommand{\intdq}[1]{\int \frac{d^4 #1}{(2\pi)^4}}

\newcommand{\fourvec}[4]{\begin{pmatrix}#1\\#2\\#3\\#4\end{pmatrix}}
\newcommand{\lambdat}[0]{\frac{g^2}{(4\pi)^3}}
\newcommand*{\vev}[1]{\left\langle #1 \right\rangle}

\newcommand*{\calJ}{\mathcal{J}}
\newcommand*{\vp}{\vec{p}}
\newcommand*{\vq}{\vec{q}}
\newcommand*{\vk}{\vec{k}}
\newcommand*{\vx}{\vec{x}}
\newcommand*{\vy}{\vec{y}}
\newcommand{\mhalfo}{\frac{1}{2}}	

\DeclareMathOperator{\Tr}{Tr}

\newcount\colveccount
\newcommand*\colvec[1]{
        \global\colveccount#1
        \begin{pmatrix}
        \colvecnext
}
\def\colvecnext#1{
        #1
        \global\advance\colveccount-1
        \ifnum\colveccount>0
                \\
                \expandafter\colvecnext
        \else
                \end{pmatrix}
        \fi
}

\newcommand{\showMyBoxes}{0}
\newcommand{\showMyBoxesTwo}{0}

\newcommand{\mybox}[1]{
\ifcase \showMyBoxes

\or
\vskip0.5em
\begin{tcolorbox}[
    breakable,
    center title,
    left=0pt,
    right=0pt,
    top=0pt,
    boxrule=0.75pt,
    bottom=0pt,
    colback=white,
    colframe=black,
    width=\linewidth,
    enlarge left by=0mm,
    boxsep=5pt,
    arc=0pt,outer arc=0pt]
#1
\end{tcolorbox}
\vskip0.5em
\fi
}

\newcommand{\myboxdark}[1]{
\ifcase \showMyBoxes

\or
\begin{tcolorbox}[
    breakable,
    center title,
    left=0pt,
    right=0pt,
    top=0pt,
    boxrule=0.75pt,
    bottom=0pt,
    colback=gray!20,
    colframe=black,
    width=\linewidth,
    enlarge left by=0mm,
    boxsep=5pt,
    arc=0pt,outer arc=0pt]
\textbf{TODO:} #1
\end{tcolorbox}
\vskip1em
\fi
}

\newcommand{\myboxdarkTwo}[1]{
\ifcase \showMyBoxesTwo

\or
\begin{tcolorbox}[
    breakable,
    center title,
    left=0pt,
    right=0pt,
    top=0pt,
    boxrule=0.75pt,
    bottom=0pt,
    colback=gray!20,
    colframe=black,
    width=1.2\linewidth,
    enlarge left by=0mm,
    boxsep=5pt,
    arc=0pt,outer arc=0pt]
\textbf{TODO 2:} #1
\end{tcolorbox}
\vskip1em
\fi
}

\newcommand{\myboxmargin}[1]{
\ifcase \showMyBoxes

\or
\marginpar{
#1
}
\fi
}

\begin{document}

\begin{titlepage}

\makeatletter

\begin{center}
{\Huge
\textbf{
\@title
}}\\

\vskip0.5cm
\LARGE
Markus Q. Huber\\

\vskip0.25cm
\large
\textit{Institut f\"ur Theoretische Physik, Justus--Liebig--Universit\"at Giessen, 35392 Giessen, Germany\\
Institute of Physics, University of Graz, NAWI Graz, Universit\"atsplatz 5, 8010 Graz, Austria}

\vskip2mm
\href{mailto:markus.huber@physik.jlug.de}{markus.huber@physik.jlug.de}

\vskip0.5cm
\makeatletter
\@date
\makeatother

\end{center}

\vskip1.6cm
\hrule width \textwidth
\vskip3mm
\emph{Abstract:}
Yang-Mills theories are an important building block of the standard model and in particular of quantum chromodynamics.
Its correlation functions describe the behavior of its elementary particles, the gauge bosons.
In quantum chromodynamics, the correlation functions of the gluons are basic ingredients for calculations of hadrons from bound state equations or properties of its phase diagram with functional methods.\newline
Correlation functions of gluons are defined only in a gauge fixed setting.
The focus of many studies is the Landau gauge which has some features that alleviate calculations.
I discuss recent results of correlation functions in this gauge obtained from their equations of motions.
Besides the four-dimensional case also two and three dimensions are treated, since the effects of truncations, viz., the procedure to render the infinitely large system of equations finite, can be studied more directly in these cases.
In four dimensions, the anomalous running of dressing functions plays a special role and it is explained how resummation is realized in the case of Dyson-Schwinger equations.\newline
Beyond the Landau gauge other gauges can provide additional insights or can alleviate the development of new methods.
Some aspects or ideas are more easily accessible in alternative gauges and the results presented here for linear covariant gauges, the Coulomb gauge and the maximally Abelian gauge help to refine our understanding of Yang-Mills theories.

\vskip3mm
\hrule width \textwidth

\vskip1.2cm


\makeatother

\end{titlepage}

\tableofcontents

\section{Introduction}

\index{quantum chromodynamics}
\index{standard model}
Quantum chromodynamics is the theory of quarks and gluons describing the strong interaction 
\cite{Marciano:1977su,Brambilla:2014jmp}.
It is part of the Standard Model of particle physics which since its formulation in the 1970's serves as the basic theoretical description of elementary particles and their interactions.
The Standard Model was successfully tested in many different ways and the search for physics 'beyond the Standard Model' is one of the driving forces in modern physics.
To be able to separate 'new' physics from Standard Model physics, it is necessary to understand the latter as good as possible including the strong interaction.

\index{confinement}
\index{chiral symmetry breaking}
\Gls{qcd} is an asymptotically free theory which means that the interaction becomes weak at high energies \cite{Gross:1973id,Politzer:1973fx}.
This makes perturbative studies reliable at high scales.
But \gls{qcd} has many nonperturbative properties as well.
The two most prominent ones are confinement and chiral symmetry breaking.
Confinement describes the fact that no free quark or gluon has ever been observed \cite{Nash:1974dw,Antreasyan:1977va,Stevenson:1978wn,Bergsma:1984yn,HidalgoDuque:2011je}.
Many different mechanisms have been proposed to explain this, see, e.g., Refs.~\cite{Alkofer:2006fu,Greensite:2003bk} for an overview.
Some of them might just be different viewpoints and the picture that will finally emerge will most likely contain aspects of several of them.
The second property refers to the breaking of an approximate symmetry of \gls{qcd}, chiral symmetry.
It is explicitly broken by the nonzero quark masses.
In addition, it is broken dynamically.
This explains observations like the low masses of some mesons, which are the Goldstone bosons of the broken symmetry, and the high masses of other bound states, which consist mostly of dynamically created mass.

\index{lattice QCD}
For the nonperturbative description of \gls{qcd} different methods exist, each with its own advantages and disadvantages.
A very prominent and successful method are Monte Carlo simulations on discretized spacetime \cite{Wilson:1974sk}.
With lattice \gls{qcd} one can successfully describe many aspects of \gls{qcd} including parts of the meson and baryon spectrum \cite{Colangelo:2010et,Gattringer:2010zz,Fodor:2012gf,Aoki:2016frl}.
Also thermodynamic quantities are described nicely for vanishing density \cite{Borsanyi:2013bia,Bazavov:2014pvz}.
When a chemical potential is introduced, however, lattice methods encounter a technical problem in form of a complex phase of the weight factor of the path integral \cite{deForcrand:2010ys}.
Besides, many state of the art calculations in lattice \gls{qcd} require a lot of human effort and large computing resources.
This makes alternative methods interesting which require less resources and do not suffer from the complex action problem.

\index{functional equations}
The approach followed in this work is functional equations.
They constitute sets of integral, differential or integro-differential equations.
They can be applied directly to the action of \gls{qcd}, so they start from first principles, but they also can be applied to effective model descriptions of it.
However, actual calculations typically introduce some modeling, as the underlying equations need to be approximated.
The determination of the reliability of such approximations is one of the challenging problems of this method.
Via such models, also effective models can be introduced into the system which allow a phenomenological successful description of some quantities, as, for example, bound states and their properties, e.g., \cite{Alkofer:2000wg,Bashir:2012fs,Eichmann:2016yit}.

In recent years, though, we have steadily progressed towards a description from first principles.
This required the extension of previous truncation schemes and testing the impact of neglected or modeled parts.
Giving up modeling and employing dynamically calculated quantities instead was an instructive process, since models sometimes had circumvented problems which had not been perceived as such.
A recent breakthrough in this regard was the calculation of the scalar and pseudoscalar glueball spectrum in Yang-Mills theory \cite{Huber:2020ngt} from a parameter-free determination of the correlation functions in Landau gauge \cite{Huber:2020keu}.
The results were in quantitative agreement with corresponding lattice results.

In this work, recent developments for functional equations are described with a focus on the equations of motion of correlation functions.
Calculations of three- and four-point functions are covered as are advancements in solving untruncated propagator equations.
Some aspects, which may initially seem only technical, are treated in detail, because they turn out to be important for self-consistent solutions of the equations.

Physical observables are by definition gauge independent, but it is often helpful to calculate them in a gauge-fixed setting.
Since the choice of gauge should not matter for the result, one has the freedom to choose a gauge that is convenient for the specific task.
Functional equations also operate in a gauge-fixed setting and the choice of gauge plays an important role to keep calculations feasible.
The most advanced studies were done in the Landau gauge, both on the technical and the conceptual level, but a handful of other gauges was used as well.
The purpose of employing alternative gauges is thereby twofold: First, some quantities are more easily accessible in certain gauges.
Second, some techniques are naturally developed in one gauge before they are transferred to other gauges.
We will encounter examples for both cases in this work.

In the second section, the basics of \gls{qcd} and its gauge fixed continuum formulation are introduced.
In Sec.~\ref{chp:methods}, a general introduction to functional methods is given.
After setting up the quantum field theoretical framework, \acrlongpl{dse}, the functional renormalization group, $n$PI effective action techniques and the Hamiltonian approach are introduced.
This section also contains a section dedicated to the discussion of computational tools which are becoming more and more important these days.
The section concludes with an overview of alternative methods to study correlation functions.

Readers familiar with functional equations and their use in \gls{qcd} might skip Sec.~\ref{chp:qcd} and Sec.~\ref{chp:methods} and proceed directly to Sec.~\ref{chp:results_YM_Landau} where developments and results in Landau gauge Yang-Mills theory are presented.
This gauge is in many respects the most accessible gauge for functional but also other methods, as explained in detail in Sec.~\ref{sec:why_LG}.
Fundamental aspects important for the future development of \acrlongpl{dse} like the treatment of spurious divergences and perturbative resummation are discussed.
Technical advances treated in this section include the inclusion of two-loop diagrams in the gluon propagator equation and the calculation of non-primitively divergent correlation functions and estimates of their impact on other correlation functions.
For completeness, also results for three-point functions are discussed.
Besides four dimensions, also two and three dimensions are considered.
In particular, in three dimensions a detailed analysis of the truncation dependence is presented.

Correlation functions in gauges other than the Landau gauge are treated in Sec.~\ref{chp:otherGauges}.
Investigations of linear covariant gauges, the maximally Abelian gauge and the Coulomb gauge are presented.
Although the Landau gauge is a special case of a linear covariant gauge, most methods need to be refined for the general case.
With several methods progress in this direction was achieved recently, so that linear covariant gauges receive again more attention.
The attractive feature is that ideas, concepts and methods can be tested with a continuous connection to the Landau gauge.
Another example of a covariant gauge is the maximally Abelian gauge.
It is of interest, because it provides access to the dual superconductor picture of confinement.
Again, similar methods can be employed as in the Landau gauge due to its covariance.
An example for a non-covariant gauge is the Coulomb gauge.
Different functional methods were used in this gauge.
An interesting case is a variational approach, which was developed specifically in this gauge and later extended to the Landau gauge.

Sec.~\ref{chp:conclusions} contains conclusions and is followed by two appendices.
They provide some integral kernels for reference and details on the analytic calculation of the two-loop diagrams in the gluon propagator \acrlong{dse}.

\section{Quantum chromodynamics and gauge fixed correlation functions}
\label{chp:qcd}

This section contains details on \gls{qcd} and its correlation functions in linear covariant gauges.
It serves as a reference for the rest of this work.
In particular, the \glspl{dse} and equations of motion from the 3PI effective action as well as the notation of all correlation functions are presented in Sec.~\ref{sec:corrFuncs}.

\subsection{Basics of QCD}

\subsubsection{The Lagrangian density of QCD}

\index{quantum chromodynamics}
\index{Yang-Mills theory}
\index{flavor}
\index{covariant derivative}
The fundamental matter fields of \gls{qcd} are quark fields.
They appear in three generations with two types each resulting in six flavors: up and down, strange and charm, bottom and top.
The masses of quarks range from a few MeV (up and down) to 173\,GeV (top).
The interaction between quarks is mediated by gluons.
The Lagrangian density of \gls{qcd} is \cite{Fritzsch:1973pi,Yang:1954ek}
\begin{align}\label{eq:LagrangianDensityQCD}
 \mathcal{L}_\text{QCD}&=\sum_{f=\{u,d,s,c,b,t\}}\overline{\psi}_f(-\slashed{D}+m )\psi_f + \mathcal{L}_{YM},\\
 \mathcal{L}_\text{YM}&=\frac{1}{2} \,\Tr\lbrace F_{\mu \nu }F_{\mu \nu} \rbrace \label{eq:actionYM},
\end{align}
where the sum is over all flavors of quark fields $\psi_f(x)$.
In the following, the flavor index will be suppressed again.
The covariant derivative $D_\mu$ contains the gluon field $A_\mu(x)$ as gauge field:
\begin{align}
 D_\mu=\partial_\mu+i\,g\,A_\mu.
\end{align}
The second term, $\mathcal{L}_\text{YM}$, contains only gluons.
Since \gls{qcd} is a non-Abelian gauge theory, the gluons interact among themselves.
In the limit of infinitely heavy quarks, only $\mathcal{L}_\text{YM}$ remains and we have a pure Yang-Mills theory \cite{Yang:1954ek}.
Note that in this work the Euclidean metric is used exclusively.

\index{field strength tensor}
\index{gauge group}
The field strength tensor $F_{\mu \nu}$ is given by
\begin{align}
F_{\mu \nu}&=\partial_\mu A_\nu -\partial_\nu A_\mu +i\,g\,[A_\mu,A_\nu].
\end{align}
The gauge field $A_\mu$ lives in an algebra defined by the hermitian generators $T^r$ of a generic gauge group.
For \gls{qcd} this is $SU(3)$.
They obey the relations
\begin{align}
[ T^r,T^s]&=i f^{rst}T^t,\\
\Tr \lbrace T^r T^s \rbrace &= T_f \delta^{rs},
\end{align}
with $T_f=1/2$ for the gauge group $SU(N)$.
The decomposition of the gauge field is
\begin{align}
 A_\mu=A_\mu^r T^r
\end{align}
and similarly for the field strength tensor:
\begin{align}
 F_{\mu \nu}&=F_{\mu \nu}^r T^r,\\
 F_{\mu \nu }^r&=\partial_\mu A_\nu^r-\partial_\nu A_\mu^r-g\,f^{rst}A_\mu^s A_\nu^t.
\end{align}
In components, the Yang-Mills Lagrangian reads
\begin{align}\label{eq:Lagrangian-YM}
 \mathcal{L}_{YM}&=\frac{1}{4} F_{\mu \nu }^rF^r_{\mu \nu}.
\end{align}

\index{covariant derivative}
\index{gauge transformation}
An important property of the Lagrangian density of \gls{qcd} is that it is invariant under gauge transformations.
In particular, both terms in $\mathcal{L}_\text{QCD}$ are gauge invariant by themselves.
A gauge transformation is given by
\begin{align}\label{eq:gauge-trans}
\psi^U(x)&=U(x)\psi(x),\\
\overline\psi^U(x)&=U^{-1}(x)\overline\psi(x),\\
A^U_\mu(x)&=U(x)\,A_\mu(x)\,U(x)^{-1}+\frac{i}{g}(\partial_\mu U(x))\,U(x)^{-1},
\end{align}
where $U(x)$ is
\begin{align}
U(x)=e^{i\,g\, \omega(x)}
\end{align}
with $\omega(x)$ the Lie algebra valued gauge parameter:
\begin{align}
 \omega(x)=\omega^r(x)T^r.
\end{align}
In infinitesimal form a gauge transformation reads
\begin{align}\label{eq:infinitesimalGT}
 \psi(x) &\rightarrow (1+i\,g\,\omega(x))\psi(x),\\
 \overline\psi(x) &\rightarrow\overline\psi(x) (1-i\,g\,\omega(x)),\\
 A_\mu^r &\rightarrow A_\mu^r+\delta A_\mu^r=A_\mu^r - \partial_\mu \omega^r -g\,f^{rst} \omega^s A_\mu^t=A_\mu^r-D_\mu^{rs} \omega^s,
\end{align}
where the covariant derivative in the adjoint representation $D_\mu^{rs}$ is defined as
\begin{align}
 D_\mu^{rs}=\delta^{rs}\partial_\mu+g\,f^{rst} A_\mu^t.
\end{align}

\subsubsection{Gauge fixing}
\label{sec:gauge_fixing}

\index{Gribov copies}
\index{gauge fixing}
\index{gauge orbit}
The Lagrangian density (\ref{eq:LagrangianDensityQCD}) cannot be used in functional equations as it stands.
Functional equations are formulated in terms of correlation functions of quarks and gluons, which are gauge dependent objects.
Consequently, a gauge must be fixed.
The standard procedure to choose one representative of a class of gauge equivalent configurations, called a gauge orbit, is realized by the Faddeev-Popov method \cite{Faddeev:1967fc}.
Although it was shown that this is insufficient to fix a gauge uniquely \cite{Gribov:1977wm,Singer:1978dk}, it is the standard method for perturbation theory and also functional equations.
For the former, this can be understood by the fact that standard perturbation theory is an expansion around $A_\mu(x)=0$.
Gauge equivalent configurations, so-called Gribov copies, however, appear for large amplitudes far away from $A_\mu(x)=0$.
Thus, they do not have any influence in perturbation theory.
Functional equations, on the other hand, are expected to be applicable also non-perturbatively and thus in the regime where Gribov copies are possibly relevant.
The role of Gribov copies for functional equations is currently an open question.
More in Gribov copies can be found in Sec.~\ref{sec:RGZ}.

\index{gauge fixing}
\index{Faddeev-Popov operator}
\index{gauge orbit}
\index{spin-statistics theorem}
Following the Faddeev-Popov method, we choose a gauge fixing functional $f[A]$.
Its purpose is to select a representative of each gauge orbit.
For now, we ignore the caveat of Gribov copies mentioned above.
A standard choice is $f[A]=\partial\cdot A$.
One can insert the condition $f[A]=0$ into the path integral by writing unity as
\begin{align}\label{eq:Delta}
1=\Delta[A]\int \mathcal{D}U \delta(f[A^U]).
\end{align}
$\mathcal{D}U$ is the integration over the gauge orbit and $\Delta[A]$ is the Jacobian for switching from variables $A$ to $U$.
It is explicitly given by
\begin{align}
 \Delta^{rs}[A]&=\det \left(\frac{\de f^r[A(x)]}{\de \omega^s(y)}\right)=:\det\,M^{rs}(x,y).
\end{align}
$M(x,y)$ is the Faddeev-Popov operator.
A determinant in the path integral can be localized using auxiliary fields called ghost fields.
They need to be scalar Grassmann fields and are thus unphysical, since they violate the spin-statistics theorem:
\begin{align}
 \det\,M^{rs}(x,y)=\int \mathcal{D}[\bar{c}c]e^{\int dx\,dy\,\bar{c}^r(x)\,M^{rs}(x,y)\,c^s(y)}.
\end{align}
For the quantization of Yang-Mills theory in the Landau gauge without ghost fields see, e.g., Ref.~\cite{Huffel:2017wxd}.

\index{gauge fixing}
The delta functional is regularized by writing it as a Gaussian distribution with a width $\xi$.
Although only the limit $\xi\rightarrow 0$ recovers the original gauge fixing condition, also the case with a non-zero width can be considered as a valid choice of gauge.
It corresponds to a weighted sampling over the complete gauge orbit.\footnote{See Sec.~\ref{sec:lattice} for more details on choosing a configuration on the gauge orbit.}
The complete gauge fixing part added to the Lagrangian density is then
\begin{align}\label{eq:Lagrangian-gaugeFixing}
 \mathcal{L}_{\text{gf}}= \frac{1}{2\xi}f[A]^2-\int dy \,\bar{c}^r(x)\,M^{rs}(x,y)\,c^s(y).
\end{align}

\glsreset{brst}
\index{BRST symmetry}
Adding $\mathcal{L}_\text{gf}$ to the Lagrangian density breaks gauge symmetry explictly.
However, there is another symmetry named \emph{\gls{brst} symmetry} after Becchi, Rouet, Stora \cite{Becchi:1975nq,Becchi:1974md} and Tyutin \cite{Tyutin:1975qk}.
It is very useful in proving renormalizability and unitarity of a theory, see, for example, refs. \cite{Kugo:1979gm,Piguet:1995er}.
For the quark and gauge fields, the corresponding transformations take the form of a gauge transformation with the gauge parameter $\omega$ replaced by a ghost field $c$.
The transformations of the ghost fields are constructed such that $\mathcal{L}_\text{gf}$ is invariant after a new field is introduced.

\index{Nakanishi-Lautrup field}
This field, called Nakanishi-Lautrup field $b$ \cite{Nakanishi:1972pt,Lautrup:1967zz}, is not dynamical and takes the role of a Lagrange multiplier for gauge fixing:
\begin{align}
\int Db^r\,e^{-\int dx \left(i b^r f^r[A] \right )}=N \delta(f^r[A]).
\end{align}
$N$ is a normalization factor.
One can then relax the gauge fixing condition $f[A]=0$ into $f[A]=i\,\xi\,b/2$:
\begin{align}
 \int Db^r\,e^{-\int dx \left(i b^r f^r[A]+ \frac{\xi}{2} b^r  b^r \right )}=N e^{-\frac1{2\xi}(f[A])^2}.
\end{align}
This corresponds to the Gaussian averaging over the gauge orbit in linear covariant gauges.

In Landau gauge the off-shell BRST transformation is:
\begin{subequations}\label{eq:BRST}
\begin{align}
s\,A_\mu^r&=-D_\mu^{rs} c^s, \\
s\,c^r&=-\mhalfo g\,f^{rst} c^s c^t, \\
s\,\bar{c}^r&=i \, b^r,\\
s\,b^r&=0.
\end{align}
\end{subequations}
The on-shell form is obtained by integrating out the Nakanishi-Lautrup field what amounts to replacing $b^r$ by $-i\,(\partial_\mu A_\mu^r)/\xi$.
The off-shell BRST transformation is nilpotent, viz., $s^2=0$.
Via the nilpotency one can fix the gauge without the need for a path integral \cite{Kugo:1981hm}.
This relies on the possibility to add any quantity that is the result of a BRST transformation, a so-called BRST exact quantity, to the Lagrangian without spoiling its BRST invariance.
We construct such a term by noting that the gauge fixing condition $f[A]$ has ghost number zero and the BRST transformation raises the ghost number by one.
Thus we introduce the factor $\bar{c}$ in front of $f[A]$ to get ghost number zero.
In the case of the Landau gauge this leads to the already known gauge fixing terms:
\begin{align}\label{eq:BRST-LG}
 \mathcal{L}_{gf}&=s(\bar{c}^r f[A]^r)=s(\bar{c}^r \partial_\mu A^r_\mu)=\nnnl
 &=i\,b^r\,(\partial_\mu A^r_\mu)-\bar{c}^r\, \partial_\mu (-D^{rs}_\mu c^s)= i\,b^r (\partial_\mu A^r_\mu)-\bar{c}^r\,M^{rs}\,c^s.
\end{align}
This method is very general and can also be employed, for example, for gauge fixing conditions including ghost fields \cite{Kugo:1981hm}.
Finally, it should be noted that the expectation value of any gauge invariant quantity remains unaffected by adding such a BRST exact form.
Thus, all physical observables are independent of the chosen gauge as required.

\subsection{Correlation functions of QCD in linear covariant gauges}
\label{sec:corrFuncs}

\index{Green function}
\index{Green's function}
This section contains general information about the correlation functions of \gls{qcd} in linear covariant gauges.\footnote{
In the literature, also the term Green's function is used for a general correlation function, although mathematically only a propagator fulfills its definition.
Both terms \emph{Green function} and \emph{Green's function} are in use, the latter being correct from the historical linguistic viewpoint.
Despite the fact that Green function corresponds to the prevalent style in modern English, Green's function is predominantly used \cite{Wright:2006}.
}
Their structures in color, Lorentz and Dirac spaces are described and their \glspl{dse} are given in untruncated form if not noted otherwise.
In addition, for the three-point functions the equations of motion from a three-loop expansion of the \gls{3pi} effective action are provided.
Their forms are very similar to those of the corresponding \glspl{dse}, and they provide a useful alternative description for these vertices.
Beyond the primitively divergent correlation functions, also details on the two-gluon-two-ghost and the four-ghost vertices are given, as they will be discussed in Sec.~\ref{chp:results_YM_Landau}.
For completeness, also the quark propagator and the quark-gluon vertex are included to provide a consistent description of full QCD.
The derivations of the \glspl{dse} and equations of motion for the 3PI effective action will be described later in Sec.~\ref{chp:methods} where details like signs and numeric prefactors will also be explained.
However, the equations are already given here to collect all the information about the description of correlation functions in one place.
In later sections, subsets of these equations and their truncations will be discussed.

Since many different correlation functions will be treated, it will be convenient to use a common notation where the field content of the correlation functions is put in the upper index.
$A$ denotes a gluon leg, $\bar{c}/c$ a ghost/anti-ghost leg, and  $\bar{q}/q$ a quark/anti-quark leg.
When the field content is clear because of the explicitly given Lorentz and color indices, the field indices may be omitted.
The tree-level expression is denoted by a superscript $(0)$.
Tab.~\ref{tab:names_tensors_dressings} contains an overview of all dressings and tensors introduced in the following.

Correlation functions with gluon legs can be split into transverse and longitudinal parts.
To be precise, there are two ways of splitting a correlation function $\Gamma$, see \cite{Eichmann:2015nra} for a detailed discussion:
\begin{align}\label{eq:trans-long}
 \Gamma=\Gamma^\text{sur}+\Gamma^\text{L}=\Gamma^\text{gauge}+\Gamma^\text{T}.
\end{align}
$\Gamma^\text{T}$ and $\Gamma^\text{L}$ are defined as the parts that vanish when projecting longitudinally and transversely, respectively:
\begin{align}
 T\cdot \Gamma &= T\cdot \Gamma^\text{gauge}+ \Gamma^\text{T} =T\cdot \Gamma^\text{sur},\\
 L\cdot \Gamma &= L\cdot \Gamma^\text{sur}+L\cdot \Gamma^\text{L} =L\cdot \Gamma^\text{gauge}.
\end{align}
$T$ is the transverse and $L$ the longitudinal projector.
In general, $\Gamma^\text{T}\neq \Gamma^\text{sur}$ and $\Gamma^\text{L}\neq\Gamma^\text{gauge}$.
$\Gamma^\text{gauge}$ is often called longitudinal, although it does not vanish when it is projected transversely.
As a consequence, information about $\Gamma^\text{sur}$ can be inferred using gauge techniques \cite{Salam:1963sa,Salam:1964zk,Delbourgo:1977jc,Delbourgo:1977hq} from longitudinal projections.
This was applied in \gls{qcd} for example in Refs.~\cite{Aguilar:2010cn,Rojas:2013tza,Aguilar:2018epe,Aguilar:2014lha,Aguilar:2016lbe,Oliveira:2018fkj,Aguilar:2018csq,Aguilar:2019jsj}.
On the other hand, the transversality of the gluon propagator in the Landau gauge leads to the closure of the parts of correlation functions \cite{Fischer:2008uz} that survive when transverse projections are applied.
Thus, $\Gamma^\text{sur}$ is sufficient in this gauge.
Keeping in mind these comments, we will now continue in agreement with the widespread use in the literature and call it transverse in the following.
It should also be noted that the choice of a basis is subject to many considerations including technical and physical ones.
For a discussion of the latter, see Ref.~\cite{Eichmann:2015nra}.

\begin{table}[tb]
\begin{center}
 \begin{tabular}{|l|c|c|}
\hline
 \textbf{Correlation function} & \textbf{Dressings} & \textbf{Tensors}\\
 \hline\hline
 Ghost propagator & $G$ & 1 \\	
 \hline
 Gluon propagator & $Z$, $Z^L$ & $P$, $L$\\
 \hline
 Quark two-point function & $A$, $B$ & $\slashed{p}$, $1$\\
 \hline
 \hline
 Ghost-gluon vertex & $D^{A\bar c c, T/L}$ & $P_{\mu\nu}(k)p_\nu$, $k_\mu$\\
 \hline
 Three-gluon vertex & $C^{AAA}_i$ & $\tau^i_{\mu\nu\rho}$\\
 \hline
 Quark-gluon vertex & $G^{A\bar q q}_i$ & $\lambda^i_{\mu}$ \\
 \hline
 \hline
 Four-gluon vertex & $F^{AAAA}_i$ & $\rho_{\mu\nu\rho\sigma}^{i,abcd}$\\
 \hline
 Two-ghost-two-gluon vertex & $D^{A A\bar c c}_i$ & $\rho_{\mu\nu}^{i,abcd}$\\
 \hline
 Four-ghost vertex & $E^{\bar c\bar c c c}_i$ & $\sigma^{i,abcd}$\\
 \hline
\end{tabular}
\caption{Overview of names for tensors and dressing functions.
The general pattern for the tensors is: Lorentz tensors $\tau$, color tensors $\sigma$, combined Lorentz and color tensors $\rho$, Dirac tensors $\lambda$.
Dressing functions contain the fields in the superscript.}
\label{tab:names_tensors_dressings}
\end{center}
\end{table}

\subsubsection{Ghost propagator}
\label{sec:cbc}

\begin{figure}[tb]
  \includegraphics[width=0.6\textwidth]{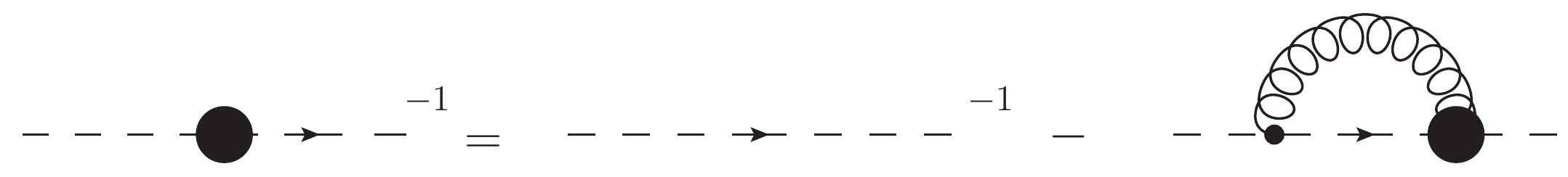}
  \vskip5mm
  \includegraphics[width=0.99\textwidth]{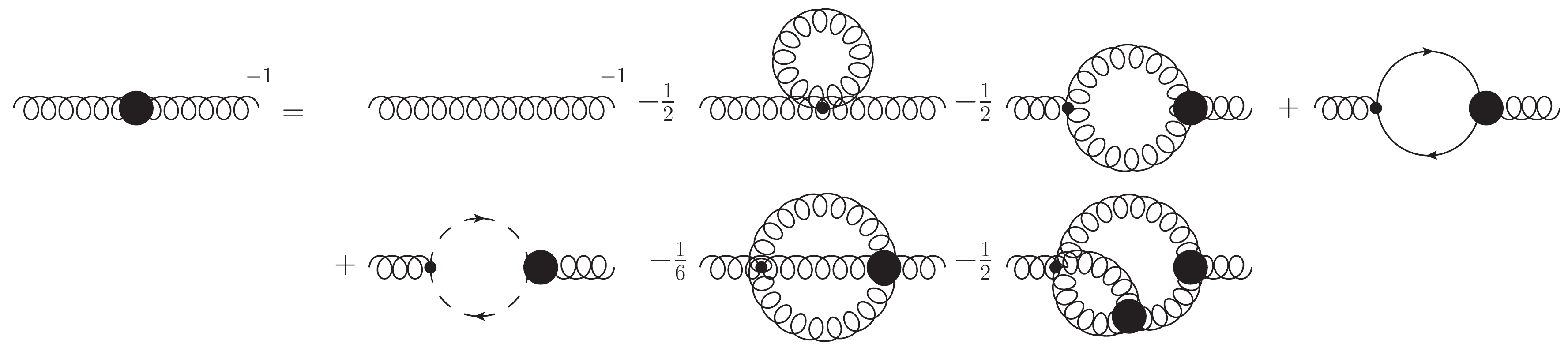}
  \vskip5mm
  \includegraphics[width=0.6\textwidth]{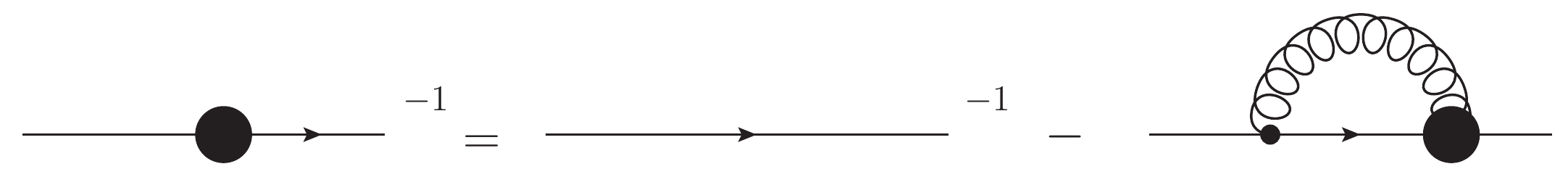}
   \begin{center}
  \caption{Top: The ghost propagator \gls{dse}.
  Center: The gluon propagator \gls{dse}.
  The loop diagrams are called tadpole, gluon loop, quark loop, ghost loop, sunset, and squint.
  Bottom: The quark propagator \gls{dse}.\newline
  Here and in other figures, internal propagators are dressed, and thick blobs denote dressed vertices, wiggly lines gluons, dashed lines ghosts and continuous lines quarks.
  }
  \label{fig:gh-gl-qu_DSEs}
 \end{center}
\end{figure}

\index{ghost propagator}
The ghost field is a scalar field and hence its propagator has one dressing function only.
The negative norm of the ghost field can be taken into account directly in the propagator via a minus sign:
\begin{align}
D^{ab}_G(p)=-\de^{ab}\frac{G(p^2)}{p^2}.
\end{align}
The tree-level corresponds to $G^{(0)}(p^2)=1$.
The full \gls{dse} of the ghost propagator is given in \fref{fig:gh-gl-qu_DSEs}.

\subsubsection{Gluon propagator}
\label{sec:AA}

\index{gluon propagator}
The gluon propagator in linear covariant gauges is uniquely split into a transverse and a longitudinal part parametrized by two dressing functions:
\begin{align}
 D^{ab}_{\mu\nu}(p)&=\de^{ab}D_{\mu\nu}(p)=\de^{ab}(D^T_{\mu\nu}(p)+D^L_{\mu\nu}(p)),\\
 D^T_{\mu\nu}(p)&=\left(g_{\mu\nu}-\frac{p_\mu p_\nu}{p^2}\right)\frac{Z(p^2)}{p^2},\\
 D^L_{\mu\nu}(p)&=\frac{p_\mu p_\nu}{p^2}\frac{Z^L(p^2)}{p^2}.
\end{align}
At the tree-level we have $Z^{(0)}(p^2)=1$ and $Z^{L,(0)}(p^2)=\xi$.
The \gls{sti} of the gluon propagator fixes the longitudinal part to be equal to the gauge fixing parameter also beyond tree-level, $Z^L(p^2)=\xi$.
The full \gls{dse} of the gluon propagator is given in \fref{fig:gh-gl-qu_DSEs}.
It is equivalent to the corresponding equation of motion of the 4PI effective action \cite{Carrington:2010qq}.

\subsubsection{Quark propagator}
\label{sec:qbq}

\index{quark propagator}
The quark propagator depends on two dressing functions which can be chosen in various ways.
A typical parametrization starts from its inverse:
\begin{align}
 (S^{ij}(p))^{-1}&=\de^{ij}(i A(p^2)\slashed{p}+B(p^2)).
\end{align}
The tree-level is $A^{(0)}(p^2)=1$ and $B^{(0)}(p^2)=m_0$ with $m_0$ the bare quark mass.
The propagator in terms of the two scalar functions $A(p^2)$ and $B(p^2)$ reads then
\begin{align}
 S^{ij}(p)&=\de^{ij}\frac{-i A(p^2)\slashed{p}+B(p^2)}{p^2 A(p^2)^2+B(p^2)^2}.
\end{align}
Two alternative parametrizations use the scalar and the vector dressing functions $\sigma_s(p^2)$ and $\sigma_v(p^2)$, respectively, or the quark renormalization function $Z_f(p^2)=1/A(p^2)$ and the quark mass function $M(p^2)=B(p^2)/A(p^2)$:
\begin{align}
 S^{ij}(p)&=\de^{ij}(-i \sigma_v(p^2)\slashed{p}+\sigma_s(p^2))=\de^{ij}\frac{Z_f(p^2)}{p^2+M(p^2)^2}\left(-i \slashed{p}+M(p^2)\right).
\end{align}
The full \gls{dse} of the quark propagator is given in \fref{fig:gh-gl-qu_DSEs}.

\subsubsection{Ghost-gluon vertex}
\label{sec:Acbc}

\index{ghost-gluon vertex}
Three-point functions depend on two independent external momenta from which one constructs three variables.
Possible sets are three squared momenta, two squared momenta and the angle between the two or even more complex combinations, see, e.g., \cite{Eichmann:2014xya}.
In the following, typically the three external momenta are used as arguments.
However, in plots for specific momentum configurations either three momentum squares or two momentum squares and the angle between the two corresponding momenta are used, e.g., $D^{A\bar cc}(k^2;p^2,q^2)$ or $D^{A\bar cc}(k^2;p^2,\alpha)$.
The first two arguments refer to the gluon and anti-ghost momenta.

The full ghost-gluon vertex can be parametrized as
\begin{align}
 \Gamma^{A\bar cc,abc}_\mu(k;p,q)=i\,g\,f^{abc}(A(k;p,q)p_\mu+B(k;p,q) k_\mu),
\end{align}
where the momentum arguments correspond to the order of the fields in the superscript.
For a discussion of the color part see Sec.~\ref{sec:AAA}.
All momenta are taken as incoming.
The bare vertex has $A^{(0)}(k;p,q)=1$ and $B^{(0)}(k;p,q)=0$.
This parametrization of the vertex contains a part that is proportional to $k_\mu$ and thus vanishes upon contraction with the transverse projector.
Contracting with a longitudinal projector, both tensors survive.
Thus, it is elucidating to parametrize the vertex as follows where a clear separation into transverse and longitudinal parts is evident:
\begin{align}
 \Gamma^{A\bar cc,abc}_\mu(k;p,q)=i\,g\,f^{abc}(D^{A\bar cc,T}(k;p,q)P_{\mu\nu}(k)p_\nu+D^{A\bar cc,L}(k;p,q) k_\mu).
\end{align}
$P_{\mu\nu}(k)=g_{\mu\nu}-k_\mu k_\nu/k^2$ is the transverse projector.

\begin{figure}[tb]
 \includegraphics[height=2.1cm]{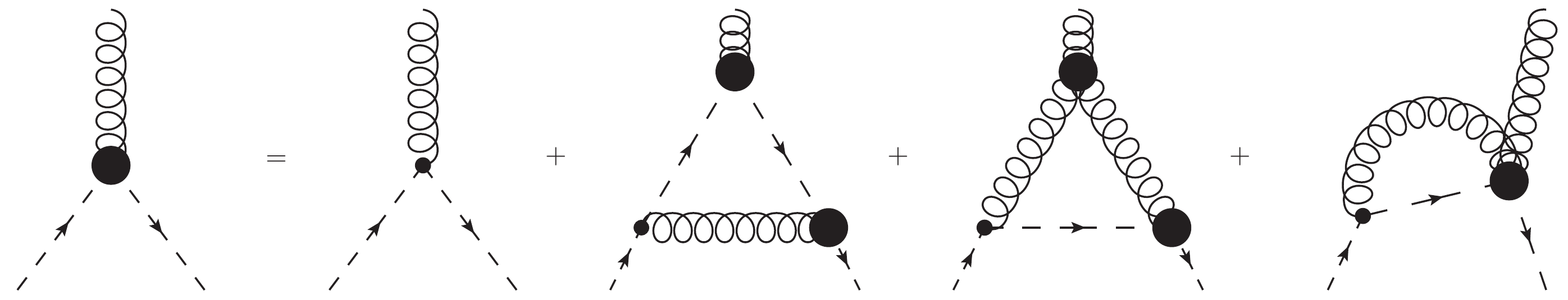}\\
 \vskip3mm
 \includegraphics[width=\textwidth]{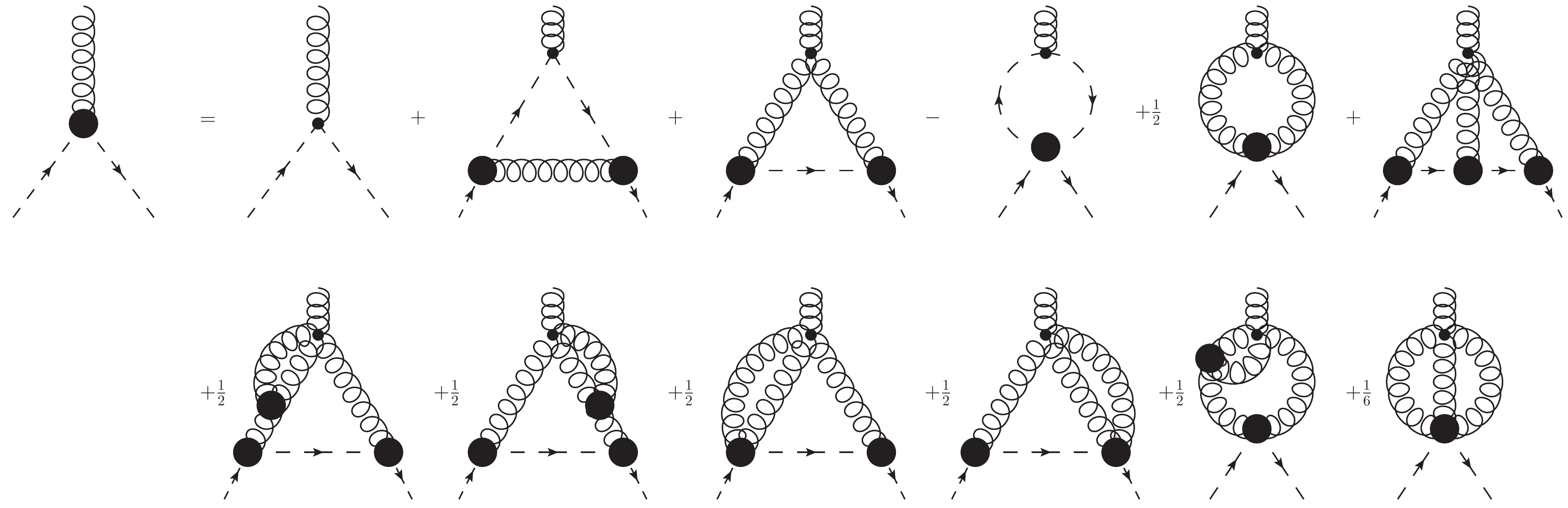}\\
 \vskip3mm
 \includegraphics[height=2.1cm]{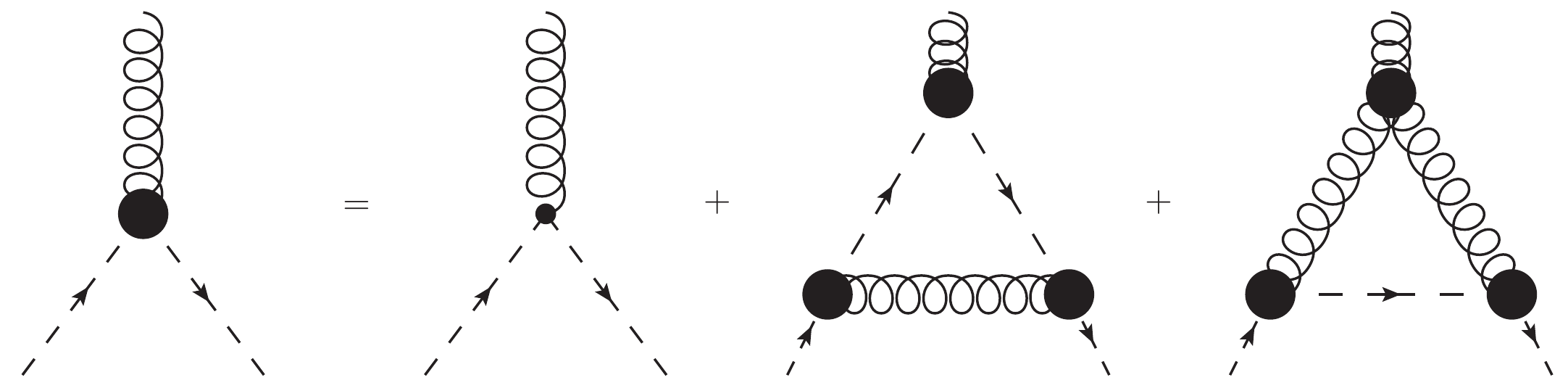}\\
 \begin{center}
 \caption{The $c$- (top) and $A$-DSEs (center) and the equation of motion from the 3PI effective action (bottom) of the ghost-gluon vertex.}
 \label{fig:ghg_DSE}
 \end{center}
\end{figure}

The ghost-gluon vertex has two different \glspl{dse} differing by the leg that is attached to the bare vertex.
They are called $A$- and $c$-DSE.
A third one, the $\bar c$-DSE, is due to the ghost--anti-ghost symmetry of the Landau gauge \cite{Alkofer:2000wg,Lerche:2002ep} equivalent to the $c$-DSE in this gauge.
The two full \glspl{dse} and the equation of motion from the 3PI effective action in a three-loop expansion are depicted in \fref{fig:ghg_DSE}.
It should be noted that the $A$- and $c$-DSE are both exact, but truncations can have different effects on them.
In contradistinction, the equation of motion from the 3PI effective action that is shown in \fref{fig:ghg_DSE} is obtained from a truncation of the effective action and thus not exact.
Differences in the description of the ghost-gluon vertex due to using different truncations are discussed in detail in Secs.~\ref{sec:res_AAcbc} and \ref{sec:YM3d_testing}.

\subsubsection{Three-gluon vertex}
\label{sec:AAA}

\index{three-gluon vertex}
\index{charge invariance}
The three-gluon vertex has 14 tensors:
\begin{align}
 \Gamma^{abc}_{\mu\nu\rho}(p,q,r)=i\,g\,f^{abc}\sum_{i=1}^{14}\tau^i_{\mu\nu\rho} C^{AAA}_i(p,q,r).
\end{align}
A possible color structure with the symmetric structure constant $d^{abc}$ is ignored here, since it would be in conflict with the charge invariance of \gls{qcd} \cite{Smolyakov:1980wq,Peccei:1998jv,Blum:2015lsa}.
In Sec.~\ref{sec:res_AAcbc} it will be explained that in three-point \glspl{dse} $d^{abc}$ can only arise from a decoupled part of the four-point functions.

The three-gluon vertex has 14 Lorentz tensors:
\begin{align}
 \Gamma^{abc}_{\mu\nu\rho}(p,q,r)=i\,g\,f^{abc}\sum_{i=1}^{14}\tau^i_{\mu\nu\rho} C^{AAA,i}(p,q,r).
\end{align}
For the Landau gauge, the four-dimensional transverse sector is sufficient due to the closure of the corresponding parts of correlation functions \cite{Fischer:2008uz}.
The resulting transverse vertex can be written as:
\begin{align}
 \Gamma^{abc,T}_{\mu\nu\rho}(p,q,r)=i\,g\,f^{abc}\sum_{i=1}^{4}\tau^i_{\mu\nu\rho} C^{AAA,T,i}(p,q,r).
\end{align}
The superscript $T$ to indicate that this is the transverse part will be suppressed whenever calculations in the Landau gauge are discussed.
An explicit transverse basis can be constructed from the naive basis, viz., the set of all possible combinations of the two independent momenta and the metric tensor with three Lorentz indices.
Upon transverse projection, six tensors survive, but only four are linearly independent.
They can be chosen as
\begin{align}
 \tau^1_{\mu\nu\rho}&=P_{\mu\mu'}(p)P_{\nu\nu'}(q)P_{\rho\rho'}(r)g_{\mu'\nu'}(p-q)_{\rho'},\\
 \tau^2_{\mu\nu\rho}&=P_{\mu\mu'}(p)P_{\nu\nu'}(q)P_{\rho\rho'}(r)g_{\mu'\rho'}p_{\nu'},\\
 \tau^3_{\mu\nu\rho}&=P_{\mu\mu'}(p)P_{\nu\nu'}(q)P_{\rho\rho'}(r)g_{\nu'\rho'}q_{\mu'},\\
 \tau^4_{\mu\nu\rho}&=P_{\mu\mu'}(p)P_{\nu\nu'}(q)P_{\rho\rho'}(r)q_{\mu'}p_{\nu'}(p-q)_{\rho'}.
\end{align}
As discussed around \eref{eq:trans-long}, this corresponds to the part of the vertex that survices when projected transversely, $\Gamma^\text{sur}$.
Alternatively, considerations of the symmetry group $S_3$ underlying the Bose symmetry of the vertex can be used to construct a basis that has specific properties under permutations of the legs \cite{Eichmann:2014xya}.
The bare vertex is given by
\begin{align}\label{eq:bare_three-gluon_vertex}
 \Gamma_{\mu\nu\rho}^{(0),abc}(p,\,q,\,r) =
   -i g f^{abc} \left[ (p-q)_\rho \delta_{\mu\nu} + (q-r)_\mu \delta_{\nu\rho} + (r-p)_\nu \delta_{\mu\rho} \right].
\end{align}

\begin{figure}[tb]
 \includegraphics[width=\textwidth]{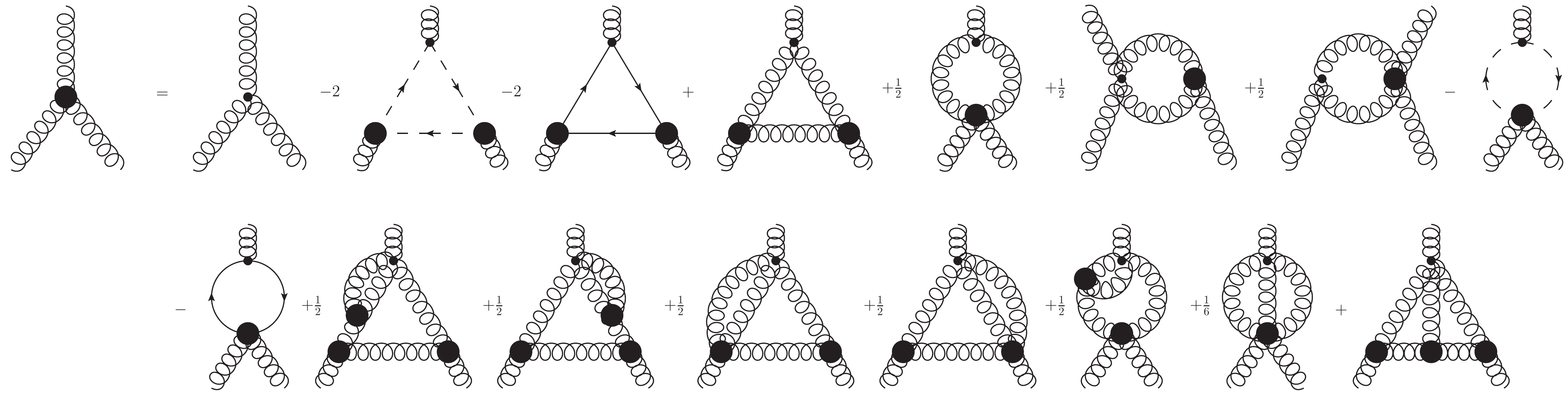}\\
 \vskip3mm
 \includegraphics[width=0.92\textwidth]{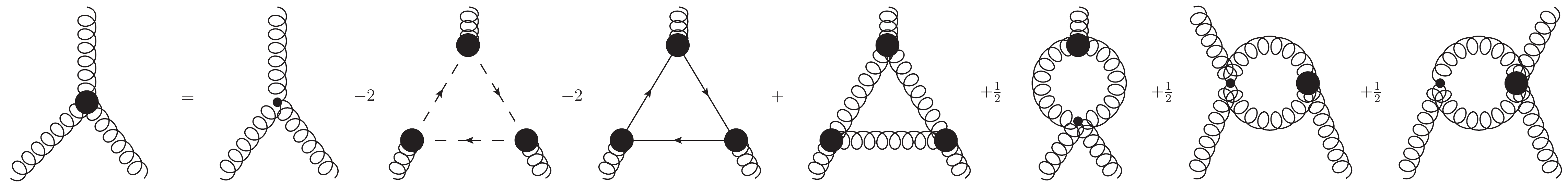}\\
 \begin{center}
 \caption{The \gls{dse} (top) of the three-gluon vertex and its equation of motion from the 3PI effective action (bottom).}
 \label{fig:3g_DSE}
 \end{center}
\end{figure}

The full \gls{dse} of the three-gluon vertex and its equation of motion from the 3PI effective action in a three-loop expansion are depicted in \fref{fig:3g_DSE}.

\subsubsection{Quark-gluon vertex}
\label{sec:AAqbq}

\index{quark-gluon vertex}
The full quark-gluon vertex has twelve tensors:
\begin{align}
 \Gamma^{A\bar qq,a,ij}_\mu(k;p,q)&=i\,g\, T^{a,ij}\sum_{l=1}^{12} \lambda_\mu^l(k;p,q) G_l(k;p,q),
\end{align}
where $T^{a,ij}$ is the generator of the gauge group.
The transverse subspace is eight-dimensional:
\begin{align}
 \Gamma^{A\bar qq,a,ij,T}_\mu(k;p,q)&=i\,g\, T^{a,ij}\sum_{l=1}^{8} \lambda_\mu^l(k;p,q) G_l(k;p,q).
\end{align}
Again, the superscript $T$ can be skipped in discussions of the Landau gauge.
The tree-level vertex is
\begin{align}
 \Gamma^{A\bar qq,a,ij,(0)}_\mu(k;p,q)&=i\,g\,T^{a,ij}\gamma_\mu.
\end{align}

\index{quark-gluon vertex}
In the calculation of the quark-gluon vertex a good choice of the basis is particularly important.
In particular, a bad choice can lead to numeric instabilities.
Various versions have been used in the literature \cite{Ball:1980ay,Hopfer:2013np,Aguilar:2014lha,Williams:2014iea,Hopfer:2014szm,Windisch:2014lce,Mitter:2014wpa,Williams:2015cvx,Cyrol:2017ewj}, among them the traditional Ball-Chiu basis \cite{Ball:1980ay} and variants thereof.

\begin{figure}[tb]
 \includegraphics[height=2.1cm]{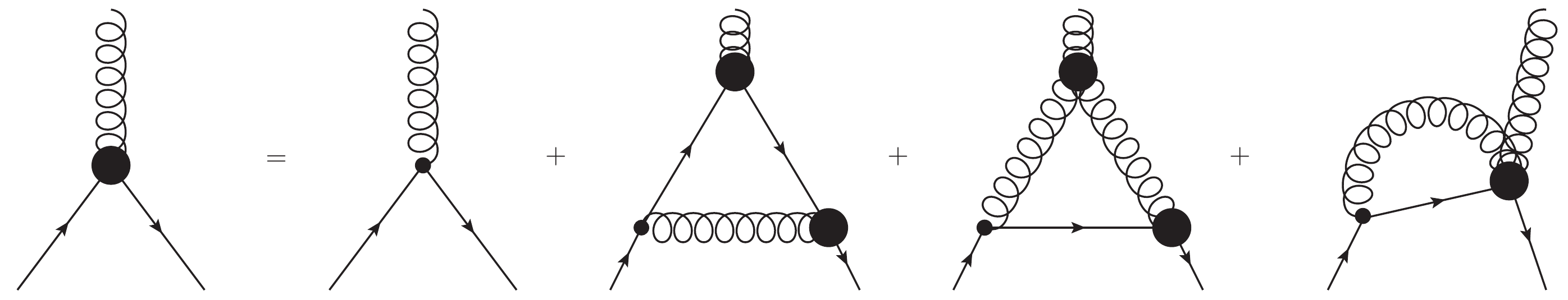}\\
 \vskip3mm
 \includegraphics[width=\textwidth]{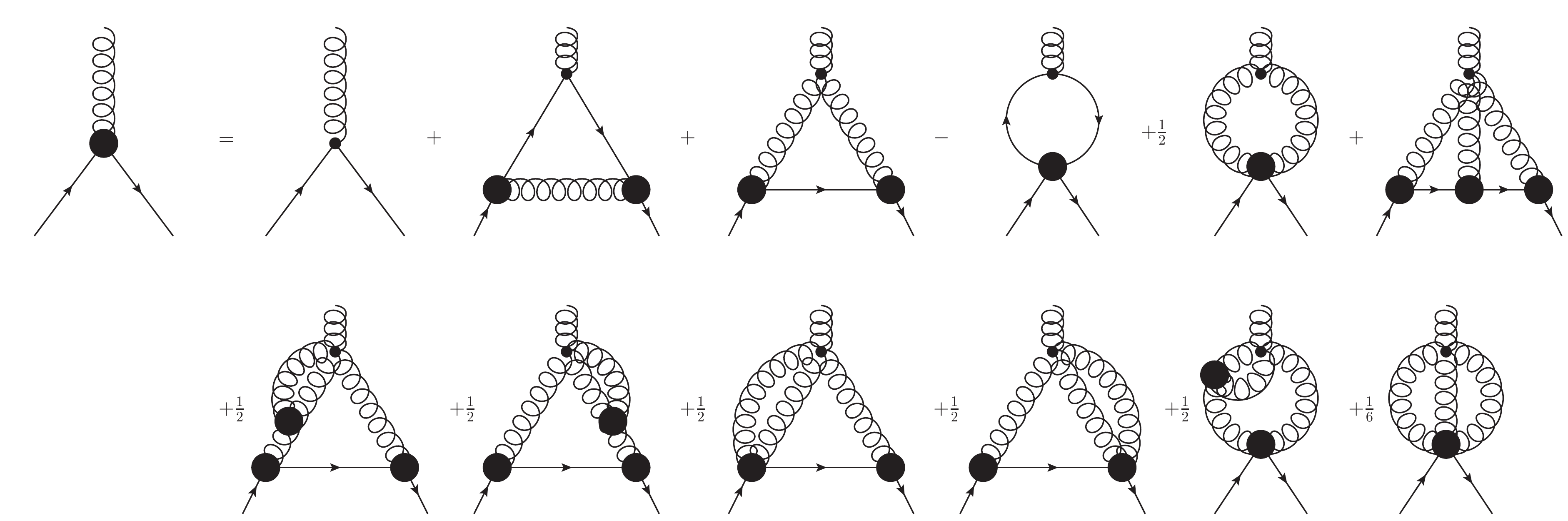}\\
 \vskip3mm
 \includegraphics[height=2.1cm]{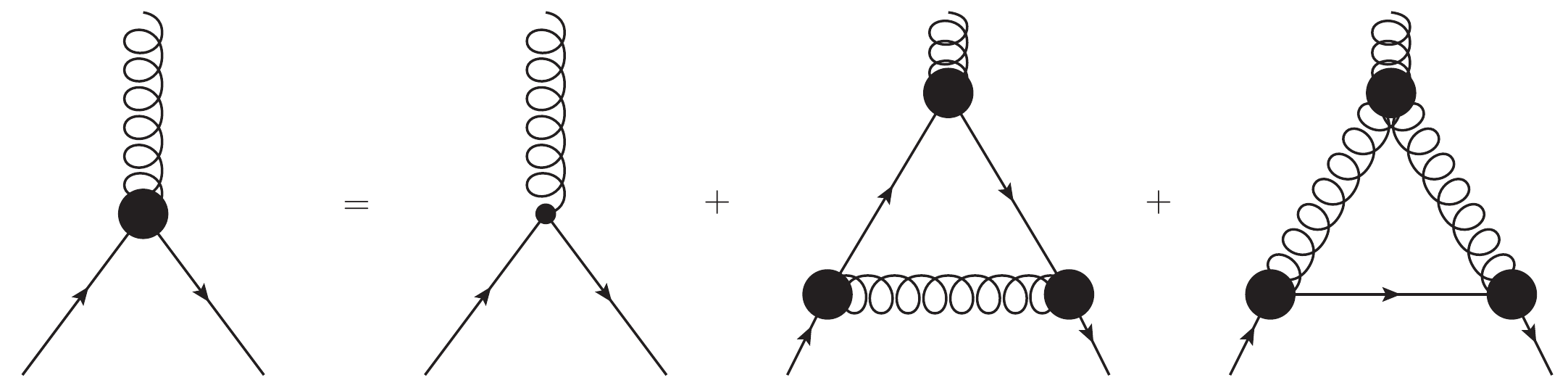}
 \begin{center}
 \caption{The $q$- (top) and $A$-DSEs (center) and the equation of motion from the 3PI effective action (bottom) of the quark-gluon vertex.}
  \label{fig:qug_DSE}
 \end{center}
\end{figure}

The diagrammatic structure of the quark-gluon vertex \glspl{dse} is the same as for the ghost-gluon vertex.
Two of its three \glspl{dse}, which are all equivalent as long as they are not truncated, are depicted in full form in \fref{fig:qug_DSE} as well as its equation of motion from the 3PI effective action in a three-loop expansion.
The $\bar q$-DSE, which is not shown, is equivalent to the $q$-DSE in the Landau gauge.

\subsubsection{Four-gluon vertex}
\label{sec:AAAA}

\index{four-gluon vertex}
With four-point functions, which have three independent momenta, the size of the tensor space increases drastically.
Writing down all tensors with four Lorentz indices constructed from the three independent momenta or the metric tensor leads to 138 tensors:
\begin{align}
 g\,g&: &3\, \text{combinations},\\
 g\,p\,p\,&:& 6\times 3^2=54\,\text{combinations},\\
 p\,p\,p\,p\,&:&  3^4=81\,\text{combinations}.
\end{align}
However, the basis is spanned by 136 tensors only, since some tensors are not linearly independent \cite{Eichmann:2015nra}.
The transverse subspace has 41 tensors \cite{Eichmann:2015nra}.

In addition, the color space is also more complicated.
With four color indices, one can construct from the basis elements of the Kronecker delta $\delta^{ab}$ and the symmetric and anti-symmetric structure constants $d^{abc}$ and $f^{abc}$, respectively, the following 15 combinations:
\begin{align}
 \de\, \de&:& 3\,\text{combinations},\\
 f\, f&:& 3\,\text{combinations},\\
 d\,d&:& 3\,\text{combinations},\\
 d\,f&:& 6\,\text{combinations}.
\end{align}
However, several identities relate tensors to each other \cite{Pascual:1980yu}:
\begin{align}\label{eq:SUN_id_fd}
 &f^{abo}d^{cdo}+f^{aco}d^{bdo}+f^{ado}d^{aco}=0 \text{ and 2 permutations},\\
 \label{eq:SUN_id_ff}
 &f^{abo}f^{cdo} = \frac{2}{N}\left(\de^{ac}\de^{bd}-\de^{ad}\de^{cb}\right)+d^{aco}d^{bdo}-d^{ado}d^{cbo}\\
  &\qquad\qquad \text{ and 2 permutations.}\nonumber
\end{align}
The Jacobi identity is a combination of the permutations of \eref{eq:SUN_id_ff}.
These identities reduce the number of independent tensors by six.
Finally, for $SU(3)$ an additional identity reduces the final number of independent tensors to eight \cite{Pascual:1980yu}:
\begin{align}
 \de^{ab}&\de^{cd}+\de^{ac}\de^{bd}+\de^{ad}\de^{bc} = 3(d^{abo}d^{cdo}+d^{aco}d^{bdo}+d^{ado}d^{bco}).
\end{align}
For $SU(2)$, there is no symmetric structure constant which reduces the number of tensors even further to three.
In the following we will restrict ourselves to $SU(3)$.
Hence, for $SU(N)$, $N>3$ one needs to generalize.

A full basis can be chosen as
\begin{align}
\begin{aligned}
 \overline\sigma_1^{abcd}& = f^{acd}f^{bde}, \\
 \overline\sigma_2^{abcd}& = \de^{ab}\de^{cd},\\
 \overline\sigma_3^{abcd}& = \de^{ac}\de^{bd}, \\
 \overline\sigma_4^{abcd}& = \de^{ad}\de^{bc}, \\
 \end{aligned}
 \qquad\qquad
 \begin{aligned}
 \overline\sigma_5^{abcd}& = f^{abe}f^{cde},\\
 \overline\sigma_6^{abcd}& = f^{abe}d^{cde},\\
 \overline\sigma_7^{abcd}& = f^{ade}d^{bce},\\
 \overline\sigma_8^{abcd}& = f^{cde}d^{abe}.
 \end{aligned}
\end{align}
Combining these tensors such that they have clear permutation symmetries would be advantageous.
For now, however, we restrict ourselves to the permutations of $a$ with $b$ and $c$ with $d$, as these symmetries are those relevant for the two-ghost-two-gluon vertex.
The full symmetrization is discussed in Ref.~\cite{Eichmann:2015nra}.
The partially symmetric basis reads
\begin{align}
\label{eq:color_basis_4p}
\begin{aligned}
 \sigma_1^{abcd}& = -2f^{acd}f^{bde}+f^{abd}f^{cde}, \\
 \sigma_2^{abcd}& = \de^{ab}\de^{cd},\\
 \sigma_3^{abcd}& = \de^{ad}\de^{bc}+\de^{ac}\de^{bd}, \\
 \sigma_4^{abcd}& = -\de^{ad}\de^{bc}+\de^{ac}\de^{bd},
\end{aligned}
\qquad\qquad
\begin{aligned}
 \sigma_5^{abcd}& = f^{abe}f^{cde},\\
 \sigma_6^{abcd}& = f^{abe}d^{cde},\\
 \sigma_7^{abcd}& = \frac{1}{2}f^{abe}d^{cde}+f^{ade}d^{bce}+\frac{1}{2}f^{cde}d^{abe},\\
 \sigma_8^{abcd}& = f^{cde}d^{abe}.
\end{aligned}
\end{align}
The symmetry properties are summarized in \tref{tab:color_tensors_symmetries}.

\begin{table}
 \begin{center}
 \begin{tabular}{c|c|c|c|c|c|c|c|c}
  \hline
   & $\sigma_1$ & $\sigma_2$ & $\sigma_3$ & $\sigma_4$ & $\sigma_5$ & $\sigma_6$ & $\sigma_7$ & $\sigma_8$\\
   \hline\hline
  $a\leftrightarrow b$ & + & + & + & - & - & - & - & +\\
  \hline
  $c\leftrightarrow d$ & + & + & + & - & - & + & - & -\\
  \hline
 \end{tabular}
 \caption{Symmetry properties of the color basis tensors for four-point functions given in \eref{eq:color_basis_4p}.}
 \label{tab:color_tensors_symmetries}
 \end{center}
\end{table}

If the symmetric structure constant is neglected, the number of independent tensors reduces to five.
Based on the relations in \eref{eq:SUN_id_ff}, one might expect that neglecting $d^{abc}$ leads to problems with an incomplete basis.
However, it was already noted in Ref.~\cite{Driesen:1997wz} that the set $\{\sigma_1,\ldots,\sigma_5\}$ closes under \gls{dse} iterations if no symmetric color part from three-point functions is taken into account.
Indeed, the sets $\{\sigma_1,\ldots,\sigma_5\}$ and $\{\sigma_6,\ldots,\sigma_8\}$ are orthogonal to each other for $SU(3)$.
The former set is called the \emph{reduced basis} in the following.
Furthermore, the second set only couples to the symmetric color part in the \glspl{dse} of three-point functions as discussed in Sec.~\ref{sec:res_AAcbc}.

The tree-level tensor of the four-gluon vertex is given by
\begin{align}
\label{eq:bare_four-gluon_vertex}
  \Gamma^{{(0)},abcd}_{\mu\nu\rho\sigma}&(p,q,r,s) =
    -g^2  \Big[\;\, 				\left(f^{acn'}f^{bdn'}-f^{adn'}f^{cbn'}\right)\delta_{\mu\nu}\delta_{\rho\sigma}\nnnl
	  & +     \left(f^{abn'}f^{cdn'}-f^{adn'}f^{bcn'}\right)\delta_{\mu\rho}\delta_{\nu\sigma}
	   +     \left(f^{acn'}f^{dbn'}-f^{abn'}f^{cdn'}\right)\delta_{\mu\sigma}\delta_{\rho\nu}\Big].
\end{align}
The full transverse vertex can be written as
\begin{align}\label{eq:AAAA}
 \Gamma^{AAAA,abcd,T}_{\mu\nu\rho\sigma}(p,q,r,s) = g^2 \sum_{k=1}^{328} \rho_{\mu\nu\rho\sigma}^{k,abcd} F^{AAAA}_{k(i,j)}(p,q,r,s),
\end{align}
where the tensors $\rho_{\mu\nu\rho\sigma}^{k,abcd}$ are given by
\begin{align}
 \rho_{\mu\nu\rho\sigma}^{k,abcd}=\sigma_i^{abcd} \tau^j_{\mu\nu\rho\sigma}
\end{align}
with $k(i,j)=41(i-1)+j$.
A full list of the Lorentz tensors $\tau^j_{\mu\nu\rho\sigma}$ is not given here.
For testing purposes, some specific dressing functions are used statically in Sec.~\ref{sec:res_fg}, for example,
\begin{align}
 G^{abcd}_{\mu\nu\rho\sigma}(p,q,r,s)&=(\de^{ab}\de^{cd}+\de^{ac}\de^{bd}+\de^{ad}\de^{bc})
 (\de_{\mu\nu}\de_{\rho\sigma}+\de_{\mu\rho}\de_{\nu\sigma}+\de_{\mu\sigma}\de_{\nu\rho}),\\
 \widetilde{P}_{\mu\nu\rho\sigma}^{abcd}(p,q,r,s)&=(\de^{ab}\de^{cd}+\de^{ac}\de^{bd}+\de^{ad}\de^{bc})
 \frac{s_\mu r_\nu q_\rho p_\sigma + r_\mu s_\nu p_\rho q_\sigma + q_\mu p_\nu s_\rho r_\sigma}{\sqrt{p^2\,q^2\,r^2\,p^2}}.
\end{align}
Two other tensors, $V_2$ and $V_3$, are constructed by transverse projection and orthonormalization of the set consisting of the tree-level tensor and these two tensors \cite{Driesen:1998xc,Cyrol:2014kca}:
\begin{align}
 &\widetilde{V}_{2,\mu\nu\rho\sigma}^{abcd}=\de^{ab}\de^{cd}\de_{\mu\nu}\de_{\rho\sigma}+\de^{ac}\de^{bd}\de_{\mu\rho}\de_{\nu\sigma}+\de^{ad}\de^{bc}\de_{\mu\sigma}\de_{\nu\rho},\\
 &\widetilde{V}_{3,\mu\nu\rho\sigma}^{abcd}=(\de^{ac}\de^{bd}+\de^{ad}\de^{bc})\de_{\mu\nu}\de_{\rho\sigma}+(\de^{ab}\de^{cd}+\de^{ad}\de^{bc})\de_{\mu\rho}\de_{\nu\sigma}+(\de^{ab}\de^{cd}+\de^{ac}\de^{bd})\de_{\mu\sigma}\de_{\nu\rho}.
\end{align}

\begin{figure}[tb]
 \includegraphics[width=\textwidth]{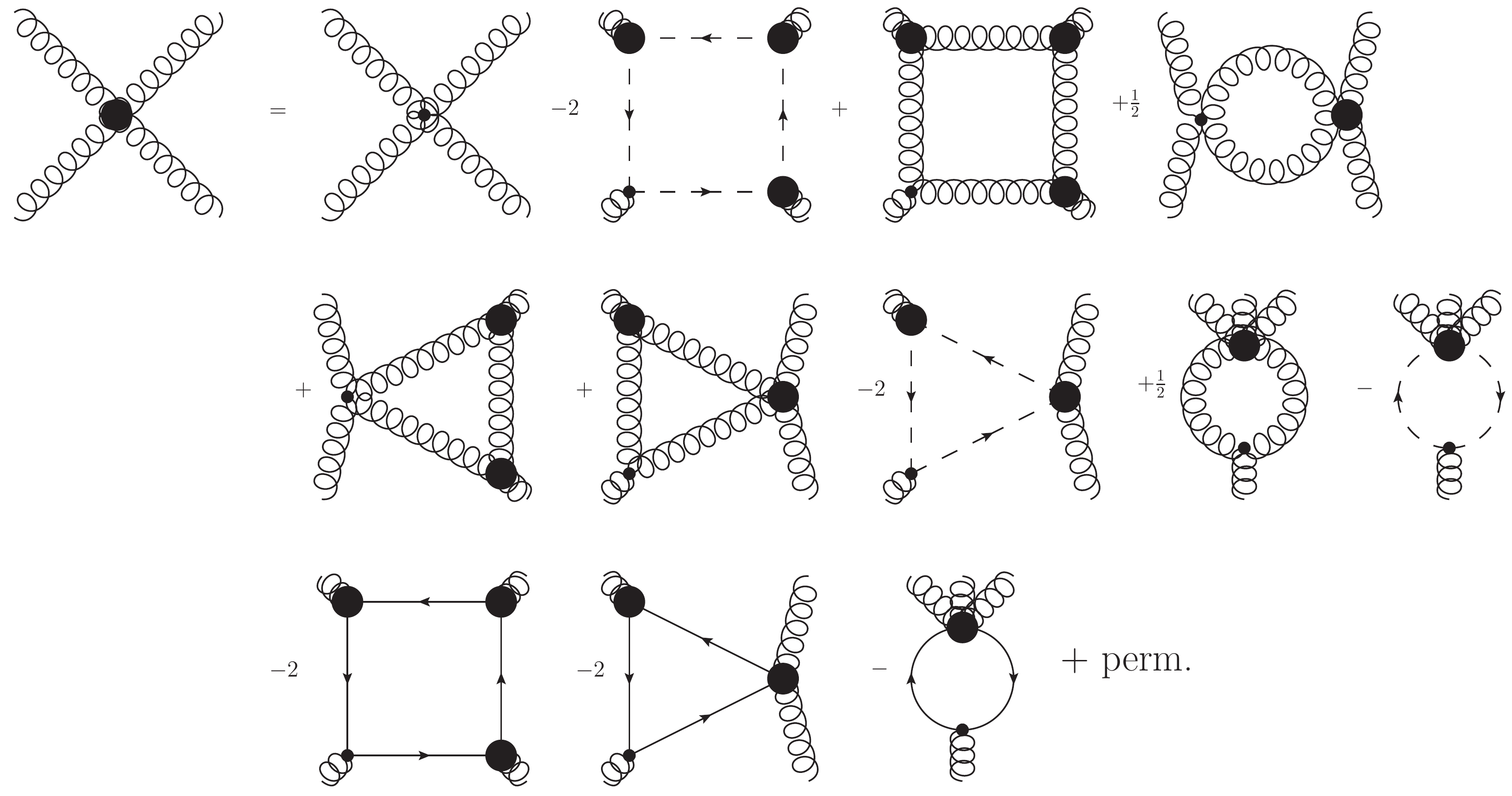}
 \begin{center}
 \caption{One-loop truncated DSE of the four-gluon vertex, i.e., two-loop diagrams are discarded.}
 \label{fig:4g_DSE}
 \end{center}
\end{figure}

Four-point functions depend on three independent momenta.
They can be chosen as follows:
\begin{align}
\label{eq:moms_4p}
  s=S\colvec{4}{0}{0}{0}{1}, \quad
  r=R\colvec{4}{0}{0}{\sin(\theta_r)}{\cos(\theta_r)}, \quad
  q=Q\colvec{4}{0}{\sin(\theta_q)\sin(\psi_q)}{\sin(\theta_q)\cos(\psi_q)}{\cos(\theta_q)\phantom{\cos(\psi_q)}}.
\end{align}
The three radial variables $S$, $R$ and $Q$ and the three angles $\theta_r$, $\theta_q$ and $\psi_q$ can then be taken to span the six-dimensional space for four-point functions.

The \gls{dse} of the four-gluon vertex truncated to one-loop diagrams is depicted in \fref{fig:4g_DSE}.
The full DSE can be inferred from the one for a scalar field theory including three- and four-point interactions given in Ref.~\cite{Carrington:2010qq}.

\subsubsection{Two-ghost-two-gluon vertex}
\label{sec:AAcbc}

\index{two-ghost-two-gluon vertex}
The two-ghost-two-gluon vertex has two Lorentz indices what simplifies its treatment in Lorentz space considerably compared to the four-gluon vertex.
In Lorentz space, the following basis is explicitly transverse and has clear (anti-)symmetry properties under the exchange of the gluon momenta:
\begin{align}
\begin{aligned}
 \tau^1_{\mu\nu}(p,q;r,s)&= t_{\mu\nu}(p,q),&
 \tau^2_{\mu\nu}(p,q;r,s)&= t_{\mu\alpha}(p,p)t_{\alpha\nu}(r,q) + t_{\mu\alpha}(p,r)t_{\alpha\nu}(q,q),\\
 \tau^4_{\mu\nu}(p,q;r,s)&= t_{\mu\alpha}(p,p)t_{\alpha\nu}(q,q),&
 \tau^3_{\mu\nu}(p,q;r,s)&= t_{\mu\alpha}(p,p)t_{\alpha\nu}(r,q) - t_{\mu\alpha}(p,r)t_{\alpha\nu}(q,q),\\
 \tau^5_{\mu\nu}(p,q;r,s)&= t_{\mu\alpha}(p,r)t_{\alpha\nu}(r,q).
\end{aligned}
\end{align}
Here, $t_{\mu\nu}(p,q)=g_{\mu\nu} p\cdot q -q_\mu p_\nu$, the two gluon momenta are $p$ and $q$ and the anti-ghost/ghost momentum is $r/s$.
The full basis is constructed as the direct product in color and Lorentz space:
\begin{align}
 \rho_{\mu\nu}^{k,abcd}=\sigma_i^{abcd} \tau^j_{\mu\nu}
\end{align}
with $k(i,j)=5(i-1)+j$.
The vertex is then written as
\begin{align}\label{eq:AAcbc}
 \Gamma^{AA\bar cc,abcd}_{\mu\nu}(p,q;r,s) = g^4 \sum_{k=1}^{40} \rho_{\mu\nu}^{k,abcd} D^{AA\bar cc}_{k(i,j)}(p,q;r,s).
\end{align}
Note that a factor of $g^4$ is put in front to account for the fact that the lowest diagrams are of this order, since there is no tree-level contribution as for the four-gluon vertex.
In Sec.~\ref{sec:res_AAcbc}, it will be explained that $25$ dressings are sufficient corresponding to the reduced set of color tensors, because the other three color tensors do not couple to the reduced set or to three-point functions.
As defined here, the dressing functions are dimensionful, because the vertex is dimensionless, but the Lorentz tensors are chosen as dimensionful.

\begin{figure}[tb]
 \begin{center}
  \includegraphics[width=0.9\textwidth]{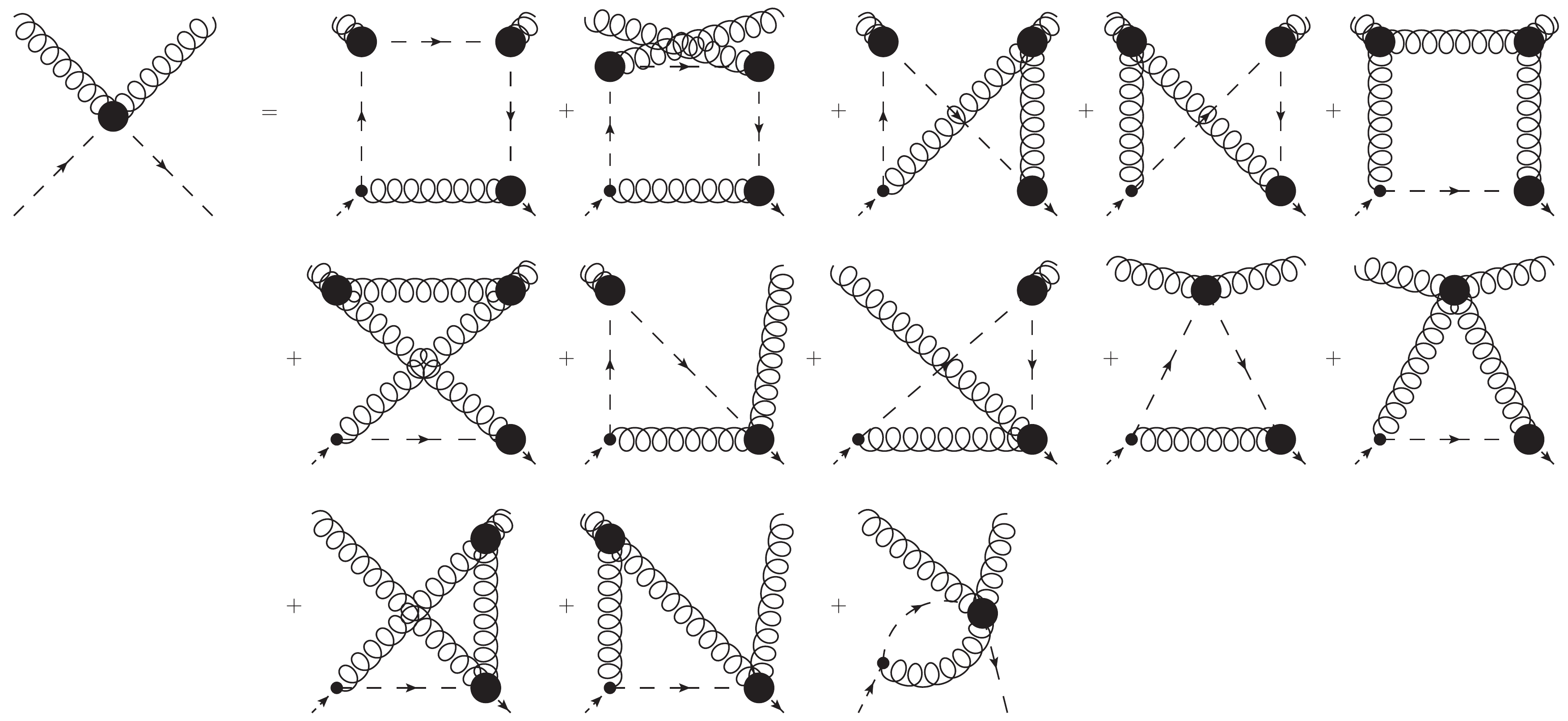}
 \end{center}
 \caption{The $c$-DSE of the two-ghost-two-gluon vertex.}
 \label{fig:2gh2gl_DSE}
\end{figure}

\myboxmargin{two-ghost-two-gluon vertex: DSE(s)}
As typical for vertices of mixed types of fields, the two-ghost-two-gluon vertex has several \glspl{dse}.
In analogy to the ghost-gluon vertex, they are called $A$-DSE and $c$-DSE based on which type of field is attached to the bare vertices.
Due to the existence of a bare four-gluon vertex, the $A$-DSE contains also two-loop terms, while the $c$-DSE has a one-loop structure.
An additional advantage of the $c$-DSE is that in contrast to the $A$-DSE, it does not contain a four-ghost vertex.
The full $c$-\gls{dse} is shown in \fref{fig:2gh2gl_DSE}.

\subsubsection{Four-ghost vertex}

\index{four-ghost vertex}
The four-ghost vertex is a comparatively simple four-point function, because it is a scalar in Lorentz space.
Thus, it features only eight tensors in total.
The full vertex is written as
\begin{align}\label{eq:cbcbcc}
 \Gamma^{\bar c\bar ccc,abcd}(p,q,r,s) = g^4 \sum_{k=1}^{8} \sigma^{k,abcd} E^{\bar c\bar ccc}_{k}(p,q,r,s).
\end{align}
As for the two-ghost-two-gluon vertex, the reduced color basis is completely decoupled from the other three tensors and thus sufficient.

\begin{figure}[tb]
 \begin{center}
  \includegraphics[width=0.9\textwidth]{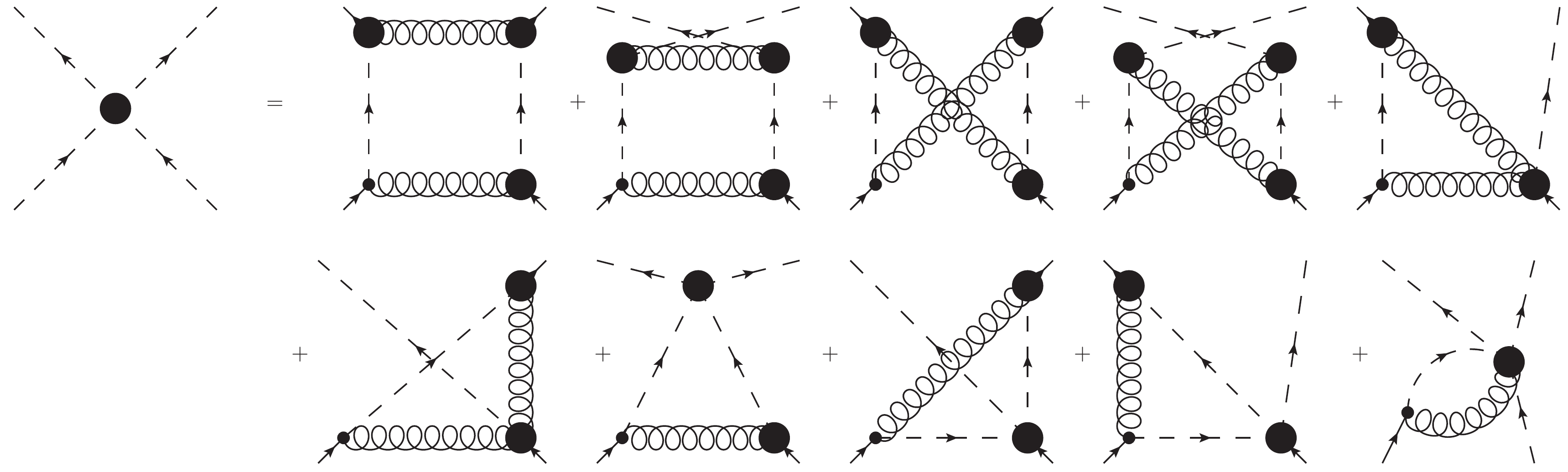}
 \end{center}
 \caption{The \gls{dse} of the four-ghost vertex.}
 \label{fig:4gh_DSE}
\end{figure}

A full \gls{dse} of the vertex is shown in \fref{fig:4gh_DSE}.
Another one, where the bare vertex is attached to the anti-ghost leg, has the same structure.

\section{Methods}
\label{chp:methods}

Many methods are used to calculate correlation functions nonperturbatively.
Sec.~\ref{sec:funcEqs} contains details on different functional equations with a focus on equations of motion from $n$PI effective actions and the functional renormalization group.
The Hamiltonian approach is also sketched.
Furthermore, an overview of available technical tools is given.
The section concludes with a short comparative summary of functional methods.
In Sec.~\ref{sec:otherMethods}, further methods are shortly discussed.

\subsection{Functional equations}
\label{sec:funcEqs}

\myboxmargin{Functional eqs.: What are they? Where are they used?}
\index{functional equations}
Functional equations relate correlation functions via integral, differential or integro--differ\-ential equations.
They come in different variants, for example, as equations of motion of correlations functions or as equations expressing underlying symmetries.
In the main part of this work only the correlations functions of \gls{qcd} are considered, but functional methods are very general and can be used in many different fields of physics, ranging from condensed matter to quantum field theory to gravity.
This section is kept general and describes the derivation of different functional equations.
Before details on specific functional equations are given, some general definitions are provided.

\index{path integral}
\index{super field}
\index{generating functional}
In the following, the path integral formalism is used as it is the most natural way to deal with functional equations.
To keep things general and to avoid cumbersome notation, a super-field $\phi_i$ is introduced where the index $i$ represents a field-type and all of its indices as well as position or momentum.
Repeated indices are summed/integrated over.
The action can then be represented as
\begin{align}\label{eq:S-phi}
 S[\phi]=&\frac{1}{2!}S_{rs}\phi_r \phi_s - \frac{1}{3!} S_{rst} \phi_r \phi_s \phi_t -
	\frac{1}{4!}S_{rstu} \phi_r \phi_s \phi_t \phi_u +\ldots.
\end{align}
For conventional reasons, minus signs were put in front of the interaction terms.
The coefficients $S_{rs}$, $S_{rst}$ and $S_{rstu}$ denote the bare two-, three and four-point functions and the statistical factors are chosen for convenience.
Here, only three- and four-point functions are included, but it is straightforward to add higher terms.

\index{path integral}
The path integral is given by
\begin{align}
Z[J]=\int D[\phi] e^{-S + \phi_i J_i}=e^{W[J]},
\end{align}
where the sources $J$ were introduced as well as the generating functional of connected correlation functions, $W[J]$.
The central object we are interested in is the effective action $\Gamma[\Phi]$ which is obtained as the Legendre transformation of $W[J]$:
\begin{align}\label{eq:effAct}
 \Gamma[\Phi]&=-W[J]+\Phi_i J_i.
\end{align}
The effective action depends on the averaged field $\Phi$ in the presence of the external source $J$:
\begin{align}
\Phi_{i}:=\left\langle \phi_{i}\right\rangle _{J}=\frac{\delta W[J]}{\delta J_{i}}=Z[J]^{-1}\int D[\phi] \phi_i e^{-S + \phi_j J_j}.
\end{align}
The inverse relation is
\begin{align}
 J_i&=\frac{\delta \Gamma[\Phi]}{\delta \Phi_i}.
\end{align}

\index{effective action}
The effective action can be expanded in $n$-point functions $\Gamma_{i_1\ldots i_n}$, also called vertex functions, around the physical ground state, which is taken here as $\Phi_\text{phys}:=\left\langle \phi_{i}\right\rangle _{J=0}=0$:
\begin{align}
 \Gamma[\Phi]=\frac{1}{2}\Gamma_{i_1 i_2} \Phi_{i_1} \Phi_{i_2} - \sum_{n>2}\frac{1}{\mathcal{N}^{i_1\ldots i_n}}\Gamma_{i_1\ldots i_n} \Phi_{i_1}\ldots \Phi_{i_n}.
\end{align}
The $\mathcal{N}^{i_1\ldots i_n}$ are symmetry factors.
The derivatives of the effective action are denoted as $\Gamma_{i_{1}\cdots i_{n}}^{J}$ where the superscript $J$ is kept to indicate that the sources are not set to zero\footnote{The negative sign is a choice of convention.
It counteracts the minus sign appearing from the derivative in \eref{eq:der-prop} below.}:
\begin{align}\label{eq:def-vertices}
\Gamma_{i_{1} i_{2}}^{J}&:=\frac{\delta\Gamma[\Phi]}{\delta\Phi_{i_{1}}\delta\Phi_{i_{2}}},\\
\Gamma_{i_{1}\cdots i_{n}}^{J}&:=-\frac{\delta\Gamma[\Phi]}{\delta\Phi_{i_{1}}\cdots\delta\Phi_{i_{n}}}, \quad n>2.
\end{align}
The coefficients in the vertex expansion correspond to
\begin{align}\label{eq:verts}
 \Gamma_{i_1\ldots i_n}=\Gamma_{i_1\ldots i_n}^{J=0}.
\end{align}

\index{propagator}
The two-point functions $\Gamma_{i_1 i_2}$ play a special role, since they are the inverse of the propagators $D_{ij}$.
For non-zero sources the relation is
\begin{align}\label{eq:super-prop}
D_{ij}^{J}:=\frac{\delta^{2}W[J]}{\delta J_{i}\delta J_{j}}=\left(\frac{\delta^{2}\Gamma[\Phi]}{\delta\Phi_{i}\delta\Phi_{j}}\right)^{-1},
\end{align}
and the physical propagator is given by
\begin{align}\label{eq:prop}
D_{ij}&:=D_{ij}^{J=0}.
\end{align}
It should be noted that \eref{eq:super-prop} is a matrix relation in field space.

\subsubsection{Dyson--Schwinger equations}
\label{sec:der_dse}

\begin{figure}[tb]
 \begin{center}
  \includegraphics[width=\textwidth]{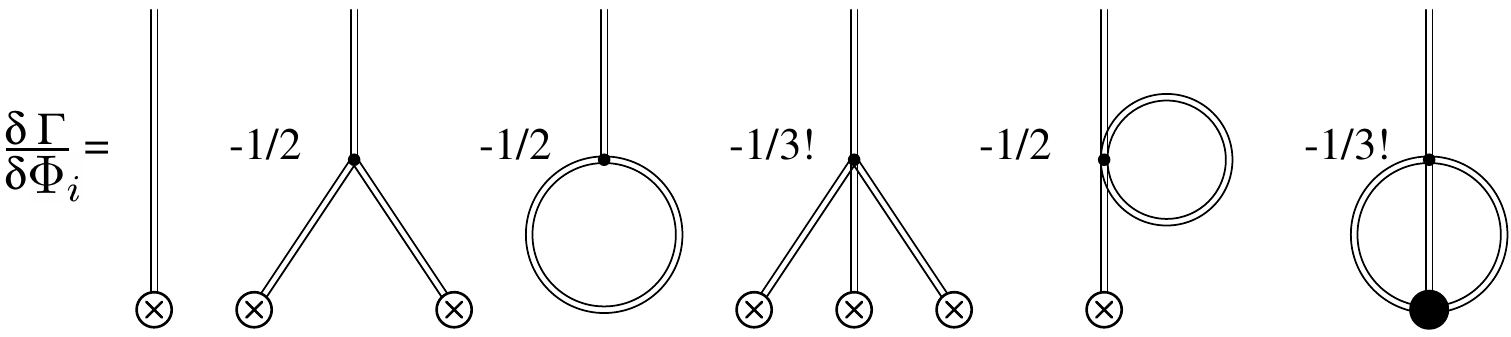}
 \end{center}
 \caption{\Gls{dse} for a generic theory with three- and four-point interactions.
 The double lines represent the super field.
 Encircled crosses denote external fields $\Phi$.}
 \label{fig:DSE_master}
\end{figure}

\begin{figure}[tb]
 \begin{center}
  \includegraphics[width=\textwidth]{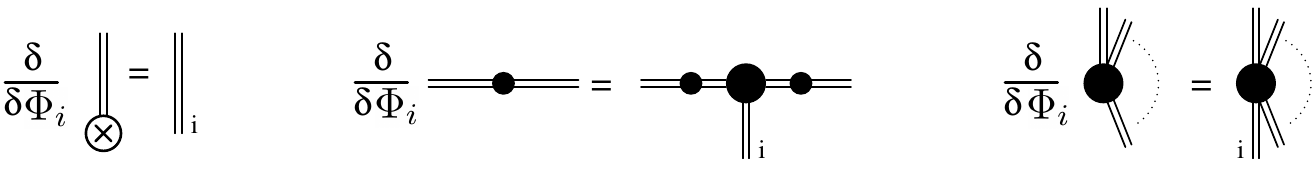}
 \end{center}
 \caption{Pictorial representation of the differentiation rules given in \eref{eq:derivatives}.
 The propagators are explicitly denoted by a blob.}
 \label{fig:DSE_diagRules}
\end{figure}

\index{correlation function, full}
\index{correlation function, connected}
Dyson-Schwinger equations are named after Julian Schwinger and Freeman Dyson who initiated the use of these equations \cite{Dyson:1949ha,Schwinger:1951ex,Schwinger:1951hq}.
Nowadays, they are most conveniently written as integral equations and derived in the path integral formalism.
\Glspl{dse} can be derived from the following total derivative:
\begin{align}\label{eq:DSE-Z}
\int D[\phi] \left( -\frac{\delta S}{\delta \phi_i} + J_i \right) e^{-S + \phi_j J_j}=\left( -\frac{\delta S}{\delta \phi'_i}\Bigg\vert_{\phi'_i=\delta/\delta J_i} +J_i \right) Z[J]=0.
\end{align}
This is the generating equation for the \glspl{dse} of full correlation functions.
To switch to connected correlation functions, the generating functional $Z[J]$ is replaced by $e^{W[J]}$ and the identity
\begin{align}
e^{-W[J]}\left(\frac{\delta}{\delta J_i}\right)e^{W[J]}= \frac{\delta W[J]}{\delta J_i}+\frac{\delta}{\delta J_i} 
\end{align}
is used.
After multiplying with $e^{-W[J]}$ from the left, \eref{eq:DSE-Z} becomes
\begin{align}\label{eq:W-DSE}
-\frac{\delta S}{\delta \phi_i}\Bigg\vert_{\phi_i=\frac{\delta W[J]}{\delta J_i}+\frac{\delta}{\delta J_i}} +J_i=0.
\end{align}
Finally, we perform a Legendre transformation to obtain the generating equation for the \glspl{dse} of \gls{1pi} correlation functions.
In the transformation, the derivative with respect to the sources becomes
\begin{align}
\frac{\delta}{\delta J_i}=\frac{\delta \Phi_j}{\delta J_i} \frac{\delta}{\delta \Phi_j}=\frac{\delta}{\delta J_i} \frac{\delta W[J]}{\delta J_j} \frac{\delta}{\delta \Phi_j}=\frac{\delta^2 W[J]}{\delta J_i \delta J_j} \frac{\delta}{\delta \Phi_j}=
D_{ij}^J \frac{\delta}{\delta \Phi_j}
\end{align}
and we finally have
\begin{align}\label{eq:DSE_master}
-\frac{\delta S}{\delta \phi_i}\Bigg\vert_{\phi_i=\Phi_i+D_{ij}^J  \, \delta/\delta \Phi_j} +\frac{\delta \Gamma[\Phi]}{\delta \Phi_i}=0.
\end{align}
An example for this equation is shown \fref{fig:DSE_master}.
From \eref{eq:DSE_master}, all \glspl{dse} of \gls{1pi} correlation functions are obtained by differentiating with respect to fields and then setting the fields to their physical values.
In the course of such derivations, the following differentiation rules are required:
\begin{subequations}\label{eq:derivatives}
\begin{align}
\frac{\delta}{\delta\Phi_{i}}\Phi_{j} & =\delta_{ij},\\
\frac{\delta}{\delta\Phi_{i}}D_{jk}^{J} & =\frac{\delta}{\delta\Phi_{i}}\left(\frac{\delta^{2}\Gamma[\Phi]}{\delta\Phi_{j}\delta\Phi_{k}}\right)^{-1}=\nnnl
 &=-\left(\frac{\delta^{2}\Gamma[\Phi]}{\delta\Phi_{j}\delta\Phi_{m}}\right)^{-1}\left(\frac{\delta^{3}\Gamma[\Phi]}{\delta\Phi_{m}\delta\Phi_{i}\delta\Phi_{n}}\right)\left(\frac{\delta^{2}\Gamma[\Phi]}{\delta\Phi_{n}\delta\Phi_{k}}\right)^{-1}=D_{jm}^{J}\Gamma_{min}^{J}D_{nk}^{J}, \label{eq:der-prop} \\
 \label{eq:GammaJ}
\frac{\delta}{\delta\Phi_{i}}\Gamma_{j_{1}\cdots j_{n}}^{J} & =-\frac{\delta\Gamma[\Phi]}{\delta\Phi_{i}\delta\Phi_{j_{1}}\cdots\delta\Phi_{j_{n}}}=\Gamma_{ij_{1}\cdots j_{n}}^{J}.\end{align}
\end{subequations}
These rules are depicted in graphical form in \fref{fig:DSE_diagRules}.

\begin{figure}[tb]
 \begin{center}
  \includegraphics[width=\textwidth]{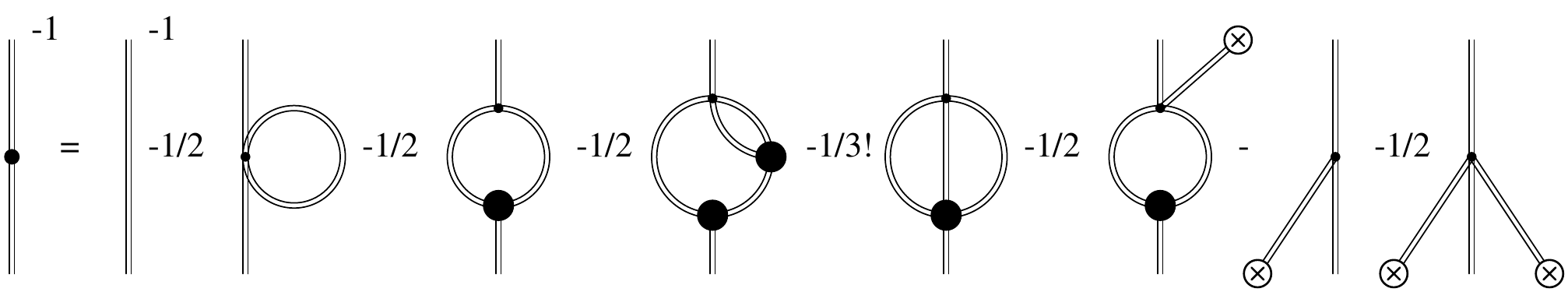}
 \end{center}
 \caption{One derivative of the generating equation (\ref{eq:DSE_master}).
 Since the sources are not yet set to zero, there are diagrams with external fields which will lead to nonvanishing contributions for higher derivatives.}
 \label{fig:DSE_2derivs}
\end{figure}

An example for a two-point function, obtained by performing one derivative of \eref{eq:DSE_master} with respect to a field, is shown in \fref{fig:DSE_2derivs}.
It is important to keep diagrams with external fields until the end.
Only then the sources are set to zero and thus the external fields take their physical values.
In the cases considered in this work, the expectation value of fields is always zero.

A special application of \glspl{dse} in gauge field theories is the combination of the pinch technique \cite{Cornwall:1981zr,Cornwall:1989gv,Binosi:2002ft,Binosi:2003rr} with the background field method \cite{Abbott:1980hw,Abbott:1981ke} which is referred to as PT-BFM \cite{Aguilar:2006gr,Binosi:2007pi,Binosi:2008qk,Binosi:2009qm}.
A special advantage of this approach in \gls{qcd} is that individual subsets of diagrams in the gluon propagator equation are fully transverse \cite{Binosi:2008qk}.
It was successfully employed for the study of propagators, e.g., \cite{Aguilar:2006gr,Aguilar:2008xm,Aguilar:2009ke,RodriguezQuintero:2010wy,Aguilar:2011ux,Aguilar:2010cn,Aguilar:2011yb,Binosi:2012sj,Aguilar:2013vaa,Aguilar:2016ock}, and three-point functions \cite{Aguilar:2013vaa,Aguilar:2016lbe,Aguilar:2018csq,Aguilar:2019jsj,Aguilar:2019kxz,Aguilar:2019uob}.

A method alternative to the one typically used in \gls{qcd} to solve \glspl{dse} is based on the homotopy analysis method and Monte Carlo sampling and was introduced in Ref.~\cite{Pfeffer:2017ufd}.
This method is relatively new and has been applied to $\phi^4$ theory only \cite{Pfeffer:2017ufd,Pfeffer:2018tkx}.

\subsubsection{Equations of motion from $n$PI effective actions}

\index{1PI effective action}
\index{$n$PI effective action}
\index{effective action}
The effective action $\Gamma[\Phi]$ is given by the Legendre transformation of $W[J]$ where $\Phi=\delta W[J]/\delta J$ is the expectation value of the field variable.
One can treat the propagator $D_{ij}$ and the proper vertices $\Gamma_{ij\ldots}$ on the same footing as the expectation value of the field by adding corresponding sources:
\begin{align}
 e^{W[J,R^{(2)},R^{(3)},\ldots]} = Z[J,R^{(2)},R^{(3)},\ldots]=\int D[\phi] e^{-S + \phi_i J_i + \frac{1}{2} R^{(2)}_{ij}\phi_i \phi_j + \frac{1}{3!} R^{(3)}_{ijk}\phi_i \phi_j \phi_k + \ldots}.
\end{align}
The $R^{(n)}_{i\ldots}$ are source terms for the propagator $D_{ij}$, defined in \eref{eq:prop}, and the vertices $\Gamma^{(n)}_{i_1 i_2 \ldots i_n}=\Gamma_{i_1 i_2 \ldots i_n}$, defined in \eref{eq:verts}, where the superscript $(n)$ was added to denote the order of a vertex.
With $W = W[J, R^{(2)}, R^{(3)}, R^{(4)}, \ldots]$ follows
\begin{align}
 \frac{\de W}{\de J_i}&=\langle \phi_i \rangle_J = \Phi_i,\\
 \frac{\de W}{\de R^{(2)}_{ij}}&=\frac{1}{2}\left( D_{ij}^J  + \Phi_i \Phi_j \right),\\
 \frac{\de W}{\de R^{(3)}_{ijk}}&=\frac{1}{6}\left( D^{(3),J}_{ijk} + D_{ij}^J \Phi_k + D_{ik}^J \Phi_j + D_{jk}^J \Phi_i  + \Phi_i \Phi_j \Phi_k \right).
\end{align}
$D^{(3),J}_{ijk}$ is the third derivative of $W$ with respect to $J$.

Performing additional Legendre transformations, the corresponding $n$PI effective action $\Gamma = \Gamma[\Phi, D, \Gamma^{(3)}, \ldots]$ is obtained:
\begin{align}\label{eq:nPI-action}
 &\Gamma =  - W + \frac{\de W}{\de J_i} J_i + \frac{\de W}{\de R^{(2)}_{ij}} R^{(2)}_{ij} + \frac{\de W}{\de R^{(3)}_{ijk}} R^{(3)}_{ijk} + \ldots.
\end{align}
From \eref{eq:nPI-action}, one can show
\begin{align}
 \frac{\de \Gamma}{\de \Phi_i} &= J_i, \qquad \frac{\de \Gamma}{\de D^{(2)}_{ij}} = \frac{1}{2} R^{(2)}_{ij}, \qquad \frac{\de \Gamma}{\de \Gamma^{(3)}_{ijk}} = -\frac{1}{6} D_{ai}D_{bj}D_{kc}R^{(3)}_{abc}.
\end{align}
For the last equation, $D^{(3)}_{ijk}=-D_{ai}D_{bj}D_{ck}\Gamma^{(3)}_{ijk}$ was used.
For vanishing sources $J_i=R^{(2)}_{ij}=R^{(3)}_{ijk}=\ldots=0$ this leads to the stationarity conditions
\begin{align}\label{eq:stationarity_conds}
 \frac{\delta \Gamma}{\de \Phi_i}=0, \qquad \frac{\delta \Gamma}{\de D_{ij}}=0, \qquad \frac{\delta \Gamma}{\de \Gamma^{(3)}_{ijk}}=0,
\end{align}
from which the equations of motion of correlations functions follow.

\index{2PI effective action}
\index{$n$PI effective action}
The effective action $\Gamma[\Phi, D]$ is called 2PI effective action as it contains only 2-particle irreducible diagrams \cite{Cornwall:1974vz}, viz., it only contains diagrams which cannot be separated by cutting two lines.
The 3PI effective action $\Gamma[\Phi, D, V^{(3)}]$ is obtained by performing an additional Legendre transformation with respect to three-point functions \cite{Berges:2004pu,Carrington:2010qq}.
It contains only 3-particle irreducible diagrams.
Effective actions with Legendre transformations up to $n$-point functions are called $n$PI effective actions, although the property of $n$-particle irreducibility does not hold any more for the 5PI effective action \cite{Carrington:2010qq}.
All effective actions are equivalent, viz.,
\begin{align}
 \Gamma[\Phi] = \Gamma[\Phi,D] = \ldots= \Gamma[\Phi, D, \Gamma^{(3)}, \ldots, \Gamma^{(n)}].
\end{align}
However, in $n$PI effective actions with $n\geq2$, propagators and vertices are not all treated on the same footing, because $m$-point functions with $m>n$ are not dressed.
Thus, only correlation functions up to $n$ legs are treated self-consistently.

\index{loop expansion}
For practical calculations, typically loop expansions of $n$PI effective actions are considered.
For a self-consistent expansion of an $n$PI effective action at least an $n$-loop expansion is necessary \cite{Berges:2004pu}.
Higher $n$PI effective actions are equivalent at the same expansion order:
\begin{align}
 \Gamma^\text{1-loop}[\Phi] &= \Gamma^\text{1-loop}[\Phi, ...],\\
 \Gamma^\text{2-loop}[\Phi, D] &= \Gamma^\text{2-loop}[\Phi, D, ...],\\
 \Gamma^\text{n-loop}[\Phi, D, \ldots, \Gamma^{(n)}] &= \Gamma^\text{n-loop}[\Phi, D, ..., \Gamma^{(m)}] \quad \forall n\leq m.
\end{align}

\begin{figure}[tb]
  \includegraphics[height=1.9cm]{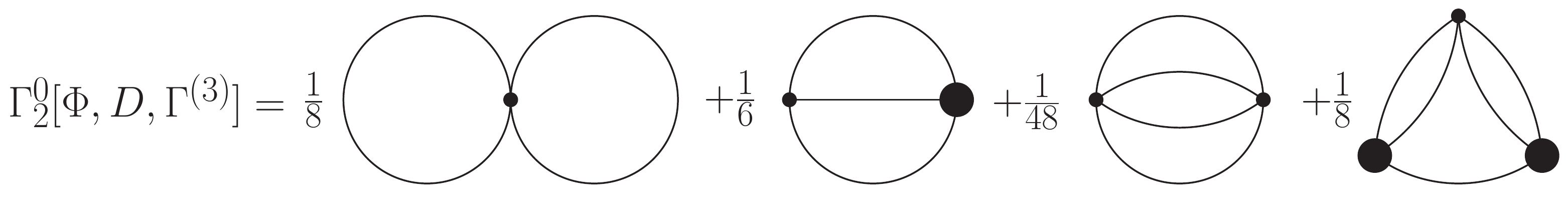}\\
  \vskip3mm
  \includegraphics[height=1.9cm]{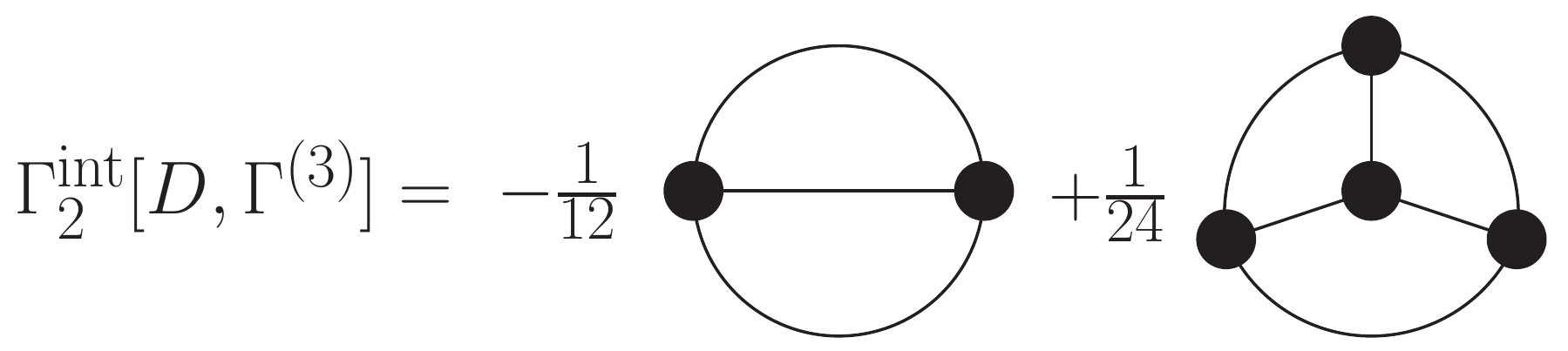}
 \caption{$\Gamma_2^0[\Phi,D,\Gamma^{(3)}]$ and $\Gamma_2^\text{int}[D, \Gamma^{(3)}]$ of the 3PI effective action at three-loop order.}
 \label{fig:Gamma2}
\end{figure}

\index{scalar theory}
As an example, the three-loop expansion of the 3PI effective action of a scalar theory with cubic and quartic interactions is considered in the following.
The generalization to \gls{qcd} can be done diagrammatically following the usual rules, viz., by endowing the fields with the appropriate indices and the diagrams with closed loops of Grassmann fields with additional minus signs.
Details for \gls{qcd} can be found in Ref.~\cite{Berges:2004pu}.
For the derivation of the 3PI effective action, it is convenient to start with the 2PI effective action \cite{Cornwall:1974vz} and perform another Legendre transformation.
The loop expansion of the 2PI effective action itself is derived using a loop expansion of the 1PI effective action \cite{Jackiw:1974cv}.
The resulting expression is \cite{Berges:2004pu,Carrington:2010qq}
\begin{align}
 \Gamma[\Phi, D, \Gamma^{(3)}] &= S[\Phi] + \frac{1}{2}\ln D^{-1}_{ii} + \frac{1}{2} S_{ij}[\Phi] D_{ij} - \Gamma_2[\Phi,D, \Gamma^{(3)}],\nnnl
 \Gamma_2[\Phi,D, \Gamma^{(3)}] &= \Gamma_2^0[\Phi,D, \Gamma^{(3)}] + \Gamma_2^\text{int}[D, \Gamma^{(3)}].
\end{align}
$S_{ij}[\Phi]$ is the field dependent inverse propagator defined as $\de^2 S[\Phi]/\de\Phi_i\de\Phi_j$.
$\Gamma_2^0$ contains bare vertices, whereas $\Gamma_2^\text{int}$ depends only on dressed quantities.
They are given by
\begin{align}
 \Gamma_2^0[\Phi,D, \Gamma^{(3)}] &= 
 \frac{1}{8}S_{ijkl}D_{ij}D_{jk}
 +\frac{1}{3!}S_{abc}D_{ai}D_{bj}D_{ck}\Gamma^{(3)}_{ijk}\nnnl
 &+\frac{1}{2\cdot4!}S_{abcd}D_{ai}D_{bj}D_{ck}D_{dl}S_{ijkl}
 +\frac{1}{8}S_{abcd}D_{ai}D_{bj}D_{ck}D_{dl}\Gamma^{(3)}_{ije}D_{ef}\Gamma^{(3)}_{klf},\\
 \Gamma_2^\text{int}[D, \Gamma^{(3)}] &= 
 -\frac{1}{2\cdot3!}\Gamma^{(3)}_{abc}D_{ai}D_{bj}D_{ck}\Gamma^{(3)}_{ijk}
 +\frac{1}{24}\Gamma^{(3)}_{abc}D_{ai}D_{bd}D_{cl}\Gamma^{(3)}_{ijk} D_{je}D_{km}\Gamma^{(3)}_{def}D_{fn}\Gamma^{(3)}_{lmn}.
\end{align}
Graphically, these expressions are depicted in \fref{fig:Gamma2}.

\begin{figure}[tb]
  \includegraphics[height=1.9cm]{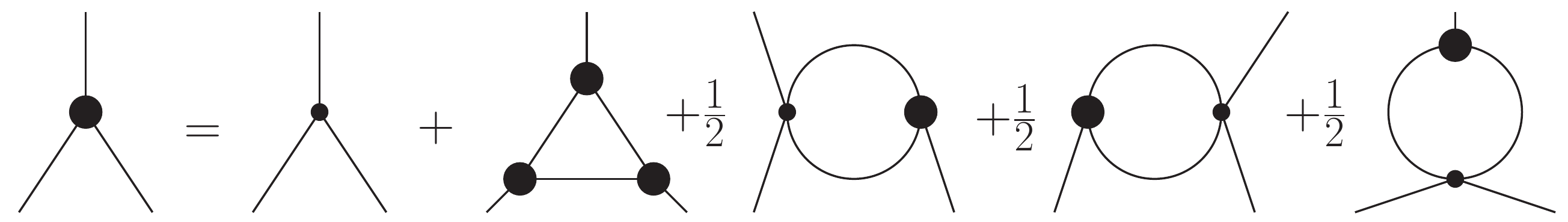}\\
 \caption{Equation of motion for the three-vertex in the three-loop expansion of the 3PI effective action.}
 \label{fig:nPI_vert}
\end{figure}

\begin{figure}[tb]
  \includegraphics[width=\textwidth]{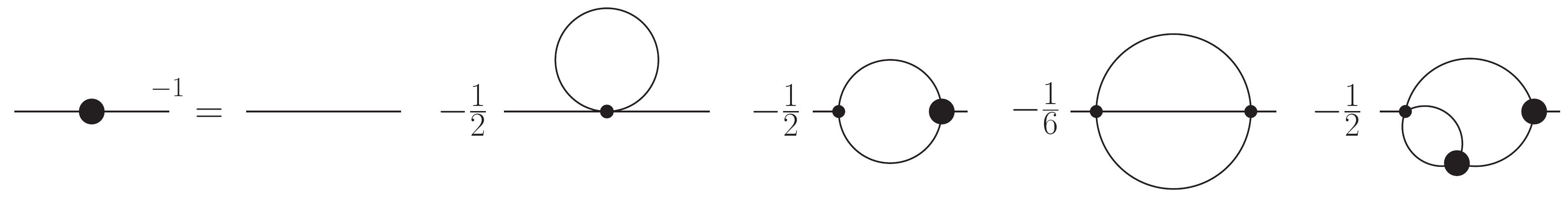}\\
 \caption{Equation of motion for the propagator in the three-loop expansion of the 3PI effective action.}
 \label{fig:nPI_prop}
\end{figure}

The equations of motion for the propagator $D$ and the vertex $\Gamma^{(3)}$ are derived from the stationarity conditions in \eref{eq:stationarity_conds}.
For the propagator one obtains
\begin{align}
 D^{-1}=D^{-1}_0 - 2 \frac{\delta \Gamma_2}{\de D},
\end{align}
where $D_0$ is the bare propagator.
Following through with the derivative one arrives at an expression that does not yet resemble the corresponding \gls{dse}.
However, it can be rewritten using the equation for the vertex $\Gamma^{(3)}$ which is shown in \fref{fig:nPI_vert}.
It is used to replace the vertex in the diagram with two dressed vertices.
This insertion corrects the prefactors of the other diagrams and leads to the equation depicted in \fref{fig:nPI_prop}.
The equation is identical to the \gls{dse} except for the four-point functions which are bare here.
This is a direct consequence of using the 3PI effective action in which only the bare four-vertex appears.

\subsubsection{Functional renormalization group}
\glsreset{frg}

\index{functional renormalization group}
The \gls{frg} is a versatile tool used in a wide range of fields including ultracold fermion gases, e.g., \cite{Diehl:2009ma,Diehl:2007th,Boettcher:2012cm}, supersymmetric models, e.g., \cite{Gies:2009az,Synatschke:2010ub,Synatschke:2009nm,Litim:2011bf,Heilmann:2012yf}, gravity, e.g., \cite{Manrique:2009tj,Reuter:1996cp,Reuter:2001ag,Codello:2007bd,Eichhorn:2010tb,Eichhorn:2009ah,Litim:2011cp,Folkerts:2011jz,Denz:2016qks,Gies:2016con,Bonanno:2017pkg,Platania:2017djo}, Higgs physics, e.g., \cite{Gies:2013fua,Gies:2014xha,Borchardt:2016xju,Gies:2017zwf} and the phase diagram of QCD, e.g.,\cite{Schaefer:2006sr,Schaefer:2006ds,Braun:2007bx,Schaefer:2008hk,Braun:2009gm,Herbst:2010rf,Pawlowski:2010ht,Fister:2011uw,Haas:2013qwp,Herbst:2013ufa,Haas:2013hpa,Tripolt:2014wra,Christiansen:2014ypa,Cyrol:2017qkl}.
This list is necessarily incomplete.
For general reviews see Refs.~\cite{Berges:2000ew,Pawlowski:2005xe,Gies:2006wv,Schaefer:2006sr}.
A short overview of the idea of the functional renormalization group is given in the following.

\index{effective average action}
The central object in the \gls{frg} is the effective average action $\Gamma^k[\Phi]$.
It introduces an artificial momentum scale $k$ that allows to interpolate between the \gls{uv} and the \gls{ir}.
In the limit $k\rightarrow \Lambda$, where $\Lambda$ is the \gls{uv} cutoff of the theory, the effective average action corresponds to the bare action at the cutoff, $\Gamma^\Lambda[\Phi] \rightarrow S$.
Lowering the scale $k$, all quantum fluctuations above $k$ are integrated out successively.
For $k\rightarrow 0$, the full quantum effective action $\Gamma[\Phi]$ is recovered.

\index{regulator}
To introduce the scale $k$, the action is modified by a regulator term:
\begin{align}
W^k[J]=\ln \,Z^k[J]=\ln \int D\phi \, e^{-S[\phi] +J_i \phi_i-\frac{1}{2}\phi_i R^k_{ij} \phi_j}.
\end{align}
The regulator function $R^k_{ij}$ needs to possess the following properties:
(1) It has to vanish for $k \rightarrow 0$ to obtain the standard effective action in this limit.
(2) It has to diverge for $k \rightarrow \infty$ so that the classical action is recovered in this limit.
(3) For small momenta $q^2<k^2$ it must be proportional to $k^2$, thus behaving like an effective mass acting as an \gls{ir} cutoff for fluctuations with small momenta.
(4) It has to vanish for large momenta $q^2>k^2$ so that it does not interfere with the high momentum behavior.

\index{effective average action}
The effective average action is obtained via a modified Legendre transformation:
\begin{align}
 \Gamma^k[\Phi]=-W^k[J]+J_i \Phi_i - \frac1{2}\Phi_i R^k_{ij} \Phi_j
\end{align}
with
\begin{align}\label{eq:def-Phi-k}
 \Phi_i=\frac{\delta W^k[J]}{\delta J_i}=\langle \phi_i \rangle_J.
\end{align}

\begin{figure}[t]
 \begin{center}
\includegraphics[width=0.3\textwidth]{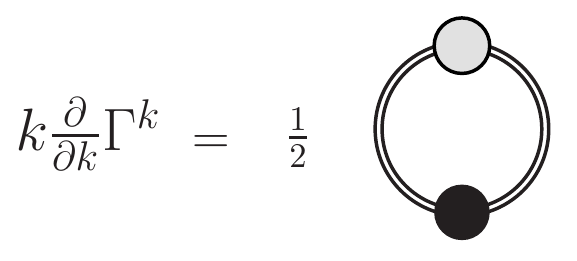}
 \caption{
 The flow equation from \eref{eq:flow_eq}.
 A gray blob denotes the regulator insertion $\partial_k R^k$.
 The dressed propagator is explicitly depicted by a black blob.}
 \label{fig:fRG_master}
\end{center}
\end{figure}

\index{flow equation}
The dependence of the effective average action on the scale $k$ is described by a flow equation:
\begin{align}\label{eq:deriv-k}
 \partial_k \Gamma^k[\Phi]=& -\partial_k W^k[J]-\frac{\delta W^k[J]}{\delta J_i} \partial_k J_i+ \frac{\partial J_i}{\partial k}  \Phi_i- \frac1{2} \Phi_i \partial_k R^k_{ij}\Phi_j=\nnnl
= & \left\langle \frac1{2} \phi_i \partial_k R^k_{ij} \phi_j \right \rangle_J -\frac1{2} \Phi_i \partial_k R^k_{ij}\Phi_j=\nnnl
= & \frac1{2}  \partial_k R^k_{ij} \left( \left \langle \phi_i  \phi_j \right\rangle_J - \Phi_i \Phi_j \right)=\nnnl
= & \frac1{2}  \partial_k R^k_{ij} D_{ij}^{k,J}.
\end{align}
Here, $\partial_k:=\partial/\partial k$ and \eref{eq:def-Phi-k} was used to cancel the second and third terms in the first line.
$\langle \phi_i \phi_j \rangle_J$ was decomposed as $D^{k,J}_{ij}+\langle \phi_i \rangle_J \langle \phi_j \rangle_J= D^{k,J}_{ij}+\Phi_i \Phi_j$.
$D^{k,J}_{ij}$ is the connected propagator in presence of the sources $J$ at the regulator scale $k$:
\begin{align}
 D^{k,J}_{ij}:=\frac{\delta^2 W^k[J]}{\delta J_i \delta J_j}.
\end{align}
Its inverse is the two-point function but with an additional contribution from the regulator $R^k$:
\begin{align}
 \de_{ij}=\frac{\de \Phi_i}{\de \Phi_j}=\frac{\de J_l}{\de \Phi_j} \frac{\de}{\de J_l} \frac{\de W^k[J]}{\de J_i}=
 \frac{\de \left(\Gamma^k[\Phi]+\frac{1}{2} \Phi_m R^k_{mn} \Phi_n \right)}{\de \Phi_j \de \Phi_l}  \frac{\de^2 W^k[J]}{\de J_l  \de J_i}=
 \left(\Gamma^{k,J}_{jl}[\Phi] + R^k_{jl} \right) D^{k,J}_{li}
\end{align}
with
\begin{align}
 \Gamma^{k,J}_{ij}:=\frac{\de^2 \Gamma^k[\Phi]}{\de \Phi_i \de \Phi_j}
\end{align}
and
\begin{align}
 J_l=\frac{\de(\Gamma^k[\Phi]+\frac{1}{2}\Phi_m R^k_{mn} \Phi_n )}{\de \Phi_l}.
\end{align}
Hence, \eref{eq:deriv-k} can also be written as \cite{Wetterich:1992yh}
\begin{align}\label{eq:flow_eq}
 \partial_k \Gamma^k[\phi]=& \frac1{2} \left(\Gamma^{k,J}_{ij}[\Phi] + R^k_{ij} \right)^{-1} \partial_k R^k_{ij}.
\end{align}
This equation is shown diagrammatically in \fref{fig:fRG_master}.
Solving the flow equation (\ref{eq:flow_eq}) corresponds to integrating out all fluctuations and going from a microscopic description, determined by the classical actions $S$, to a macroscopic description.
It should be noted that the trajectory in theory space but not the endpoint depends on the regulator function $R$.
However, this equation cannot be solved exactly and the necessary approximations lead to a regulator dependence of the endpoint.

\begin{figure}[t]
 \begin{center}
\includegraphics[width=0.68\textwidth]{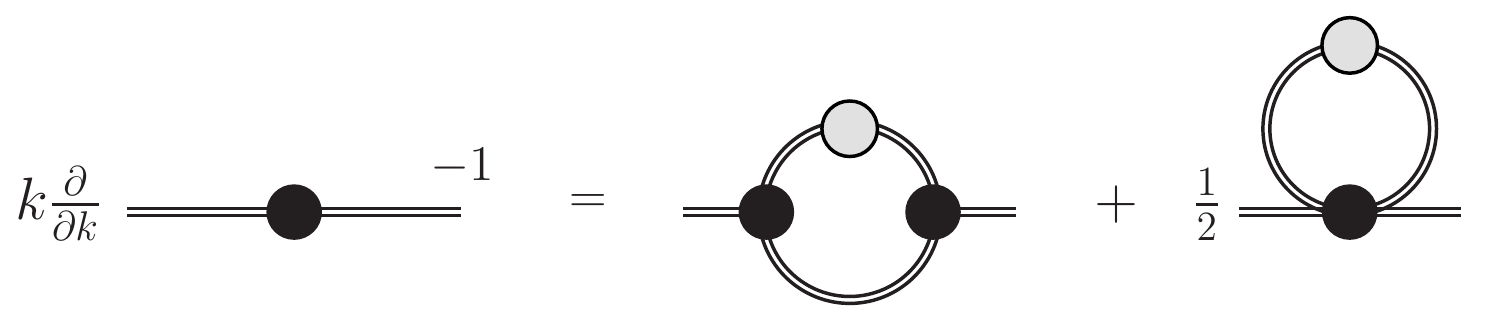}
 \caption{
 The flow equation for the two-point function.
 A black blob denotes a dressed $n$-point function.
 Propagators are all dressed.}
 \label{fig:fRG_prop}
\end{center}
\end{figure}

From \eref{eq:flow_eq}, flow equations for all correlation functions can be derived by applying derivatives with respect to fields.
To this end, the following differentiation rules are required:
\begin{subequations}\label{eq:derivatives-RG}
\begin{align}
\frac{\de D_{ij}^{k,J}}{\de \Phi_l}&=\frac{\de\left(\Gamma^{k,J}_{ij} + R^k_{ij} \right)^{-1}}{\de \Phi_l}= D^{k,J}_{im} \Gamma^{k,J}_{mln} D^{k,J}_{nj},\\
\frac{\delta}{\delta\Phi_{i}}\Gamma^{k,J}_{j_{1}\cdots j_{n}} & =-\frac{\delta\Gamma^k[\Phi]}{\delta\Phi_{i}\delta\Phi_{j_{1}}\cdots\delta\Phi_{j_{n}}}=\Gamma^{k,J}_{ij_{1}\cdots j_{n}}.
\end{align}
\end{subequations}
Again, the external sources are set to zero at the end and $D^{k,J=0}$ and $\Gamma^{k,J=0}$ correspond to the propagators and vertices of the theory for fixed $k$.
For Grassmann fields, the expected minus signs arise directly from their anticommutativity.
As an example, consider the two-point function of a scalar theory.
Two derivatives yield
\begin{align}\label{eq:FRGE-2p}
 \partial_k& \Gamma^{k,J}_{ij}=\frac{\de^2}{\de \Phi_i \de\Phi_j} \partial_k \Gamma^k[\Phi]=\nnnl
 &=\frac{\de^2}{\de \Phi_i \de\Phi_j}\frac1{2} \left(\Gamma^{k,J}_{mn} + R^k_{mn} \right)^{-1} \partial_k R^k_{mn}= \frac{1}{2} \frac{\de}{\de \Phi_i}  D^J_{mr} \Gamma^{k,J}_{rjs} D^J_{sn} \partial_k R^k_{mn} =\nnnl
 &=  D^J_{mt} \Gamma^{k,J}_{tiu} D^J_{ur} \Gamma^{k,J}_{rjs} D^J_{sn} \partial_k R^k_{mn} +
  \frac{1}{2}  D^J_{mr} \Gamma^{k,J}_{irjs} D^J_{sn} \partial_k R^k_{mn}.
\end{align}
The flow equation is obtained by setting the external source $J=0$.
The resulting integro-differential equation is depicted in \fref{fig:fRG_prop}.

As is clear from the flow equation of the effective average action, flow equations only contain one-loop diagrams.
From a technical point of view, this is very convenient.
At the same time, for an $n$-point function all vertices up to order $n+2$ appear.
For example, in \gls{qcd} the flow equation for the quark propagator contains tadpole diagrams with quark-gluon, quark-quark and quark-ghost scattering kernels in addition to the diagram with the quark-gluon vertices.
This is in marked contrast to the quark propagator \glspl{dse} which contains only a single loop diagram.

A technical advantage of the \gls{frg} is that renormalization is implemented automatically via the regulator function.
It makes the integrals \gls{uv} finite and thus no extrapolation of correlation functions in the \gls{uv} is required as is often the case otherwise.
Solving a flow equation requires to calculate the integrals at fixed scale $k$ and solve the differential equation in $k$.
This adds an additional layer of complexity.
However, often flow equations are considered under approximations that allow certain simplifications.

\subsubsection{Hamiltonian approach}
\label{sec:Hamilton}

The Hamiltonian approach is shortly discussed here as it has some similarities with other functional methods.
In contrast to the other sections, here solely its application to Yang-Mills theory is discussed.
It uses the canonical quantization in the Weyl gauge, viz., $A_0^a=0$.
The residual gauge freedom in form of time-independent gauge transformations is fixed to the Coulomb gauge, $\partial_i A_i=0$.
Resolving Gauss's law for the longitudinal part of the momentum operator leads to an extra term in the Hamiltonian, the so-called Coulomb Hamiltonian.
The final Hamiltonian depends only on transverse gauge fields.
The method was extended to Landau gauge in Ref.~\cite{Quandt:2013wna}.

Correlation functions are calculated from the corresponding vacuum expectation values:
\begin{align}\label{eq:vev_Hamilton}
\vev{K[A]} = \int D[ A] \: \calJ_A \: \lvert \varPsi[A]\rvert^2 \: K[A] ,
\end{align}
where $K[A]$ is a polynomial of fields.
The functional integral runs over all configurations in Coulomb gauge.
$\varPsi[A]$ is the vacuum wave functional and $\calJ_A=\det(G_A^{-1})$ is the Faddeev-Popov determinant of Coulomb gauge with the Faddeev-Popov operator given by\footnote{In Sec.~\ref{sec:gauge_fixing}, the Faddeev-Popov determinant was called $\Delta[A]$ and the Faddeev-Popov operator $M(x,y)$.
The notation with $\calJ_A$ and $G_A^{-1}$ is used in some literature on the Coulomb gauge.}
\begin{align}\label{eq:FP_Coulomb}
G_A^{-1}{}^{ab}(\vx,\vy) = \bigl( - \delta^{ab} \partial^2 - g f^{abc} A_i^c(\vx) \partial_i \bigr) \delta(\vx-\vy).
\end{align}
The vacuum wave functional cannot be obtained exactly by solving a functional Schr\"odinger equation with the exception of $1+1$ dimensions \cite{Reinhardt:2008ij}.
Hence, one uses an ansatz.
It is convenient to rewrite the square modulus of the vacuum wave functional as
\begin{align}
 \lvert \varPsi[A]\rvert^2 = \exp(-S[A]).
\end{align}
The functional integral in \eref{eq:vev_Hamilton} has then strong similarities with the standard path integral formulation.
Indeed, one can interpret $S[A]$ as an action and make the ansatz \cite{Campagnari:2010wc}
\begin{align}
 S[A] = \omega A^2 + \frac{1}{3!} \: \gamma_3 \, A^3 + \frac{1}{4!} \: \gamma_4 \, A^4.
\end{align}
For $\gamma_3=\gamma_4=0$, the integral is purely Gaussian \cite{Schutte:1985sd,Szczepaniak:2001rg,Feuchter:2004mk}.
The coefficients $\omega$, $\gamma_3$ and $\gamma_4$ are variational kernels that have to be determined by minimization of the vacuum energy $\vev{H_\mathrm{YM}} = E[\omega,\gamma_3,\gamma_4] $ \cite{Campagnari:2010wc}.
They are given explicitly in Sec.~\ref{sec:Coulomb}.

\begin{figure}[tb]
\begin{center}
\includegraphics[width=\linewidth]{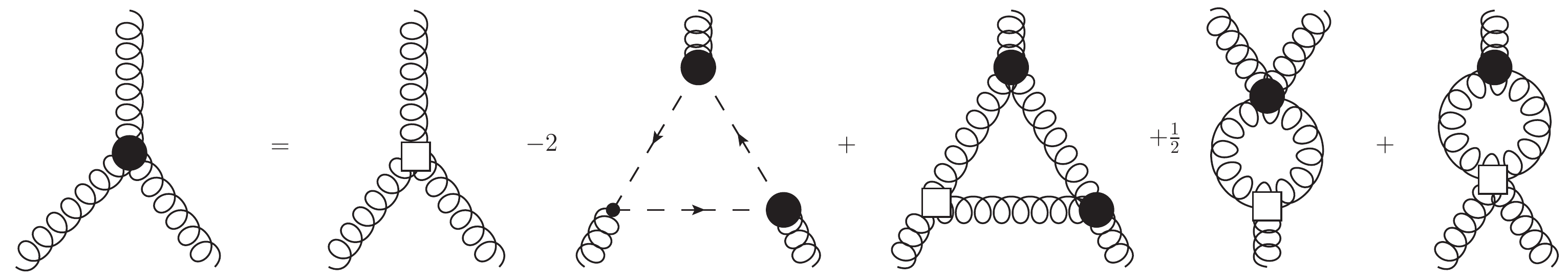}
\caption{Truncated equation for the three-gluon vertex.
Bose symmetrization in the implementation is taken into account by including only one swordfish diagram with a four-gluon variational kernel but a modified prefactor.
Empty boxes represent variational kernels.}
\label{fig:dse_coulomb_3g}
\end{center}
\end{figure}

\begin{figure}[tb]
\includegraphics[height=2.6cm]{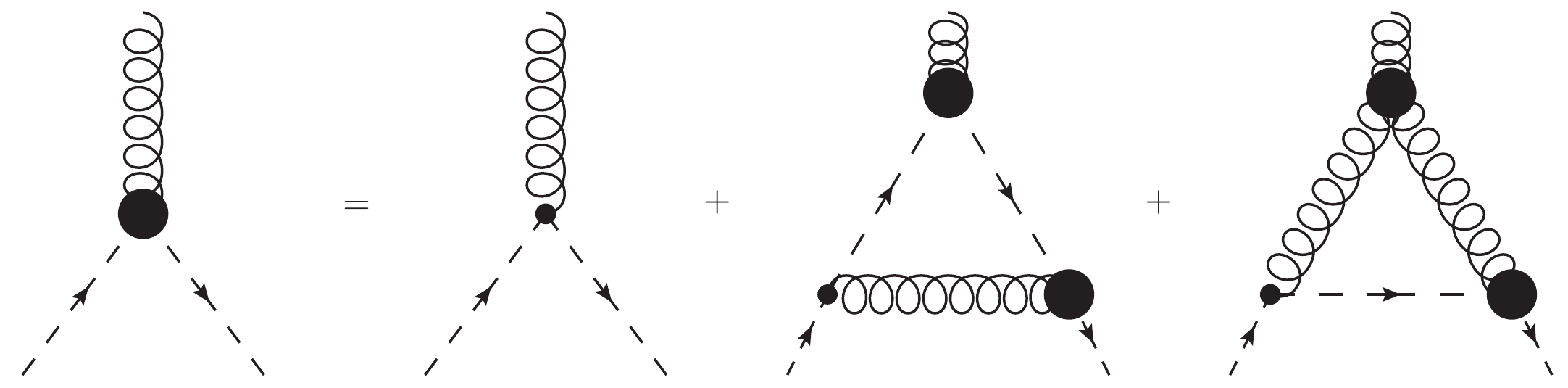}\\
\vskip3mm
\includegraphics[height=2.6cm]{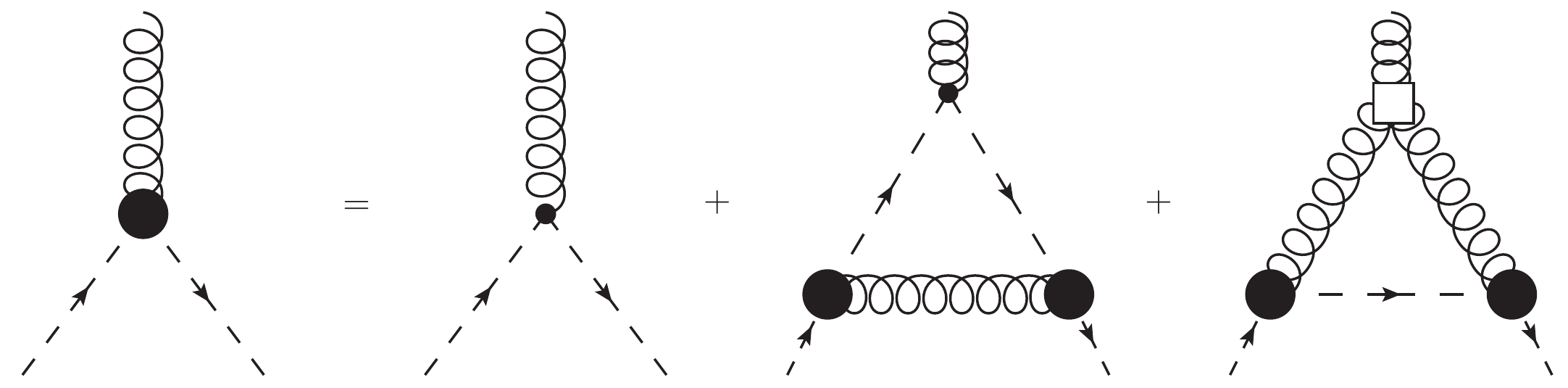}
\begin{center}
\caption{The truncated equations for the ghost-gluon vertex arising from operator identities.
The equation with the bare vertex attached to the ghost leg is equivalent to the equation at the top.}
\label{fig:dse_coulomb_ghg}
\end{center}
\end{figure}

In analogy with the standard path integral, one can directly derive equations relating the different correlation functions from the identity
\begin{align}
 0 = \int D[A] \: \frac{\delta}{\delta A} \left( \calJ_A \: e^{-S[A]} \: K[A]\right) .
\end{align}
The truncated equation for the three-gluon vertex is shown in \fref{fig:dse_coulomb_3g}.
For correlation functions involving ghost legs, one can also start from the inversion of the Faddeev-Popov operator given in \eref{eq:FP_Coulomb}, multiplies it with the appropriate number of gluon fields and takes the expectation value, see Ref.~\cite{Campagnari:2011bk} for details.
For the ghost-gluon vertex, for example, this leads as in the case of \glspl{dse} to two different equations.
They are depicted in \fref{fig:dse_coulomb_ghg} in truncated form.

\subsubsection{Technical tools}
\label{sec:tools}

Part of the progress with functional equations in recent years was driven by new tools that were partially developed specifically for use with functional equations.
The development of such tools became necessary with the systems of equations growing in size over the years.
This was made possible by the increase in available computing power but also the improved conceptual understanding of the equations.

The calculation of a particular correlation function from a functional equation consists of two steps, the derivation of the equation and the actual process of solving it.
For both steps dedicated programs exist which can also interface with each other.
Advantageously, some of these programs were made publicly available and can be used by everyone.
Of course, it depends on the specific problem if it is worth spending the time to learn how to use these programs.
However, it should be clear that beyond a certain complexity of the problem automatizations become necessary.

\index{DoDSE}
The first publicly available program for the derivation of functional equations was \textit{DoDSE} which is short for 'Derivation Of DSEs' \cite{Alkofer:2008nt}.
It is a \textit{Mathematica} \cite{Wolfram:2004} package that can derive \glspl{dse} and represent them graphically.
The output are \glspl{dse} in a symbolic form that does not refer to any specifics of the fields beyond their type.
A single index contains all physical specifics like color, flavor or component similar to the notation used in Sec.~\ref{sec:der_dse}.
The commutation and anticommutation properties of the fields are taken into account as well.
To use the abstract output in numeric calculations, it has first to be translated manually into the full algebraic form and then the contractions of indices have to be performed.
Finally, these expressions have to be put into the numeric program.

\index{DoFun}
\index{FORM}
\index{computer algebra system}
While \textit{DoDSE} represented the first step towards automatization and was capable of dealing with large systems of equations like arising for the \gls{mag} \cite{Huber:2009wh} or the \gls{gz} action \cite{Huber:2009tx}, the lack of output usable for numeric calculations constituted a major bottleneck.
With the upgrade to version 2.0, a possibility was added to transform the symbolic output into algebraic form.
To this end, the user needs to supply the corresponding Feynman rules.
A tool to derive such rules from a given action was also added.
The definition of Feynman rules also provides control over which parts of a correlation function to include.
To contract the indices of the algebraic expression only rudimentary functions are provided like handling Kronecker deltas.
The user is free to choose other available programs to perform such contractions of which there is a fair choice, e.g., \textit{FORM} \cite{vanRitbergen:1998pn,Vermaseren:2000nd,Kuipers:2012rf,Kuipers:2013pba,Ruijl:2017dtg}, \textit{FeynCalc} \cite{Mertig:1990an,Shtabovenko:2016sxi,Shtabovenko:2016whf}, \textit{HEPMath} \cite{Wiebusch:2014qba} or \textit{TRACER} \cite{Jamin:1991dp}.
The program \textit{FORM} has a long tradition in high energy physics calculations as it is very efficient in corresponding calculations due to its specialization.
However, using it requires learning the programming language.
The \textit{Mathematica} package \textit{FormTracer} \cite{Cyrol:2016zqb} makes the features of \textit{FORM} relevant for functional calculations, viz., contractions of Lorentz, color and Dirac indices, directly available in \textit{Mathematica} while still having a very flexible syntax.

\index{DoFun}
Version 2.0 of \textit{DoDSE} added the derivation of functional flow equations \cite{Huber:2011qr}.
Hence the name was changed to \textit{DoFun}, 'Derivation Of FUNctional equations'.
The inclusion of flow equations expanded the applicability of \textit{DoFun} to gravity \cite{Denz:2016qks} and effective \gls{qcd} models including the \gls{njl} and quark-meson models \cite{Braun:2017srn,Braun:2018bik,Alkofer:2018guy,Eser:2018jqo,Braun:2019aow,Divotgey:2019xea}.
It was also used for \gls{qcd} and Yang-Mills theory \cite{Alkofer:2008dt,Huber:2009wh,Huber:2009tx,Fister:2010yw,Macher:2011ys,Alkofer:2011pe,Huber:2012zj,Huber:2012kd,Blum:2014gna,Mitter:2014wpa,Huber:2014isa,Cyrol:2014kca,Huber:2014tva,Braun:2014ata,Rennecke:2015eba,Huber:2015ria,Huber:2016tvc,Cyrol:2016tym,Huber:2016xbs,Cyrol:2017qkl,Contant:2017gtz,Cyrol:2017qkl,Cyrol:2017ewj,Pawlowski:2017gxj,Huber:2017txg,Corell:2018yil,Leonhardt:2019fua,Braun:2019aow,Hajizadeh:2019qrj,Contant:2019lwf,Huber:2020keu}, the Thirring model \cite{Janssen:2012pq} and for calculations of spectral functions in the $O(N)$ model \cite{Strodthoff:2016pxx}.

In version 3.0 \cite{Huber:2019dkb}, the workflow was made more flexible by implementing stricter rules for the definition of fields and the code was partially restructured for easier maintainance.
In addition, the derivation of equations for composite operators was included and the code was put on a public \textit{GitHub} repository, \href{https://github.com/markusqh/DoFun}{https://github.com/markusqh/DoFun}.
Bugfixes and updates are also made accessible via this webpage.
At the time of writing, the current version number is 3.0.1.
The documentation of \textit{DoFun} is contained in \textit{Mathematica}'s Documentation Center.

\index{CrasyDSE}
For solving functional equations numerically, a wide range of programs can be used, e.g., \textit{C}, \textit{C++}, \textit{Fortran}, \textit{Python} or \textit{Julia}.
Various packages can be used to deal with standard operations like integration or interpolation.
Technical details on solving \glspl{dse} can be found, e.g., in Refs.~\cite{Atkinson:1997tu,Maas:2005xh,Huber:2011xc,Sanchis-Alepuz:2017jjd}.
A framework for handling \glspl{dse} is provided by \textit{CrasyDSE} \cite{Huber:2011xc} written in \textit{C++}.
The name is an acronym for 'Computation of RAther large SYstems of DSEs'.
It does not rely on any particular packages to make it as autonomous as possible.
Hence it can be deployed easily on different systems without worrying about dependencies.
Naturally, this comes at the price of not being as fast as some dedicated packages.
\textit{CrasyDSE} consists of separate modules for integration, interpolation and handling \glspl{dse} which can be used independently from each other.
A useful tool contained in \textit{CrasyDSE} is a \textit{Mathematica} package that provides the functionality to create kernel files from expressions in \textit{Mathematica}.
This can be used directly with the output of \textit{DoFun}, but it works with general expressions.

\index{FORM}
\textit{FORM} \cite{vanRitbergen:1998pn,Vermaseren:2000nd,Kuipers:2012rf,Kuipers:2013pba,Ruijl:2017dtg} was already mentioned as a tool for contracting indices.
In addition, it provides an optimization routine \cite{Kuipers:2013pba,Ruijl:2017dtg} that brings large expressions into a form that not only reduces their size but also speeds up their evaluation considerably.
For large kernels as appearing, for example, for four-point functions \cite{Huber:2016tvc,Huber:2017txg}, such optimization routines are extremely helpful.
Expressions can be passed on from \textit{Mathematica} to \textit{FORM}, optimized with it and then be read back in to export them to dedicated kernel files, e.g., with \textit{CrasyDSE}.
Alternatively, \textit{FormTracer} also provides access to this \textit{FORM} feature.

In summary, recent years have seen an increased use of automated tools for deriving and solving functional equations.
Such tools are helpful not only to reduce errors and allow an efficient treatment, but given the size and complexity of modern truncations their use has become obligatory in many cases and their importance will most likely grow.
It should also be noted that public availability and exposition of technical details increase the accessibility of the field and the trust in the results by researchers from other fields.
Thus, any effort to make tools and programs public in the future is welcome.

\subsubsection{Summary: Similarities and differences of functional approaches}

All of the functional equations described above share some similarities.
They can be represented with Feynman diagrams of dressed and, in some cases, bare quantities.
The Feynman diagrams correspond to integrals, but also differential forms of the equations exist, e.g., \cite{Rivers:1988pi}.

An interesting difference lies in the role of bare quantities.
While in \glspl{dse} every diagram contains one bare vertex, flow equations contain only dressed vertices.
Thus, although some diagrams look very similar except for one bare/dressed vertex, there are also diagrams appearing only in one set of equations.
In the equations of motion of $n$PI effective actions, bare quantities appear either because the corresponding vertex has more than $n$ legs, or because a resummation takes place that leads to the same diagrams as for the \gls{1pi} effective action.
In the Hamiltonian approach, some bare vertices are replaced by variational kernels which are determined separately by a variational principle.
The variational kernels thus contain already more nonperturbative information than the standard bare vertices.
Tab.~\ref{tab:summary_funEqs} contains an overview of the differences between equations of motion and flow equations.

\begin{table}[tb]
\begin{center}
\begin{tabular}{|l|l|l|l|}
\hline
 & DSEs & $n$PI & FRG \\
\hline\hline
 Effective action & $\Gamma[\phi]$ & $\Gamma[\phi, D, R^{(3)}, \ldots, R^{(n)}]$ & $\Gamma^k[\phi]$\\
\hline
 Loops & fixed & loop expansion possible & 1\\
\hline
 Bare vertices & one per diagram & yes & none \\
\hline
 Remarks & integrated RGEs &    & differential DSEs,\\
    & & & regulator \\
\hline
\end{tabular}
\caption{Comparison between \glspl{dse}, equations of motion from $n$PI effective actions and flow equations.}
\label{tab:summary_funEqs}
\end{center}
\end{table}

\Glspl{dse} and equations of motion of $n$PI effective actions are naturally very similar on the technical level, since \glspl{dse} are nothing else than the equations of motion of the \gls{1pi} effective action.
Thus, the same techniques can be used to solve them.
The equations of the Hamiltonian approach have a similar structure as well.
However, the appearance of variational kernels instead of bare vertices requires to rethink the renormalization procedure, since variational kernels can have a different momentum structure than the bare vertices, see, e.g., Ref. \cite{Huber:2014isa}.
Flow equations, finally, can be handled in a similar fashion for fixed \gls{rg} scale $k$, but in addition the solution of the differential equation in $k$ is required.
The appearance of the regulator has advantages as far as regularization is concerned.
However, it also complicates real-time calculations by introducing additional poles \cite{Pawlowski:2015mia}.

\subsection{Other methods}
\label{sec:otherMethods}

\Gls{qcd} correlation functions are studied with a range of methods.
Their complementarities, e.g., analytic vs. numerical, continuous vs. discrete spacetime, are useful and sometimes lead to additional benefits.
In this section, methods other than functional methods are shortly reviewed to sketch the landscape of approaches used for studying nonperturbative \gls{qcd}.

\subsubsection{Monte Carlo simulations on the lattice}
\label{sec:lattice}

\myboxmargin{lattice simulations in general and their success}
\index{lattice calculations}
Lattice \gls{qcd}, viz., Monte Carlo simulations of \gls{qcd} on a discretized spacetime \cite{Wilson:1974sk}, are very successful in describing many aspects of \gls{qcd} e.g., \cite{Colangelo:2010et,Gattringer:2010zz,Fodor:2012gf,Borsanyi:2013bia,Bazavov:2014pvz,Aoki:2016frl}.
It relies on making spacetime discrete and finite by reducing it to a lattice on a four-dimensional torus.
The quark fields live on the sites of the lattice and the gauge fields on the links.
This formulation is very convenient from the conceptual point of view, because the \gls{uv} regularization via the lattice spacing $a$ does not break gauge symmetry.
In addition, physical quantities, viz., gauge independent quantities, can be calculated directly without fixing a gauge.

However, to make contact with other nonperturbative methods it is useful to fix the gauge nevertheless.
Then, quantities like correlation functions of elementary fields can be calculated and directly compared.
Unfortunately, the gauge fixing procedure is not unique due to the Gribov problem mentioned in Sec.~\ref{sec:gauge_fixing}.
This directly affects lattice calculations as different algorithms to fix the gauge can be used.
Based on the specific way to choose a gauge copy, 'different' gauges can be defined which are equivalent on the perturbative level.
For example, in case of the Landau gauge, choosing the first copy found is known as minimal Landau gauge and choosing the copy with the lowest norm of the gauge configuration as absolute Landau gauge.
Many other choices are possible, see, e.g., \cite{Zwanziger:1993dh,Henty:1996kv,Cucchieri:1997dx,Silva:2004bv,Bogolubsky:2005wf,Bogolubsky:2007bw,Bornyakov:2008yx,Maas:2008ri,Maas:2009ph,Maas:2009se,Bornyakov:2010nc,Sternbeck:2012mf,Vandersickel:2012tz,Maas:2013vd}.
Averages over the full gauge orbits constitute also a valid choice \cite{Hirschfeld:1978yq,Neuberger:1986xz,Parrinello:1990pm,Fachin:1991pu,Fachin:1993qg,Kalloniatis:2005if,vonSmekal:2007ns,vonSmekal:2008en,vonSmekal:2008es,Mehta:2009zv,Maas:2012ct,Serreau:2012cg,Serreau:2013ila}, but this approach has not been realized yet with lattice methods.
Results for correlations functions are available for propagators, e.g., \cite{Cucchieri:1999sz,Maas:2007uv,Maas:2008ri,Maas:2009se,Maas:2009ph,Maas:2007af,Maas:2011se,Cucchieri:2016jwg,Cucchieri:2003di,Cucchieri:2006tf,Cucchieri:2007rg,Cucchieri:2008qm,Cucchieri:2009zt,Cucchieri:2011ig,Bornyakov:2011fn,Maas:2017csm,Bloch:2003sk,Furui:2003jr,Bogolubsky:2005wf,Silva:2005hb,Cucchieri:2006za,Cucchieri:2007md,Sternbeck:2007ug,Oliveira:2007dy,Cucchieri:2007zm,Cucchieri:2008fc,Gong:2008td,Oliveira:2008uf,Bornyakov:2008yx,Bogolubsky:2009dc,Bornyakov:2009ug,Oliveira:2009eh,Pawlowski:2009iv,Oliveira:2012eh,Bornyakov:2013ysa,Maas:2014xma} and three-point functions \cite{Cucchieri:2004sq,Ilgenfritz:2006he,Cucchieri:2008qm,Sternbeck:2016ltn,Athenodorou:2016oyh,Cucchieri:2016jwg,Maas:2017csm,Boucaud:2017obn} in two, three and four dimensions at zero and nonvanishing temperatures.

These results are very useful for comparisons with the results from functional methods.
However, care must be taken in such comparisons.
In lattice studies questions of infinite volume and continuum limits are not always fully clarified.
In addition, it is not clear which gauge fixing prescription on the lattice corresponds to which solution of functional equations.
In some cases, the two approaches were also combined.
For example, fits to gluon propagator results can be used as input in functional equations at nonvanishing temperatures \cite{Fischer:2010fx,Fischer:2011mz,Fischer:2012vc,Fischer:2013eca,Fischer:2014ata,Eichmann:2015kfa,Contant:2017gtz,Contant:2017onc}.

\myboxmargin{gauge fixing on the lattice}

\myboxmargin{Gribov problem}

\subsubsection{(Refined) Gribov-Zwanziger framework}
\label{sec:RGZ}

\glsreset{fmr}

\myboxmargin{gauge fixing a la Gribov-Zwanziger}
\index{Gribov-Zwanziger framework}
As mentioned in Sec.~\ref{sec:gauge_fixing}, it is not possible to fix a gauge uniquely in the continuum.
For example, many gauges can be implemented by introducing a delta functional of a gauge fixing functional or a Gaussian smearing of it together with the Jacobian, the Faddeev-Popov operator.
However, this always leaves some remnant gauge copies.
Gribov suggested a way to alleviate the situation \cite{Gribov:1977wm} by restricting the integration in field configuration space on the gauge fixing hypersurface to a smaller region defined as the region where the Faddeev-Popov operator is positive.
This region is nowadays known as first Gribov region, bounded by the Gribov horizon.
At the boundary, the first eigenvalue of the Faddeev-Popov operator becomes zero and directly beyond the horizon the Faddeev-Popov determinant is negative.
Where the second eigenvalue becomes negative and the determinant becomes positive again, the second Gribov horizon is crossed from the second to the third Gribov region.
However, even if the restriction to the first Gribov region can be implemented, there are still gauge copies \cite{vanBaal:1991zw}.

\myboxmargin{fundamental modular region}
\index{fundamental modular region}
Formally, one can define a copy-free region called the \gls{fmr} \cite{vanBaal:1991zw}.
It is contained within the first Gribov region and shares some of its boundary.
Some of its boundary points are identified which makes the \gls{fmr} a nonlocal and highly nontrivial object.
It is not known how to fix the gauge to this region in the continuum.
On the lattice, finding the absolute minimum is also not feasible, since this minimization problem is of the spin-glass type.
However, one can approximate the search by taking the lowest minimum of the gauge fixing functional found in a (finite) sampling of the gauge orbit.
This is known as the absolute Landau gauge.

\myboxmargin{first Gribov region, gauge fixing a la Gribov}
\index{Gribov region, first}
For the Landau gauge, the first Gribov region is defined by
\begin{align}
 \Omega:= \{A\, | \, \partial_\mu A_\mu=0, \, M >0 \}.
\end{align}
$M$ is the Faddeev-Popov operator which is related to the ghost propagator as
\begin{align}
 D^{ab}_{\bar{c}c}(k)=-\frac{\delta^{ab}}{k^2}c_{\bar{c}c}(k)=-( M^{-1} )^{ab}.
\end{align}
Gribov's idea to restrict the integration in field configuration space to the region where $M$ is positive relied on parametrizing the ghost propagator as \cite{Gribov:1977wm}
\begin{align}
 D^{ab}_{\bar{c}c}(k)=-\frac{\delta^{ab}}{k^2}\frac1{1-\sigma(k,A)}.
\end{align}
It can be shown that $\sigma(k,A)$ increases with decreasing $k$.
Thus, it is sufficient to demand $\sigma(0,A)<1$.
This relation is known as no-pole condition.
The ghost-self-energy can be calculated as a series in the external field $A$.
In the path integral, the no-pole condition can be implemented via a Heaviside functional as
\begin{align}
 \theta(1-\sigma(0,A))=\int_{-\infty}^{\infty} \frac{d\beta}{2\pi\,i}\frac{e^{i\,\beta(1-\sigma(0,A))}}{\beta-i\,\epsilon}.
\end{align}
Using the saddle-point method, the integral over $\beta$ can be evaluated.
$\beta$ then takes a certain value $\gamma$ which is also known as Gribov parameter.
It has the dimension of mass and must be calculated separately.

\glsreset{gz}
The generalization to all orders shows that one can add the so-called horizon condition to the Lagrangian density that implements the restriction to the first Gribov horizon \cite{Zwanziger:1989mf}:
\begin{align}\label{eq:hor-func3}
 h(x)=\gamma^4\,g^2 \int dx dy f^{abc} A^c_\mu(x)\left( M^{-1}\right )^{ad}(x,y) f^{dbe} A^e_\mu(y).
\end{align}
This expression can be localized by introducing four new fields forming a \gls{brst} quartet leading to the so-called \gls{gz} action \cite{Zwanziger:1992qr}.
The resulting action has a gluon propagator that vanishes at tree-level \cite{Zwanziger:1991gz}.
The ghost dressing function at one-loop level is \gls{ir} divergent \cite{Zwanziger:1992qr,Zwanziger:1993dh}.
Among other things, the breaking of the standard \gls{brst} symmetry \cite{Dudal:2009xh}, the definition of a nonperturbative \gls{brst} transformation for this action \cite{Kondo:2009qz,Sorella:2009vt} or the construction of physical operators \cite{Baulieu:2009ha,Sorella:2010fs,Dudal:2010vn} were also investigated.

\glsreset{rgz}
A generalization of the \gls{gz} action takes into account the existence of several condensates \cite{Dudal:2008sp}.
This is known as the \gls{rgz} action.
The form of the action depends on details of the considered condensates \cite{Dudal:2008sp,Gracey:2010cg,Dudal:2011gd}, but it is always chosen such that the gluon propagator at tree-level and the ghost dressing function at one-loop level are \gls{ir} finite.
The condensates are difficult to calculate dynamically.
Thus, they are typically determined by fits to lattice data \cite{Cucchieri:2011ig}.
Such results can then be used in further calculations as input in analytic form.
For example, glueball masses \cite{Dudal:2010cd,Capri:2011ki,Dudal:2013wja}, the Polyakov loop \cite{Canfora:2015yia} or the topological susceptibility \cite{Dudal:2017kxb} were calculated.
The method, originating in Landau gauge, was extended also to the maximally Abelian gauge \cite{Capri:2005tj,Dudal:2006ib,Capri:2006cz,Capri:2007hw,Capri:2008ak} and linear covariant gauges \cite{Capri:2015ixa,Capri:2015nzw,Capri:2016aif,Capri:2017bfd}.

\subsubsection{Massive extensions of Yang-Mills theory}
\label{sec:massiveYM}

The fact that lattice simulations find a non-zero and finite value for the gluon propagator at zero momentum has motivated several model studies that contain a mass term for the gluon in the Lagrangian.
This term is sometimes considered only on a phenomenological level, but there are also approaches with a physical motivation.
The refined Gribov-Zwanziger framework, mentioned in Sec.~\ref{sec:RGZ}, belongs to the latter class, ascribing the mass term to the existence of certain condensates arising from the restriction of the integration in the path integral to the first Gribov region.

\index{massive Yang-Mills theory}
Another approach that links a gluon mass to the Gribov problem leads to a massive extension of Yang-Mills theory \cite{Serreau:2012cg,Serreau:2013ila} in form of the Curci-Ferrari model \cite{Curci:1976bt}.
In contrast to the refined Gribov-Zwanziger framework, which contains many nonperturbative condensates, only one additional parameter appears in form of a mass term for the gluon.
With perturbative one-loop and two-loop calculations, an effective description of the nonperturbative regime is obtained, e.g., for two- \cite{Tissier:2010ts,Tissier:2011ey,Pelaez:2014mxa,Pelaez:2017bhh,Gracey:2019xom} and three-point functions \cite{Pelaez:2013cpa,Pelaez:2015tba}.
Besides the coupling, the gluon mass is also a free parameter which can be fixed by fitting to lattice data.
The analysis was also extended to two loops \cite{Gracey:2014dna}.
This model was also successfully applied to studies at nonzero temperatures and densities \cite{Reinosa:2013twa,Reinosa:2014ooa,Reinosa:2014zta,Reinosa:2015gxn,Reinosa:2015oua,Reinosa:2016iml,Maelger:2017amh}.
Other studies of massive extensions of Yang-Mills theory include \cite{Weber:2011nw,Weber:2012vf,Weber:2012tp,Weber:2014lxa,Siringo:2014lva,Siringo:2015gia,Machado:2016cij,Weber:2016biv,Siringo2016,Siringo2016a,Comitini:2017zfp,Siringo:2018uho}.

\section{Correlation functions of Landau gauge Yang-Mills theory}
\label{chp:results_YM_Landau}

\glsreset{mag}

\myboxmargin{correlations functions of Yang-Mills theory}
\index{Yang-Mills theory}
The correlation functions of Yang-Mills theory have been studied with various methods ranging from phenomenological modeling to studies from first principles.
While consensus has been reached in some questions, there are also some open issues pending further investigation.
In particular, although several methods agree qualitatively and even quantitatively, there are still some subtle issues concerning details of gauge fixing.

In this section, the status of results from \glspl{dse} is reviewed.
First, an overview of the Landau gauge is given.
In the subsequent section, Yang-Mills theory in four dimensions is discussed.
A particular focus lies on testing truncation dependences and clarifying some aspects which are important for a self-contained solution.
The cases of three and two dimensional Yang-Mills theory are then investigated in Secs.~\ref{sec:YM3d} and \ref{sec:YM2d}, respectively.
These theories are not only interesting by themselves, but they also allow insights on the general structure of functional equations and their truncations.

\subsection{Landau gauge Yang-Mills theory}
\label{sec:why_LG}

\myboxmargin{Advantages of Landau gauge: lowest number of fields and correlation functions and relevant dressing functions, nonrenormalization of ghost-gluon vertex}
\index{Landau gauge}
The Landau gauge is the gauge investigated best with functional methods.
It is, technically speaking, more accessible than other gauges and was always a preferred gauge also for other methods.
This allowed useful comparisons and complementary combinations of methods.
Quite generally, it is also advantageous that the Landau gauge has the lowest number of 'terms' possible.
This refers, on the one hand, to the number of primitively divergent correlation functions.
On the other hand, the physically relevant transverse correlation functions form a closed system \cite{Fischer:2008uz} and the longitudinal ones do not need to be computed.
Another set of functional equations, the \glspl{sti}, constrains the longitudinal part only.
Turning to other gauges, more dressings, for instance, the longitudinal ones in linear covariant gauges, or even more fields and interaction terms are needed.
An example for the latter case is the maximally Abelian gauge where the diagonal and off-diagonal parts of all fields are treated separately thus doubling the field content.

\myboxmargin{MOM scheme dependence on the vertex}
\index{MOM scheme}
It should also be mentioned that the Landau gauge has some interesting perturbative properties.
Very often, Feynman gauge is used for perturbative calculations due to the simpler structure of the gluon propagator.
However, when \gls{mom} renormalization schemes are used, it turns out that the dependence on which vertex is used to define the coupling is weakest in the Landau gauge \cite{Celmaster:1979km}.

\myboxmargin{Landau gauge and hadron phenomenology}
\index{hadrons}
For studies of hadrons using bound state equations, the Landau gauge is also a standard choice.
A vast list of calculations using effective interactions exists, see Refs.~\cite{Bashir:2012fs,Eichmann:2016yit} for recent reviews, but also results directly from calculated correlation functions were obtained, e.g., \cite{Williams:2015cvx,Huber:2020ngt}.

\myboxmargin{Landau gauge: role of ghost-gluon vertex}
\index{ghost-gluon vertex}
Historically, the nonrenormalization of the ghost-gluon vertex played an important role in the investigation of the Landau gauge.
The observation that his vertex is finite in the Landau gauge is attributed to Taylor and thus known as Taylor's nonrenormalization theorem \cite{Taylor:1971ff,Marciano:1977su}.
Using an approximated \gls{sti}, one can show that in the limit of the ghost momentum going to zero the vertex is bare.
Variations for deriving this can be found, e.g., in \cite{Taylor:1971ff,Marciano:1977su,Boucaud:2005ce,Alkofer:2004it,Boucaud:2011ug,Boucaud:2011eh,Huber:2012kd}.
Although it can only be shown for this special limit, it was often used as justification to employ a bare ghost-gluon vertex as ansatz.
This was the entry point for many calculations of propagators.
For the ghost propagator it is the only vertex that is needed and in the one-loop truncated gluon propagator only the three-gluon vertex remains to be specified.
In dedicated calculations of the ghost-gluon vertex using lattice, functional and other methods, it was later found that the ghost-gluon vertex indeed shows only a quantitative deviation from a bare vertex \cite{Cucchieri:2004sq,Schleifenbaum:2004id,Ilgenfritz:2006he,Cucchieri:2008qm,Boucaud:2011eh,Huber:2012zj,Huber:2012kd,Pelaez:2013cpa,Aguilar:2013xqa,Huber:2013xb,Mintz:2017qri,Aguilar:2018csq,Maas:2019ggf,Huber:2020keu}.
This explained the success of using this vertex ansatz.
Investigations of other gauges, with the exception of the Coulomb gauge where the ghost-gluon vertex shows a similar behavior \cite{Schleifenbaum:2006bq,Campagnari:2011bk,Huber:2014isa}, are aggravated by the fact that corresponding simplifications are not known.

\myboxmargin{Landau gauge: role of lattice calculations}
\index{lattice calculations}
Another factor contributing to the widespread use of the Landau gauge in functional calculations is its accessibility with lattice methods which can thus be used for comparisons.
The calculation of correlation functions with lattice methods has a long history itself.
A breakthrough was achieved when finally lattice calculations were able to probe propagators also in the \gls{ir} regime \cite{Cucchieri:2007md,Cucchieri:2008fc,Sternbeck:2007ug,Bogolubsky:2009dc}.
Since then, the corresponding results in the midmomentum regime are often used as benchmarks for other methods.
Propagators were studied heavily in two \cite{Maas:2007uv,Maas:2008ri,Maas:2009se,Maas:2009ph,Maas:2007af,Cucchieri:2011ig,Maas:2011se,Cucchieri:2016jwg,Maas:2017csm,Maas:2019ggf}, three \cite{Cucchieri:1999sz,Cucchieri:2003di,Cucchieri:2006tf,Maas:2007uv,Cucchieri:2007rg,Maas:2007af,Maas:2008ri,Cucchieri:2008qm,Maas:2009se,Maas:2009ph,Cucchieri:2009zt,Cucchieri:2011ig,Bornyakov:2011fn,Maas:2011se,Maas:2017csm,Maas:2019ggf} and four dimensions \cite{Bloch:2003sk,Furui:2003jr,Bogolubsky:2005wf,Silva:2005hb,Cucchieri:2006za,Cucchieri:2007md,Sternbeck:2007ug,Oliveira:2007dy,Cucchieri:2007rg,Cucchieri:2007zm,Cucchieri:2008fc,Gong:2008td,Oliveira:2008uf,Bornyakov:2008yx,Bogolubsky:2009dc,Bornyakov:2009ug,Oliveira:2009eh,Pawlowski:2009iv,Cucchieri:2009zt,Cucchieri:2011ig,Maas:2011se,Oliveira:2012eh,Duarte:2016iko,Cucchieri:2016jwg,Boucaud:2017ksi,Maas:2017csm,Duarte:2017wte,Boucaud:2018xup,Biddle:2018dtc,Maas:2019ggf} and some results for vertices are available as well \cite{Cucchieri:2004sq,Ilgenfritz:2006he,Cucchieri:2008qm,Sternbeck:2016ltn,Athenodorou:2016oyh,Sternbeck:2017ntv,Boucaud:2017obn,Maas:2019ggf}.
Also for lattice calculations of correlation functions the Landau gauge is investigated best with only very limited results available beyond this gauge.
However, some issues like gauge fixing ambiguities due to the Gribov problem are not fully settled yet and still actively investigated, e.g., \cite{Maas:2011se,Sternbeck:2012mf,Maas:2017csm,Maas:2019ggf} and references therein.

\myboxmargin{decoupling vs. scaling}
\index{decoupling solution}
\index{scaling solution}
An aspect of correlation functions of Yang-Mills theory that is most likely related to the unresolved Gribov issue is the existence of several solutions.
Historically, the first calculation that solved the ghost and the gluon propagators together \cite{vonSmekal:1997is,vonSmekal:1997vx} found a solution that is characterized by power laws of all dressing functions \cite{Alkofer:2004it,Fischer:2006vf,Huber:2007kc,Alkofer:2008bs,Fischer:2009tn}.
The gluon propagator was found to be \gls{ir} vanishing and the ghost dressing function \gls{ir} enhanced.
This is in qualitative agreement with predictions in relation to the confinement scenario by Kugo and Ojima \cite{Kugo:1979gm,Kugo:1995km} and in the \gls{gz} framework.
The exponents of the corresponding power laws,
\begin{align}\label{eq:powerLaws}
 Z(p^2)\propto (p^2)^{\de_A}, \quad G(p^2) \propto (p^2)^{\de_c},
\end{align}
were shown to be related by \cite{vonSmekal:1997is,vonSmekal:1997vx,Fischer:2009tn}
\begin{align}
 \ka=-\de_c=\frac{\de_A}{2}.
\end{align}
The value obtained for $\ka$ was $0.595353$ \cite{Zwanziger:2001kw,Lerche:2002ep,Huber:2012zj}.
Lattice calculations at that time could not reach low enough momenta to confirm that.
Only ten years later, first calculations on large enough lattices were made that showed a direct conflict with this behavior \cite{Cucchieri:2007md,Cucchieri:2008fc,Sternbeck:2007ug,Bogolubsky:2009dc}.
What was found was an \gls{ir} finite gluon propagator and an \gls{ir} finite ghost dressing function.
The solution of this type was named massive or decoupling solution and the earlier one scaling solution.
Soon after these lattice results, solutions with this behavior were also found with other methods \cite{Dudal:2008sp,Boucaud:2008ji,Boucaud:2008ky,Aguilar:2008xm,Fischer:2008uz,Alkofer:2008jy}.
However, the idea of the dynamic generation of a gluon mass goes back to the 80's \cite{Cornwall:1981zr} and was investigated previously, for example, in Refs.~\cite{Aguilar:2006gr,Aguilar:2004sw}.
The current status is that both types of solutions can be obtained with functional and other continuum methods while on lattice only the decoupling solution is found.
In particular, with functional equations it is possible to find an infinite number of decoupling solutions and the scaling solution corresponds to the endpoint with an infinite gluon mass \cite{Fischer:2008uz}.
If all these solutions are physically equivalent, they must necessarily lead to the same results for physical quantities.
This was tested explicitly for scalar and pseudoscalar glueballs for which a range of different solutions was tested as input for their bound state equations, but no change in the spectrum was found \cite{Huber:2020ngt}.
It should be noted that only the scaling solution possesses an intact standard \gls{brst} symmetry \cite{Kugo:1979gm} and this symmetry is not realized in lattice calculations \cite{Neuberger:1986vv,Neuberger:1986xz}.

The fact that for the Landau gauge many results from lattice simulations exist also led to this gauge being a preferred gauge for many other methods.
For example, models with a gluon mass term fix the value of this mass by fits to lattice results, e.g., \cite{Tissier:2010ts,Tissier:2011ey,Pelaez:2014mxa,Siringo:2014lva,Gracey:2019xom}.
Such approaches are discussed in Sec.~\ref{sec:massiveYM}.
In the \gls{rgz} framework, originating in restricting the integration in the path integral to the first Gribov region \cite{Gribov:1977wm,Zwanziger:1991gz,Zwanziger:1992qr}, see Sec.~\ref{sec:RGZ}, also fits to lattice data are often used, e.g., \cite{Cucchieri:2011ig,Cucchieri:2016jwg}, since a self-consistent determination of the corresponding quantities is difficult \cite{Dudal:2011gd}.

Beyond the vacuum, correlation functions have also been studied best in the Landau gauge.
At vanishing density and nonzero temperature, lattice studies are less abundant than in the vacuum but still provide useful guidelines, e.g., \cite{Fischer:2010fx,Maas:2011ez,Cucchieri:2012nx,Silva:2013maa,Fister:2014bpa,Silva:2016onh,Silva:2016msq}.
At nonzero density there are currently no lattice results for correlation functions of \gls{qcd} due to the complex action problem.
However, results for the gauge group $SU(2)$ exist \cite{Boz:2013rca,Hajizadeh:2017ewa,Boz:2018crd,Bornyakov:2020kyz}.
With functional methods, studies range from using effective actions originally designed for hadron phenomenology, e.g., \cite{He:2007zzg,Xin:2014ela,Gao:2016qkh}, to combinations with lattice methods, e.g., \cite{Fischer:2011mz,Fischer:2012vc,Muller:2013pya,Fischer:2014ata,Eichmann:2015kfa,Contant:2017gtz,Contant:2019lwf,Isserstedt:2019pgx,Gunkel:2019xnh}, to pure functional studies, e.g., \cite{Maas:2005hs,Fister:2011uw,Huber:2013yqa,Quandt:2015aaa,Huber:2016xbs,Cyrol:2017qkl,Contant:2018zpi,Hajizadeh:2019qrj}.
The most advanced pure functional studies calculated propagators \cite{Cyrol:2017qkl,Quandt:2015aaa,Hajizadeh:2019qrj} and even vertices \cite{Huber:2013yqa,Cyrol:2017qkl}.
For a recent review on the \gls{dse} approach to the phase diagram of \gls{qcd} see Ref.~\cite{Fischer:2018sdj}.
Also other methods rely on the Landau gauge for studies of the phase diagram of \gls{qcd}, e.g., \cite{Reinosa:2015gxn,Canfora:2015yia,Reinosa:2016iml}.

Finally, in the Landau gauge also results on the analytic structures of its propagators are available from different sources, e.g., \cite{Alkofer:2003jj,Fischer:2005en,Krassnigg:2009gd,Strauss:2012dg,Dorkin:2013rsa,Dudal:2013yva,Cornwall:2013zra,Windisch:2016iud,Lowdon:2017gpp,Lowdon:2017uqe,Lowdon:2018mbn,Cyrol:2018xeq,Kaptari:2018vkk,Lowdon:2018uzf,Kaptari:2018vkk,Dudal:2019gvn}.
As far as the Yang-Mills propagators are concerned, the only direct functional calculation for complex momenta up to now was done in this gauge \cite{Strauss:2012dg}.
These results were used for solving bound state equations for the scalar and pseudoscalar glueballs \cite{Sanchis-Alepuz:2015hma}.
For further calculations of glueballs using bound state equations see Refs.~\cite{Meyers:2012ka,Souza:2019ylx,Kaptari:2020qlt,Huber:2020ngt}.
The direct calculation of the analytic structures of propagators is possible with functional methods.
However, there are technical obstacles related to calculating dressing functions for complex momenta \cite{Windisch:2013dxa,Pawlowski:2015mia}. 
To overcome them, various techniques were developed \cite{Maris:1995ns,Fischer:2008sp,Fischer:2005en,Krassnigg:2009gd,Windisch:2012zd,Windisch:2012sz,Windisch:2013dxa,Dorkin:2013rsa,Windisch:2013mg}, but they have not been applied to state-of-the-art truncations yet.
More details are given in Sec.~\ref{sec:analytic_structure}.

Since other gauges are investigated even less than the Landau gauge, one can consider the situation in the Landau gauge as the state of the art of gauge fixed nonperturbative studies of \gls{qcd}.
And despite the plethora of studies using the Landau gauge there is still much work to be done.
There are some conceptual questions that need to be answered, for example, what role Gribov copies play and how different ways to fix the Landau gauge on the lattice are mirrored in functional equations.
In recent years, studies with functional methods have made some big steps forward and provided interesting results on the question of the impact of truncations.
Yet, when it comes to studies of the phase diagram, most results still rely on input from lattice results.
Obtaining reliable results at nonvanishing density requires to repeat calculations in a more complicated setting and possibly extend the truncation setup.
Such comprehensive studies should be the goal for the coming years.

\glsreset{dse}

\subsection{Correlation functions from Dyson--Schwinger equations}
\label{sec:YM4d}

\index{Yang-Mills theory}
Yang-Mills theory, being the limit of \gls{qcd} with infinitely heavy quarks and describing the gluonic part of \gls{qcd}, has been studied intensively with functional methods.
For quite some time, it was assumed that the gluonic part of \gls{qcd} constitutes the main obstacle to use functional methods for nonperturbative studies of \gls{qcd}.
The reason was that for a long time no quantitative results could be obtained and even the qualitative behavior in the \gls{ir} was debated.
The latter issue seems to be resolved nowadays, and the former is most likely close to a resolution.
Results from the \gls{frg} give good reasons to believe that \cite{Cyrol:2016tym} and results from equations of motion presented here and in Ref.~\cite{Huber:2020keu} also point in this direction.
In this context it should be noted, though, that the matter sector of \gls{qcd} turned out not to be as simple as expected for a long time.
Recent studies indicate that in order to desert the successfully employed modeling in the quark sector and proceed to a self-contained solution, some unexpected subtleties still have to be resolved, see, e.g., \cite{Alkofer:2008tt,Hopfer:2013np,Windisch:2014lce,Hopfer:2014szm,Williams:2014iea,Mitter:2014wpa,Aguilar:2014lha,Williams:2015cvx,Blum:2016fib,Binosi:2016wcx,Aguilar:2016lbe,Cyrol:2017ewj,Aguilar:2018epe,Aguilar:2019uob}.

\index{resummation}
\index{two-loop diagrams}
\index{quadratic divergences}
In this section several aspects important for a self-contained description from \glspl{dse} are discussed.
Some of them have been an obvious item on the to-do list of \gls{dse} practitioners, for example, the inclusion of \emph{two-loop diagrams in the gluon propagator}.
Others, like the issue of \emph{quadratic divergences} or the correct \emph{perturbative \gls{rg} resummation}, are only becoming important now, as their impact is only noticeable in new truncations.
Secs.~\ref{sec:spurDivs}, \ref{sec:resummation} and \ref{sec:twoLoop} treat these points.
The final section, Sec.~\ref{sec:res_verts}, contains results for \emph{three- and four-point functions}.
In particular, the study of non-primitively divergent correlation functions in Sec.~\ref{sec:res_AAcbc} is of interest with regard to the importance of higher correlation functions in functional equations.

\subsubsection{Spurious divergences}
\label{sec:spurDivs}

\index{spurious divergences}
\index{quadratic divergences}
\gls{qcd} is a renormalizable quantum field theory.
Employing the usual power counting, the superficial degree of divergence of correlation functions can be determined \cite{Pascual:1984,Ryder:1996qf}.
The ghost and gluon propagators turn out to have a superficial degree of divergence of two (quadratic).
All other correlation functions have a lower degree.
However, the ghost two-point function must be proportional to the external momentum $p^2$.
Thus, any calculation automatically leads to a separation of this factor and the effective degree of divergence is zero (logarithmic) \cite{Pascual:1984}.
The gluon two-point function, on the other hand, has two tensors, $g_{\mu\nu}$ and $p_\mu p_\nu$.
For the latter, the same argument as for the ghost propagator applies, but for the former a quadratic divergence remains.
Here, gauge symmetry comes to the rescue.
It entails via an \gls{sti} that the longitudinal part of the gluon propagator does not receive any quantum corrections and equals the tree-level.
Only the transverse part of the propagator gets dressed where one can pull out a factor $p^2 g_{\mu\nu}-p_\mu p_\nu$.
And again the effective degree of divergence gets reduced to zero.
Also for three-point functions the effective degrees of divergence are zero.

\index{quadratic divergences}
\index{dimensional regularization}
\index{cutoff regularization}
So in \gls{qcd} all divergences are logarithmic.
However, when calculating the gluon propagator explicitly, it turns out that it has quadratic divergences when a hard \gls{uv} cutoff is employed.
This also happens for other regularizations.
Their common property is that they break gauge covariance.
A notable exception to this is dimensional regularization \cite{'tHooft:1972fi,Bollini:1972ui}, which is widely employed for perturbative calculations.
Unfortunately, its applicability to numeric calculations is difficult at best.
It was tested for logarithmic divergences in \gls{qed} \cite{Schreiber:1998ht,Gusynin:1998se} and for the Lippmann-Schwinger equation \cite{Phillips:1999bf}.
However, the regularization of power law divergences has not been attempted yet.
In Ref.~\cite{Phillips:1999bf} it was noted that the employed technique relies on being able to directly take the limit $\epsilon=(4-d)/2 \rightarrow 0$.
Power law divergences appear as poles in $\epsilon\neq0$ and the techniques must then be adapted accordingly.

Since dimensional regularization can (currently) not be applied to deal with the quadratic divergences of the gluon propagator, one has to find alternative ways.
As it is numerically most convenient, one typically uses an $O(4)$ invariant Euclidean cutoff $\Lambda$.
In analytic calculations, one can then discard the part of the result that contains the quadratic cutoff dependence.
In numeric calculations, this cannot be done directly.
In addition, the question of mixing between high and low momentum regimes arises, as we will see below.

\index{Brown-Pennington projector}
Based on the observation that the gluon propagator corrections must be proportional to $p^2 g_{\mu\nu}-p_\mu p_\nu$ and the individual parts do not scale differently, Brown and Pennington argued that it would be sufficient to project onto $p_\mu p_\nu$, since the quadratic divergences only reside in the part proportional to $g_{\mu\nu}$ \cite{Brown:1988bn}.
This can be achieved with what has become known as the Brown-Pennington projector.
It is a special case of the generalized projector
\begin{align}
 P^{\zeta}_{\mu\nu}(p)=g_{\mu\nu}-\zeta\frac{p_\mu p_\nu}{p^2}. \label{eq:Pzeta}
\end{align}
Acting on the two parts of the transverse tensor with this projector yields
\begin{align}
 P^\zeta_{\mu\nu}(p)g_{\mu\nu}&=d-\zeta,\\
 P^\zeta_{\mu\nu}(p)p_\mu p_\nu&=1-\zeta.
\end{align}
The Brown-Pennington projector corresponds to $P^{d}_{\mu\nu}(p)$ for which the first projection vanishes.
While the quadratic divergences can be discarded with this projector, some new problems arise \cite{Lerche:2002ep,Huber:2012zj}.
First of all, when acting with $P^\zeta_{\mu\nu}(p)$ on the right-hand side of the gluon propagator \gls{dse}, only $\zeta=1$, which corresponds to the transverse projector $P_{\mu\nu}$, ensures the absence of the longitudinal parts of the vertices.
This would not be a problem if the longitudinal parts were known.
However, we want to avoid calculating them.
Furthermore, a projection with $\zeta\neq1$ introduces a spurious parameter dependence \cite{Lerche:2002ep}.
To see this, the ghost-gluon vertex is written as (neglecting a part proportional to the gluon momentum for now)
\begin{align}
 \Gamma^{A\bar{c}c,abc}_\mu(k;p,q)=i\,g\,f^{abc} \left( \eta \,p_\mu - \hat{\eta}\, q_\mu\right).
\end{align}
$\eta$ and $\hat \eta$ are parameters related by $\eta+\hat\eta=1$.
A possible dressing of the vertex is taken to be one here. 
This approximation is sufficient in the \gls{uv} and the arguments below would not be changed by a more general dressing.
Th employed parametrization reflects that one can always shift contributions from the ghost to the anti-ghost legs due to the transversality of the Landau gauge:
\begin{align}
 P_{\mu\nu}(p+q)p_\mu=-P_{\mu\nu}(p+q)q_\mu.
\end{align}
Plugging this vertex into the gluon propagator \gls{dse} and projecting with $P^\zeta_{\mu\nu}(p)$, the kernel of the ghost loop becomes with $x=p^2$ and $y=q^2$ \cite{Huber:2014tva}
\begin{align}
  K_{Z}^{gh,\zeta}(p,q)&=\frac{x^2 
\left(\zeta - 2 - 4\, \eta\hat\eta \, (\zeta-1)  \right)
+2 x (y+z)-\zeta  (y-z)^2}{12 x^2 y z}.
\end{align}
Appendix~\ref{sec:app_kernels_LG} contains the full expressions for the \gls{dse}.
One can see that only for $\zeta=1$ the dependence on $\eta$ vanishes and for $\zeta\neq1$ an undesired dependence on $\eta$ remains.
Hence, only $\zeta=1$ will be used in this work for calculations.
Nevertheless, it will be kept explicitly in some analytic expressions.

The divergences in the gluon propagator \gls{dse} can be identified by analyzing the \gls{uv} behavior of the equation.
To this end, one approximates all dressing functions by their \gls{uv} expressions and discards the external momentum in their arguments, since we are interested in the case $p<q$.
Then, all angle integrals can be performed.
For the one-loop diagrams one obtains
\begin{align}\label{eq:Z_UV}
 \frac{1}{Z_\text{UV}(p^2)}&= Z_3 - Z_4\frac{N_c\,g^2}{64\pi^2}\frac{1}{x}\,\int_x^{\Lambda^2} dy \left(3 (\zeta -4)\right) Z_\text{UV}(y)\nnnl
& +\widetilde{Z}_1\frac{N_c\,g^2}{192\pi^2}\,\int_x^{\Lambda^2} dy \frac{x
   \left(\zeta-2-4 \, \eta \hat\eta \, (\zeta-1)\right)-(\zeta -4) y}{x y} G_\text{UV}(y)^2\nnnl
& +Z_1\frac{N_c g^2}{384 \pi^2}\int_x^{\Lambda^2} dy \frac{7 x^2+12  (-4+\zeta ) y^2-2 x y (24+\zeta ) }{ x y^2}C^{AAA}_\text{UV}(y) Z_\text{UV}(y)^{2}.
\end{align}
$Z_\text{UV}$ and $G_\text{UV}$ are the dressing functions $Z$ and $G$ of the gluon and ghost propagators, respectively, in the \gls{uv}.
The dressing function of the three-gluon vertex in the \gls{uv} is denoted by $C^{AAA}_\text{UV}$.
The renormalization constants that appear are $Z_3^{1/2}$ for the gluon field, $\widetilde{Z}_3^{1/2}$ for the ghost field, $Z_1$ for the three-gluon vertex, $\widetilde{Z}_1$ for the ghost-gluon vertex and $Z_4$ for the four-gluon vertex.
The quadratically divergent part is proportional to $\zeta-4$.

To remove this part from the gluon propagator equation, various methods have been used in the past.
Besides the Brown-Pennington projector, also modifications of the integrands \cite{Fischer:2002eq,Fischer:2002hn,Sampaio:2005pc,Maas:2005hs,Cucchieri:2007ta,Ferreira:2011cv,Huber:2012kd}, modifications of the vertices \cite{Fischer:2008uz}, fitting the momentum \cite{Fischer:2005en,Fischer:2014ata} or the cutoff dependence \cite{Huber:2016tvc}, or additional counterterms \cite{LlanesEstrada:2012my,Meyers:2014iwa,Huber:2012zj,Huber:2017txg,Huber:2020keu} were used.
In the combination of the pinch technique with the background field method, so-called seagull identities get rid of the quadratic divergences \cite{Aguilar:2009ke,Aguilar:2016vin}.
Also in the functional renormalization group spurious divergences are introduced by the regulator.
In practical calculations, it is overcome by fine-tuning the initial value of the gluon mass at the initial cutoff scale \cite{Cyrol:2016tym}.
For a more detailed overview see Ref.~\cite{Huber:2014tva}.

For simple truncations, viz., using models for the vertices and only one-loop diagrams in the gluon propagator \gls{dse}, the subtraction terms can be calculated analytically \cite{Huber:2014tva}.
In the \gls{uv}, all dressings $Z_i(p^2)$ take their perturbative form.
Here, we take for this the one-loop resummed expression discussed in detail in Sec.~\ref{sec:resummation}:
\begin{align}\label{eq:dressings_UV_gen}
 D_i(s)\left(1+\omega \ln\frac{p^2}{s}\right)^{\gamma_i}=D_i(s)(\omega\, t_{p})^{\gamma_i}.
\end{align}
$\gamma_i$ is the corresponding anomalous dimension and $t_{p}=\ln\left(p^2/\LQ\right)$ with $\Lambda_\text{QCD}=s\,e^{-1/\omega}$.
$s$ is any perturbative scale and $\omega=\omega(s)$ is given by $11\,N_c\,\alpha(s)/12/\pi=\beta_0 g^2(s)$ with $\alpha(s)$ the coupling.
One can then perform the quadratically divergent integral in \eref{eq:Z_UV}:
\begin{align}\label{eq:spurDivIntegral}
\int_{x_1}^{\LL} dy \left(\omega\,
t_{y}\right)^{\gamma}= I(\LL,\gamma)-I(x_1,\gamma) =\LQ (-\omega)^{\gamma}\, \big( \Gamma\left(1+\gamma, -t_{\Lambda}\right) - \Gamma\left(1+\gamma,-t_{1}\right)\big).
\end{align}
$\Gamma$ is the incomplete Gamma function.
Its arguments are
\begin{align}
t_\Lambda=\ln\left(\LL/\LQ\right), \qquad t_{1}=\ln \left(x_1/\LQ\right),
\end{align}
where $x_1$ is the lower boundary of the integral.
$x_1$ can be chosen equal to $\LQ$.
This corresponds to the Landau pole of the employed parametrization of the perturbative dressings.
Using an alternative parametrization, one can show that if $x_1$ is chosen as the lowest possible value, the results for both parametrizations agree \cite{Huber:2014tva}.
For this choice of $x_1$, the second term in \eref{eq:spurDivIntegral} only cancels the imaginary part of the first part and leaves the real part untouched.
Rewriting the incomplete Gamma functions into a convenient representation, the final subtraction coefficient is
\begin{align}\label{eq:C_sub}
C_\mathrm{sub}&:=\LQ b\,\omega^{2\delta}\sum_{n=0}^{\infty}\frac{(t_{\Lambda})^{1+2\delta+n}}{n!(1+2\delta+n)}.
\end{align}
$\delta$ is the anomalous dimension of the ghost propagator and $b$ is a coefficient that depends on the details of the vertex models.
It should be noted that the result is not proportional to $\LL$ but to $\LQ$.
The dependence on the cutoff is more intricate and completely encoded in the incomplete Gamma functions.
This is a direct consequence of resummation in four-dimensional Yang-Mills theory.
If we had used the leading tree-level contributions of the dressings, the result would be proportional to $\LL$.

\begin{figure}[tb]
 \includegraphics[width=0.49\textwidth]{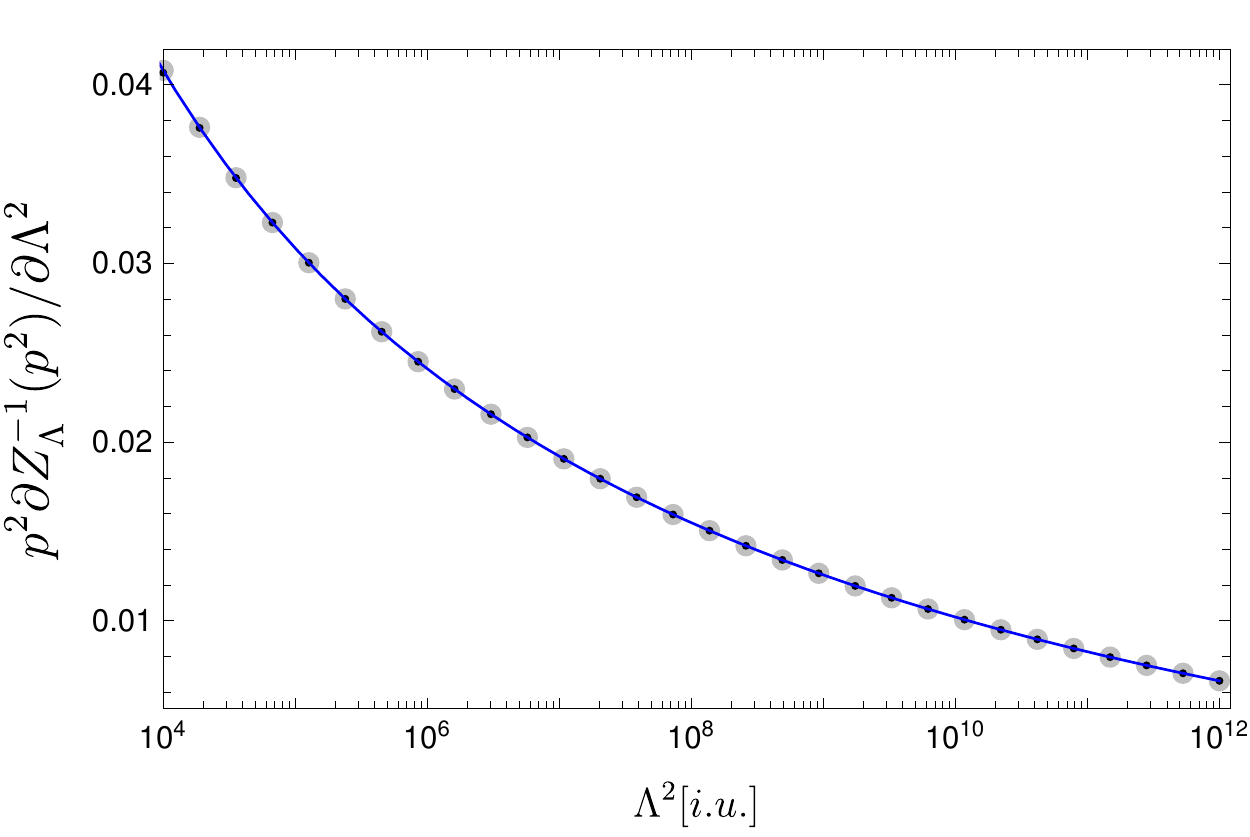}
 \includegraphics[width=0.49\textwidth]{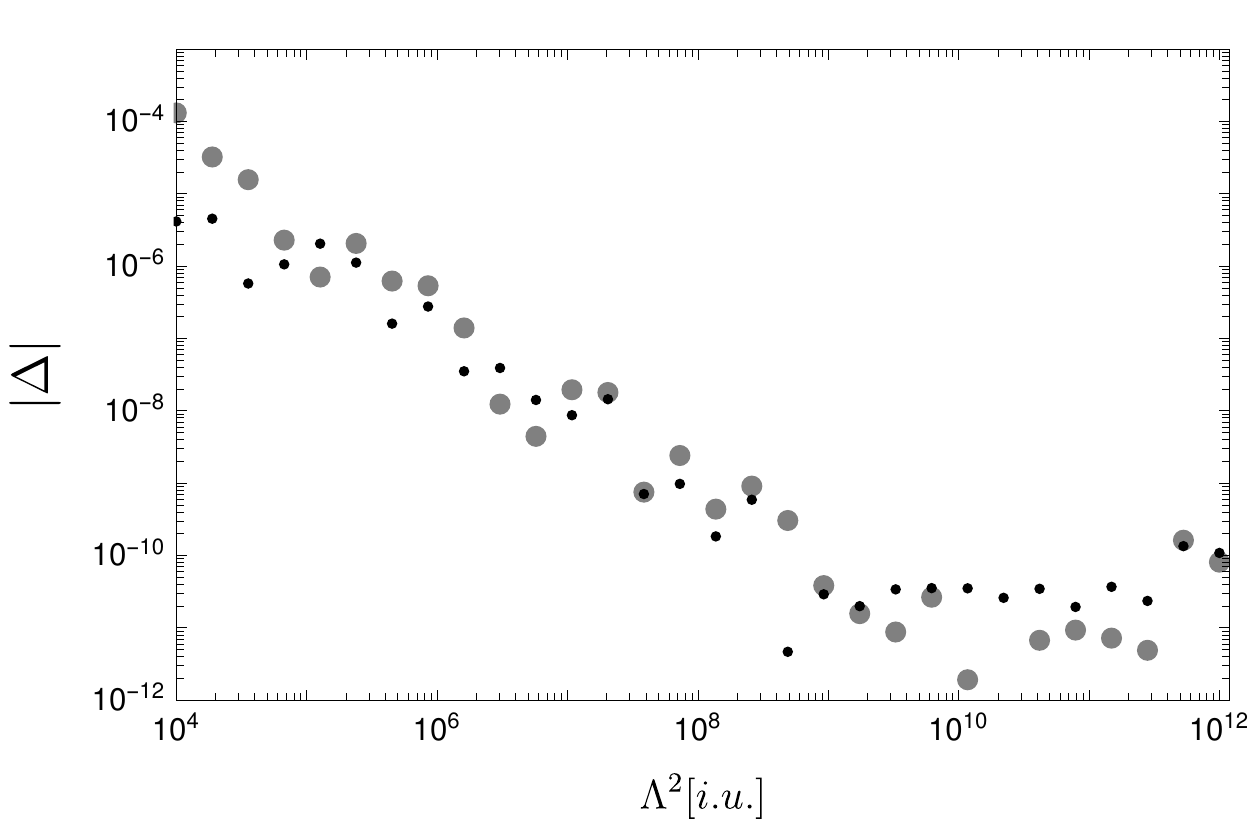}
 \caption{
 Left: Cutoff dependence of the full solution $Z_\Lambda(p^2)$ \cite{Huber:2014tva}.
 The big gray (small black) dots represent results for the largest (smallest) external momentum used in the calculations.
 The solid line shows the corresponding analytic expression.
 The cutoff $\Lambda$ is given in arbitrary internal units.
 Right: Absolute values of differences $\Delta$ in the left plot between smallest and largest momentum results (thick, gray dots), as well as numerical (smallest momentum) result and analytic
 expression (thin, black dots).} 
 \label{fig:spurDivCutoffDep}
\end{figure}

With this subtraction term, the quadratically divergent part can be subtracted from the gluon self-energy by adding $-C_\text{sub}/p^2$.
Assuming that all dressings behave like in \eref{eq:dressings_UV_gen}, $b$ reads for the ghost and gluon loops of \eref{eq:Z_UV}
\begin{align}
b = \frac{N_c g^2}{64\pi^2}\left(G(s)^2-6C^{AAA}(s)Z(s)^{2}\right).
\end{align}
Note that \gls{rg} improvement terms as discussed in Sec.~\ref{sec:resummation_oneLoop} are not taken into account in this expression and still must be added appropriately.
To test the precision of this method, the derivative with respect to the cutoff can be calculated which must agree with the original integrand.
For an example calculation, this is shown in \fref{fig:spurDivCutoffDep}.
It should be noted that a precise determination of the coefficient $b$ is important and that already tiny changes have an effect, see, for example, \fref{fig:YM3d_varCsub} below.

Unfortunately, this method of calculating the subtraction term is not feasible when dynamic vertices or two-loop diagrams are included.
In that case, a convenient method is to introduce a mass counterterm for the gluon \cite{Meyers:2014iwa}.
Although a mass term for the gluon is forbidden by gauge invariance of the Yang-Mills action, a counterterm can be introduced once a gauge is fixed \cite{Collins:2008re}.
One can then explicitly calculate this counterterm with a second renormalization condition.
In practical calculations it is easiest to fix the gluon propagator at a low momentum value.
Writing the gluon propagator \gls{dse} as
\begin{align}
 Z^{-1}(x)=Z_3+\Sigma(x)-\frac{C_\text{sub}}{x},
\end{align}
one can determine the renormalization constant $Z_3$ by choosing a fixed value for $Z(x_s)$:
\begin{align}
 Z_3 = Z^{-1}(x_s)-\Sigma(x_s)+\frac{C_\text{sub}}{x_s}.
\end{align}
$C_\text{sub}$ is determined by demanding that the propagator $D(x)=Z(x)/x$ has a fixed value at $x_m$:
\begin{align}
 C_\text{sub}=\frac{x_m x_s}{x_s-x_m}\left(\Sigma(x_m)-\Sigma(x_s)\right)+\frac{x_m x_s}{x_s-x_m} Z^{-1}(x_s)-\frac{ x_s}{x_s-x_m}D^{-1}(x_m).
\end{align}
This method is employed for the two-loop calculation in Sec.~\ref{sec:twoLoop}.
It should be noted that results can depend on this counterterm.
However, for a large enough truncation, this dependence vanishes and the original number of relevant counterterms is restored \cite{Huber:2020keu}.

\subsubsection{Resummed perturbative behavior}
\label{sec:resummation}

\myboxmargin{perturbative behavior of DSEs}
\index{perturbation theory}
In many studies of the \gls{qcd} propagators using \glspl{dse} from the last 20 years, the one-loop resummed perturbative behavior is recovered.
This means that the dressing functions behave at high momenta like a logarithm with a power known as anomalous dimension.
The usual way to extract this behavior is solving the corresponding renormalization group equations which are based on the invariance of bare correlation functions under a change of the renormalization scale.
A one-loop \gls{rg} calculation leads to this anomalous running, whereas direct perturbative one-loop calculations yield only pure logarithms.
However, the corresponding coefficients contain the anomalous dimensions.
This is related to a general structure in perturbation theory that allows to predict certain aspects already from lower orders \cite{Collins:2008re}.

\myboxmargin{renormalization group}
\index{renormalization group}
Unfortunately, this structure can be violated by truncations of \glspl{dse}.
For example, it is in general not possible to solve a \gls{dse} self-consistently to a fixed order in perturbation theory.
Self-consistency means here that the expression on the left-hand side of the equation is the same as on the right-hand side.
A perturbative calculation lacks this consistency, as the right-hand side is calculated with a fixed input and the result is assigned to the left-hand side.
A \gls{dse}, when solved self-consistently, has contributions up to infinite order.
In the simplest case, a solution can be obtained by iteration.
In this case, the left-hand side is repeatedly inserted on the right-hand side until it does not change any more.
In every iteration step, a subset of higher order perturbative contributions is included.
The best we can say perturbatively about the solution of a \gls{dse} is that it contains all contributions up to a certain order and some contributions beyond that.

\paragraph{Resummation in Dyson--Schwinger equations}
\label{sec:resummationInDSEs}

\myboxmargin{Resummed behavior in DSEs}
\index{resummed perturbation theory}
\index{perturbation theory}
So how can the anomalous running be obtained from a \gls{dse}?
As the name resummed perturbative behavior suggests, its origin is in the resummation of certain diagrams.
The renormalization group is a natural way to realize this resummation, but without exposing the individual contributions beyond one-loop.
This is due to the aforementioned structure of perturbation theory.
To perform resummation in \glspl{dse}, it is necessary to actually calculate these diagrams.
The resulting series in the coupling constant can partially be resummed.
To see how this works it is easiest to start from the other end.
The one-loop resummed expression for a propagator dressing function reads
\begin{align}\label{eq:dressings_UV}
  Z(p^2)=Z(\mu^2)\left(1+ g^2 \beta_0\, \ln{\frac{ p^2}{\mu^2}}\right)^{\gamma}.
\end{align}
The anomalous dimension is denoted by $\gamma$.
$\mu$ is the renormalization scale and $\beta_0$ the first coefficient of the $\beta$-function.
A series expansion yields
\begin{align}
  \left(1+g^2 \beta_0\, \ln{\frac{ p^2}{\mu^2}}\right)^{\gamma} = 1+\gamma\beta_0 g^2 \ln{\frac{ p^2}{\mu^2}} +\frac{1}{2}\gamma(\gamma-1)\left(\beta_0 g^2\right)^2 \ln^2{\frac{ p^2}{\mu^2}} + \mathcal{O}(g^6).
\end{align}
Thus, the resummation contains all terms which contain a logarithm of the same power as $g^2$.

\begin{figure}[tb]
 \begin{center}
  \includegraphics[height=2cm]{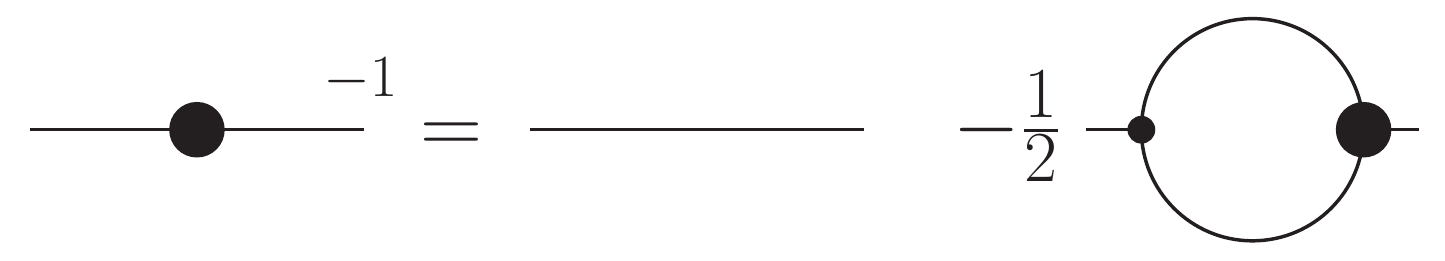}
  \\
  \includegraphics[height=2.5cm]{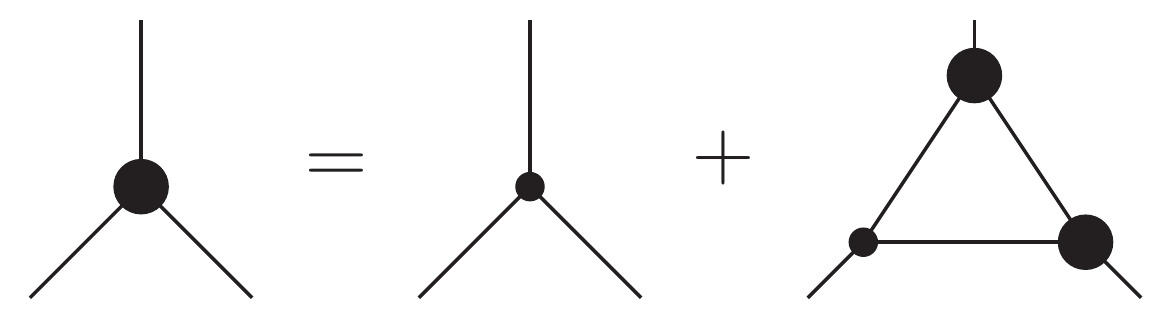}
 \end{center}
 \caption{The \glspl{dse} of the propagator and the vertex of $\varphi^3$ theory.
 Small black dots represent bare vertices with a renormalization constant $Z_1$.}
 \label{fig:phi3_DSEs}
\end{figure}

\begin{figure}[tb]
 \begin{center}
  \includegraphics[width=\textwidth]{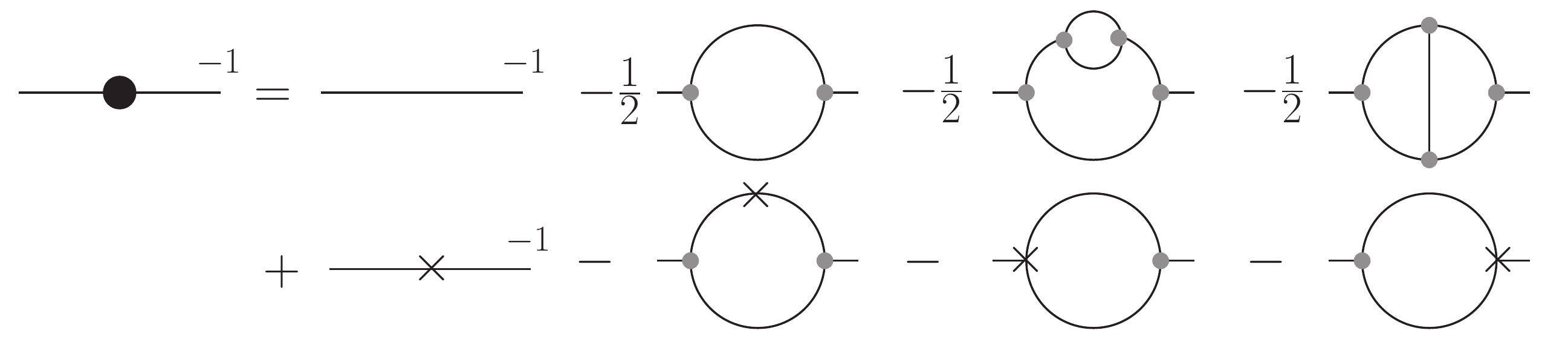}
 \end{center}
 \caption{The propagator of $\varphi^3$ theory up to two-loop level.
 The counterterms, denoted by crosses, are explicitly shown in the second row.
 Small gray dots denote bare vertices without a renormalization constant.}
 \label{fig:phi3_prop_pert}
\end{figure}

\myboxmargin{example: resummation in phi cubed theory}
\index{$\varphi^3$ theory}
To illustrate how such terms are generated, we consider as an example a scalar theory in six dimensions with a cubic interaction:
\begin{align}
 \mathcal{L}=\varphi(-\partial^2+m^2)\varphi +\frac{g}{3!}\varphi^3.
\end{align}
The propagator \gls{dse} is shown in \fref{fig:phi3_DSEs}.
The self-energy has one diagram at the perturbative one-loop level.
At two-loop level, there are two diagrams, one from the one-loop vertex correction and one from the one-loop correction of the internal propagators.
The corresponding diagrams are depicted in \fref{fig:phi3_prop_pert}.

\index{dimensional regularization}
\index{self-energy}
\myboxmargin{calculation of $\varphi^3$ one-loop diagram}
For simplicity, the following calculations are done in dimensional regularization considering $d=6-2\epsilon$ dimensions and with $m=0$.
The calculation of the one-loop diagram yields \cite{Muta:1998}
\begin{align}\label{eq:phi3_prop_ol}
 \Pi^{(1)}(p^2,\mu^2)&=g^2 \left(\frac{\mu^2}{p^2}\right)^{3-\frac{d}{2}}p^2\frac{\Gamma\left(2-\frac{d}{2}\right)\Gamma\left(\frac{d}{2}-1\right)^2}{2(4\pi)^{\frac{d}{2}}\Gamma(d-2)}\nonumber\\
 &=\lambdat \frac{p^2}{12} \Bigg( \frac{1}{\eps} - \frac{8}{3} -\gamma_E -\ln\left(\frac{p^2}{4\pi \mu^2}\right) \nnnl
 &\quad+\frac{\eps}{36} \left( 18 \ln \left(\frac{p^2}{ \mu^2} \right) + \ldots\right)  + \mathcal{O}(\epsilon^2) \Bigg).
\end{align}
Here and in the following, the numbers in parentheses in the superscript refer to the loop order.
$\gamma_E$ is the Euler-Mascheroni constant.
The terms at order $\epsilon$ denoted by the dots will not be relevant in the following.
We will also need the one-loop result for the vertex up to order $\eps^{-1}$:
\begin{align}
 \Gamma^{(1)}(p^2,q^2,k^2) = \lambdat \frac{1}{2\epsilon} + \mathcal{O}(\eps^0).
\end{align}
Using the \gls{ms} scheme to fix the finite parts, the counterterms at one-loop level are:
\begin{align}
 \Pi ^{(1),\text{ct}}&= -\lambdat \frac{1}{12\eps},\\
 \Gamma ^{(1),\text{ct}}&= -\lambdat \frac{1}{2\eps}.
\end{align}

\myboxmargin{calculation of $\varphi^3$ two-loop diagrams}
\index{two-loop diagrams}
The two two-loop diagrams yield \cite{Muta:1998}
\begin{align}
 \Pi^{(2)}_a(p^2,\mu^2)&=\left(\lambdat\right)^2 \frac{p^2}{144} \Bigg( -\frac{1}{\eps^2} + \frac{1}{6\eps}\left( 12 \gamma_E-43 +12 \ln \left(\frac{p^2}{4\pi \mu^2}\right) \right)\nnnl
 &\quad- \frac{1}{36}\left(72 \ln^2\left(\frac{p^2}{\mu^2}\right) + \ldots \right) + \mathcal{O}(\eps) \Bigg),\\
 \Pi^{(2)}_b(p^2,\mu^2)&=  \left(\lambdat\right)^2 \frac{p^2}{24} \Bigg( \frac{1}{\eps^2} - \frac{2}{\eps}\left( \gamma_E-3 +  \ln \left(\frac{p^2}{4\pi \mu^2}\right) \right)\nnnl
 &\quad + \frac{1}{6}\left(12 \ln^2\left(\frac{p^2}{ \mu^2}\right) + \ldots \right) + \mathcal{O}(\eps) \Bigg).
\end{align}
$\Pi^{(2)}_a(p^2,\mu^2)$ is the propagator correction and $\Pi^{(2)}_b(p^2,\mu^2)$ the vertex correction depicted by the third and fourth diagrams, respectively, in the right-hand side in \fref{fig:phi3_prop_pert}.
As expected, terms proportional to $\ln^2 p^2/\mu^2$ appear in the finite parts.
They will contribute at order $g^4$ to the resummation.

\index{overlapping divergences}
Renormalization beyond one-loop order requires that subdivergences are subtracted before the overall divergences are subtracted.
At two-loop level, this is achieved by adding diagrams with appropriate counterterm insertions as shown in the second row of \fref{fig:phi3_prop_pert}.
For \glspl{dse}, the counterterms enter automatically at the right places and they do not need to be added explicitly, since one is working with renormalized quantities only.
In this way, also overlapping divergences are automatically dealt with correctly in \glspl{dse}.
As an example, consider the overlapping divergence arising from the second perturbative two-loop diagram shown in \fref{fig:phi3_prop_pert}.
It corresponds to a correction of the vertex and is also known as kite diagram.
It contains two subdivergences which overlap and require separate counterterms.
These are shown as the last two diagrams in \fref{fig:phi3_prop_pert}.
In the propagator \gls{dse}, \fref{fig:phi3_DSEs}, these counterterms enter differently:
One is automatically included directly as the bare vertex comes with the renormalization constant of the vertex which is $Z_1=1-\Gamma^{\text{ct}}$.
The second one is contained in the dressed vertex.
Its \gls{dse} contains as lowest order term also the renormalization constant $Z_1$ and thus the appropriate counterterm.

\index{counterterms}
\myboxmargin{order $g^4$ terms from counterterms}
The one-loop diagrams containing the one-loop counterterms are of order $g^4$.
The required integrals can be calculated from the expressions we already encountered, since they correspond to the one-loop diagram with additional momentum independent factors:
\begin{align}
 \Pi^{(2),\text{ct}}_a&=-2 \Pi^{(1),\text{ct}} \Pi^{(1)}\\
 \Pi^{(2),\text{ct}}_b&=2 \Gamma^{(1),\text{ct}} \Pi^{(1)}.
\end{align}
The minus sign for $\Pi^{(2),\text{ct}}_a$ stems from the fact that the counter term for the propagator at one-loop order is $-\Pi^{(1),\text{ct}}$, because it is the inverse of the two-point function.
The one-loop counterterms in the \gls{ms} scheme contain only the terms of order $\epsilon^{-1}$.
Thus, since we are here interested in the finite parts, we need the one-loop self-energy terms of order $\epsilon$.
The resulting expressions contain the following squared logarithms:
\begin{align}
 \Pi^{(2),\text{ct}}_a&=\left(\lambdat\right)^2 \frac{p^2}{144} \ln^2\left(\frac{p^2}{\mu^2}\right) + \ldots\\
 \Pi^{(2),\text{ct}}_b&=-\left(\lambdat\right)^2 \frac{p^2}{24} \ln^2\left(\frac{p^2}{\mu^2}\right) + \ldots.
\end{align}
Note that the disappearance of subdivergences, which is required for renormalizability, is reflected in the fact that all terms containing $\eps^{-1}\ln p^2/\mu^2$ cancel out in the sum of all expressions.

\index{resummation}
\myboxmargin{resumming all contributions}
Finally, if we sum all relevant terms from \fref{fig:phi3_prop_pert} and drop a factor of $p^2$, we obtain
\begin{align}\label{eq:phi3_prop_pert_resum}
1-&\frac{1}{12}\lambdat \ln\left(\frac{p^2}{\mu^2}\right)  + \left(\lambdat\right)^2 \left( -\frac{1}{72} +\frac{1}{12} + \frac{1}{144} - \frac{1}{24} \right) \ln^2\left(\frac{p^2}{\mu^2}\right)=\nnnl
1-&\frac{1}{9}\frac{3}{4}\lambdat \ln\left(\frac{p^2}{\mu^2}\right) + \left(\lambdat\right)^2 \frac{1}{2}\left(-\frac{1}{9}\right) \left( -\frac{1}{9}-1 \right) \left(\frac{3}{4}\right)^2 \ln^2\left(\frac{p^2}{\mu^2}\right)=\nnnl
1-&\gamma \beta_0 \lambdat \ln\left(\frac{p^2}{\mu^2}\right) + \frac{1}{2}(-\gamma)(-\gamma-1)\left(\lambdat\right)^2 \beta_0^2 \ln^2\left(\frac{p^2}{\mu^2}\right).
\end{align}
The prefactors were rearranged and expressed in terms of $\beta_0=3/4/(2\pi)^3$ and $\gamma=1/9$.
Eq.~(\ref{eq:phi3_prop_pert_resum}) corresponds to the first terms of the Taylor series of
\begin{align}\label{eq:phi3_resummed}
 \left(1+\beta_0 g^2 \ln\left(\frac{p^2}{\mu^2}\right)\right)^{-\gamma}.
\end{align}
This, however, is the one-loop solution to the renormalization group equation for the two-point function,
\begin{align}
 \left( \mu \frac{\partial}{\partial \mu} + \beta(g)\frac{\partial}{\partial g}-2 \gamma(g)\right) \Gamma_2(p,\mu,g)=0.
\end{align}
Thus, \eref{eq:phi3_resummed} corresponds to the one-loop resummed expression for the inverse propagator.

\myboxmargin{resum diagrams in DSEs}
\index{resummation}
The resummed expressions thus contains all one-loop contributions and a subset of the higher terms.
To realize the resummation in a \gls{dse} calculation, all contributions required for each order must be calculated.
In particular, the renormalization constants have to be included correctly.
In a one-loop truncation of the \gls{qcd} propagator \glspl{dse}, already many terms of order $g^4 \ln^2(p^2/\mu^2)$ and higher are contained via the nonperturbative propagators and possibly even the employed vertex models.
However, in the gluon propagator \gls{dse}, the two-loop diagrams are required as they contain a sizable contribution to the second order.
Higher terms can be included systematically, although this has been done only up to two-loop level up to now \cite{Huber:2017txg}.
It turns out that this is sufficient and approximates the one-loop resummed behavior very well as discussed in Sec.~\ref{sec:twoLoop}.
In particular, \fref{fig:YM4d_props_tl} in that section shows that the \gls{uv} behavior of the propagator dressing functions recovers the pure perturbative one-loop resummed behavior quite well.

\paragraph{RG improvement in propagator equations: Yang-Mills propagators and other examples}
\label{sec:resummation_oneLoop}

\index{ghost propagator}
\index{gluon propagator}
In the remainder of this section it is discussed how the one-loop resummed behavior can be obtained in one-loop truncations.
Most truncations do not contain two-loop terms for reasons of technical complexity.
However, having the correct one-loop resummed perturbative behavior is advantageous and can be obtained by modifying the equations.
This is done in many calculations and can be considered as part of the employed modeling of the missing parts.
The modifications are derived by enforcing the self-consistency of a \gls{dse} in the perturbative regime.
One possiblity to realize this is to interpolate between a given vertex model and an expression with the correct \gls{uv} behavior \cite{Pennington:2011xs}.
However, since the necessary modification is part of the modeling anyway, once can include it directly in the vertex models. 
To see how this works, we consider the system of ghost and gluon propagator \glspl{dse} by following \cite{Fischer:2002eq,Fischer:2003zc,Huber:2012kd}.
The dressing functions of the ghost and gluon propagators, $G(p^2)$ and $Z(p^2)$, respectively, can be parametrized as follows at the one-loop resummed level:
\begin{subequations}
\label{eq:GZ-UV}
\begin{align}
 G(p^2)&=G(s)\left(1+\omega \ln\left(\frac{p^2}{s}\right) \right)^\delta,\label{eq:G-UV}\\
 Z(p^2)&=Z(s)\left(1+\omega \ln\left(\frac{p^2}{s}\right) \right)^\gamma.\label{eq:Z-UV}
\end{align}
\end{subequations}
Here, $\omega$ is used as, for now, generic coefficient of the logarithm.
Also the anomalous dimensions $\gamma$ and $\delta$ are not specified yet.
$s$ is a perturbative momentum scale.

\index{ghost propagator}
\index{gluon propagator}
The kernels of the propagator equations are given in Appendix~\ref{sec:app_kernels_LG}.
We are interested in high external momenta $p^2$.
The integrals in the \glspl{dse} are dominated then by the region $p^2<q^2<\Lambda^2$, where $\Lambda$ is a \gls{uv} cutoff.
In this regime, the momentum dependence of all dressing functions is dominated by the loop momentum $q$.
Hence, we approximate all dressing functions by ignoring the external momentum in their arguments, e.g., $G((p+q)^2)$ is replaced by $G(q^2)$, see also above \eref{eq:Z_UV}.
This allows calculating all three angle integrals analytically:
\begin{align}
 \frac{1}{G(p^2)}&\rightarrow \tilde{Z}_3+\widetilde{Z}_1\frac{N_c\,g^2}{64\pi^2}\,\int_x^{\Lambda^2} dy \frac{x-3y}{ y^2} Z(y)G(y)D^{A\bar cc}(y) \\
 \frac{1}{Z(p^2)}&\rightarrow Z_3+\widetilde{Z}_1\frac{N_c\,g^2}{192\pi^2}\,\int_x^{\Lambda^2} dy \frac{-y(\zeta-4)+x(\zeta-2)}{ x\,y} G(y)G(y)D^{A\bar cc}(y)\nnnl
& +Z_1\frac{N_c g^2}{384 \pi^2}\int_x^{\Lambda^2} dy \frac{7 x^2+12 y^2 (-4+\zeta )-2 x y (24+\zeta ) }{ x y^2}Z(y)^{2} C^{AAA}(y) .
\end{align}
Here, $x=p^2$ and $y=q^2$ and $\zeta$ is a parameter from the projector kept for convenience.
For the dressed ghost-gluon and three-gluon vertices only their tree-level structures, which dominate in the \gls{uv}, were taken into account.
They were dressed with the dressing functions $D^{A\bar cc}$ and $C^{AAA}$ (depending only on the dominant scale), respectively, and the coupling $g$ was explicitly split off.
Also the vertex dressings are taken to depend only on the large momentum scale $q$ with a form similar to Eqs.~(\ref{eq:GZ-UV}) with anomalous dimensions $\delta_\text{ghg}$ and $\delta_\text{3g}$.
From counting the powers in $y$ in the integrals, the quadratic and logarithmic parts can be identified.
The quadratic terms are discussed in Sec.~\ref{sec:spurDivs}.
Here we are only interested in the logarithmic behavior and discard the other terms.
Note that perturbatively the $\zeta$ dependence cancels for the logarithmic terms between the ghost and the gluon loops.
Plugging in the perturbative expressions for the dressings, the radial integral can also be performed:
\begin{align}\label{eq:prop_integrals_UV}
 &G(s)^{-1}\left(1+\omega \ln\left(\frac{x}{s}\right) \right)^{-\delta}=\tilde{Z}_3-\widetilde{Z}_1\frac{3N_c\,g^2 D^{A\bar cc}(s)G(s)Z(s)}{64\pi^2(1+\gamma+\delta+\gamma_\text{ghg})\omega}\nnnl
 &\qquad\times\Bigg[\left(1+\omega \ln\left(\frac{\Lambda^2}{s}\right)\right)^{1+\gamma+\delta+\gamma_\text{ghg}}-\left(1+\omega \ln\left(\frac{x}{s}\right)\right)^{1+\gamma+\delta+\gamma_\text{ghg}}\Bigg]  \\
 &Z(s)^{-1}\left(1+\omega \ln\left(\frac{x}{s}\right) \right)^{-\gamma}=Z_3+\widetilde{Z}_1\frac{\left(\zeta-2\right)N_c\,g^2 D^{A\bar cc}(s)G(s)^2}{192\pi^2(1+2\delta+\gamma_\text{ghg})\omega}\nnnl
 &\qquad\times\Bigg[\left(1+\omega \ln\left(\frac{\Lambda^2}{s}\right)\right)^{1+2\delta+\gamma_\text{ghg}}-\left(1+\omega \ln\left(\frac{x}{s}\right)\right)^{1+2\delta+\gamma_\text{ghg}}\Bigg]\\
 &\quad + Z_1\frac{\left(-(24+\zeta)\right)N_c\,g^2 C^{AAA}(s)Z(s)^2}{192\pi^2(1+2\gamma+\gamma_\text{3g})\omega}\nnnl
 &\qquad \times\Bigg[\left(1+\omega \ln\left(\frac{\Lambda^2}{s}\right)\right)^{1+2\gamma+\gamma_\text{3g}}-\left(1+\omega \ln\left(\frac{x}{s}\right)\right)^{1+2\gamma+\gamma_\text{3g}}\Bigg].
\end{align}
The first terms in the brackets contain the cutoff dependence and are canceled by the renormalization constants.
Of interest are the second terms.
Self-consistency requires that the exponents of the logarithms are the same on the left- and right-hand sides.
The dressings at the point $s$ are also discarded for the moment.
Normally one discards these terms by choosing appropriate renormalization conditions \cite{Fischer:2002eq}.
We will see below that this is actually not necessary.

\index{ghost-gluon vertex}
\index{three-gluon vertex}
The ghost-gluon vertex is \gls{uv} finite.
Thus, its anomalous dimension is zero, $\gamma_\text{ghg}=0$, and was only kept for illustration purposes, as was the renormalization constant $\widetilde{Z}_1$, which will be chosen as $1$ in the following.
Matching the exponents in the ghost propagator \gls{dse} yields
\begin{align}
 2\delta+\gamma+1=0,
\end{align}
which is a sum rule known from the perturbative renormalization group.
The ghost-loop in the gluon propagator \gls{dse} leads to the same relation.
If the exponents should match also for the gluon loop, the anomalous dimension of the three-gluon vertex must be $\gamma_\text{3g}=2(\delta-\gamma)$.
Plugging this in and setting $Z_1=1$, there are three equations (the exponents and the two coefficients in the two equations) from which one can determine $\omega$, $\delta$ and $\gamma$ as
\begin{align} 
 \omega&= \frac{11N_c}{3}\frac{g^2}{(4\pi)^2}=g^2\beta_0,\\
 \de &=-\frac{9}{44},\\
 \gamma&=-\frac{13}{22}.
\end{align}
The first coefficient of the Yang-Mills $\beta$ function, $\beta_0=11N_c/3/(4\pi)^2$ was used to highlight that $\omega$ corresponds to the expression used in \eref{eq:dressings_UV}.

\index{three-gluon vertex}
Two things need to be emphasized with regard to the preceding calculation.
First, the anomalous dimension of the three-gluon vertex is not $2(\delta-\gamma)$.
Second, taking into account also the dressings at $s$, the solution does not look so simple anymore and depends on the relative sizes of $G(s)$, $Z(s)$, $D^{A\bar cc}(s)$, and $C^{AAA}(s)$.
On the other hand, the approximation of taking only the loop momentum as argument of the dressing functions is not problematic.
It actually turns out to work very well, as can be seen in the analysis of the quadratically divergent part which is extremely sensitive to even small deviations \cite{Huber:2014tva}.

\index{three-gluon vertex}
So how can one fix the analysis?
The correct anomalous dimension of the three-gluon vertex is actually only half of the one above, viz., $\gamma_\text{3g}=\delta-\gamma=17/44$.
One could thus think of replacing the dressed three-gluon vertex by the squared three-gluon vertex.
This indeed leads to self-consistency also of the gluon propagator \gls{dse}.
The same can even be done for the ghost-gluon vertex, as it does not change the exponents.
It also cures the second problem by changing the coefficients as explained below.

\index{running coupling}
\index{MiniMOM scheme}
It is known that from every vertex one can extract an expression for the running coupling \cite{Alles:1996ka,Alkofer:2004it,Eichmann:2014xya}, for example:
\begin{subequations}
\label{eq:couplings}
\begin{align}
 \alpha_\text{ghg}(p^2)&=\alpha(\mu^2)\left(D^{A\bar cc}(p^2)\right)^2G^2(p^2)Z(p^2),\\
 \alpha_\text{3g}(p^2)&=\alpha(\mu^2)\left(C^{AAA}(p^2)\right)^2Z^3(p^2),\\
 \alpha_\text{4g}(p^2)&=\alpha(\mu^2)F^{AAAA}(p^2)Z^2(p^2).
\end{align}
\end{subequations}
Here, $\alpha(\mu^2)=g^2/4\pi^2$.
For a review on the \gls{qcd} running coupling see Ref.~\cite{Deur:2016tte}.
The definitions above should be contrasted with the so-called \textit{MiniMOM} \cite{vonSmekal:1997is,vonSmekal:2009ae} or Taylor coupling \cite{Boucaud:2008gn}:
\begin{align}
 \alpha_\text{MM}(p^2)&=\alpha(\mu^2)G^2(p^2)Z(p^2).
\end{align}
It agrees with $\alpha_\text{ghg}(p^2)$ for a bare ghost-gluon vertex.
Its advantage is that the scale parameter $\Lambda_\text{QCD}$ can be determined from two-point functions alone.
A range of studies exists for different numbers of quark families \cite{Sternbeck:2007br,Boucaud:2008gn,Sternbeck:2010xu,Blossier:2010ky,Blossier:2011tf,Blossier:2012ef,Sternbeck:2012qs,Blossier:2013ioa,Zafeiropoulos:2019flq}.
From the momentum dependent \glspl{sti} it follows that the couplings (\ref{eq:couplings}) must agree in the perturbative regime.
This entails
\begin{align}
 \left(C^{AAA}(p^2)\right)^2Z^2(p^2)=\left(D^{A\bar cc}(p^2)\right)^2G^2(p^2).
\end{align}
To make the coefficients of the two diagrams agree so that the \gls{sti} can be used, we square the vertex dressings.
$\omega$ then becomes
\begin{align} 
 \omega&= \frac{11N_c}{12}\alpha(\mu^2)\left(D^{A\bar cc}(s)\right)^2 G^2(s) Z(s)\nnnl
 &=\frac{11N_c}{12}\alpha(\mu^2)\left(C^{AAA}(s)\right)^2  Z^3(s)\\
 &=\frac{11N_c}{12\pi}\alpha_\text{ghg}(s).
\end{align}

\index{RG improvement}
\index{three-gluon vertex}
The only term that is not accounted for is the renormalization constant of the three-gluon vertex, $Z_1$.
The doubling of the vertex introduces an additional factor which modifies the behavior under changes of the renormalization scale.
To maintain multiplicative renormalizability, we must counterbalance this change.
By construction, the new term behaves under changes of the renormalization scale exactly as $Z_1$.
Thus, we do not need it and discard it.
There are two ways to do this:
One can include a term $1/Z_1$ in the vertex model or interpret the additional term as a momentum dependent renormalization constant \cite{vonSmekal:1997vx}.
The renormalization constant is related to the renormalization constants of the propagators via an \gls{sti}:
\begin{align}\label{eq:sti_Z1}
 Z_1=\frac{Z_3}{\widetilde{Z}_3}.
\end{align}
The renormalization constants $\widetilde{Z}_3$ and $Z_3$ run with the negative anomalous dimension, as can be seen from \eref{eq:prop_integrals_UV}.
Thus the 'anomalous dimension' of $Z_1$ is indeed $\delta-\gamma$.
However, it runs with the cutoff, not with the momentum.
Since the extra terms fix the \gls{rg} running of the dressings, they are sometimes referred to as \gls{rg} improvement terms \cite{Huber:2012kd}.

\index{FRG}
It is interesting to compare these modifications with the flow equations of the functional \gls{rg}.
In these equations all vertices are dressed and no renormalization constants appear.
Thus, dressing the bare vertices in \glspl{dse} seems like a natural choice.
However, there is no one-to-one correspondence between a one-loop truncation with all vertices dressed and the complete equation.

\begin{figure}[tb]
 \includegraphics[width=0.49\textwidth]{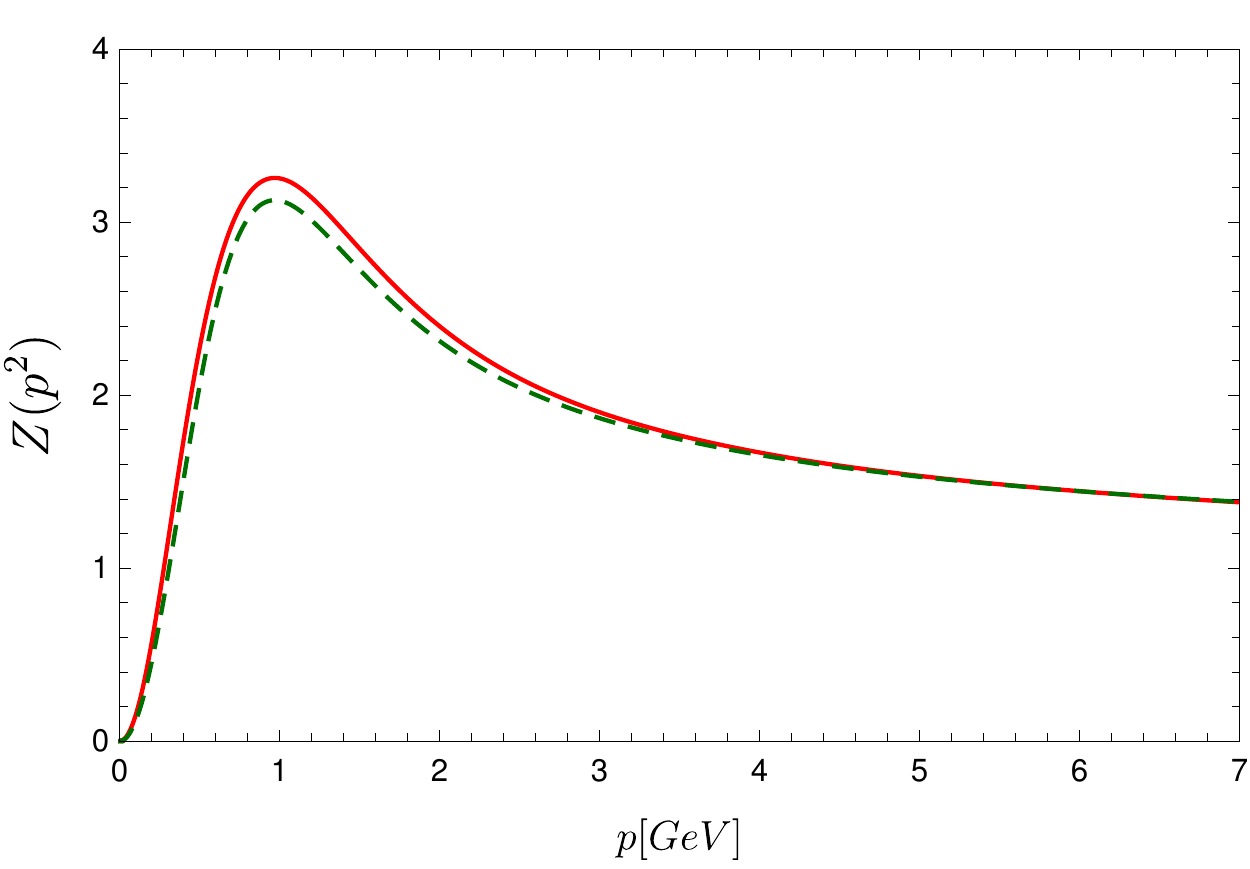}
 \includegraphics[width=0.49\textwidth]{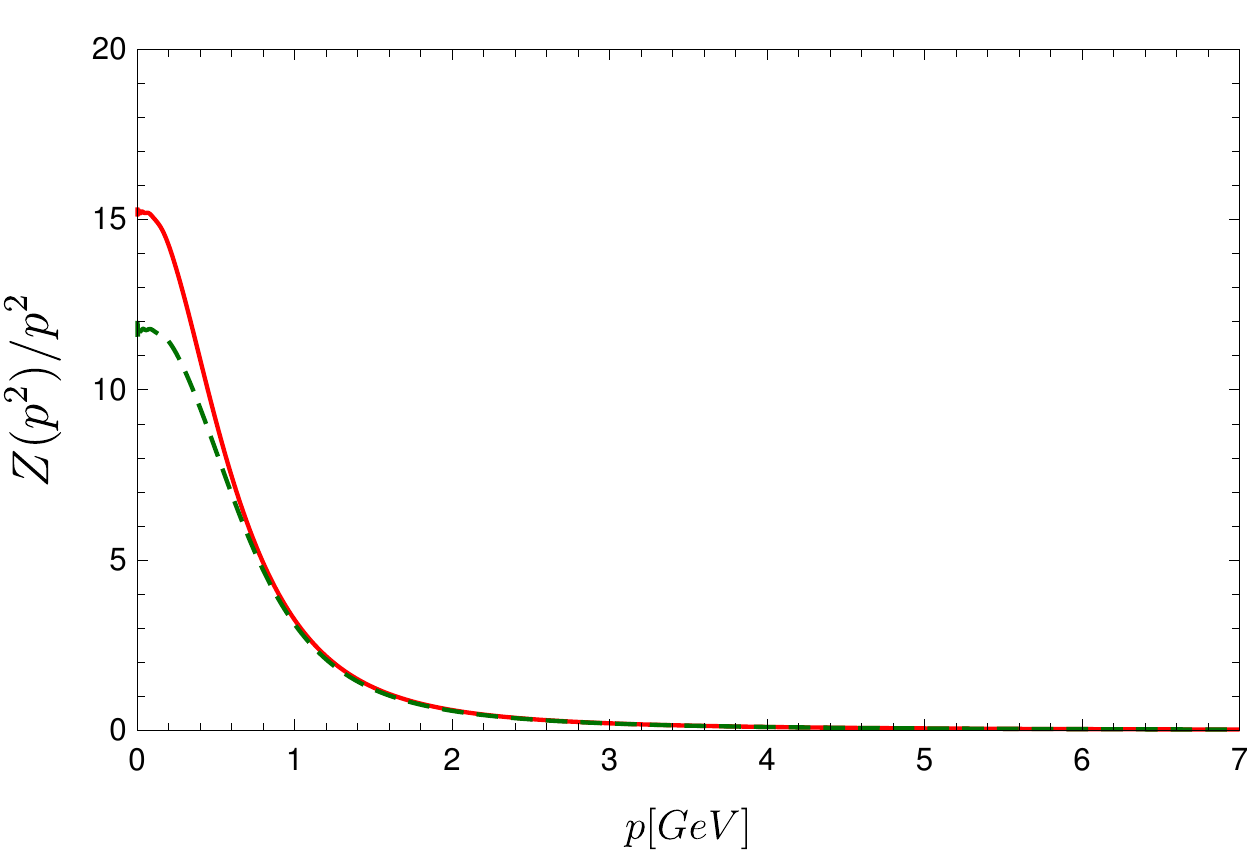}
 \caption{The gluon dressing function and propagator calculated with $C^{AAA}_{RG}$ from \eref{eq:DAAA_RG1} (red, continuous) and with $C^{AAA}_{RG}$ from \eref{eq:DAAA_RG2} (green, dashed) \cite{Huber:2014tva}.}
 \label{fig:YM4d_gl_RGI}
\end{figure}

\index{RG improvement}
\index{three-gluon vertex}
While \gls{rg} improvement terms are useful to get a self-consistent \gls{uv} behavior, they also introduce problems, because the connection to the nonperturbative regime is not clear.
There are various ways to continue the \gls{uv} running to the nonperturbative regime and unfortunately results depend quantitatively on this choice.
For the propagators, this was explicitly studied in \cite{Huber:2014tva}.
Fig.~\ref{fig:YM4d_gl_RGI} shows an example where two different versions of \gls{rg} improvement terms were used.
One version uses
\begin{align}\label{eq:DAAA_RG1}
 C^{AAA}_\text{RG}(p^2)&= G(p^2)^\alpha Z(p^2)^\beta,
\end{align}
where $\alpha$ and $\beta$ are determined such that the anomalous dimension of the expression corresponds to the one of the three-gluon vertex.
The remaining parameter is fixed such that the vertex becomes \gls{ir} finite.
For the scaling solution one obtains then $\alpha=-2-6\delta=-17/22$ and $\beta=-1-3\delta=-17/44$ and for a decoupling solution $\alpha=3+1/\delta=-17/9$ and $\beta=0$.
This vertex model leads to the correct logarithmic running, but in the coefficients the factors of $G(s)$ and $Z(s)$ do not combine as they should.
However, it turns out that the inconsistency between the ghost and the gluon loops is not problematic in practical calculations, probably due to the fact that the gluon loop is perturbatively much larger then the ghost loop.

\index{RG improvement}
\index{three-gluon vertex}
A second variant reads
\begin{align}\label{eq:DAAA_RG2}
 C^{AAA}_{RG}(p^2)&= \frac{G_{RG}(p^2)}{Z_{RG}(p^2)}.
\end{align}
It is motivated by the \gls{sti} \eref{eq:sti_Z1} and thus the factors of $G(s)$ and $Z(s)$ combine as they should.
The employed propagator dressings have the perturbative form with the Landau pole regularized:
\begin{align}
 G_{RG}(x):=G(s)\left(\omega\,\ln \left(a_{gh}+\frac{x}{\Lambda^2_\mathrm{QCD}} \right)\right)^\delta,\\
 Z_{RG}(x):=Z(s)\left(\omega\,\ln \left(a_{gl}+\frac{x}{\Lambda^2_\mathrm{QCD}} \right)\right)^\gamma,
\end{align}
with
\begin{align}
 a_{gl}=e^{\frac{Z(s)^{-1/\gamma}}{\omega}}, \quad
 a_{gh}=e^{\frac{G(s)^{-1/\delta}}{\omega}}.
\end{align}
Fig.~\ref{fig:YM4d_gl_RGI} shows the influence of the choice of the \gls{rg} improvement term.
It is large enough that this effect has to be taken into account in modern truncations.
For vertices such \gls{rg} improvement terms were used in the past \cite{Blum:2014gna,Eichmann:2014xya,Cyrol:2014kca} as well.
For the three-gluon vertex it was found that the \gls{rg} improvement terms have a sizable impact also on the nonperturbative regime \cite{Eichmann:2014xya}.

\index{$\varphi^3$ theory}
\myboxmargin{RG improvement for $\varphi^3$ theory}
The usage of \gls{rg} improvement terms works also in other cases.
Let me illustrate this with the previous example of $\varphi^3$ theory in six dimensions.
The propagator dressing function behaves asymptotically as
\begin{align}\label{eq:phi3_prop_uv}
 Z(p^2)=Z(s)\left(1+\omega \ln \frac{p^2}{s}\right)^{\gamma}.
\end{align}
We calculated in Sec.~\ref{sec:resummationInDSEs} that $\gamma=1/9$ and $\omega=\beta_0 g^2$ with $\beta_0=3/4/(2\pi)^3$.
The \gls{dse} of the propagator has the form depicted in \fref{fig:phi3_DSEs}:
\begin{align}
 p^2 Z(p^2)^{-1}=p^2 Z_3-\Sigma(p^2).
\end{align}
$Z_3$ is the field renormalization constant.
The integral of the self-energy reads
\begin{align}
 \Sigma(p^2)=\frac{g^2}{2}Z_1\int \frac{d^6q}{(2\pi)^6} \frac{Z(q^2)Z((p+q)^2)\Gamma_3(q,-p-q,p)}{q^2(p+q)^2}.
\end{align}
The coupling $g$ was explicitly pulled out from the dressed three-point function.
For the propagators we use now \eref{eq:phi3_prop_uv} and for the dressed three-point function we use a perturbative model that contains an \gls{rg} improvement term and discards the renormalization constant $Z_1$ in the self-energy:
\begin{align}
 \Gamma_3(p,q,r)=\frac{\Gamma_3(s)^2}{Z_1}\left( 1+\omega  \ln \frac{\overline p^2}{s}\right)^{2\gamma_3}.
\end{align}
$\overline{p}^2$ is given by $(p^2+q^2+r^2)/2$ and the anomalous dimension of the vertex is $\gamma_3$.
For large loop momenta $q$ we approximate the arguments of the dressing functions by their dominant part and obtain
\begin{align}
 \Sigma(p^2)=\frac{g^2}{2}\int \frac{d^6q}{(2\pi)^6} \frac{Z(s)^2\Gamma_3(s)^2}{q^2(p+q)^2}\left( 1+\omega  \ln \frac{q^2}{s}\right)^{2\gamma+2\gamma_3}.
\end{align}
Four angle integrals can be performed trivially:
The fifth one can be done analytically as well, but the cases $p<q$ and $p>q$ have to be distinguished.
We are only interested in the former case:
\begin{align}
\Sigma(p^2) \rightarrow& \frac{g^2}{6(2\pi)^4} \int_{p^2}^{\Lambda^2} dq^2\,q^4\pi\frac{p^2-3q^2}{8q^4}\frac{1}{q^2}Z(s)^2\Gamma_3(s)^2\left( 1+\omega \ln \frac{q^2}{s}\right)^{2\gamma+2\gamma_3}\nnnl
 \rightarrow&-\frac{g^2}{12(4\pi)^3}p^2 \int_{p^2}^{\Lambda^2} dq^2\frac{1}{q^2}Z(s)^2\Gamma_3(s)^2\left( 1+\omega \ln \frac{q^2}{s}\right)^{2\gamma+2\gamma_3}\nnnl
 =&-\frac{g^2}{12(4\pi)^3} p^2 Z(s)^2\Gamma_3(s)^2\frac{\left( 1+\omega \ln \frac{\Lambda^2}{s}\right)^{2\gamma+\gamma_3+1}-\left( 1+\omega \ln \frac{p^2}{s}\right)^{2\gamma+2\gamma_3+1} }{\omega (1+2\gamma+2\gamma_3)}.
\end{align}
In the second step only the logarithmically divergent part was kept.
The renormalization constant $Z_3$ is determined by the expression depending on the cutoff $\Lambda$.
The second part must match the inverse dressing function on the left-hand side.
From the exponents we obtain
\begin{align}
 1+3\gamma+2\gamma_3=0.
\end{align}
Matching the coefficients yields
\begin{align}
 1= -\frac{1}{12(4\pi)^3} \frac{g^2 Z(s)^3\Gamma_3(s)^2}{\omega (1+2\gamma+\gamma_3)} = \frac{1}{12(4\pi)^3} \frac{g^2 Z(s)^3\Gamma_3(s)^2}{\omega \gamma}
\end{align}
from which we infer
\begin{align}
 \omega=\frac{g^2}{(4\pi)^3} Z(s)^3\Gamma_3(s)^2\frac{1}{12\gamma} =  \alpha(s)\frac{1}{12\gamma}=g^2 \beta_0
\end{align}
with the running coupling $\alpha(p^2)=g^2 Z(s)^2 \Gamma_3(s)^2 /(4\pi)^3 $.
In the Yang-Mills case the anomalous dimension could be calculated, because we used $\gamma_\text{ghg}=0$.
Here, we also need some external information and took the anomalous dimension $\gamma=1/9$ from the one-loop coefficient in \eref{eq:phi3_prop_pert_resum}.

\index{quark propagator DSE}
\myboxmargin{quark propagator DSE and RGI: Fischer}
Similar mechanisms to obtain the one-loop resummed behavior have also been employed in the quark sector.
A version that is very close to the one explained above employs for the quark-gluon vertex a model that runs like the ghost dressing function squared.
As in the previous examples, the self-consistency of the quark propagator equation using this ansatz can be shown by counting the powers of the logarithms.
For simplicity we consider only the chiral case.
The loop diagram contains a gluon propagator (anomalous dimension $\gamma$), a quark propagator ($\gamma_\text{q}$) and a quark-gluon vertex ($\gamma_\text{qg}$).
The integration leads to another power of the logarithm so that one ends up with $\gamma+\gamma_\text{q}+\gamma_\text{qg}+1$ as exponent of the logarithm on the right hand side.
Since the anomalous dimension of the quark propagator wave function renormalization is zero, one can see that $\gamma_\text{qg}=2\delta$, where $\delta$ is the anomalous dimension of the ghost propagator, leads to the same relation as in the Yang-Mills sector: $\gamma+2\delta+1=0$.
The case where the anomalous dimensions of the integrand add up to $-1$ as here is actually not well behaved, but a more detailed analysis, including also the quark mass function, shows that the general statements made here hold \cite{Fischer:2003rp,Fischer:2003zc}.
The true anomalous dimension of the quark-gluon vertex, however, is $\gamma_\text{qg}=\delta$.
Again we see that an effective \gls{uv} dressing of the bare vertex is necessary to enforce perturbative self-consistency at the one-loop level.
In practical calculations of the quark propagator \gls{dse}, the modification of the anomalous dimension is realized, for example, by introducing an additional ghost dressing function in the model for the quark-gluon vertex, e.g., \cite{Fischer:2003rp,Fischer:2003zc,Aguilar:2010cn}, or by directly changing the anomalous dimension from $\delta$ to $2\delta$ \cite{Fischer:2009gk}.
The latter option was often used for calculations at nonvanishing temperatures and/or densities, e.g., \cite{Fischer:2011mz,Fischer:2012vc,Fischer:2013eca,Fischer:2014ata,Eichmann:2015kfa,Contant:2016ndj,Contant:2017gtz,Contant:2017onc,Mitter:2017iye,Contant:2019lwf}.

\index{quark propagator DSE}
\index{Maris-Tandy model}
\myboxmargin{quark propagator DSE and RGI: Maris-Tandy}
An alternative version to fix the one-loop resummed behavior was already used much earlier in form of the Maris-Tandy interaction.
It also has a term that repairs the \gls{uv} behavior at one-loop level \cite{Maris:1999nt,Maris:1997tm}.
This interaction replaces the gluon propagator and the quark-gluon vertex dressing functions.
The \gls{uv} term of the interaction reads
\begin{align}
 \mathcal{G_\text{UV}}(k^2)&=\frac{4\pi^2 \gamma_m \left(1-e^{-\frac{k^2}{\Lambda_t^2}}\right)}{\frac{1}{2}\ln\left(e^2-1+\left(1+\frac{k^2}{\Lambda_{\text{QCD}}^2}\right)^2\right)}.
\end{align}
At high momenta it behaves as
\begin{align}
 \frac{4\pi^2 \gamma_m }{\ln\left(\frac{k^2}{\Lambda_{\text{QCD}}^2}\right)}.
\end{align}
However, the combination of gluon propagator and quark-gluon vertex should run with the exponent $\gamma+\gamma_\text{qg}=\gamma+\delta=-1-\delta$.
Again we see that a missing $\delta$ was added.
With this interaction, the integrand has the same asymptotic behavior as in the previous case and the anomalous dimension of the quark wave function vanishes.
Although this approach can be employed also at nonvanishing temperature \cite{Blank:2010bz}, the \gls{uv} part is typically ignored in such studies.

\subsubsection{Two-loop diagrams in the gluon propagator DSE}
\label{sec:twoLoop}

\myboxmargin{introduction: why are two-loop terms important -> RG and non-perturbatively}
\index{resummed perturbation theory}
\index{two-loop diagrams}
\index{gluon propagator}
The \glspl{dse} of \gls{qcd} are two-loop exact, viz., only one- and two-loop diagrams appear.\footnote{The reason lies in the structure of the bare action of \gls{qcd} that has only three- and four-point interactions.
Higher interactions would lead to higher loop diagrams.}
At fixed order of perturbation theory, the hierarchy of diagrams is clear.
For example, in the gluon propagator \gls{dse} the contributions at order $O(g^4)$ stem from the explicit two-loop diagrams and the one-loop corrections of the propagators and vertices in the one-loop diagrams.
However, as discussed in Sec.~\ref{sec:resummation}, they contribute to the one-loop \textit{resummed} perturbative behavior.
Here, it should only be noted that (some) two-loop diagrams are required to obtain the one-loop resummed perturbative behavior.
In any case, in the perturbative regime there is a clear hierarchy of diagrams and one-loop diagrams are more important than two-loop diagrams.
In the non-perturbative regime, on the other hand, the role of two-loop terms is a priori unclear, since no obvious ordering scheme exists.
Hence, two-loop diagrams can become quantitatively more important than one-loop diagrams.

\myboxmargin{non-perturbative studies of two-loop diagrams: analytic, scaling solution}
\index{two-loop diagrams}
\index{scaling solution}
Unfortunately, two-loop diagrams are conceptually and numerically more difficult to calculate and corresponding studies are scarce.
In \glspl{dse} they come in two types called sunset and squint diagrams, see Appendix~\ref{sec:app_twoLoop}.
Analytically, the importance of two-loop diagrams in the non-perturbative regime was analyzed for the scaling type of solutions in the Landau gauge \cite{Fischer:2006vf,Huber:2007kc,Alkofer:2008jy,Fischer:2009tn,Huber:2009tx,Huber:2010cq} and the \gls{mag} \cite{Huber:2008ea,Huber:2009wh,Huber:2010ne,Huber:2011fw}.
Such analyses were made possible by the combination of the functional \gls{rg} and \glspl{dse} as initially introduced in \cite{Fischer:2006vf} or under the assumption of a convergent skeleton expansion \cite{Huber:2007kc,Alkofer:2008jy}.
A general structure, which was worked out in Ref.~\cite{Huber:2009wh,Huber:2010ne}, emerges that explains why in the Landau gauge diagrams with the highest number of ghost propagators are dominant in the \gls{ir}.
This holds provided no cancellations appear, which, at least at the level of the propagator equations, would be very difficult to realize in this specific case.
The corresponding analysis can also be applied to the Gribov-Zwanziger action \cite{Huber:2009tx}.
Due to the mixing of fields at the two-point level the algebraic analysis yields four possible solutions at first, two of which could be discarded right away \cite{Huber:2009tx}.
A third one was discarded later numerically \cite{Huber:2010cq}, leaving as only solution the one which is equivalent to the solution from the Faddeev-Popov action.

\myboxmargin{MAG and IR solution}
\index{scaling solution}
\index{two-loop diagrams}
\index{maximally Abelian gauge}
The \gls{mag} constitutes a nice example where the hierarchy of diagrams is inverted in the \gls{ir}.
A scaling analysis revealed that certain sunset diagrams yield the dominant contributions in the \gls{ir} with the corresponding squint diagrams possibly contributing as well \cite{Huber:2009wh}.
The generalized method developed in Ref.~\cite{Huber:2009wh} to determine the hierarchy of diagrams at low momenta was necessary to find this solution, since previously typically only the one-loop diagrams had been taken into account.
More details on the \gls{mag} can be found in Sec.~\ref{sec:mag}.

\myboxmargin{non-perturbative studies of two-loop diagrams: numerical solutions}
\index{two-loop diagrams}
\index{squint diagram}
\index{sunset diagram}
Numerical studies of two-loop diagrams were initially limited by the available computing power.
A first inclusion via approximating one of the two loops was realized in \cite{Bloch:2003yu}.
Full numerical calculations were done later \cite{Mader:2013ru,Hopfer:2014th,Mader:2014qca,Meyers:2014iwa,Huber:2016tvc,Huber:2017txg,Huber:2020keu}.
In the Landau gauge, an early finding was that the sunset diagram leads only to a very small change \cite{Mader:2013ru}.
The squint diagram, on the other hand, has a larger impact as was shown in the case of four \cite{Hopfer:2014th,Meyers:2014iwa} and three dimensions \cite{Huber:2016tvc}.
In Ref.~\cite{Meyers:2014iwa}, models for the vertices were used and their couplings were fitted to match lattice results of the propagators.
In the latter case of Ref.~\cite{Huber:2016tvc}, dynamically calculated vertices were employed that agree well with available lattice results.\
A solution in four dimensions with partially coupled vertices was obtained in \cite{Huber:2017txg} and the full system of primitively divergent correlation functions was solved in Ref.~\cite{Huber:2020keu}.

\myboxmargin{two-loop diagrams and divergences}
\index{two-loop diagrams}
\index{spurious divergences}
The analysis of two-loop diagrams is aggravated by the fact that analytic expressions are more difficult to obtain.
A calculation using power laws for all the dressing functions as required for scaling type solutions can be realized using the \gls{ndim} \cite{Halliday:1987an,Dunne:1987am,Dunne:1987qb,Ricotta:1990nd}.
The case of two-loop diagrams is treated, e.g., in Ref.~\cite{Suzuki:2000us}.
For the specific case of massless propagators and including a tensorial structure it was explicitly worked out in Ref.~\cite{Huber:2010ne}.
As in one-loop diagrams, spurious divergences appear in the two-loop diagrams which are best exposed when using a hard \gls{uv} cutoff.
However, in most perturbative studies, dimensional regularization is employed which cannot be used to identify these divergent terms.
The sunset diagram was worked out for a \gls{uv} cutoff in \cite{Mader:2013ru,Mader:2014qca}.
The general method, which can also be applied to the squint diagram, is described in Appendix~\ref{sec:app_twoLoop}.
The corresponding results are discussed below.
A numeric solution for the gluon propagator concludes the discussion of two-loop diagrams.

\paragraph{Analytic calculation of two-loop diagrams in the gluon propagator Dyson--Schwinger equation}

\myboxmargin{two-loop diagrams calculation analytically}
\index{resummed perturbation theory}
The two-loop diagrams of the gluon propagator \gls{dse}, see \fref{fig:gh-gl-qu_DSEs}, can be calculated analytically in the perturbative regime.
The involved part are the angle integrations.
With the method described in Appendix~\ref{sec:app_twoLoop}, one can derive the following expressions for the squint and sunset integrals.
Only the tree-level structure of the vertices is taken into account and all dressing functions have been set to one, so these are the perturbative results of leading order:
\begin{align}
 I&_\text{squint}(p^2)=
 \int_{p^2}^{\Lambda^2}dy_1\int_{y_1}^{\Lambda^2}dy_2\frac{(15y_2-y_1)(-6y_1(\zeta-4)+p^2(9+\zeta))}{24576 \pi^4 p^2 y_1 y_2^2}\nnnl
 &+\int_{p^2}^{\Lambda^2}dy_1\int_0^{y_1}dy_2\frac{(9y_1+5y_2)(-6y_1(\zeta-4)+p^2(9+\zeta))}{24576 \pi^4 p^2 y_1^{3}}\nnnl
 &+\int_{0}^{p^2}dy_1\int_0^{y_1}dy_2\frac{-(9_1^2+5y_2)(3p^2(\zeta-11)+2y_1\zeta)}{24576 \pi^4 p^6 y_1}\nnnl
 &+\int_{0}^{p^2}dy_1\int_{y_1}^{y_1}\frac{y_1(y_1-15y_2)(3p^2(\zeta-11)+2y_1\zeta)}{24576 \pi^4 p^6 y_2^2},\\
 I&_\text{sunset}(p^2)=
 \int_{p^2}^{\Lambda^2}dy_1\int_{y_1}^{\Lambda^2}dy_2\frac{p^2(\zeta-1)-(y_1-21 y_2)(\zeta-4)}{8192 \pi^4 p^2 y_2^2}\nnnl
 &+\int_{p^2}^{\Lambda^2}dy_1\int_0^{y_1}dy_2\Bigg(\frac{3y_1^3(21y_1-y_2)(\zeta-4) + 5p^4 y_2(y_1- y_2)\zeta }{24576 \pi^4 p^2 y_1^{5}}\nnnl
 &+\frac{ p^2 (9y_1+5y_1 y_2^2 \zeta-2y_1^2 y_2(\zeta+6)}{24576 \pi^4 p^2 y_1^{5}}\Bigg)\nnnl
 &+\int_{0}^{p^2}dy_1\int_0^{y_1}dy_2\frac{3(y_1 \zeta+3p^2(2\zeta-9))}{8192 \pi^4 p^6}\nnnl
 &+\int_{0}^{p^2}dy_1\int_{y_1}^{\Lambda^2}dy_2\frac{y_1(-4y_1^2\zeta + 13 y_1 y_2 \zeta + p^2(y_1(9+4\zeta)+2y_2(-126+25\zeta)))}{24576 \pi^4 p^6 y_2^2}.
\end{align}
The projector from \eref{eq:Pzeta} was used.
$y_i=q_i^2$ with the $q_i$ being the loop momenta.
The quadratic divergences can be identified as the terms proportional to $\zeta-4$ as in the one-loop case.
The remaining radial integrals can be done directly:
\begin{align}
 I&_\text{squint}(p^2)=\frac{(\zeta -20) p^2}{49152 \pi ^4 \Lambda^2}-\frac{13 (\zeta -4) \Lambda^2}{2048 \left(\pi ^4 p^2\right)}\nnnl
 &\quad+\frac{700 \zeta +45 (\zeta +9) \ln ^2\left(\frac{p^2}{\Lambda^2}\right)-12 (34 \zeta -9) \ln \left(\frac{p^2}{\Lambda^2}\right)-1836}{147456 \pi ^4},\\
 I&_\text{sunset}(p^2)=\frac{\zeta  p^2}{73728 \pi ^4 \Lambda^2}+\frac{-1423 \zeta +108 (5 \zeta -21) \ln \left(\frac{p^2}{\Lambda^2}\right)+5562}{442368 \pi ^4}+\frac{41 (\zeta -4) \Lambda^2}{8192 \pi ^4 p^2}.
\end{align}
Of direct relevance is the squared logarithm in the result for the squint diagram which is important for the resummation of logarithms.
The sunset diagram does not yield such a term and thus does not contribute to the perturbative one-loop resummation at two-loop order.

\paragraph{Two-loop complete calculation of the Yang-Mills propagators}

\myboxmargin{two-loop solution of propagator equations: models}
Being able to solve the two-loop diagrams in the gluon propagator \gls{dse}, one can solve the untruncated system of propagator \glspl{dse} of Yang-Mills theory.
The vertices still need to be specified.
For the ghost-gluon vertex, a bare vertex should be sufficient.
However, since it is not problematic and actually advantageous for using the obtained propagator solutions as input for the vertex equations in Sec.~\ref{sec:res_AAcbc}, the ghost-gluon vertex will be included in a one-configuration approximation, see Sec.~\ref{sec:res_AAcbc} for details.
The four-gluon vertex is taken as bare.
This vertex appears here only in the sunset diagram, which, however, is quantitatively not as important as the squint diagram.
For the three-gluon vertex only the tree-level is dressed with the model function
\begin{align}
 C^{AAA}&(p^2,q^2,r^2)=\frac{G(\overline{p}^2)}{Z(\overline{p}^2)}\frac{\overline{p}^2}{\overline{p}^2+\Lambda_s^2}.
\end{align}
$\overline{p}^2$ is given by $(p^2+q^2+r^2)/2$.
This is a simple model with the correct \gls{uv} behavior.
The reason for choosing the structure $G(p^2)/Z(p^2)$ is motivated by the \gls{uv} analysis of the propagator \glspl{dse}, see Sec.~\ref{sec:resummation}.
However, since this term is IR divergent, a damping function is added to tame it in the IR.
The damping scale is chosen as $\Lambda^2_s= 1.54\,\text{GeV}^2$.
Note that neglecting non-tree-level dressings should be irrelevant for the perturbative resummation, since such dressings decay with a power law and thus do not yield terms proportional to $g^4 \ln^2 p^2/\mu^2$.

\index{tadpole diagram}
\myboxmargin{two-loop solution of propagator equations: tadpole and spurious divergences}
One comment needs to be made about the tadpole diagram.
Its role was scrutinized in Ref.~\cite{Huber:2014tva}.
While a small quantitative impact could not be excluded, it is clear that it behaves like a constant over $p^2$.
Thus, in most variants to subtract quadratic divergences it will be absorbed directly without any possibility to influence the results \cite{Huber:2014tva}.
This is also the case for the method employed here where a second renormalization condition for the gluon propagator is used, $D(0)=15.54\,\text{GeV}^{-2}$.
Any tadpole contribution will be absorbed in this condition and the tadpole is thus discarded.

\index{MiniMOM scheme}
\myboxmargin{renormalization constants}
The employed renormalization scheme is the \emph{MiniMOM} scheme \cite{vonSmekal:1997vx,vonSmekal:2009ae}, viz. the propagators are renormalized with a momentum subtraction.
The specific renormalization conditions are $G(0)=10$ for the ghost and $Z(x_s)=1$, with $x_s=7720\,\text{GeV}^2$, for the gluon.
The momentum subtraction scheme entails that the renormalization constants of the ghost and the gluon propagators, $\widetilde Z_3$ and $Z_3$, respectively, are fixed implicitly and only once a solution is obtained, they can be calculated exactly.
However, the renormalization constants of the three- and four-gluon vertices, $Z_1$ and $Z_4$, respectively, are required as prefactors of the gluon loop and the two-loop diagrams.
They are obtained from the corresponding STIs:
\begin{align}\label{eq:Z1Z4}
 Z_1=\frac{Z_3}{\widetilde Z_3}, \qquad Z_4=\frac{Z_3}{\widetilde Z_3^2}.
\end{align}
As a starting point, the solution of a one-loop truncated system of propagators is used.
This gives access to values for $Z_1$ and $Z_4$ which are reasonable close to their final values so that the system can be solved with fixed values for $Z_1$ and $Z_4$.
From the solution, their new values are calculated and the system is solved again.
This meta iteration process is continued until $Z_1$ and $Z_4$ do not change anymore.
In these calculations, the value for the renormalized coupling was chosen as $\alpha(\mu^2)=0.05$, since it is advantageous for the convergence of the resummed series to have a small value.
However, $\alpha(\mu^2)$ must be large enough that the scale $s$ is inside the considered momentum region.

\begin{figure}[tb]
  \begin{center}
  \includegraphics[width=0.48\textwidth]{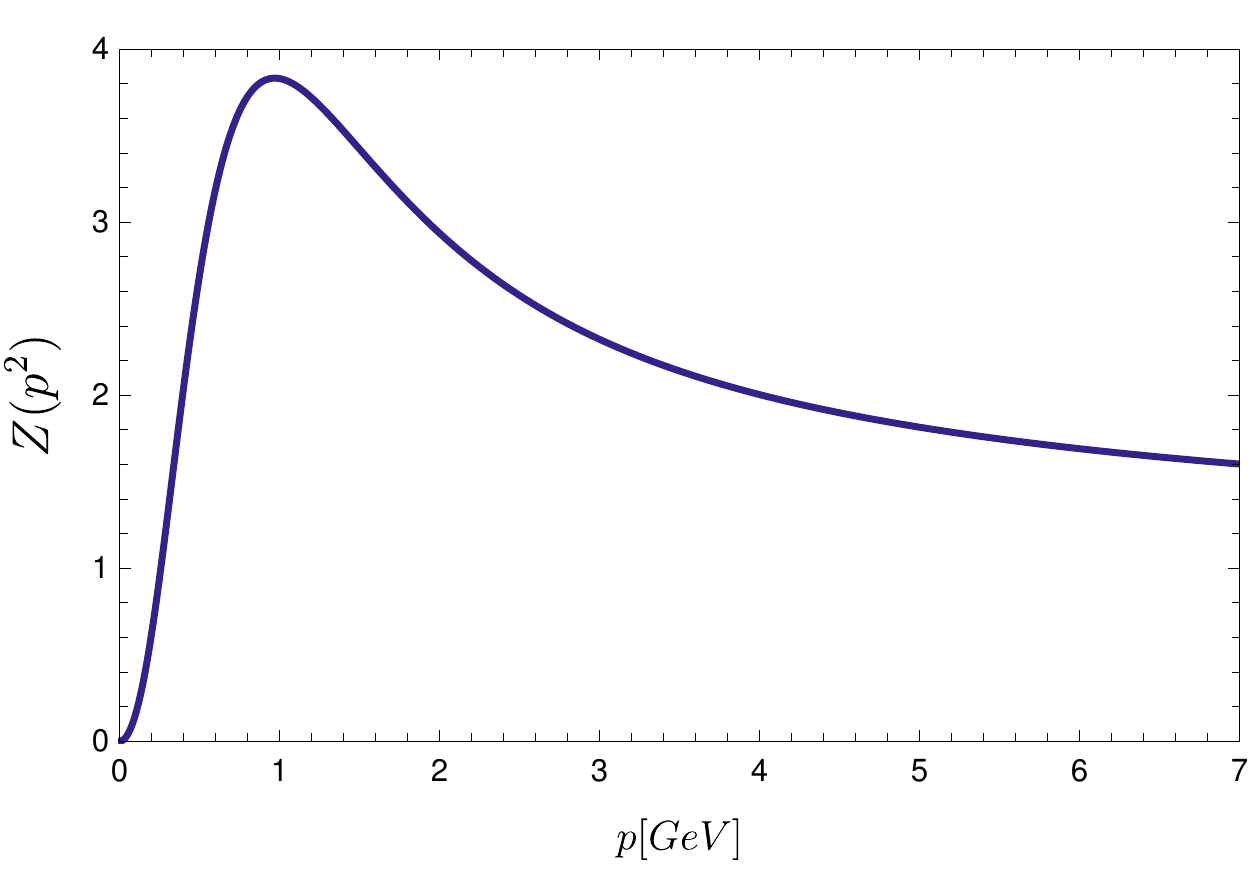}
  \hfill
  \includegraphics[width=0.48\textwidth]{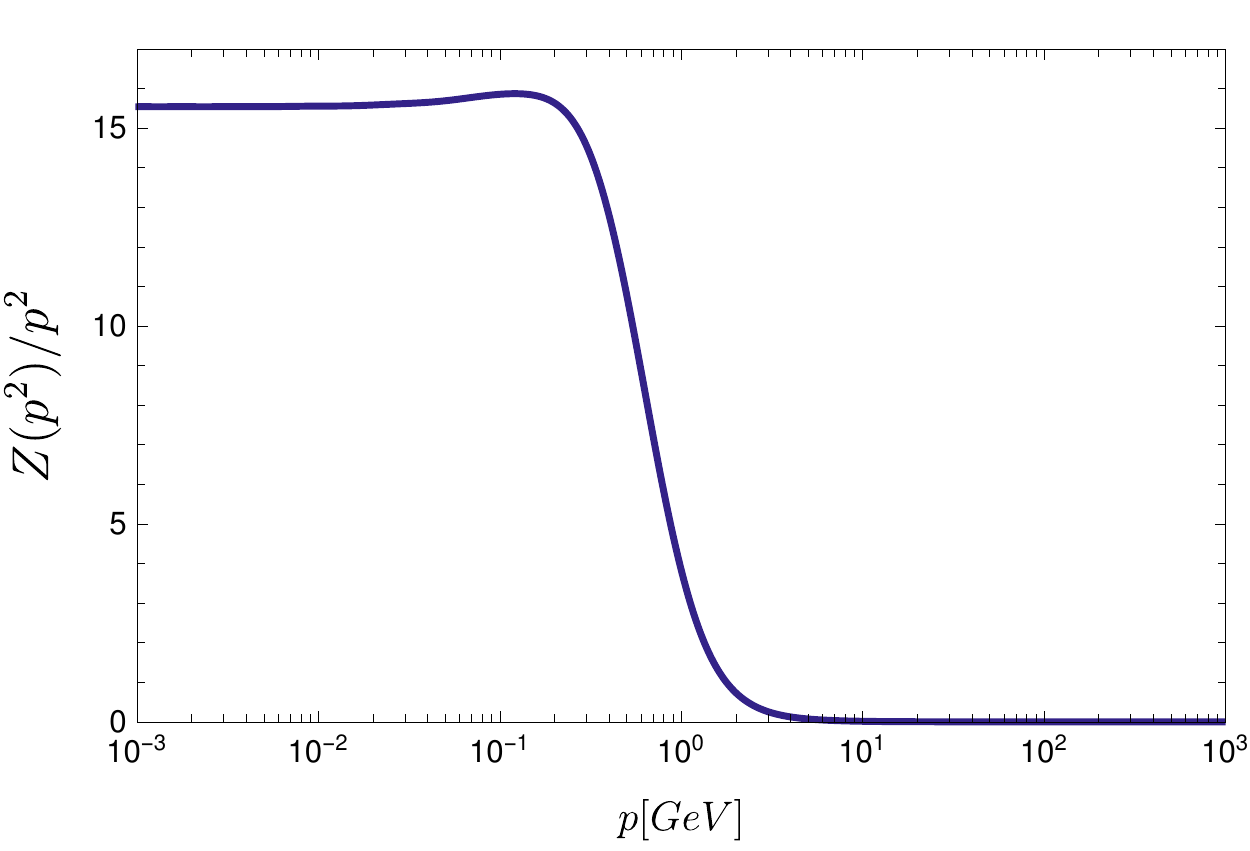}\\
  \includegraphics[width=0.48\textwidth]{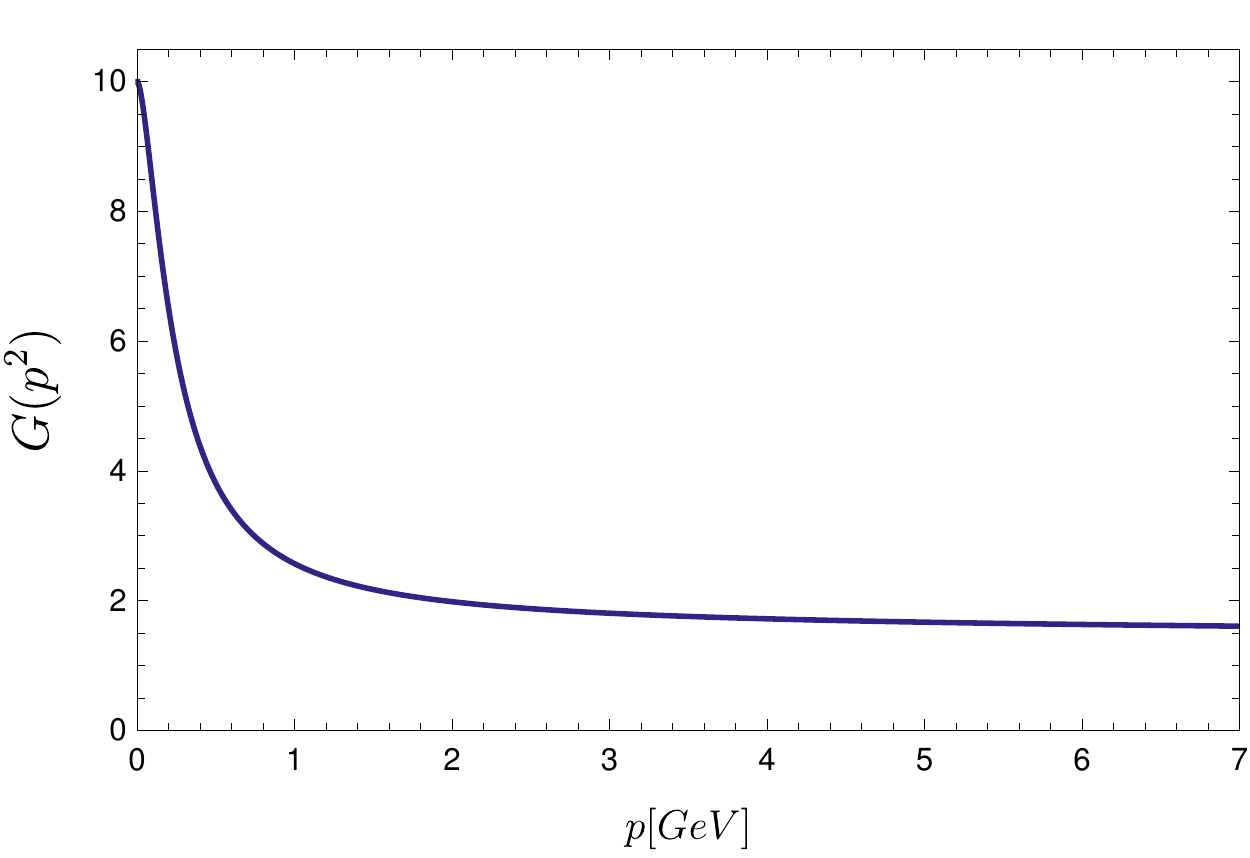}
  \hfill
  \includegraphics[width=0.48\textwidth]{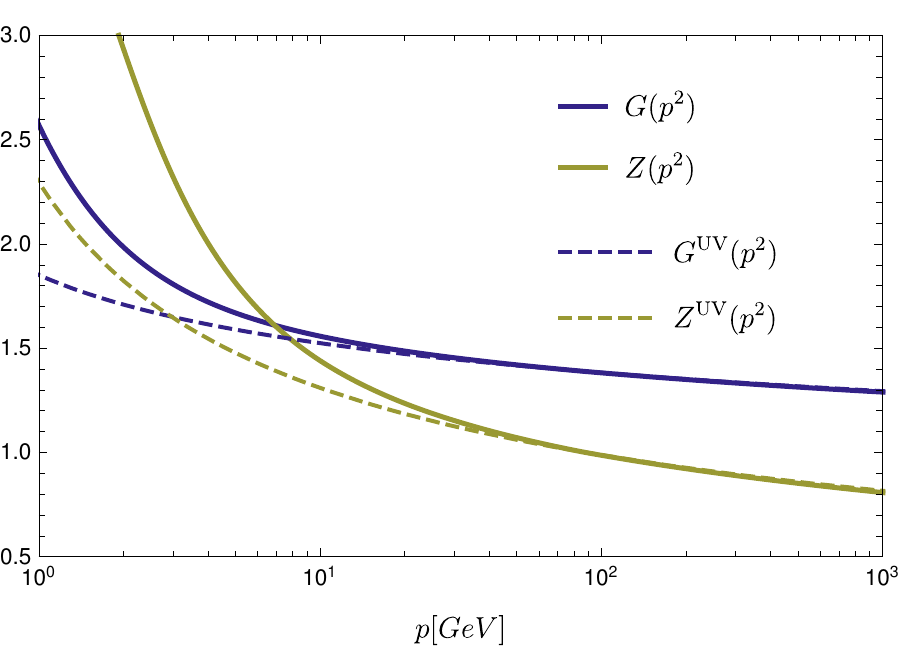}
  \caption{\label{fig:YM4d_props_tl}
  The solutions of the Yang-Mills propagator \glspl{dse} including the two-loop diagrams \cite{Huber:2017txg}.
  The top row shows the gluon dressing function (left) and the gluon propagator (right), the bottom row the ghost dressing function (left) and the propagator dressing functions in the \gls{uv} compared to the perturbative one-loop resummed behavior (right).}
 \end{center}
\end{figure}

\myboxmargin{two-loop solution of propagator equations: results, resummed behavior}
\index{two-loop diagrams}
\index{resummation}
The results for the propagators are shown in \fref{fig:YM4d_props_tl}.
The two most important consequences of including the two-loop diagrams concern the \gls{uv} behavior of the propagators and the quantitative changes in the midmomentum regime.
A comparison of the obtained dressing functions with the analytic expression for the resummed one-loop behavior is also shown in \fref{fig:YM4d_props_tl}.
The plot shows that the solutions respect the expected behavior.
It should be stressed, though, that only up to order $g^4$ all contributions are included.
Beyond that, many contributions are included, for example, nested propagator self-energy insertions.
Missing contributions are, for instance, contributions of order $g^4$ from the ghost-gluon vertex.
Nevertheless, the agreement is quite good and is sufficient for practical purposes.

\myboxmargin{two-loop solution of propagator equations: results, midmomentum}
\index{two-loop diagrams}
\index{squint diagram}
\index{sunset diagram}
As expected, the inclusion of two-loop diagrams has an impact on the midmomentum regime.
The more sizable contribution comes from the squint diagram, whereas the sunset diagram is relatively small.
This can be tested by discarding the latter.
The dressing functions do not change much then.

The present calculation still depends on models for the three- and four-gluon vertex.
For a future complete analysis, full expressions for these vertices need to be used including non-tree-level dressings.
The status of corresponding calculations of vertices is discussed in Sec.~\ref{sec:res_verts}.

\subsubsection{Analytic structure of the gluon propagator}
\label{sec:analytic_structure}

\index{analytic structure}
The analytic structure of the gluon propagator is of interest for several reasons.
In bound state equations the correlation functions of the constituent particles are accessed at complex values of the momenta, see, e.g., \cite{Alkofer:2000wg,Bashir:2012fs,Eichmann:2016yit,Sanchis-Alepuz:2017jjd}.
Futhermore, the spectral function of a particle provides information about its spectrum.
And finally, real-time quantities like transport coefficients require knowledge of the real-time properties of particles.

\index{Wick rotation}
Functional equations are typically solved after a Wick rotation to Euclidean  metric is performed, viz., the time coordinate is rotated to the imaginary axis as $t\rightarrow -i\,t_E$, where the subscript $E$ indicates now Euclidean time.
In momentum space, this corresponds to $p_0\rightarrow i\,p_{0,E}$.
This entails $p^2\rightarrow -p_E^2$.
Calculations are then performed for real $p_E^2>0$ where no poles or branch cuts exist.

\index{analytic continuation}
To undo the Wick rotation, Euclidean results need to be continued analytically to $k_E^2<0$.
This is an ill-posed inverse problem and always requires some form of bias.
Possible reconstruction methods are, for example, the maximum entropy method \cite{Jarrell:1996rrw,Asakawa:2000tr,Ding:2012sp,Nickel:2006mm,Mueller:2010ah,Qin:2010pc,Haas:2013hpa,Qin:2013ufa,Christiansen:2014ypa,Gao:2014rqa,Rothkopf:2016luz,Ilgenfritz:2017kkp,Fischer:2017kbq}, fits motivated by analytic results or phenomenology \cite{Brandt:2015sxa,Brandt:2015aqk,Ding:2014dua,Cyrol:2018xeq}, Pad\'e approximants \cite{Vidberg:1977}, the Schlessinger point method \cite{Schlessinger:1968spm,Tripolt:2017pzb}, the Tikonov regularization \cite{Dudal:2013yva}, or the Backus-Gilbert method \cite{Brandt:2015sxa}.

\index{Cauchy contour method}
\index{shell method}
\index{contour deformation method}
Alternatively, it is possible to solve the equations for complex momenta.
However, this complicates any calculations considerably as care has to be taken which regions of the complex plane are probed in the integration.
Poles or branch cuts constitute obstructions that have to be avoided.
Methods suitable for this task include the shell method \cite{Maris:1995ns,Fischer:2008sp}, the Cauchy contour method \cite{Fischer:2005en,Fischer:2008sp,Krassnigg:2009gd}, and others \cite{Dorkin:2013rsa}.
A very flexible method is based on contour deformation.
Unfortunately, it is also very intricate and was thus not used extensively up to now.
It was already successfully applied in QED \cite{Maris:1995ns}, QCD \cite{Alkofer:2003jj,Strauss:2012dg,Weil:2017knt} and for the scalar glueball propagator in the Born approximation \cite{Windisch:2012sz}.
The challenge is to deform the contour of the radial integration in $q^2$ such that any non-analyticities in the complex plane are avoided.
Such non-analyticities can also be created by the angle integration \cite{Maris:1995ns,Alkofer:2003jj,Windisch:2013dxa}.
For massless particles the contour deformation can be done along rays starting at the origin, see Ref.~\cite{Strauss:2012zz} for details.
This ensures that the branch cut created by the angle integration is avoided via the only possible point $q^2=p^2$, where $p$ is the complex external momentum.
The integration path can then be continued outside the branch cut to the \gls{uv} cutoff.

\index{spectral function}
\index{Cauchy's integral formula}
\index{Stieltjes transformation}
The analytic structure of a propagator is directly reflected in the spectral function.
In general, one expects a branch cut for time-like momenta starting at a threshold value $s_0$ and possibly poles.\footnote{The possible existence of a term proportional to the Dirac delta functional is discussed in Refs.~\cite{Lowdon:2017uqe,Lowdon:2018mbn,Lowdon:2018uzf}.
}
As a direct consequence of Cauchy's integral formula the propagator can then be written as
\begin{align}
 D(p^2)=\int_{s_0}^\infty ds \frac{\widetilde{\rho}(s)}{p^2+s} + \sum_{i}\frac{R_i}{p^2+m_i^2},
\end{align}
where the sum is over the potential pole contributions with residues $R_i$ and masses $m_i$ which are either real with $m_i^2<s_0$ or complex.
If there is only a real pole at $-m^2$, this expression can be rewritten as
\begin{align}\label{eq:KaellenLehmann}
 D(p^2)=\int_{0}^\infty ds \frac{\rho(s)}{p^2+s},
\end{align}
with
\begin{align}
 \rho(s)=\widetilde{\rho}(s)+R\delta(s-m^2)
\end{align}
and $\rho(s)=0$ for $s<s_0$.
Eq.~(\ref{eq:KaellenLehmann}) has the form of a Stieltjes representation for the propagator \cite{Widder:1971,Debnath:2007}.
The inversion operation is formally known but includes infinitely many derivatives \cite{Widder:1971,Debnath:2007,Dudal:2010wn}.
The difficulties in inverting this expression also become obvious when rewriting it as a double Laplace transform:
\begin{align}
 D(p^2)=\int_0^\infty du\, e^{-u\,p^2}\int_0^\infty ds\, e^{-u\,s}\rho(s).
\end{align}

\index{K\"all\'en-Lehmann representation}
\index{reflection positivity}
\index{Osterwalder-Schrader axioms}
\index{positivity violation}
\index{Schwinger function}
Eq.~(\ref{eq:KaellenLehmann}) has a physical interpretation as a sum of free propagators with density $\rho(s)$.
This is called K\"all\'en-Lehmann spectral representation.
It requires $\rho(s)$ to be nonnegative so that the field is part of the physical state space of the theory.
This is not necessarily the case for gauge theories, but \eref{eq:KaellenLehmann} can be used even if $\rho(s)$ contains negative contributions.
Turning the argument around, negative norm contributions mean that the particle cannot describe an asymptotic phyiscal particle and is thus confined.
This is a sufficient but not a necessary condition for confinement.
Mathematically, it is encoded in the axiom of \emph{reflection positivity} for Euclidean quantum field theory as formulated by Osterwalder and Schrader \cite{Osterwalder:1974tc,Osterwalder:1973dx}.
It boils down to calculating the Schwinger function defined as the time Fourier transformation of the momentum space propagator for vanishing spatial momentum:
\begin{align}
 \Delta(t)=D(t,\vec{p}=\vec{0})=\int_{-\infty}^\infty \frac{dp_0}{2\pi} e^{-i\,p_0 \,t}D(p_0, \vec{0})=\frac{1}{\pi}\int_0^\infty dp_0 \cos(p_0\,t)D(p_0,\vec{0}).
\end{align}
If the propagator can be written as in \eref{eq:KaellenLehmann}, this is the Laplace transform of the spectral density:
\begin{align}
 \Delta(t)=\int_0^\infty d\omega e^{-\omega\,t}\rho(\omega^2)
\end{align}
Clearly, if one can show that $\Delta(t)$ is negative somewhere, then $\rho(s)$ must also contain negative contributions.
If the propagator vanishes at zero momentum, as it does for a solution of the scaling type, positivity violation is obvious from
\begin{align}
 D(p^2=0)=\int d^4x \,D(x)=\int dt \,\Delta(t)=0.
\end{align}
From the derivative of \eref{eq:KaellenLehmann} with respect to $p^2$, one can infer that a non-monotonous propagator also violates positivity.
Such a behavior is seen in lattice and functional results in three dimensions, e.g., \cite{Bornyakov:2013ysa,Maas:2014xma,Cucchieri:2016jwg,Huber:2016tvc,Corell:2018yil}.
In four dimensions, the peak in the gluon propagator is only very small and not yet confirmed with lattice calculations.
Functional methods, on the other hand, find such a peak, e.g., \cite{Fischer:2008uz,Aguilar:2013vaa,Cyrol:2016tym,Cyrol:2017ewj,Huber:2017txg,Huber:2020keu} so that one can conclude without explicit calculation of the Schwinger function that positivity is violated also for decoupling type solutions.
The existence of a maximum also has consequences for the spectral dimension of the gluon propagator, which is the dimension felt by the propagator.
If the propagator has a maximum, it is reduced from four to one \cite{Kern:2019nzx}.

\begin{figure}[tb]
  \begin{center}
  \includegraphics[width=0.48\textwidth]{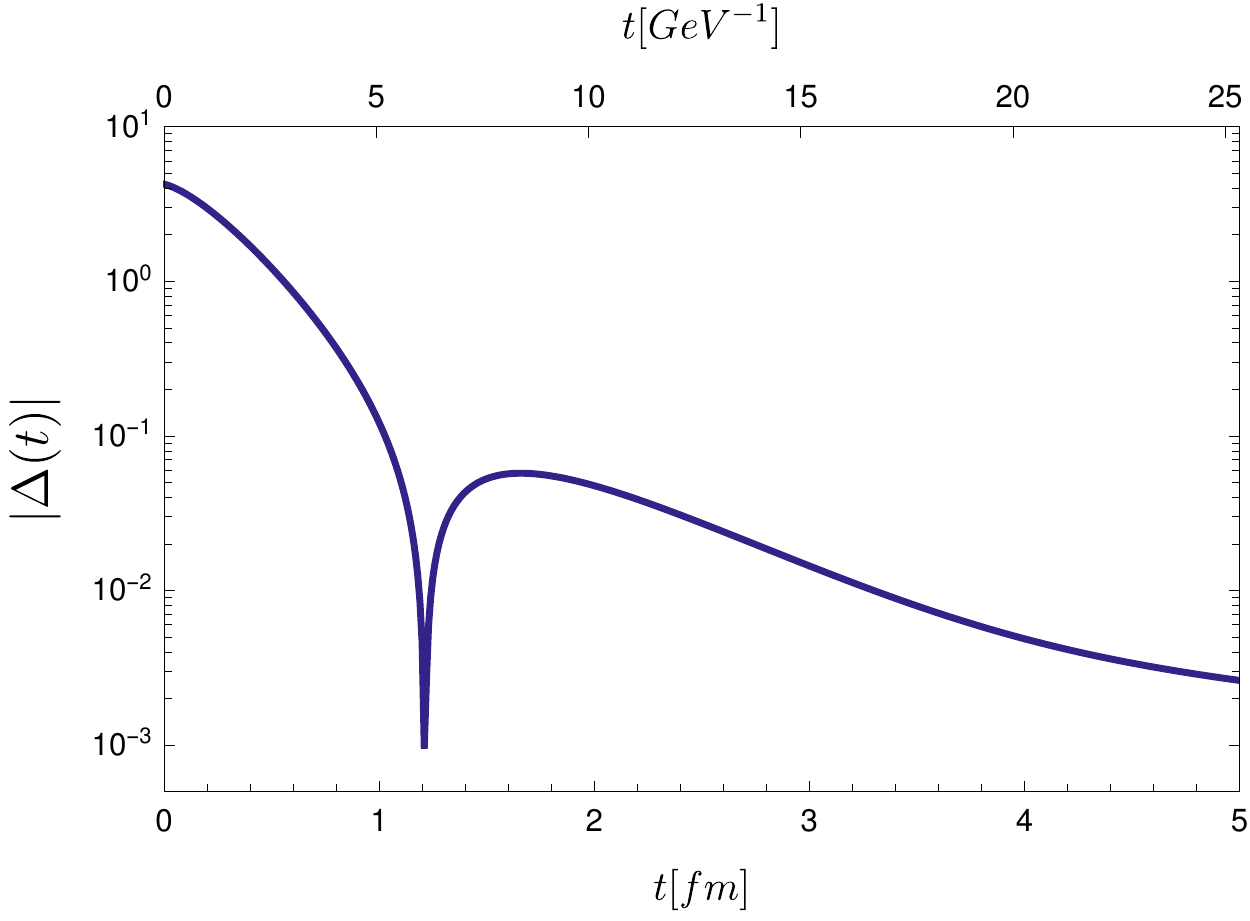}
  \hfill
  \includegraphics[width=0.48\textwidth]{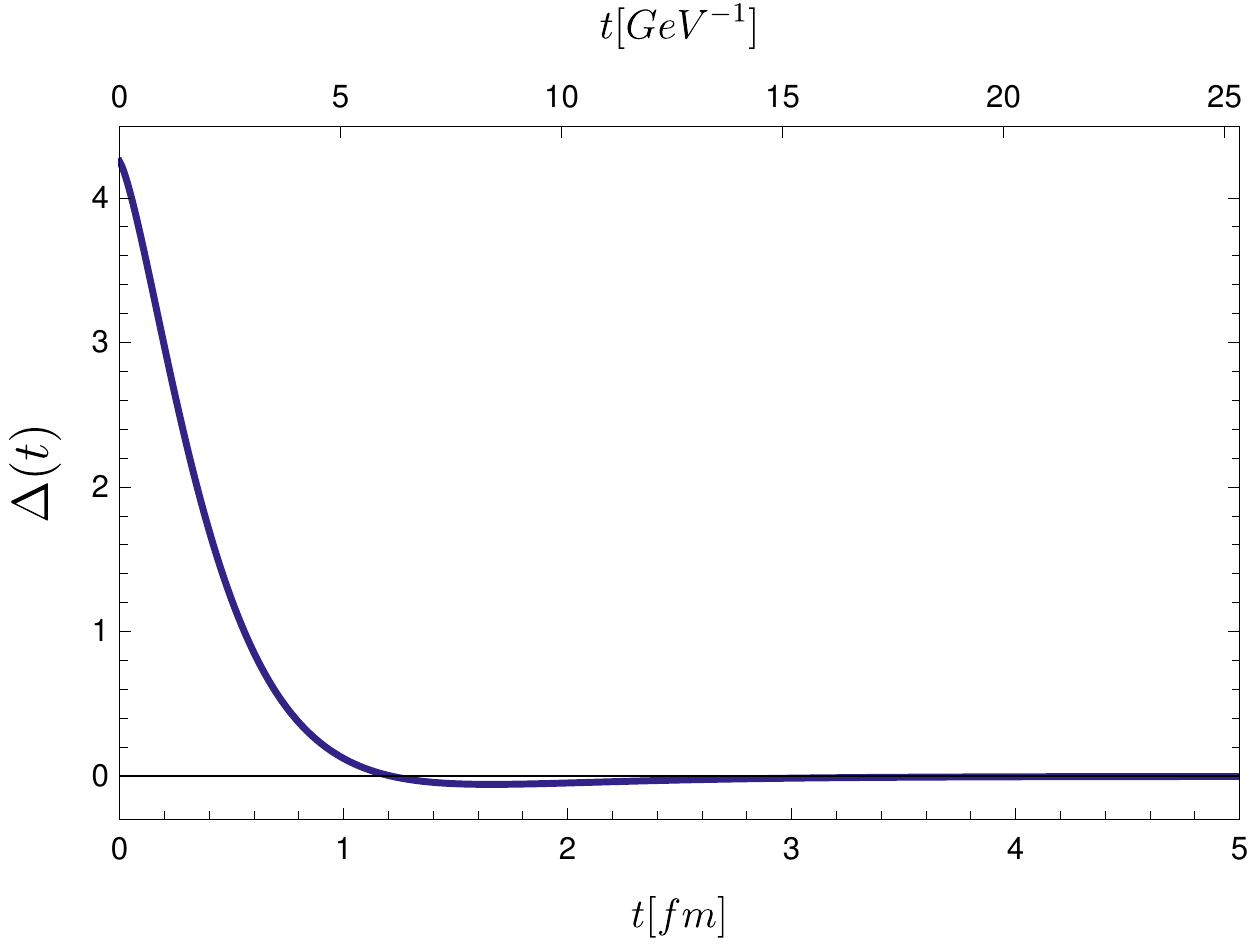}
  \caption{\label{fig:YM4d_gl_Schwinger}
  The Schwinger function of the gluon propagator of \fref{fig:YM4d_props_tl}.}
 \end{center}
\end{figure}

\index{Schwinger function}
The Schwinger function can be calculated from Euclidean data alone and thus provides an easy means of testing for positivity violation.
For the gluon propagator, the positivity violation in the Schwinger function was explicitly calculated with lattice and functional methods, e.g., \cite{vonSmekal:1997vx,Alkofer:2003jj,Cucchieri:2004mf,Maas:2004se,Bowman:2007du,Fischer:2008uz,Maas:2011se,Huber:2013xb,Huber:2020keu}.
The Schwinger function of the gluon propagator from \fref{fig:YM4d_props_tl} is shown in \fref{fig:YM4d_gl_Schwinger}.
Positivity violation is observed at a typical scale of about $1\,\text{fm}$.
For quarks, on the other hand, positivity violation is more subtle.
In particular, it depends on the employed quark-gluon interaction and also the quark masses, see, for example, Refs.~\cite{Alkofer:2003jj,Fischer:2018sdj}.

\subsubsection{Vertices of Yang-Mills theory}
\label{sec:res_verts}

In this section, results for the primitively divergent vertices of Yang-Mills theory, viz., the ghost-gluon, three-gluon and four-gluon vertices, as well as for two non-primitively divergent vertices, the two-ghost-two-gluon and four-ghost vertices, are discussed.

\paragraph{Ghost-gluon vertex}
\label{sec:res_ghg}

\index{ghost-gluon vertex}
Three-point functions have for a long time only been modeled.
This worked quite well for the ghost-gluon vertex, because using a bare vertex is close enough to the correct vertex that it is sufficient for many calculations.
Only once quantitative precision is required, its quantum corrections need to be added.
First semi-perturbative calculations using fits for the propagators recovered the qualitative behavior of the vertex actually quite well \cite{Schleifenbaum:2004id}.
Later dynamical calculations only showed some quantitative changes \cite{Huber:2012kd,Aguilar:2013xqa}.
The full momentum dependence of the vertex is known up to two loops \cite{Gracey:2014mpa}.
Further pertubative calculations can be found, for example, in Refs.~\cite{Chetyrkin:2000dq,Chetyrkin:2000fd,Davydychev:1996pb,Davydychev:1997vh,Davydychev:1998aw,Gracey:2011vw}.

\begin{figure}[tbp]
 \begin{center}
 \includegraphics[width=0.49\textwidth]{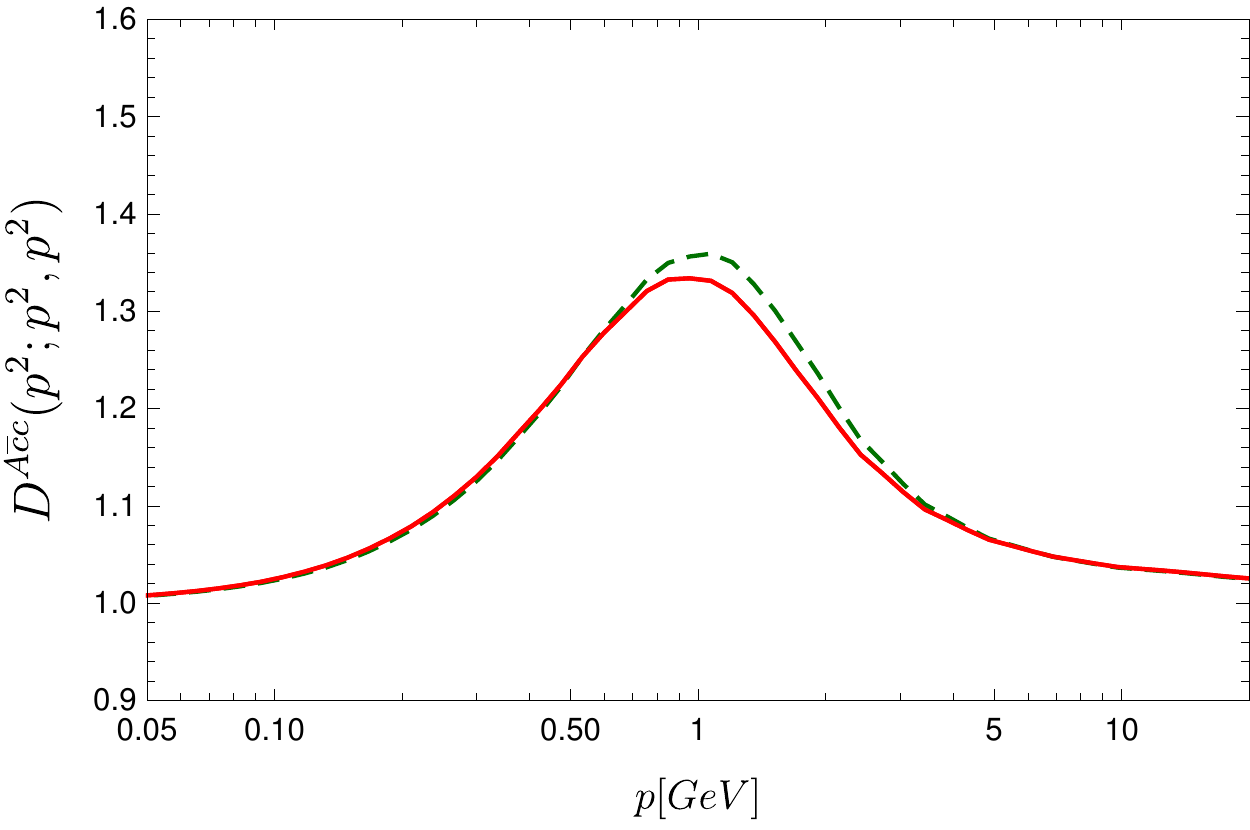}
 \hfill
 \includegraphics[width=0.49\textwidth]{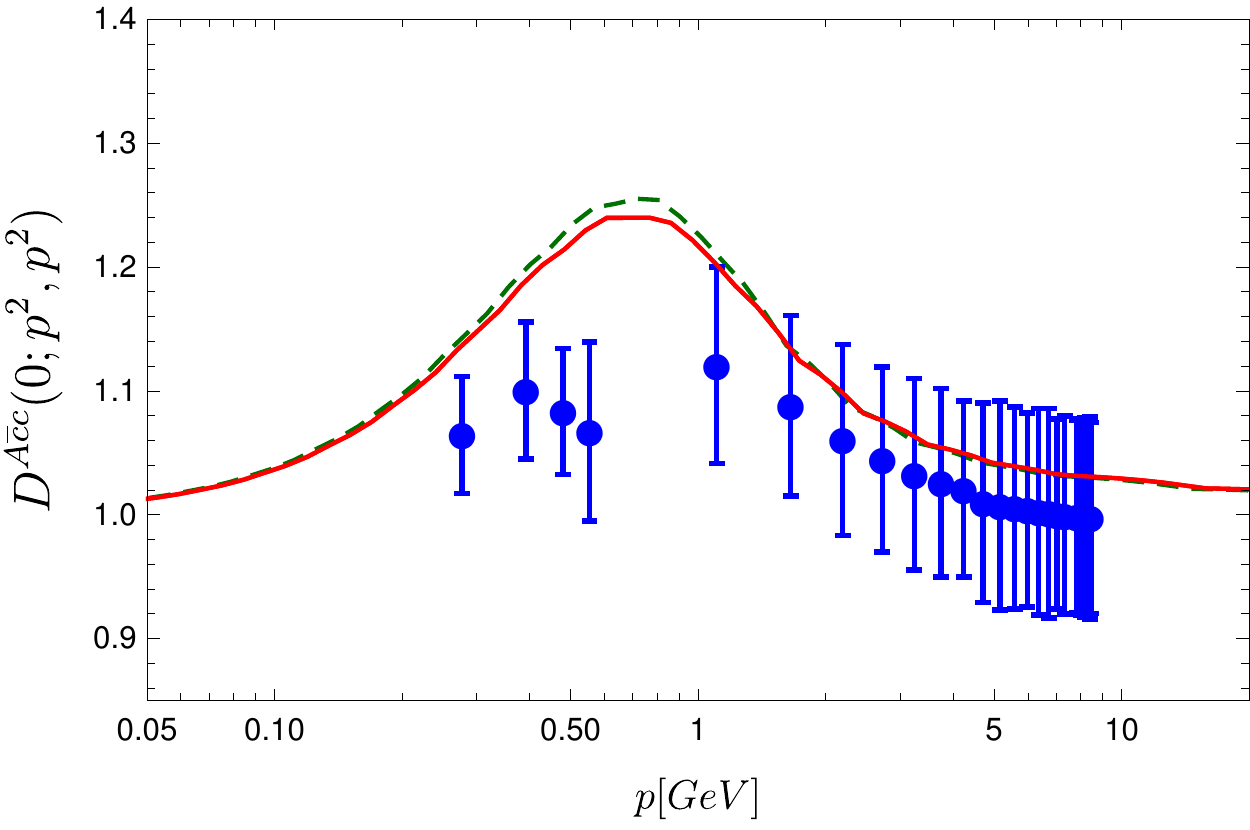}
 \caption{The ghost-gluon vertex at the symmetric point (left) and the soft gluon configuration (right) \cite{Huber:2012kd}.
 The latter is compared to lattice results from Ref.~\cite{Sternbeck:2006rd} (blue circles: $N=48$ at $\beta=6$).
 The two solutions shown were obtained with different models for the three-gluon vertex.
 }
 \label{fig:ghg_compL}
 \end{center}
\end{figure}

\begin{figure}[p]
 \begin{center}
 \includegraphics[width=0.48\textwidth]{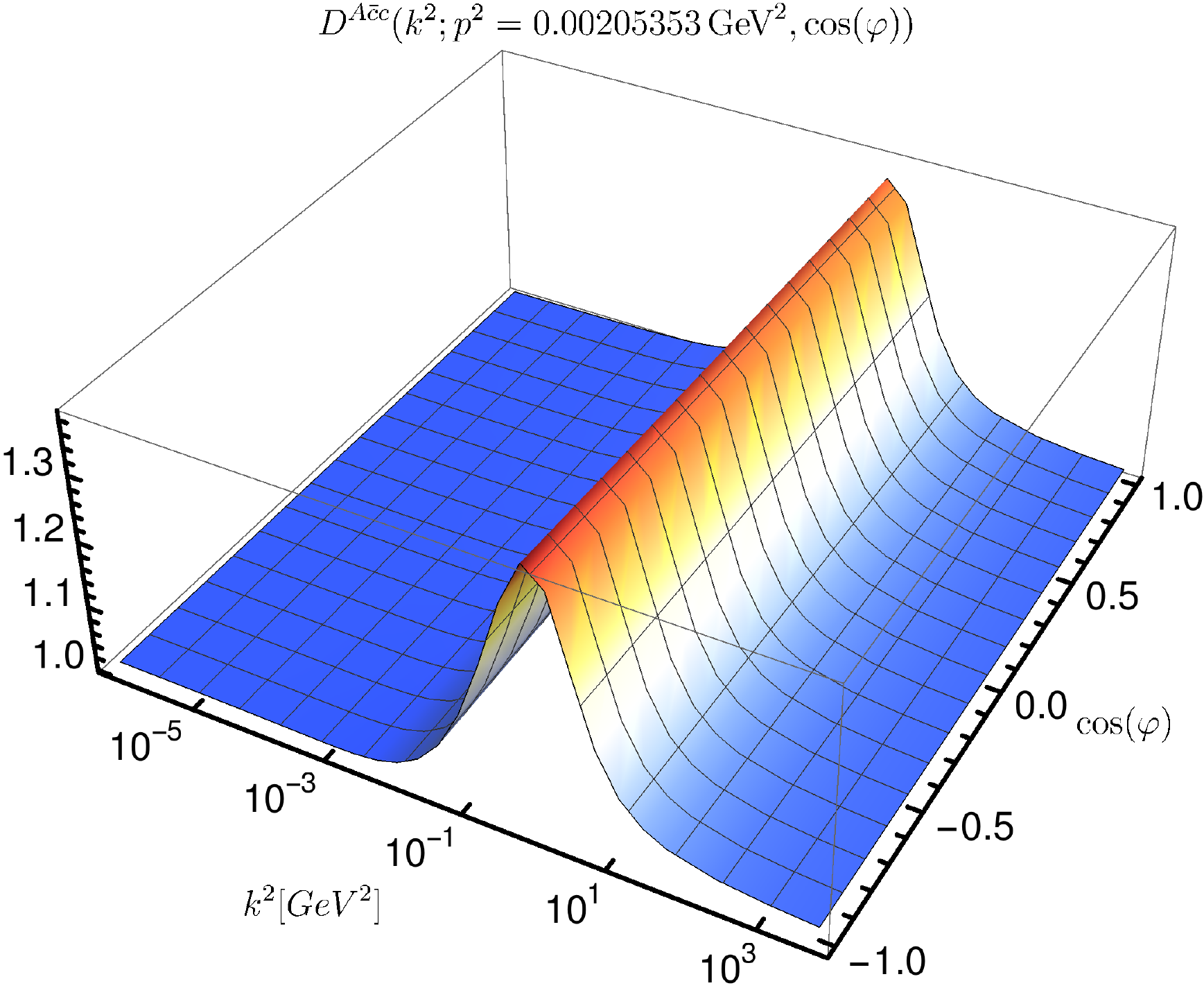}
 \hfill
 \includegraphics[width=0.48\textwidth]{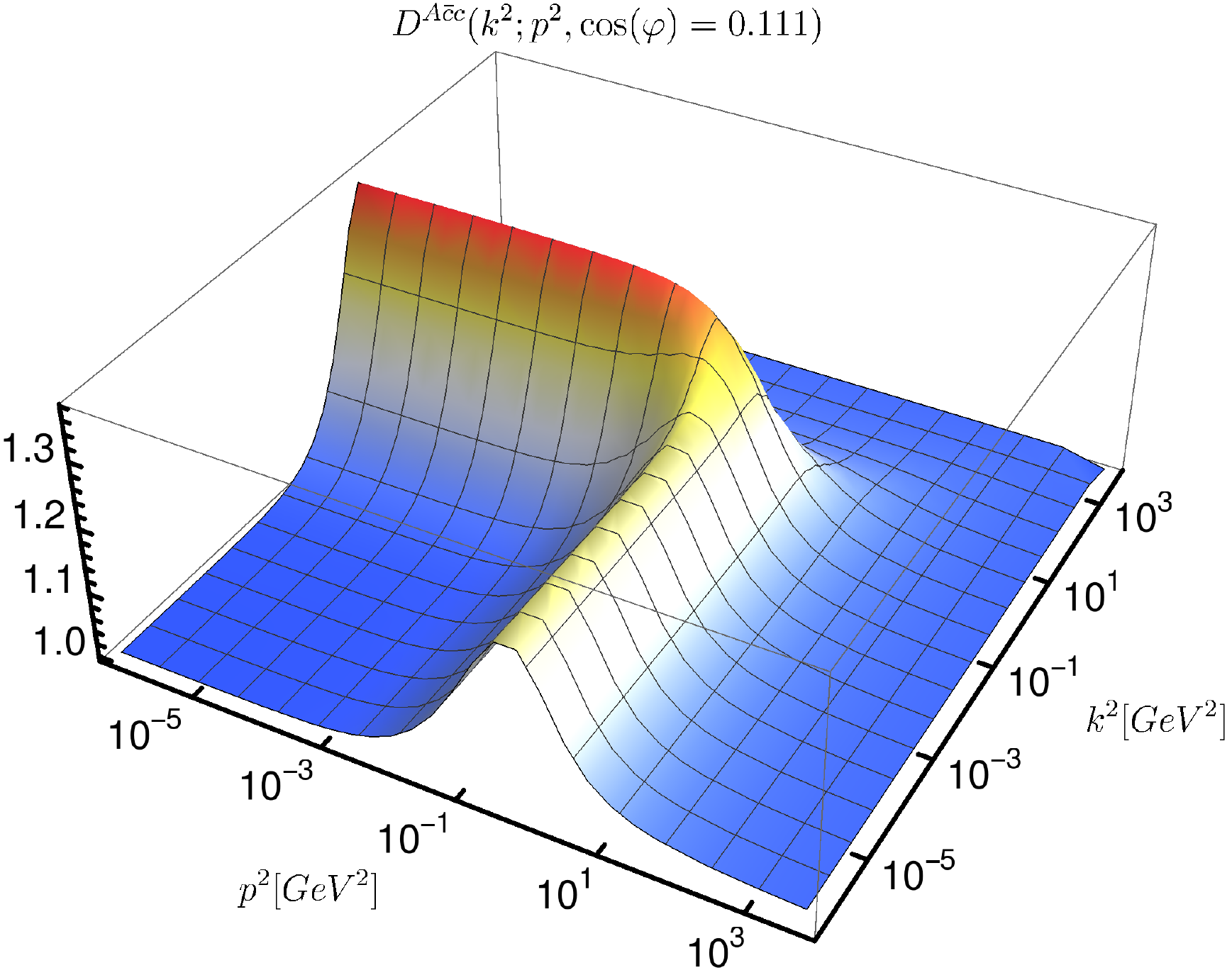} 
 \vskip3mm
 \includegraphics[width=0.48\textwidth]{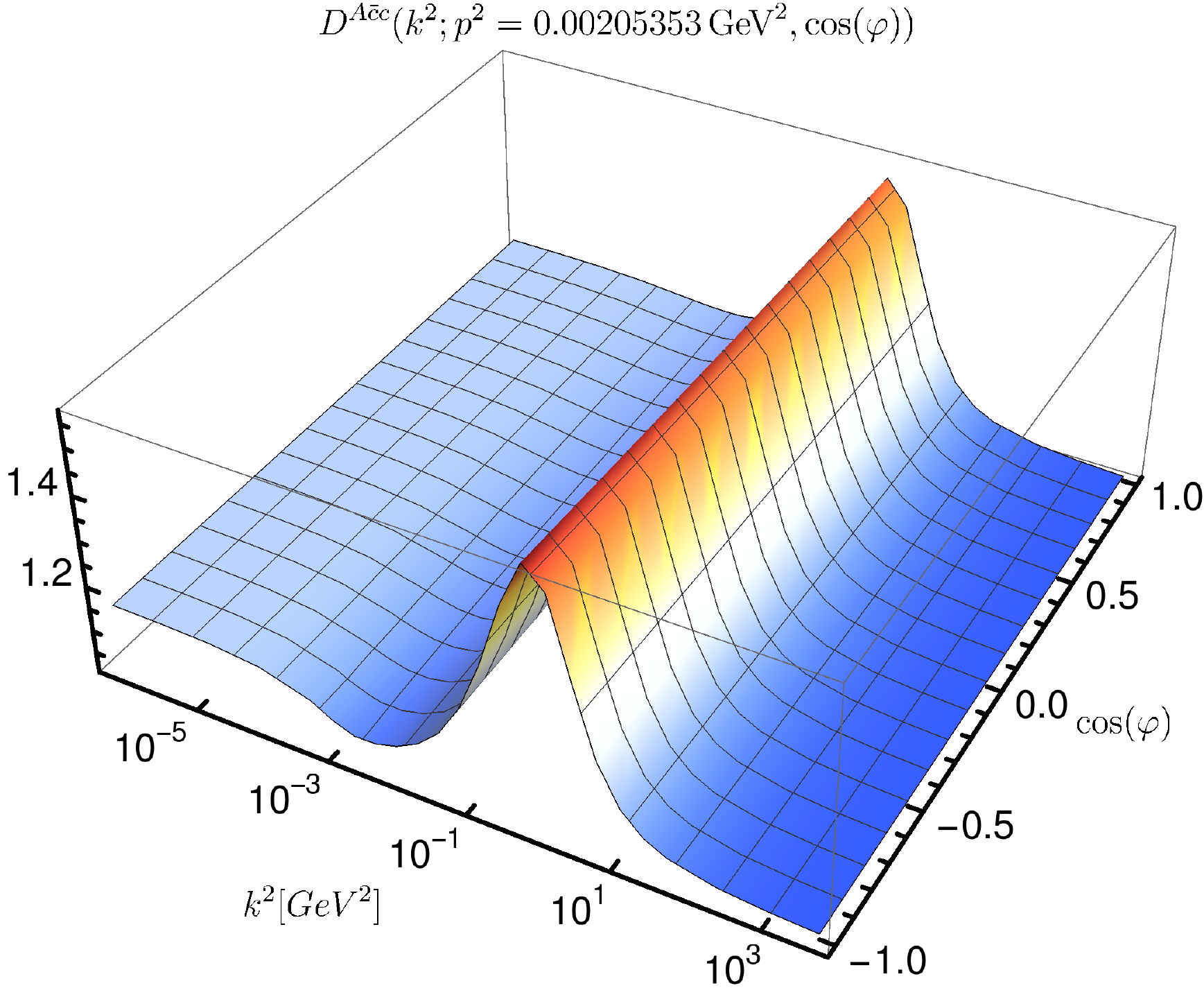}
 \hfill
 \includegraphics[width=0.48\textwidth]{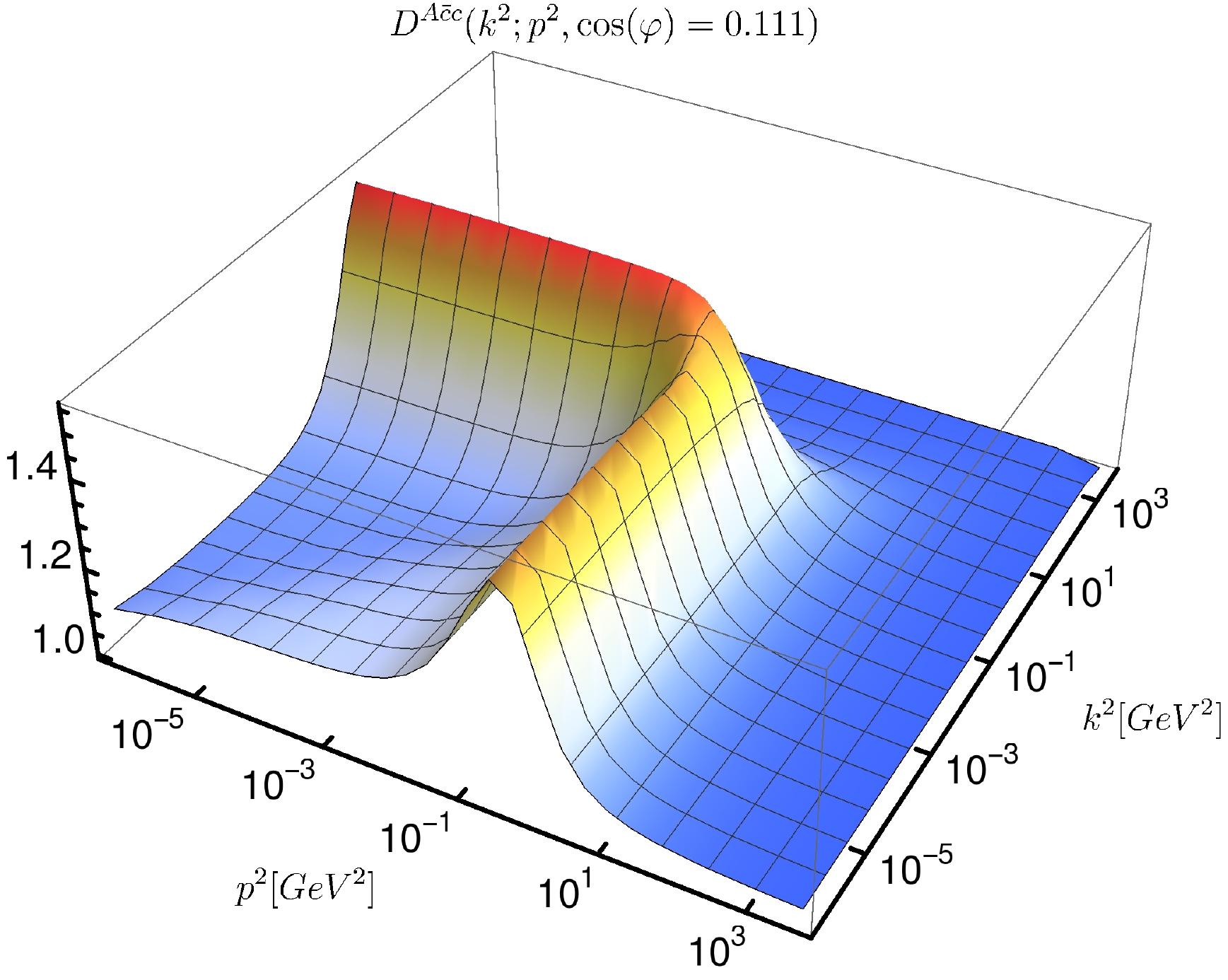} 
 \caption{Selected momentum configurations of the ghost-gluon vertex for a decoupling (top) and a scaling solution (bottom) \cite{Huber:2012kd}.
 Fixed momentum (left) or angle (right) as indicated at the top of the plots.}
 \label{fig:ghg_3d}
 \end{center}
\end{figure}

\index{ghost-gluon vertex}
The results from a combined calculation of propagators and the ghost-gluon vertex are shown in \fref{fig:ghg_3d}.
A comparison to lattice results is presented in \fref{fig:ghg_compL}.
The general features of available lattice results are reproduced, viz., there is a bump in the midmomentum regime and the dressing goes down to one at zero momentum.
A noteworthy anomaly concerns the scale of the bump for the gauge group $SU(2)$ \cite{Cucchieri:2008qm}.
All continuum results have the bump at a lower scale than the lattice results \cite{Huber:2012kd,Aguilar:2013xqa,Pelaez:2013cpa,Mintz:2017qri,Huber:2017txg,Aguilar:2018csq,Huber:2020keu}.

\paragraph{Three-gluon vertex}
\label{sec:res_tg}

\index{three-gluon vertex}
The three-gluon vertex is a crucial ingredient for the solution of the gluon propagator \gls{dse} as discussed in Sec.~\ref{sec:resummation}.
An important property of initial models was a \gls{uv} behavior that leads to a consistent \gls{uv} behavior of the gluon propagator \gls{dse} \cite{Fischer:2002eq}.
Lattice calculations in lower dimensions indicated later \cite{Cucchieri:2006tf,Cucchieri:2008qm} that the tree-level dressing of the three-gluon vertex could become negative in the \gls{ir}.
In first exploratory continuum studies it was found that the ghost triangle indeed becomes negative \cite{Schwenzer:2008pc}.
The first full calculation was then done for two dimensions and the zero crossing was confirmed \cite{Huber:2012zj}.
By now this feature was found in many approaches, e.g., \cite{Pelaez:2013cpa,Aguilar:2013vaa,Blum:2014gna,Huber:2014bba,Eichmann:2014xya,Vujinovic:2014fza,Alkofer:2014taa,Williams:2015cvx,Huber:2017txg,Aguilar:2019uob,Aguilar:2019jsj,Huber:2020keu}.
Also lattice methods find evidence of it \cite{Duarte:2016ieu,Athenodorou:2016oyh,Sternbeck:2016ltn,Boucaud:2017obn}, although the low position of the crossing makes a definite confirmation difficult.
Continuum methods show that the vertex diverges logarithmically in four dimensions \cite{Pelaez:2013cpa,Aguilar:2013vaa,Blum:2014gna,Eichmann:2014xya,Huber:2017txg,Aguilar:2019jsj,Huber:2020keu}.
The possibility of additional kinematic \gls{ir} singularities, viz., only one momentum goes to zero, was ruled out for the decoupling solutions \cite{Alkofer:2008jy,Fischer:2009tn}.
For the scaling solution they can be present, although they are weaker than the overall \gls{ir} ones \cite{Alkofer:2008jy,Alkofer:2008dt,Huber:2008mq,Fischer:2009tn}.
It should also be noted that the creation of a gluon mass gap is related to a potential \gls{ir} irregularity of the vertex \cite{Cornwall:1981zr,Boucaud:2007hy,Cyrol:2016tym}.
Perturbatively the full momentum dependence of the vertex is known up to two-loop order \cite{Gracey:2014mpa}.
More perturbative studies can be found, for example, in Refs.~\cite{Celmaster:1979km,Brandt:1985zz,Chetyrkin:2000dq,Chetyrkin:2000fd,Davydychev:1996pb,Davydychev:1997vh,Davydychev:1998aw,Gracey:2011vw,Gracey:2014mpa}.
String inspired methods for the decomposition of the vertex \cite{Ahmadiniaz:2012ie} are also known \cite{Ahmadiniaz:2012ir,Ahmadiniaz:2012xp,Ahmadiniaz:2016qwn,Ahmadiniaz:2012tf}.

\begin{figure}[tb]
\begin{center}
 \includegraphics[width=0.49\textwidth]{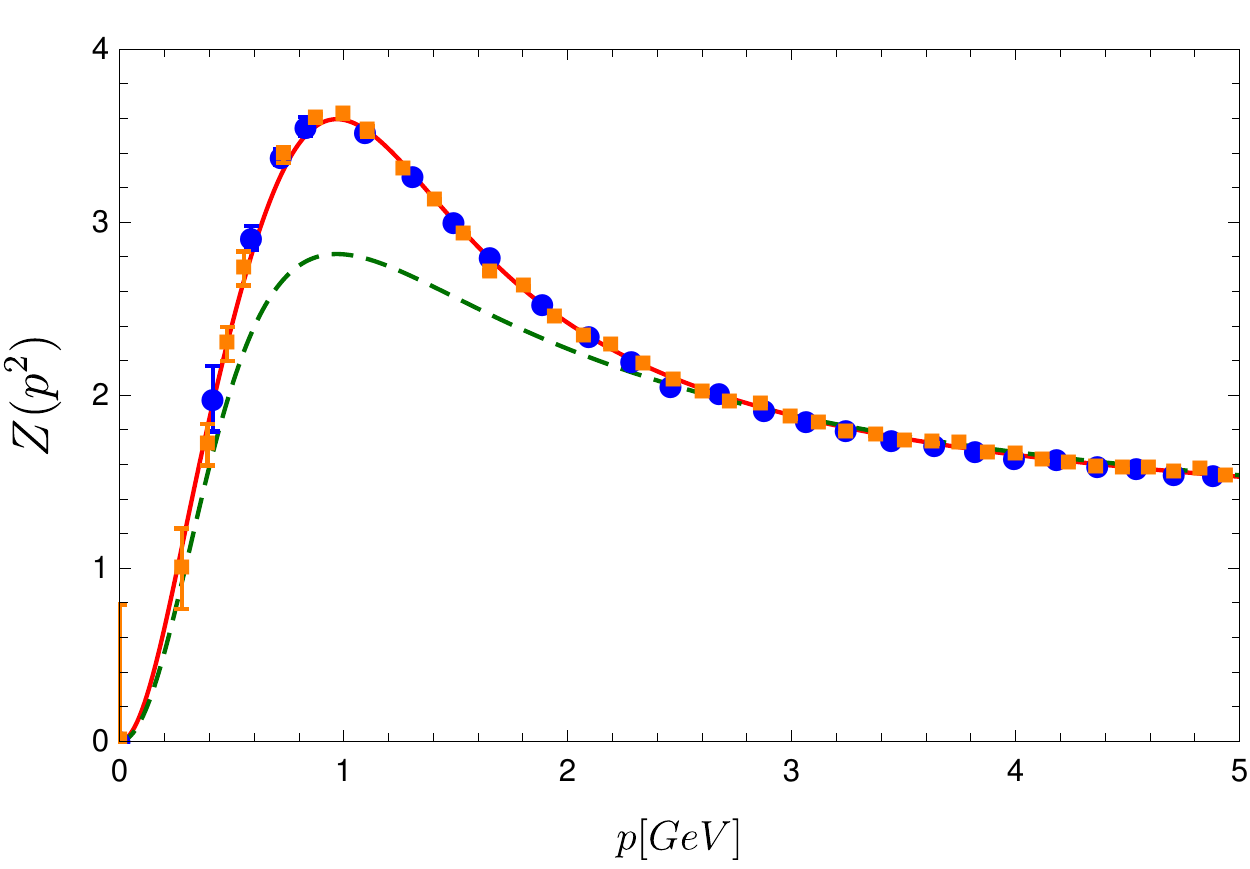}
 \hfill
 \includegraphics[width=0.49\textwidth]{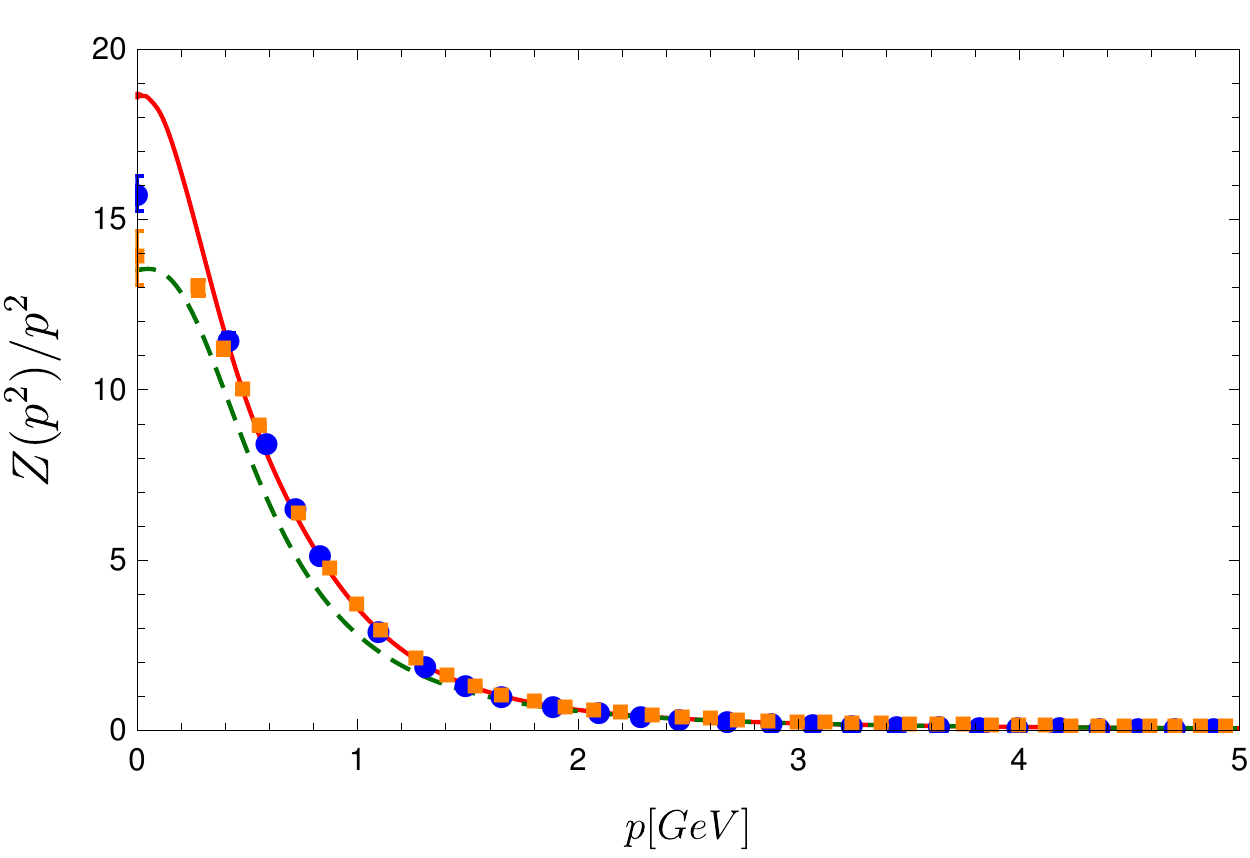}\\
 \caption{The gluon dressing function (left) and propagator (right) obtained with the effective three-gluon vertex model \eref{eq:3g_eff} with $\Lambda_\text{3g}=2.1\,\text{GeV}$ and a dynamic ghost-gluon vertex compared to lattice results \cite{Huber:2012kd}.
 The red/continuous line was obtained with the optimized effective three-gluon vertex.
 For comparison the green/dashed line is shown, which was obtained with the three-gluon vertex model of Ref.~\cite{Fischer:2002eq} and a bare ghost-gluon vertex.
 Lattice data is for $\beta=6$ and lattice sizes of $L=32$ (blue circles) and $L=48$ (orange squares) \cite{Sternbeck:2006rd}.}
 \label{fig:gl_eff3g}
\end{center}
\end{figure}

\begin{figure}[tb]
\begin{center}
 \includegraphics[width=0.49\textwidth]{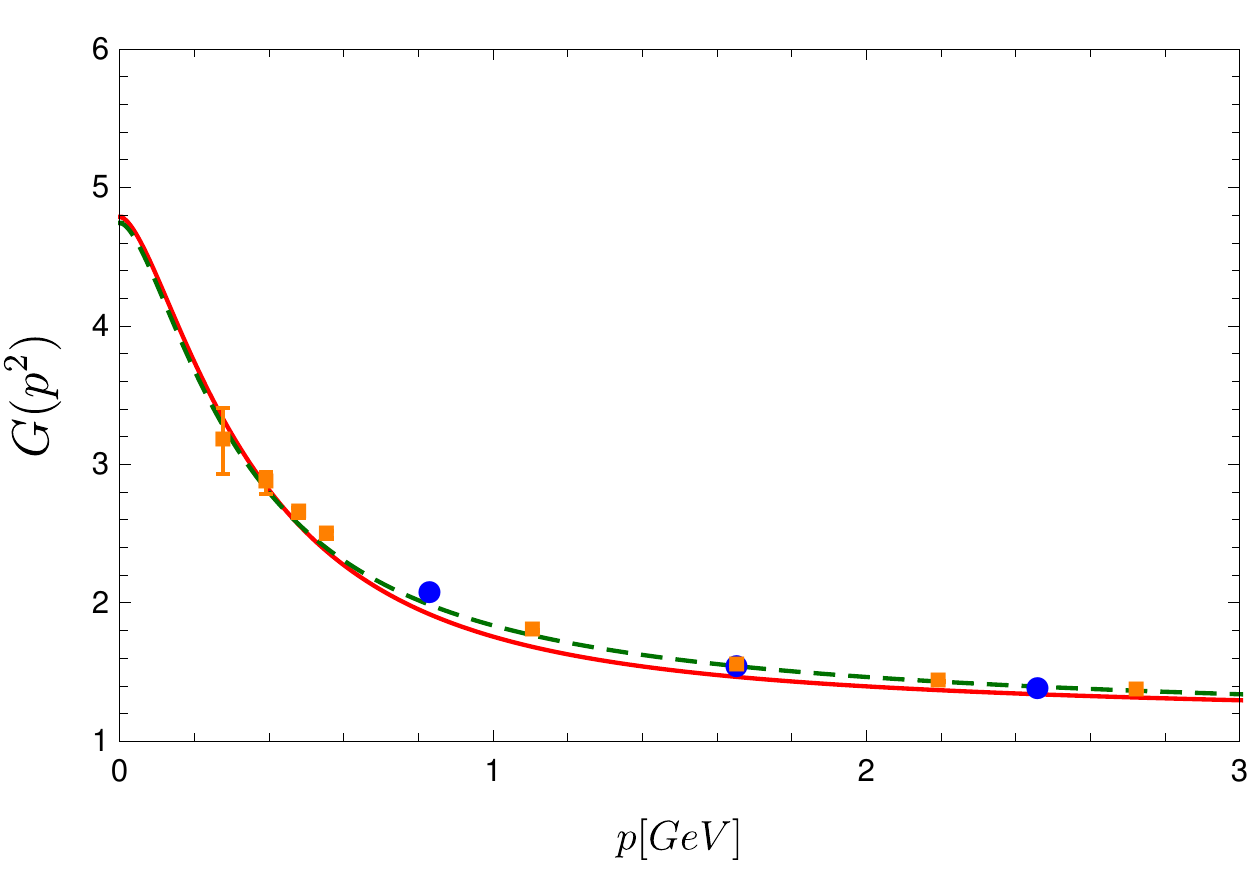}
 \caption{The ghost propagator corresponding to \fref{fig:gl_eff3g} \cite{Huber:2012kd}.}
 \label{fig:gh_eff3g}
\end{center}
\end{figure}

\index{three-gluon vertex}
A model that implements this zero crossing is given by \cite{Huber:2012kd}
\begin{align}\label{eq:3g-new}
 C^{AAA}(p,q,-p-q)=D^{AAA,IR}(p,q,-p-q)+D^{AAA,UV}(p,q,-p-q),
\end{align}
where the \gls{uv} part is (with $x=p^2$, $y=q^2$ and $z=(p+q)^3$)
\begin{align}\label{eq:3g-UV}
D^{AAA,UV}(p,q,-p-q)=G\left(\frac{x+y+z}{2}\right)^{\alpha}Z\left(\frac{x+y+z}{2}\right)^{\beta}
\end{align}
and the \gls{ir} part
\begin{align}\label{eq:3g_IR}
 D^{AAA,IR}(p,q,-p-q)=h_\text{IR} \,G(x+y+z)^{3}(f^{3g}(x)f^{3g}(y)f^{3g}(z))^4,
\end{align}
with the damping factors
\begin{align}
 f^{3g}(x):=\frac{\Lambda^2_\text{3g}}{\Lambda_\text{3g}^2+x}.
\end{align}
The exponents $\alpha$ and $\beta$ can be determined such as to reproduce the correct anomalous running and make the vertex finite at zero momentum.
For decoupling one obtains $\alpha=3+1/\de=-17/9$ and $\beta=0$ and for scaling $\alpha=-2-6\de=-17/22$ and $\beta=-1-3\de=-17/44$, see also below \eref{eq:DAAA_RG1}.
The parameter $h_\text{IR}$ is typically chosen as $-1$.
The scale $\Lambda_\text{3g}$ can then be used to move the zero crossing.
To obtain the correct anomalous dimension of the gluon propagator in a one-loop truncation, the final expression used for the three-gluon vertex is (see Sec.~\ref{sec:resummation_oneLoop} for details)
\begin{align}\label{eq:3g_eff}
 \Gamma^{AAA}(p,q,-p-q)=\frac{1}{Z_1}C^{AAA}(p,q,-p-q) \,D^{AAA,UV}(p,q,-p-q).
\end{align}
It turned out that this model can be used to obtain a gluon propagator in good agreement with lattice results.
However, it was clear that with the choice of parameters required for that agreement, the model can only be considered an effective model, because the zero crossing was at momenta so high that there was a clear disagreement with lattice results.
Nevertheless, results from this calculation, which are shown in Figs.~\ref{fig:gl_eff3g} and \ref{fig:gh_eff3g}, proved as useful input for subsequent studies of vertices.

\begin{figure}[tb]
 \includegraphics[width=\textwidth]{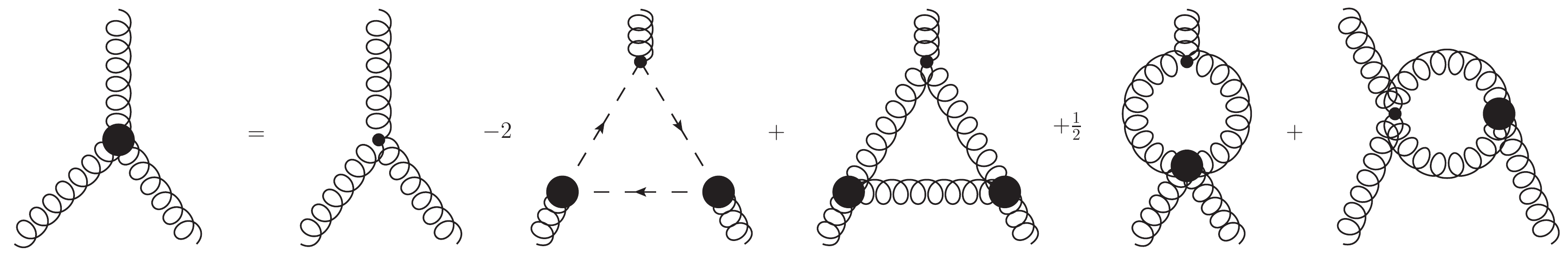}\\
 \begin{center}
 \caption{The truncated \gls{dse} of the three-gluon vertex.
 Due to manual symmetrization of the solution, modified prefactors of the swordfish diagrams are used and only two instead of three appear.}
 \label{fig:3g_DSE_trunc}
 \end{center}
\end{figure}

\begin{figure}[tb]
\begin{center}
 \includegraphics[width=0.47\textwidth]{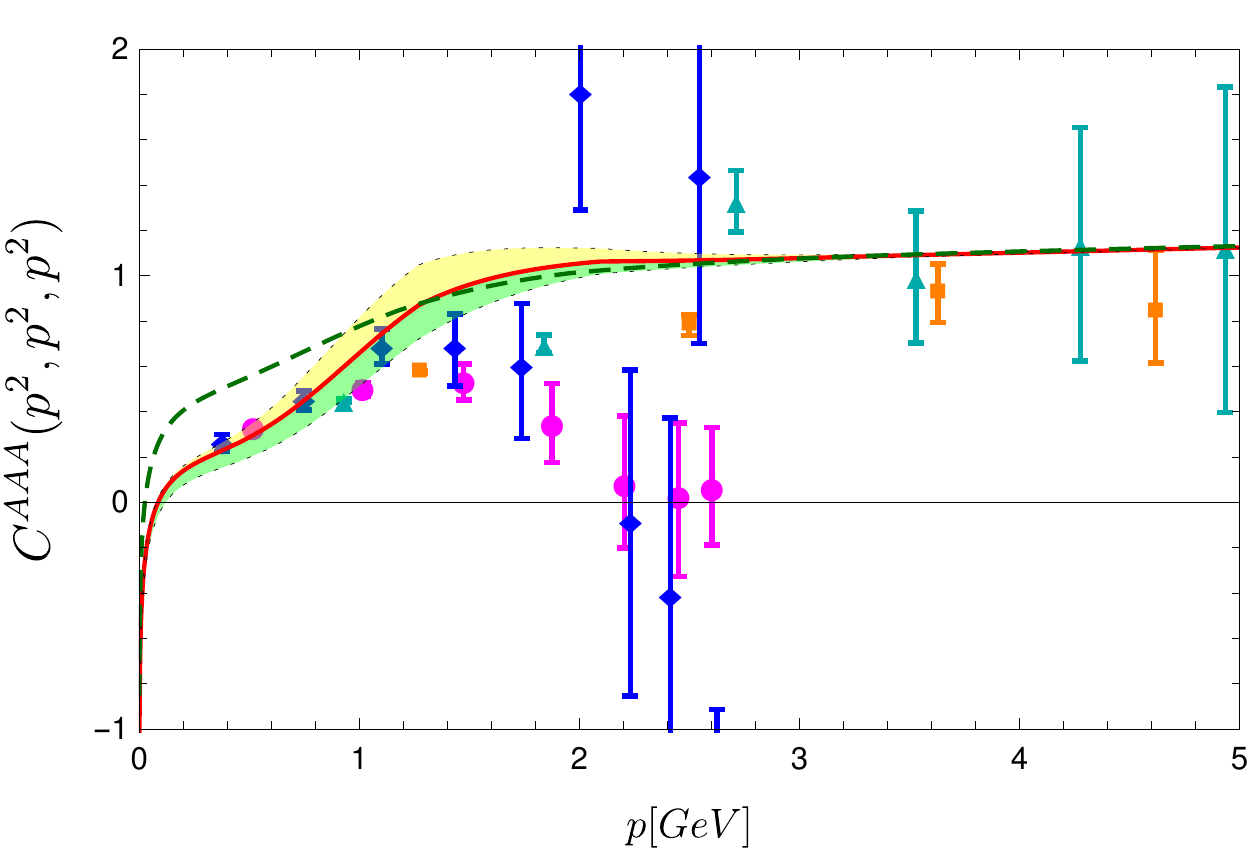}
 \hfill
 \includegraphics[width=0.47\textwidth]{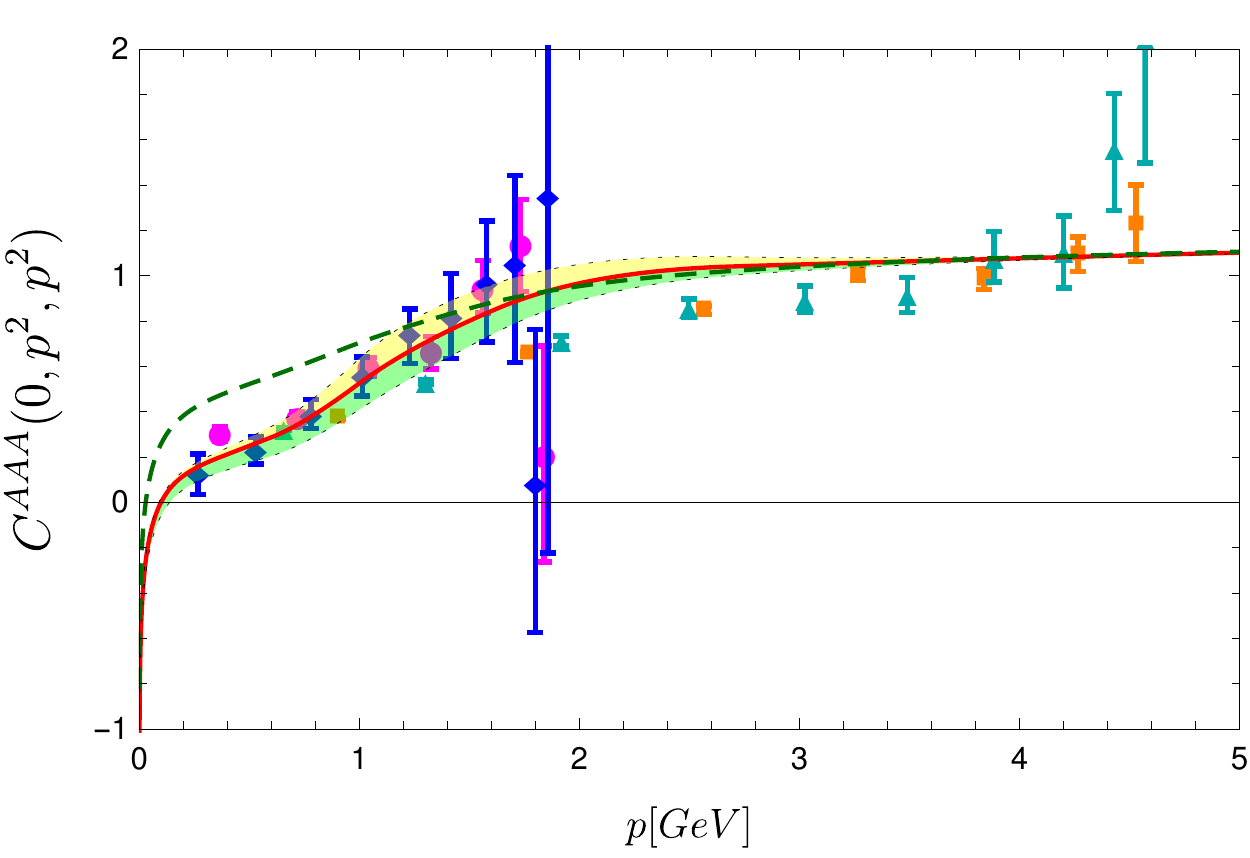}\\
 \includegraphics[width=0.47\textwidth]{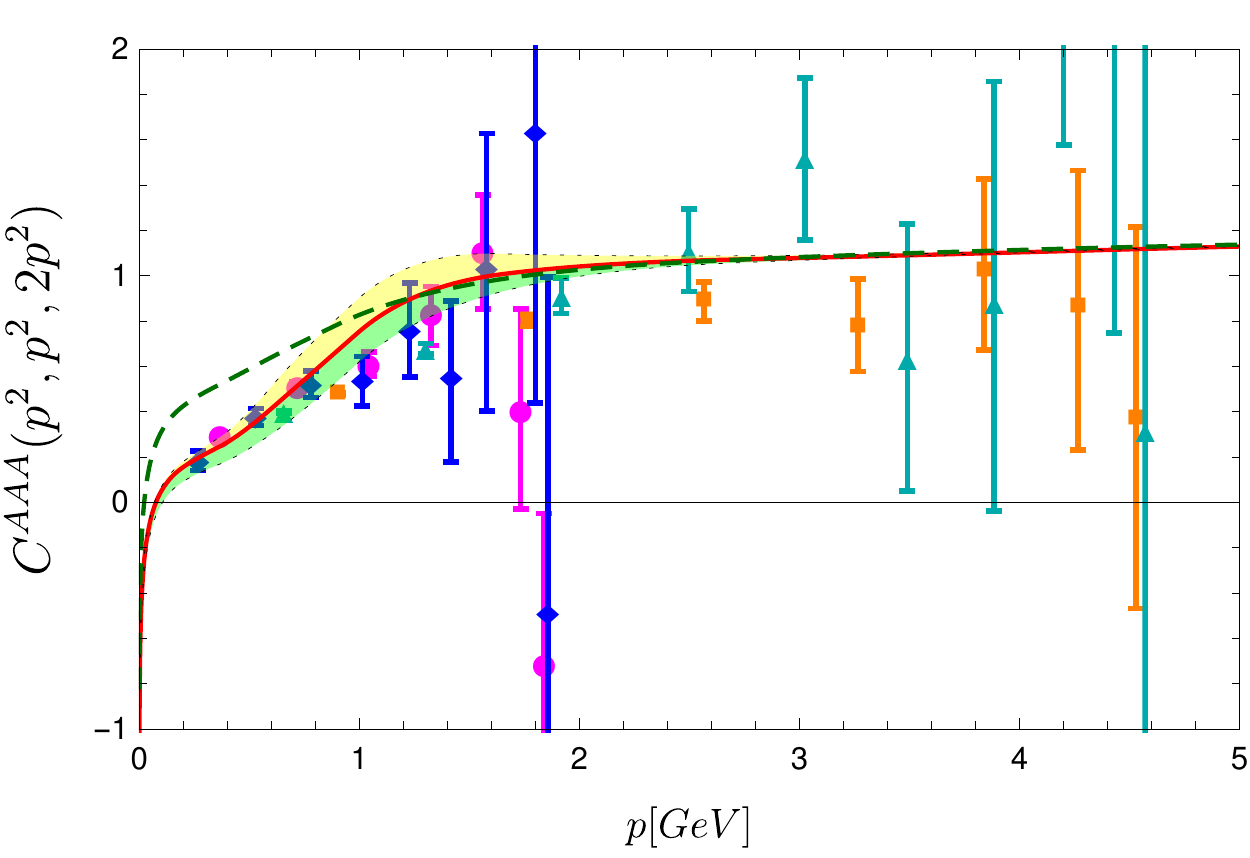}
 \hfill
 \includegraphics[width=0.47\textwidth]{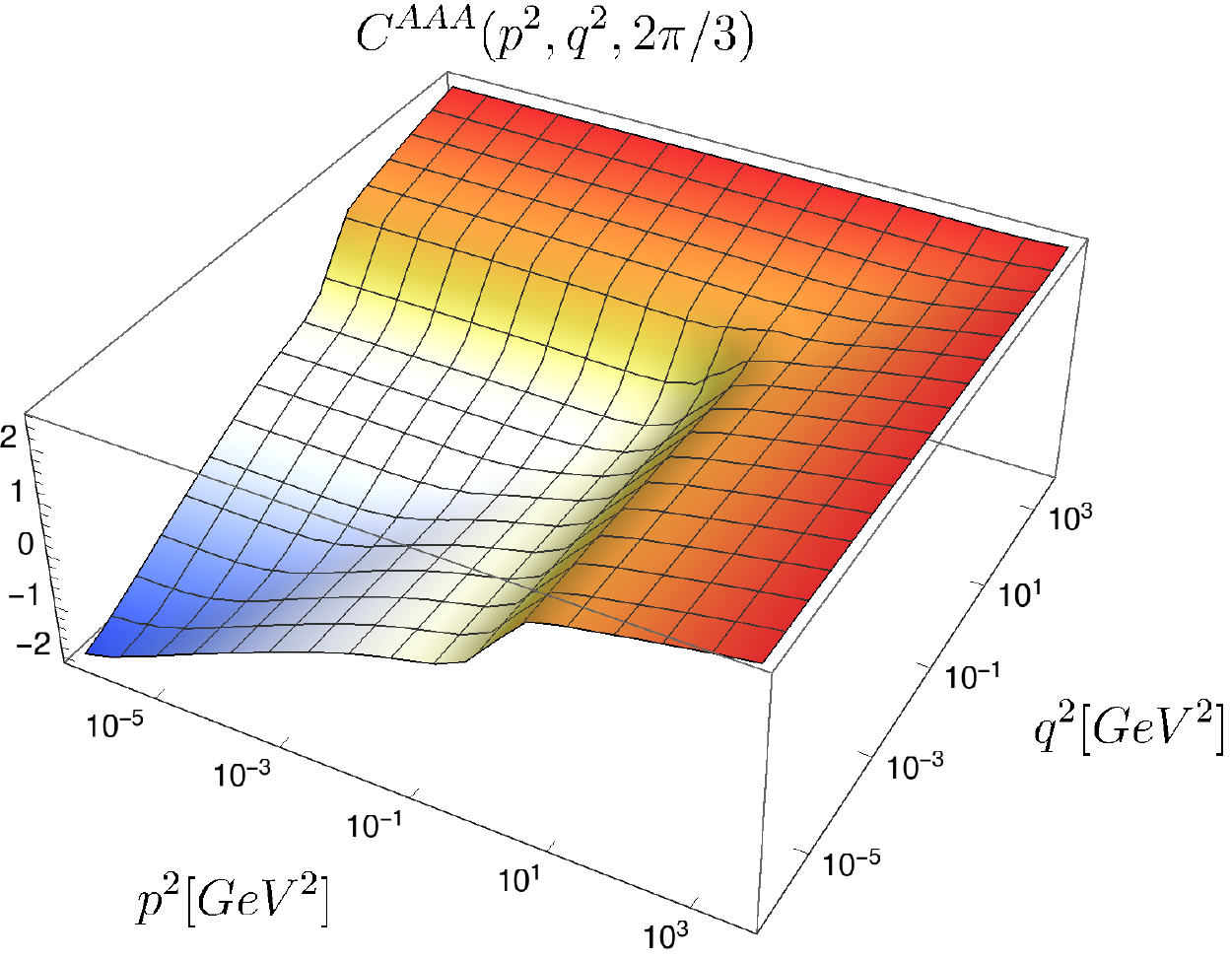}
 \caption{Three-gluon vertex dressing function with restricted kinematics as indicated in the plot labels \cite{Blum:2014gna}.
 For comparison, lattice data from Ref.~\cite{Cucchieri:2008qm} is shown.
 Different colors/symbols refer to different values of $\beta\in \{2.2, 2.5\}$ and different lattice sizes $1.4\,\text{fm}<L<4.7\,\text{fm}$.
 Solid red line: Standalone solution with the parameters $a=1.5$ and $b= 1.95$ GeV$^2$ in the four-gluon vertex model \eref{eq:4g-model}.
 Upper (yellow) band: Variation with $b$ down to $ 1.46$ GeV$^2$.
 Lower (green) band: Strengths up to $a=2$.
 Green dashed line: Solution of fully coupled system ($a=1.5$, $b=1.94$ GeV$^2$).
 The lower right plot shows the momentum dependence for a fixed angle $\varphi = 2\pi/3$.}
 \label{fig:3g_res}
\end{center}
\end{figure}

\index{three-gluon vertex}
First dynamical calculations of the three-gluon vertex were done in Refs.~\cite{Blum:2014gna,Eichmann:2014xya}.
Using a modern form of the gauge technique, the longitudinal parts of the three-gluon vertex were also calculated \cite{Aguilar:2019jsj}.
All these calculations confirmed the existence of the zero crossing.
A comparison with lattice results for different kinematic configurations is shown in \fref{fig:3g_res} \cite{Blum:2014gna}.
The employed truncation is shown in \fref{fig:3g_DSE_trunc}.
An important step when solving the equation is the manual symmetrization of the results.
Since the Bose symmetry of the vertex is broken by the truncation, one has to restore it by hand by averaging over all three distinct permutations of the external legs.
When this procedure is used, one can discard one swordfish diagram and multiply another one with $2$ as indicated in \fref{fig:3g_DSE_trunc}.

The results are shown in form of a band that was obtained by varying the parameters of the employed four-gluon vertex model given by
\begin{align}\label{eq:4g-model}
F^{AAAA}(p, q, r, s) =(a \tanh(b/\bar{p}^2)+1)\,  G\left(\bar{p}^2\right)^{\alpha_{4g}} Z\left(\bar{p}^2\right)^{\beta_{4g}}
\end{align}
where $\bar{p}^2=(p^2+q^2+r^2+s^2)/2$ and $a$ determines the additional IR interaction strength and $b$ the momentum scale of its onset.
The exponents $\alpha_{4g}$ and $\beta_{4g}$ are determined from the leading anomalous dimension of the four-gluon vertex, $\gamma_{4g}=2/11$, and from the requirement that the vertex approaches a constant value in the IR.
This yields $\alpha_{4g}=-8/9$ and $\beta_{4g}=0$ for decoupling and $\alpha_{4g}=-4/11$ and $\beta_{4g}=-2/11$ for scaling. 
The qualitative agreement of the tree-level dressing with lattice results is quite good and it was shown in Ref.~\cite{Eichmann:2014xya} that the other three dressing functions are smaller than the tree-level one.
The vertex results from \fref{fig:3g_res} were also fed back into the gluon propagator equation.
To fix the \gls{uv} behavior, the gluon loop was amended by \eref{eq:3g-UV} as \gls{rg} improvement term.
Assuming that the vertex results are close enough to the true form of the vertex, one can in this way estimate the importance of the missing two-loop diagrams in the gluon propagator \gls{dse}.
The gap of the results to the lattice results can be interpreted as an estimate of the missing contributions of the two-loop diagrams \cite{Blum:2014gna}.
In Sec.~\ref{sec:twoLoop}, the importance of two-loop diagrams was confirmed by a direct calculation.

\paragraph{Four-gluon vertex}
\label{sec:res_fg}

\index{two-ghost-two-gluon vertex}
\index{decoupling solution}
\index{scaling solution}
\index{four-gluon vertex}
The first truncations that were realized for the ghost-gluon and the three-gluon vertex \glspl{dse} involved models for the four-gluon vertex and set the other four-point functions to zero \cite{Schleifenbaum:2004id,Huber:2012zj,Huber:2012kd,Aguilar:2013xqa,Blum:2014gna,Eichmann:2014xya,Williams:2015cvx,Huber:2017txg}.
A motivation for discarding the latter is that the corresponding diagrams are of order $g^4$ and thus suppressed in the \gls{uv}.
The diagrams with a four-gluon vertex, on the other hand, contribute to the leading perturbative order.
It was also found that this vertex plays a role for the convergence of the three-gluon vertex \gls{dse} \cite{Blum:2014gna,Eichmann:2014xya,Huber:2017txg}.
Thus, a more detailed analysis of the four-gluon vertex is of interest.
Up to now this vertex was investigated with its \gls{dse} \cite{Kellermann:2008iw,  Kellermann:2009rh,Binosi:2014kka,Cyrol:2014kca,Huber:2015fna,Huber:2016tvc,Huber:2016hns,Huber:2017txg,Huber:2019wxx,Huber:2020keu}, the \gls{frg} \cite{Cyrol:2016tym,Cyrol:2017ewj,Corell:2018yil} and perturbatively up to one-loop for various kinematic configurations \cite{Brandt:1985zz,Pascual:1980yu,Gracey:2014ola,Gracey:2017yfi}, in particular the symmetric point \cite{Pascual:1980yu,Gracey:2014ola,Gracey:2017yfi}.
String inspired methods for the decomposition of the vertex \cite{Ahmadiniaz:2012ie} are also known \cite{Ahmadiniaz:2013rla,Ahmadiniaz:2016qwn}.

\begin{figure}[tb]
\begin{center}
 \includegraphics[width=0.48\textwidth]{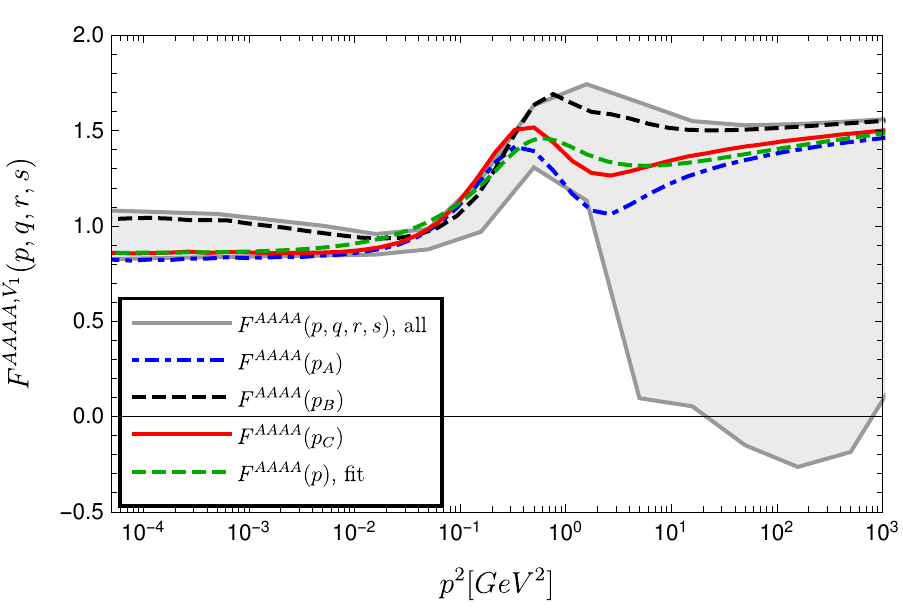}
 \includegraphics[width=0.48\textwidth]{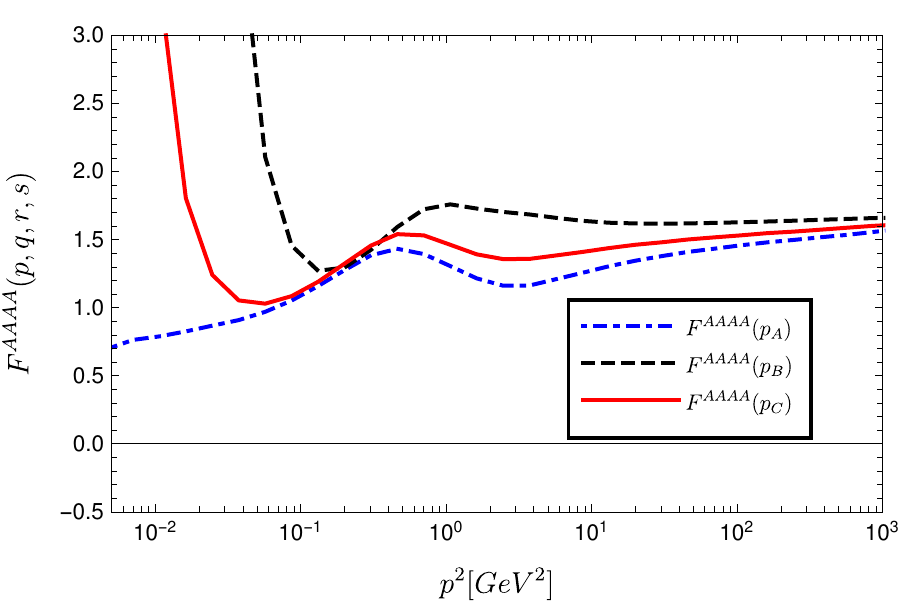} 
 \caption{Four-gluon vertex dressing function for three momentum configurations \cite{Cyrol:2014kca}.
 The angle dependence is indicated by the colored gray area.
 The dashed line corresponds to a fit \cite{Cyrol:2014kca}.
 Left: Decoupling solution.
 Right: Scaling solution.}
 \label{fig:4g_3confs}
\end{center}
\end{figure}

\begin{figure}[tb]
\begin{center}
 \includegraphics[width=0.45\textwidth]{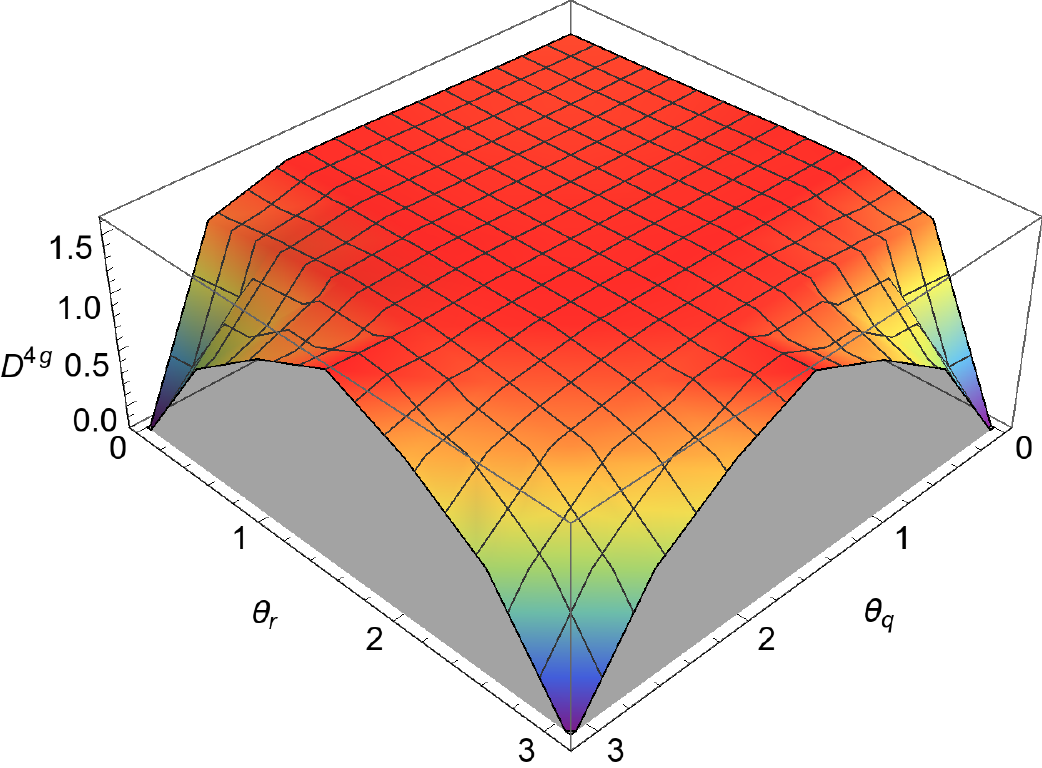}
 \includegraphics[width=0.45\textwidth]{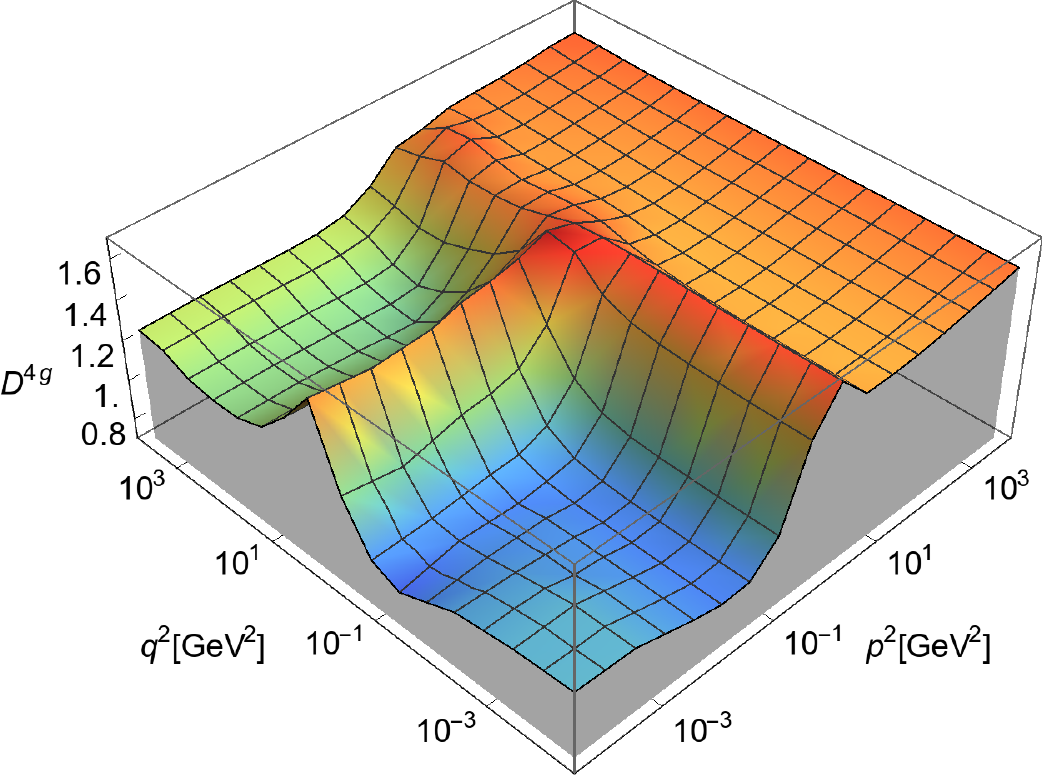}
 \caption{Left: The angular dependence of the dressing function for four variables fixed:
 The squared momenta and the angle \(\psi_q\) are kept constant: \(S^2=R^2=Q^2=160\,\GeV^2\), \(\psi_q=0\).
 The shown configuration corresponds to the point with the largest angle dependence found.
 Right: Momentum dependence of the dressing function: \(S^2=R^2=p^2\), \(Q^2=q^2\), \(\theta_r=\theta_q=\psi_q=\pi/2\).
 Both plots are for decoupling \cite{Cyrol:2014kca}.}
 \label{fig:4g_3d}
\end{center}
\end{figure}

\begin{figure}[tb]
\begin{center}
 \includegraphics[width=0.48\textwidth]{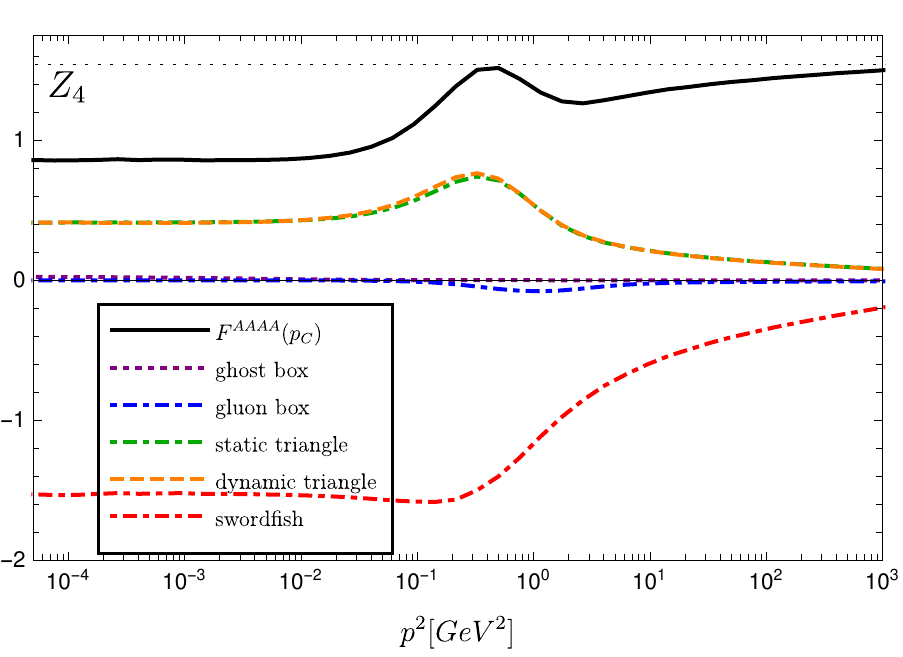}
 \includegraphics[width=0.48\textwidth]{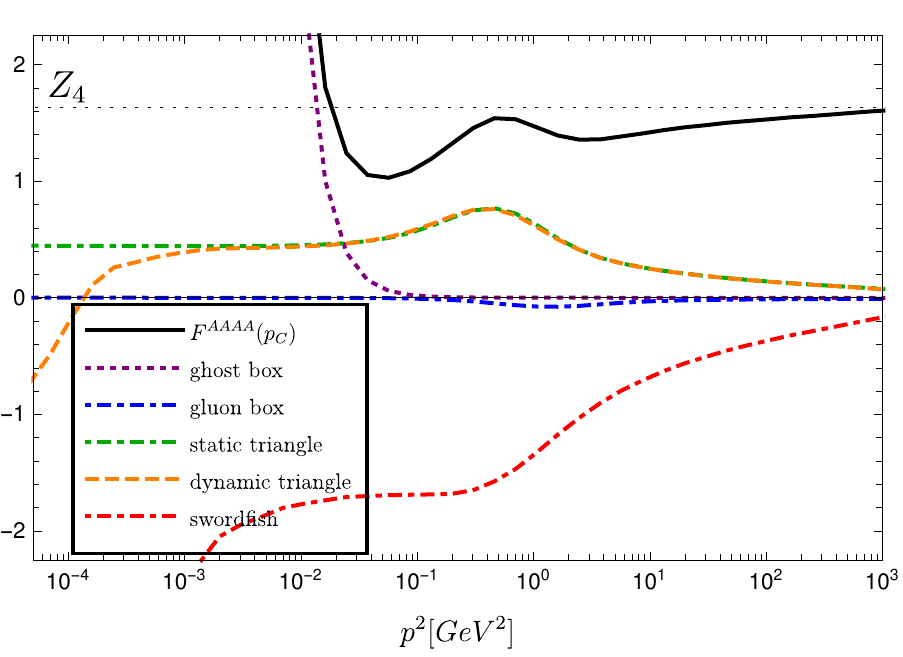}
 \caption{Contributions of individual diagrams to the tree-level dressing function for a specific momentum configuration for the decoupling (left) and scaling solutions (right) \cite{Cyrol:2014kca}.}
 \label{fig:4g_diagrams}
\end{center}
\end{figure}

\index{four-gluon vertex}
The four-gluon vertex is the most complicated four-point function, because it not only has four color indices but also four Lorentz indices as detailed in Sec.~\ref{sec:AAAA}.
Its \gls{dse} has 60 diagrams.
39 of them are two-loop and 20 are one-loop diagrams.
For the presently employed truncation, all of the former are dropped as are all one-loop diagrams containing the two-ghost-two-gluon vertex (3 diagrams) or five-point functions (2 diagrams).
This leaves 15 one-loop diagrams and the tree-level diagram.
The impact of the discarded diagrams with two-ghost-two-gluon vertices is discussed in Sec.~\ref{sec:res_AAcbc}.
However, the Bose symmetry of the four-gluon vertex allows reducing the number of diagrams to be calculated even further.
If the complete kinematic dependence of a diagram is known, one can extract also the information about the diagrams obtained by permutations of legs.
Thus, instead of calculating similar diagrams several times, only a single calculation is necessary.
In total, five diagrams need to be calculated as can be seen in \fref{fig:4g_DSE}, where also the other one-loop diagrams are shown.

\index{four-gluon vertex}
\index{Bose symmetry}
The employed truncation violates Bose symmetry.
This can be repaired by averaging over all four distinct four-gluon vertex \glspl{dse}.
Finally, there are twelve diagrams of each type which correspond to all possible permutations and the \gls{dse} is schematically calculated as
\begin{align}
  & D^{\text{4g}}(p,\,q,\,r,\,s)=\\
  &\quad Z_4 + \frac{1}{4} \Big[      L(p,\,q,\,r,\,s) +      L(q,\,r,\,s,\,p) +     L(r,\,s,\,p,\,q) +      L(s,\,p,\,q,\,r) +\\&\quad\quad
    L(p,\,r,\,s,\,q) +      L(q,\,s,\,p,\,r) +      L(r,\,p,\,q,\,s) +      L(s,\,q,\,r,\,p) +\\&\quad\quad
    L(p,\,s,\,q,\,r) +      L(q,\,p,\,r,\,s) +      L(r,\,q,\,s,\,p) +      L(s,\,r,\,p,\,q)
  \Big].
\end{align}
Here, $L(p,q,r,s)$ is the transverse projection of the one-loop diagrams $\Lambda^{abcd}_{\mu\nu\rho\sigma}$ contracted with the tree-level tensor $\Gamma^{{(0)},abcd}_{\mu'\nu'\rho'\sigma'}$.
The normalization is chosen such that the tree-level corresponds to one:
\begin{equation}
L(p,\,q,\,r,\,s) := 
\frac{
	\Lambda^{abcd}_{\mu\nu\rho\sigma} \; P^\text{T}_{\mu\mu'} \, P^\text{T}_{\nu\nu'} \, P^\text{T}_{\rho\rho'} \, P^\text{T}_{\sigma\sigma'} \, \Gamma^{{(0)},abcd}_{\mu'\nu'\rho'\sigma'}
}{
	\Gamma^{{(0)},efgh}_{\alpha\beta\gamma\delta} \; P^\text{T}_{\alpha\alpha'} \, P^\text{T}_{\beta\beta'} \, P^\text{T}_{\gamma\gamma'} \, P^\text{T}_{\delta\delta'} \, \Gamma^{{(0)},efgh}_{\alpha'\beta'\gamma'\delta'}
}.
\end{equation}
As another approximation, only the tree-level tensor out of the 41 transverse dressing functions is considered.

\index{four-gluon vertex}
For solving the four-gluon vertex \gls{dse} in this approximation, fixed input given by the propagators from Refs.~\cite{Huber:2012kd,Huber:2014tva}, a bare ghost-gluon vertex and the three-gluon vertex from \cite{Blum:2014gna,Cyrol:2014kca} is used.
Both decoupling and scaling type solutions are considered and the results are shown in \fref{fig:4g_3confs} for three different kinematic configurations.
The shaded area indicates the angle dependence which seems to be quite large.
Fig.~\ref{fig:4g_3d}, however, shows that the angle dependence is in general very mild and only at the boundaries a strong dependence can be observed.
As can be seen in \fref{fig:4g_3confs}, the dressing is very close to the tree-level.
Thus, it is interesting to note that the contributions of the individual diagrams are not small by themselves but they largely cancel out as can be inferred from the results of the individual diagrams shown in \fref{fig:4g_diagrams}.

\begin{figure}[tb]
\begin{center}
 \includegraphics[width=0.48\textwidth]{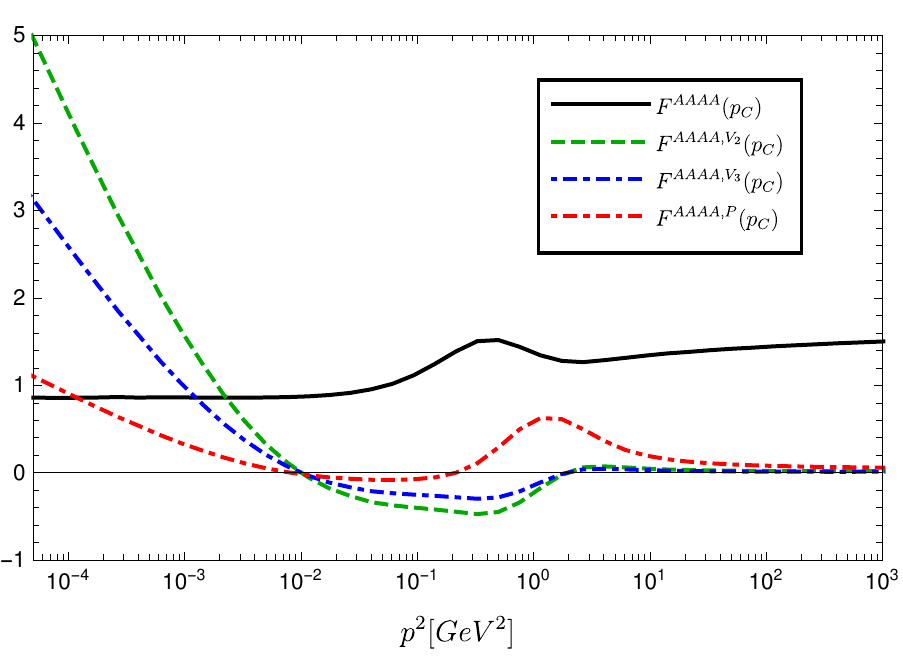}
 \caption{Comparison of the dressing functions $D^{\text{4g}}$, $D^{\text{4g},V_2}$, $D^{\text{4g},V_2}$ and $D^{\text{4g},P}$ for a specific momentum configuration \cite{Cyrol:2014kca}.}
 \label{fig:4g_dressings}
\end{center}
\end{figure}

Since the four-gluon vertex has many dressings, it is interesting to investigate if some of them are sizable.
For a first impression, one can evaluate alternative projections of the four-gluon vertex \gls{dse} using the obtained results as input.
The result for three projections is shown in \fref{fig:4g_dressings}.
As can be seen, they are not larger than the corrections to the tree-level projection.
Thus, within the tested set of tensors, the tree-level tensor is dominant due to its constant contribution from the bare vertex.

\paragraph{Two-ghost-two-gluon vertex}
\label{sec:res_AAcbc}

\myboxmargin{Why are we interested in the AAcbc? Where does it appear?}
\index{two-ghost-two-gluon vertex}
Landau gauge Yang-Mills theory has two non-primitively divergent four-point functions, the two-ghost-two-gluon and the four-ghost vertices.
For the scaling type solution it is known that diagrams with the two-ghost-two-gluon vertex can belong to the class of \gls{ir} leading diagrams \cite{Huber:2007kc}.
Neglecting them does not change the scaling laws themselves, but in a quantitative study, the contributions of these diagrams might be important.
If this is true for scaling, one might expect that for decoupling they might play a similar role.
Thus, a detailed study of their influence on lower correlations functions is warranted.
Perturbatively this vertex was studied in Ref.~\cite{Gracey:2017yfi} and with its \gls{dse} in Ref.~\cite{Huber:2017txg}.

\index{two-ghost-two-gluon vertex}
The two-ghost-two-gluon vertex, being non-primitively divergent, does not appear in the Lagrangian density.
Thus, its lowest perturbative order is one order higher than that of the four-gluon vertex.
Its impact in the midmomentum regime is hardly studied due to the technical complexity of corresponding calculations.
However, its structure reveals a few simplifications that alleviate its study.
This also includes the fact that its \glspl{dse} can be truncated such that almost all contributions are retained as discussed below.

\myboxmargin{Which DSE for AAcbc?}
\index{two-ghost-two-gluon vertex}
As the ghost-gluon vertex, the two-ghost-two-gluon vertices has formally three different \glspl{dse}.
The two with the ghost and anti-ghost legs attached to the bare vertex can be transformed into each other by exchanging ghost and anti-ghost.
They yield identical equations.
The third equation, with the gluon leg attached to the bare vertices, is the more complex equation due to the appearance of two-loop diagrams.
Thus, the equation with the ghost attached to the bare vertices is used here, see \fref{fig:2gh2gl_DSE}.
The employed truncation discards only one diagram, namely the one containing a five-point function.

\myboxmargin{basis for AAcbc}
\index{two-ghost-two-gluon vertex}
The study of the two-ghost-two-gluon vertex is complicated by the absence of a tree-level tensor which is a good first approximation of the full tensor bases in case of the three- and the four-gluon vertices \cite{Eichmann:2014xya,Cyrol:2014kca,Huber:2016hns}.
A possible guide could be the fact that the anti-symmetric structure constants contracted with the vertex in the corresponding diagrams of the ghost-gluon and three-gluon vertex \glspl{dse} annihilate the color part symmetric with respect to the two corresponding legs.
However, in the generation of the relevant dressing functions the complete color space may play a role.
Thus, the full basis as given in \eref{eq:AAcbc} is used.

\myboxmargin{Calculational details of the two-ghost-two-gluon vertex}
\index{two-ghost-two-gluon vertex}
The kinematic dependence of the two-ghost-two-gluon vertex is approximated by a single momentum scale.
Given the size of angle dependences observed in three-point functions, see, for example, \fref{fig:ghg_3d}, the error induced in the two-ghost-two-gluon vertex, which is smaller in magnitude than the three-point functions, is expected to be small.
The advantage of this approximation is that calculations are much faster compared to calculations of the full kinematic dependence with six variables \cite{Cyrol:2014kca}.
The chosen kinematic configuration is the average of the squared momenta:\footnote{The same variable name $\overline{p}$ is used for all $n$-point functions, since it is clear from the context the average of which momentum squares is meant.}
\begin{align}
 \overline{p}^2=\frac{p^2+q^2+r^2+s^2}{4}.
\end{align}
To calculate the 40 individual dressing functions $D^{AA\bar cc}_k(\overline p^2)$, the \gls{dse} of the two-ghost-two-gluon vertex is contracted with the basis tensors yielding 40 scalar equations.
Due to the non-orthogonal basis, the dressing functions are extracted by multiplication with the corresponding rotation matrix $R^{-1}$, where $R$ is given by
\begin{align}
 R^{kl}=  \rho_{\mu\nu}^{k,abcd} \rho_{\mu\nu}^{l,abcd}.
\end{align}

\myboxmargin{behavior of the AAcbc}
\index{two-ghost-two-gluon vertex}
The chosen basis leads to dimensionful dressing functions.
For plotting, it is advantageous to work with the dimensionless dressing functions $\overline{D}^{AA\bar cc}_k(\overline p^2)$ instead by multiplying with appropriate power laws:
\begin{align}
 \overline{D}^{AA\bar cc}_k(\overline p^2)=\frac{D^{AA\bar cc}_k(\overline p^2)}{(\overline p^2)^{d_k}}.
\end{align}
The $d_k$ are the dimensions of the dressings $D^{AA\bar cc}_k(\overline p^2)$ given by the negative dimensions of the basis tensors.
Thus, the quantities $\overline{D}^{AA\bar c c}$ reflect the actual momentum behavior that enters in other equations.
Power laws with the dimensions of the dressings are also used for the \gls{uv} extrapolations.
The suppression from these power laws is crucial to avoid spurious \gls{uv} contributions.

\begin{figure}
  \includegraphics[width=0.47\textwidth]{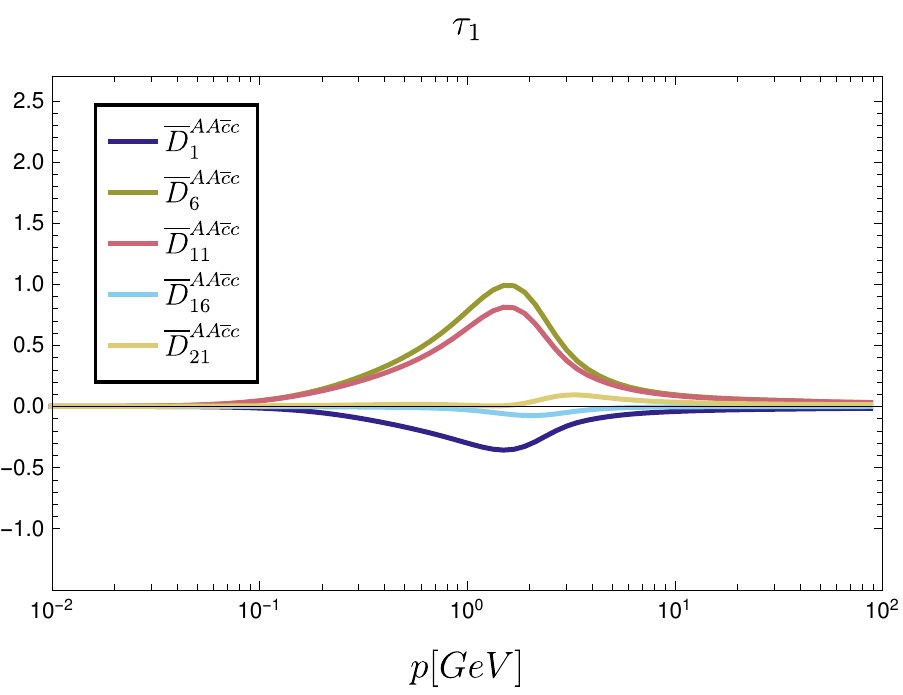}\hfill
  \includegraphics[width=0.47\textwidth]{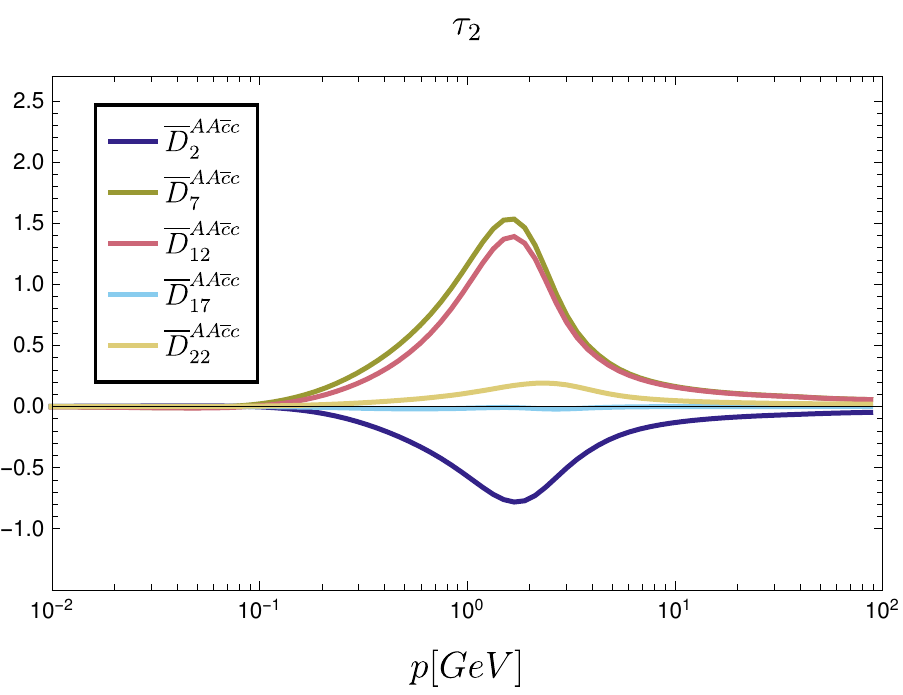}\\
  \vskip5mm
  \includegraphics[width=0.47\textwidth]{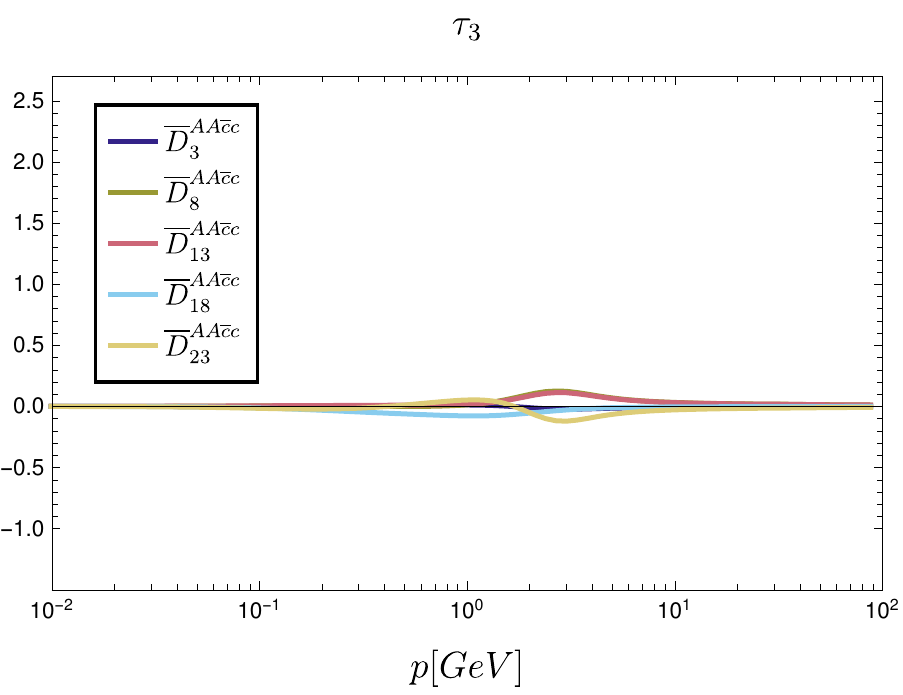}\hfill
  \includegraphics[width=0.47\textwidth]{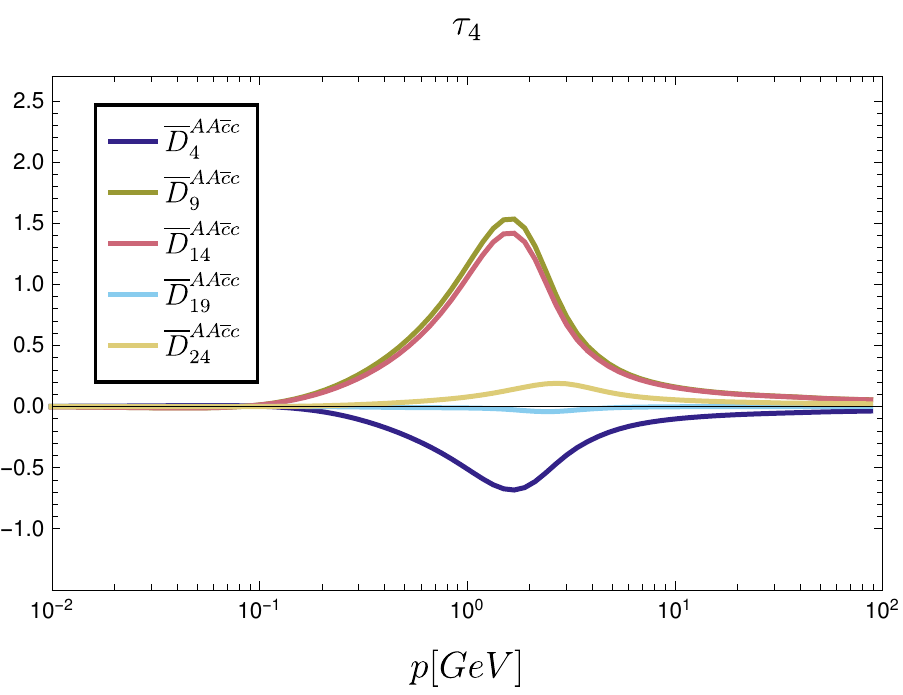}
  \vskip5mm
  \includegraphics[width=0.47\textwidth]{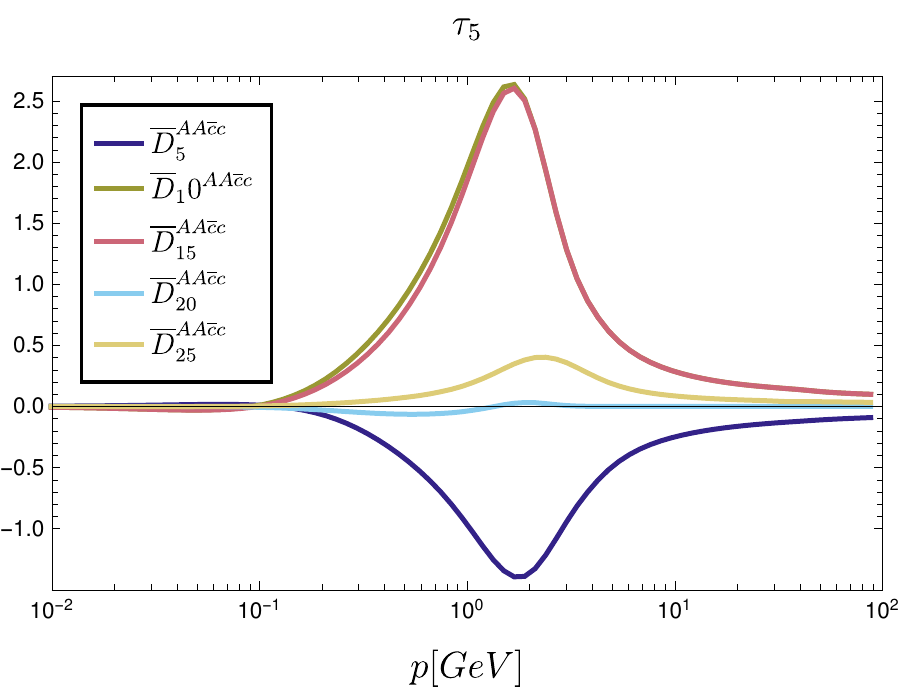}\hfill
  \hfill
  \includegraphics[width=0.47\textwidth]{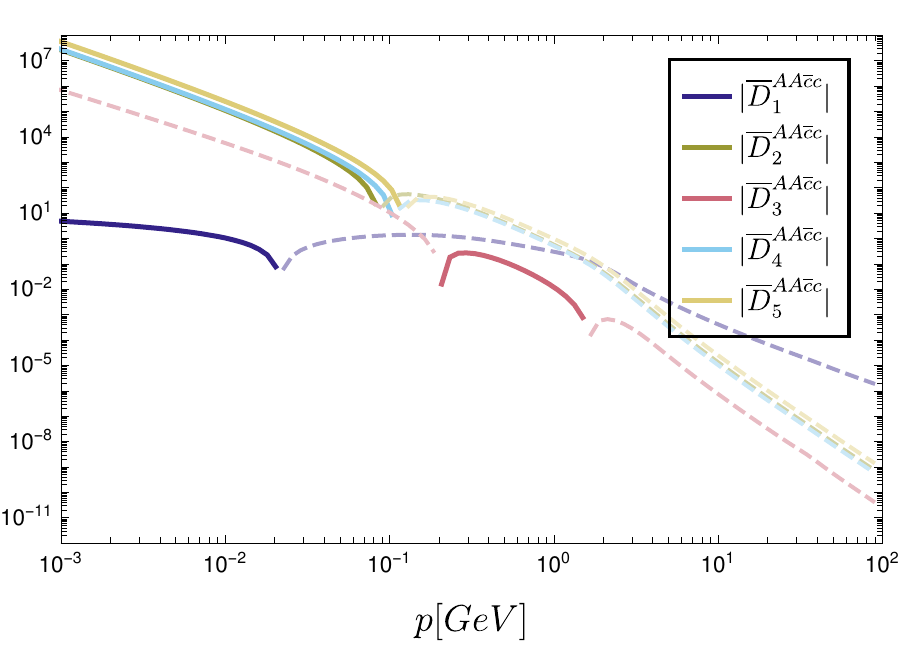}
  \begin{center}
  \caption{\label{fig:YM4d_AAcbc}
  Results for the two-ghost-two-gluon vertex \cite{Huber:2017txg}.
  Each plot shows the results corresponding to one Lorentz tensor $\tau_i$.
  The dressing functions related to the color tensors $\sigma_6$, $\sigma_7$ and $\sigma_8$ are zero.
  The lower right image shows selected dressing functions in a double-logarithmic plot to expose the power law behavior.
  Negative values are depicted dashed.}
 \end{center}
\end{figure}

\myboxmargin{Results for the AAcbc}
\index{two-ghost-two-gluon vertex}
The two-ghost-two-gluon vertex is calculated together with the ghost-gluon, three-gluon and four-gluon vertices for which also one-configuration approximations are used.
In their \glspl{dse}, the diagrams with the two-ghost-two-gluon vertex were included.
The propagators are taken as fixed input and correspond to the results shown in \fref{fig:YM4d_props_tl}.
In this context it should be noted that the dynamic inclusion of the ghost-gluon vertex for the results of \fref{fig:YM4d_props_tl} is important, as otherwise the three-gluon vertex did not converge.
This could be an artifact of the employed truncation and the sensibility of the three-gluon vertex \gls{dse} on the employed input in general.
The results for the two-ghost-two-gluon vertex are shown in \fref{fig:YM4d_AAcbc} in form of the dimensionless dressing functions $\overline{D}^{AA\bar cc}_k(\overline p^2)$.
The behavior of the dressing functions in the \gls{uv} shows that the leading behavior is, as expected, in agreement with the power laws from the dimensional analysis.
In the \gls{ir}, the dressing functions are power law suppressed compared to the \gls{uv}.
This is expected from a power counting analysis of the corresponding diagrams:
Every gluon dressing function, behaving like $p^2$ in the \gls{ir}, suppresses the integral.
The power law behavior is exposed in the lower right image of \fref{fig:YM4d_AAcbc} for a few dressing functions.
The other five images show the 25 dressing functions corresponding to the reduced set of color tensors, see Sec.~\ref{sec:AAAA}, grouped by their Lorentz parts.
The scales on the ordinate have been chosen the same for easier comparison.
Clearly, the largest dressing functions are those belonging to $\tau_5^{abcd}$ followed by those belonging to $\tau_2^{abcd}$ and $\tau_4^{abcd}$.
The group belonging to $\tau_3^{abcd}$ is basically negligible.
Also the dressing function corresponding to the color tensors $\sigma_4$ and $\sigma_5$ are very small.
These two are anti-symmetric under exchange of the two gluon legs, see \tref{tab:color_tensors_symmetries}.

\begin{figure}[tb]
 \includegraphics[width=0.45\textwidth]{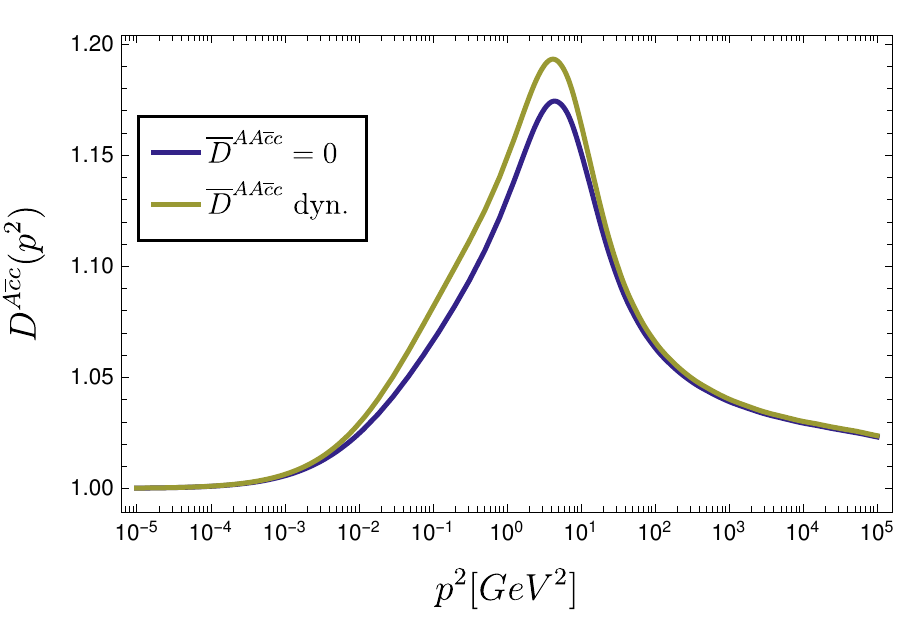}
 \hfill
 \includegraphics[width=0.45\textwidth]{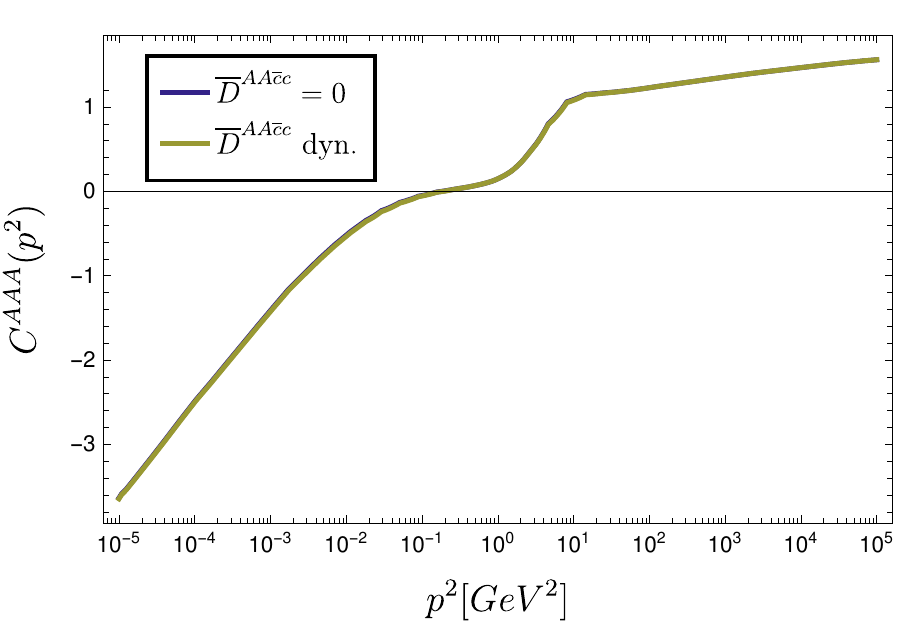}\\
 \includegraphics[width=0.45\textwidth]{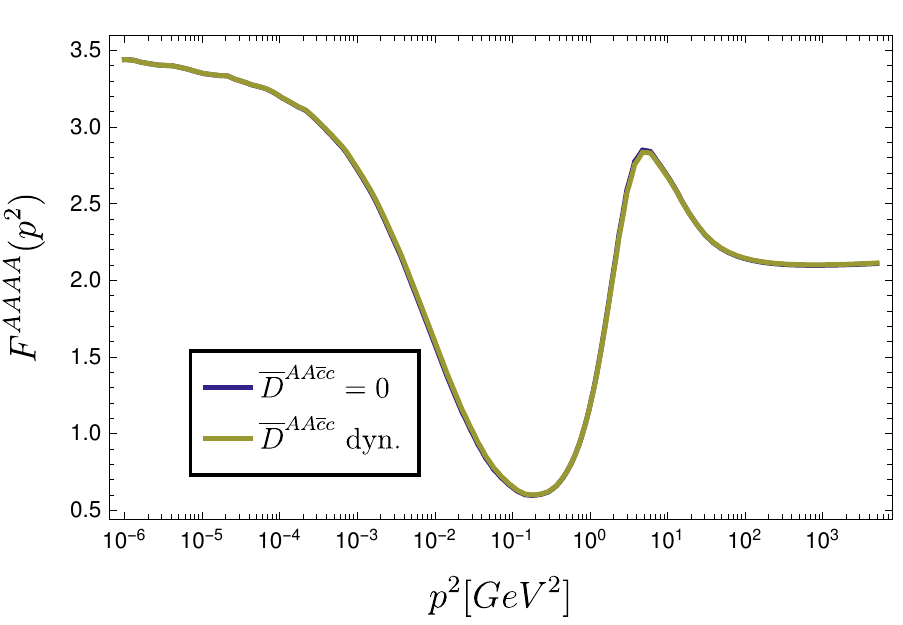}
 \hfill
 \includegraphics[width=0.48\textwidth]{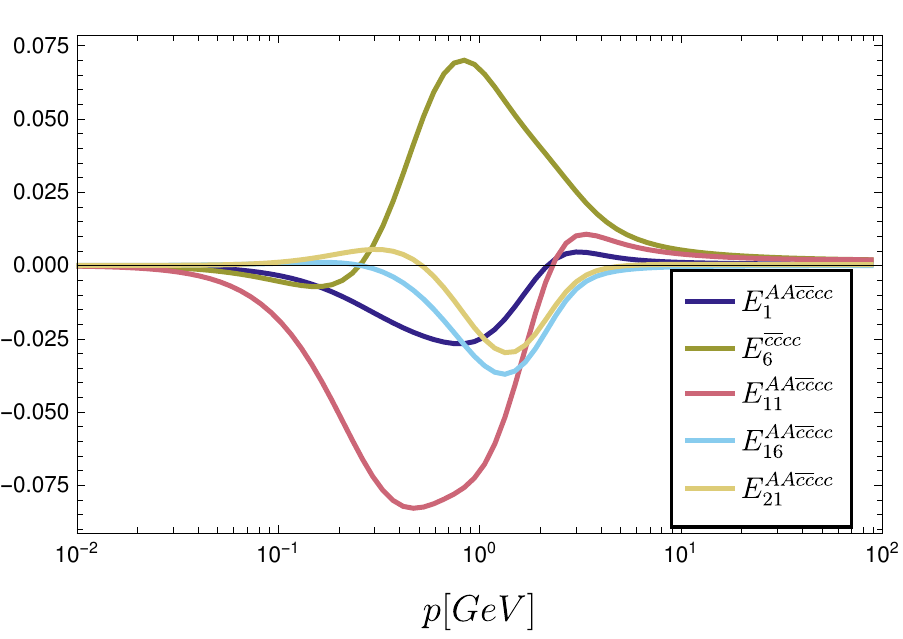}
 \caption{Comparison of results for the ghost-gluon (top left), three-gluon (top right) and four-gluon (bottom left) vertices obtained from truncations with and without the two-ghost-two-gluon vertex \cite{Huber:2017txg}.
 Results for the four-ghost vertex (bottom right) \cite{Huber:2017txg}.
 The dressing functions $E_6$, $E_7$ and $E_8$ are compatible with zero.
 }
 \label{fig:YM4d_ghgTgFg_influenceAAcbc+cbcbcc}
\end{figure}

The dressings of the remaining three color tensors can be considered separately, because they decouple from the reduced set of color tensors.
Projecting with these three tensors, only the diagrams with a two-ghost-two-gluon vertex remain.
In this vertex, only these three color tensors remain so that this part of the vertex does not couple with the rest.
It should be noted that this part of the vertex only contributes to the color symmetric part of the three-point functions, which, however, should be zero as discussed in Sec.~\ref{sec:AAA}.
In the present case, the equation reduces in this color subspace to a homogeneous linear equation.
For such an equation the trivial solution of all dressings equal to zero always exists.
The attempt to find an alternative solution by starting from a simple ansatz always led to the trivial solution.
A nontrivial solution cannot be excluded at this point, but it would be irrelevant due to the complete decoupling from the rest of the system of equations.

\myboxmargin{influence of AAcbc on other correlation functions}
\index{ghost-gluon vertex}
\index{three-gluon vertex}
\index{four-gluon vertex}
\index{two-ghost-two-gluon vertex}
To assess the influence of the two-ghost-two-gluon vertex on other correlation functions, the same system as above can be solved without this vertex.
The results for the ghost-gluon and the three-gluon vertices are shown in \fref{fig:YM4d_ghgTgFg_influenceAAcbc+cbcbcc}.
Most remarkably, the three-gluon vertex does not change.
This can be traced back to the color structure.
In the three-gluon vertex \gls{dse} the gluon legs of the two-ghost-two-gluon vertex are contracted with an antisymmetric structure constant.
Consequently, all contributions which are symmetric in the color indices of the two gluon legs vanish and only contributions from the color tensors $\sigma_4$ and $\sigma_5$ remain.
However, \fref{fig:YM4d_AAcbc} shows that these are the dressing functions which are very small.
They cannot have any substantial influence on the three-gluon vertex.
There could be indirect effects via the dependence of the other vertices on the two-ghost-two-gluon vertex, but as it turns out that they are not influenced largely either, this effect is tiny.
In the four-gluon vertex the effect is also very small.
Only the ghost-gluon vertex shows some difference, which, however, is $1.7\%$ at most.

\paragraph{Four-ghost vertex}

\myboxmargin{Why is the AAcbc more important than the cbcbcc?}
\index{four-ghost vertex}
The second non-primitively divergent four-point function is the four-ghost vertex.
It was studied perturbatively in Ref.~\cite{Gracey:2017yfi} and with its \gls{dse} in Ref.~\cite{Huber:2017txg}.
This vertex does not appear in the three-gluon vertex \gls{dse}.
For the ghost-gluon vertex, it appears only in the $A$-DSE.
However, having the two-ghost-two-gluon vertex available, one can use the $c$-DSE for the ghost-gluon vertex which is complete then and thus the preferred choice.
At the four-point level, the four-ghost vertex does neither appear in the four-gluon vertex nor the $c$-DSE for the two-ghost-two-gluon vertex.
Hence, the four-ghost vertex does not couple back within this truncation.

\myboxmargin{Results 4gh}
\index{four-ghost vertex}
The four-ghost vertex can thus be calculated a posteriori.
Its setup is similar to the two-ghost-two-gluon vertex calculation.
Its \gls{dse} is depicted in \fref{fig:4gh_DSE}.
Only one diagram is dropped in the truncation, namely the one with a five-point function.
The full tensor basis, in this case consisting of eight color tensors, is employed, but the momentum dependence is approximated by a single momentum configuration.
The result for the reduced color basis is shown in \fref{fig:YM4d_ghgTgFg_influenceAAcbc+cbcbcc}.
The other three color tensors lead also in this case to a homogeneous linear equation for which only the vanishing solution is found.

\subsection{Correlation functions of three-dimensional Yang-Mills theory}
\label{sec:YM3d}

\myboxmargin{Why YM3d? high temperature, lattice, continuum}
\myboxmargin{references, Gribov copy counting}
\index{Yang-Mills theory! in three dimensions}
The study of Yang-Mills theory in three dimensions is interesting for several quite different reasons.\footnote{All results in this section were obtained for the gauge group $SU(2)$, since all lattice results in three dimensions shown here are for this gauge group.
However, within the employed truncations, the solutions are invariant under rescaling $g^2 N_c$.}
First of all, the high temperature limit of four-dimensional Yang-Mills theory is a three-dimensional Yang-Mills theory with an adjoint Higgs field which is the leftover of the fourth component of the gluon field.
This was studied, for example, in Ref.~\cite{Maas:2004se}.
On the other hand, there are technical reasons as well for such studies.
In lattice studies, three and also two dimensions were studied to reach lower momenta.
For example, propagators were calculated in \cite{Cucchieri:2003di,Cucchieri:2004mf,Cucchieri:2008qm,Maas:2008ri,Maas:2009se,Maas:2009ph,Cucchieri:2006tf,Cucchieri:2008fc,Cucchieri:2007rg,Cucchieri:2009zt,Cucchieri:1999sz,Cucchieri:2011ig,Bornyakov:2011fn,Maas:2010qw,Maas:2011se, Bornyakov:2013ysa,Maas:2014xma,Maas:2015nva,Cucchieri:2016jwg,Maas:2017csm,Maas:2019ggf} and there are a few studies of vertices as well \cite{Cucchieri:2006tf,Cucchieri:2008qm,Maas:2019ggf}.
Studies of the Gribov problem are also simpler in lower dimensions \cite{Maas:2008ri,Maas:2009se,Maas:2009ph,Bornyakov:2011fn,Maas:2011se,Bornyakov:2013ysa,Maas:2015nva,Maas:2017csm}.
While the problem of reaching the deep \gls{ir} does no exist for continuum studies, three dimensions are of interest as well, since they have a simpler \gls{uv} behavior than four dimensions.
This alleviates the treatment of spurious divergences as will be explained below.
Further previous studies of the correlation functions of three-dimensional Yang-Mills theory were done with functional methods in \cite{Maas:2004se,Huber:2007kc,Alkofer:2007hc,Alkofer:2008dt,Aguilar:2010zx,Aguilar:2013vaa,Cornwall:2015lna,Huber:2016hns,Corell:2018yil}, with the (refined) \gls{gz} framework in \cite{Dudal:2008rm} and with a massive extension of Yang-Mills theory in \cite{Tissier:2010ts,Weber:2011nw,Tissier:2011ey,Pelaez:2013cpa}.

\myboxmargin{What will I do in this section?}
In this section I will present results for Yang-Mills theory in three dimensions obtained from the equations of motion of the \gls{1pi} and the 3PI effective actions.
The employed truncation does not contain any model parameters and is fully self-contained.
This allows studying modifications of this truncation to test its stability.
Furthermore, the hierarchy of diagrams in the gluon propagator \gls{dse} is identified.

In the following, the differences between three- and four-dimensional Yang-Mills theory will be explained.
Results for the primitively divergent correlation functions will be presented in Sec.~\ref{sec:YM3d_fullResults}.
In Sec.~\ref{sec:YM3d_testing} the stability of the employed truncation will be discussed and Sec.~\ref{sec:YM3d_disc} contains a discussion of the results.

\subsubsection{Specifics of three-dimensional Yang-Mills theory}
\label{sec:YM3d_specifics}

\myboxmargin{no resummation: two-loop diagrams easier to add to the whole}
\myboxmargin{finiteness of three dimensions}
\index{anomalous dimension}
\index{UV divergence}
In four dimensions, Yang-Mills theory is only logarithmically divergent due to gauge symmetry.
Lowering the dimension by one, the theory becomes finite.
One-loop diagrams behave then like $g^2/p$, which can directly be inferred from the fact that the coupling has a mass dimension of $1/2$.
In addition, the \gls{uv} finiteness leads to a trivial behavior under the renormalization group and, consequently, the absence of anomalous dimensions.
This difference in the \gls{uv} behavior makes three dimensions somewhat easier to treat numerically.

\myboxmargin{spurious divergences in YM3d}
\index{regularization}
\index{cutoff}
\index{regulator function}
\index{gauge covariance! breaking of}
As discussed in detail in Sec.~\ref{sec:spurDivs}, the standard regularizations employed by functional methods break gauge covariance which entails the appearance of spurious divergences.
In three dimensions, these spurious divergences are of degree one.
As a remnant of resummation in four dimensions, also logarithmic divergences appear.
This can be seen by considering that the one-loop \gls{uv} behavior of the propagator dressings is of the form $1+c\,g^2/p$ \cite{Jackiw:1980kv}:
\begin{align}
 G(p^2)=1+\frac{g^2\,N_c}{16p},\quad
 Z(p^2)=1+\frac{11g^2\,N_c}{64p}.
\end{align}
In contrast to four dimensions, where the dressing functions behave logarithmically, the dressing functions in three dimensions become constant at asymptotically high momenta.
Plugging the one-loop result into a one-loop expression, the second part lowers the degree of divergence by one leading to a logarithmic divergence proportional to $g^4$.
For dimensional reasons, it is obvious that no divergences can arise from higher orders in $g$ since the mass dimension of $g$ must be canceled by a corresponding momentum.
This reflects the absence of resummation in three dimensions.

\begin{figure}[tb]
 \begin{center}
  \includegraphics[width=0.48\textwidth]{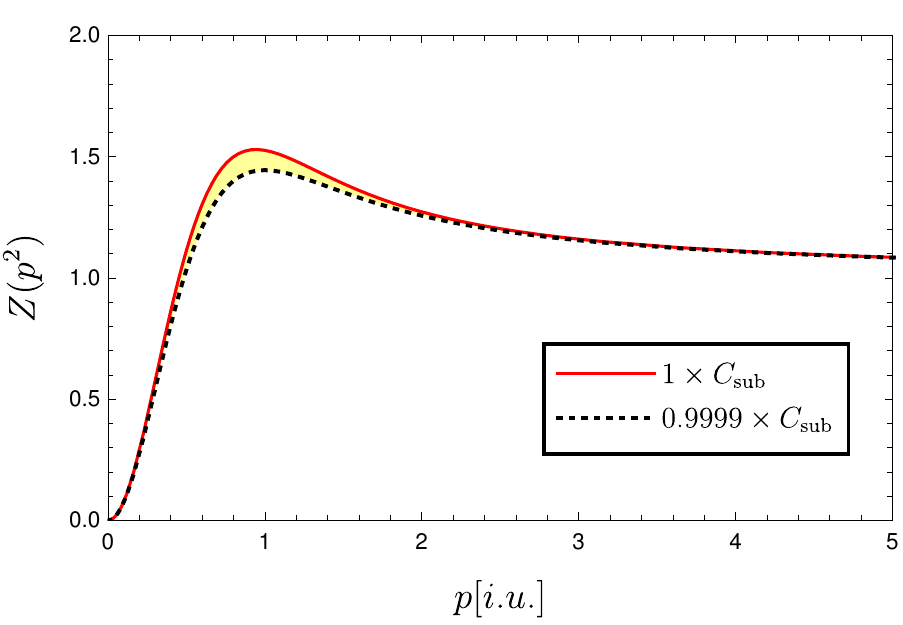}
  \hfill
  \includegraphics[width=0.48\textwidth]{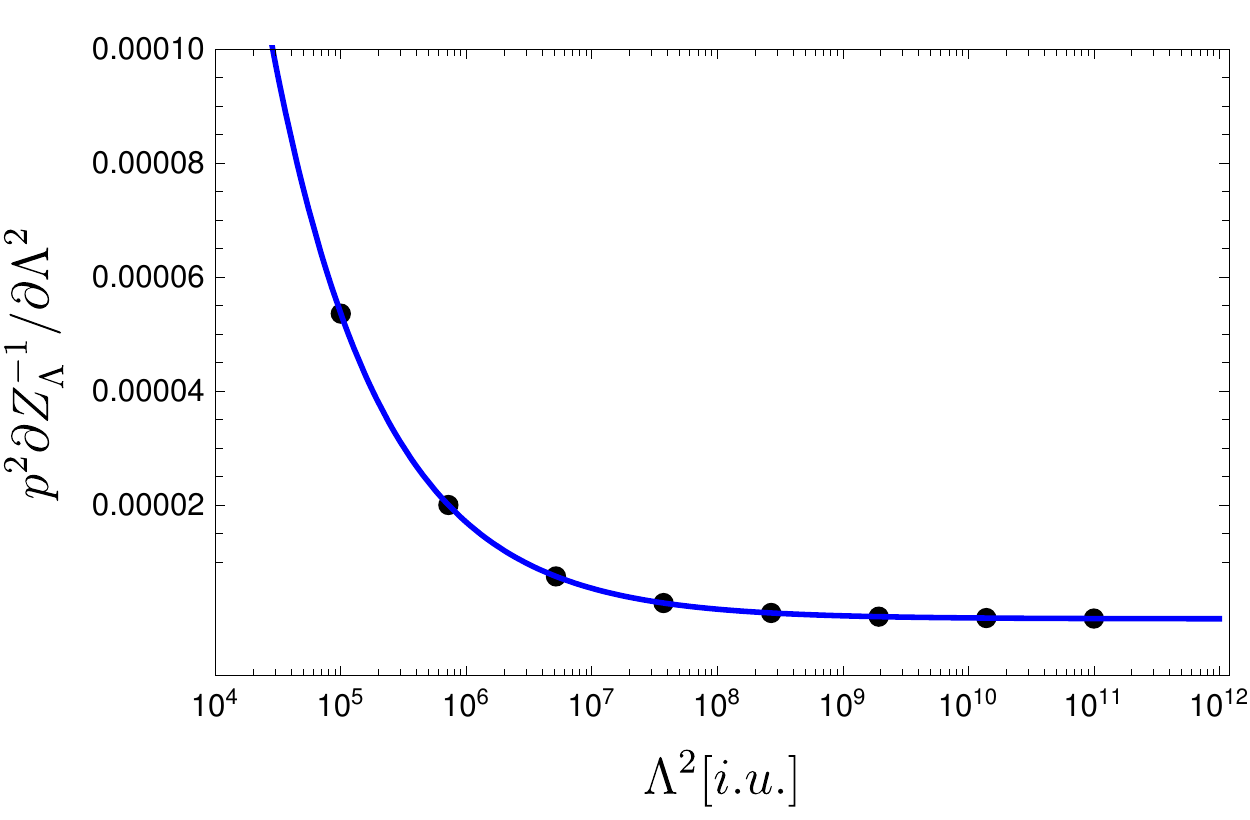}
  \caption{\label{fig:YM3d_varCsub}Left: Gluon dressing function calculated from the system of ghost and gluon propagators with a bare ghost-gluon vertex and a modeled three-gluon vertex with the correct and a rescaled value for $C_\text{sub}$ \cite{Huber:2016tvc}.
  Right: Cutoff dependence of the right-hand side of the gluon propagator DSE.
  The dots correspond to calculated values, and the line corresponds to the fit function.}
 \end{center}
\end{figure}

\myboxmargin{treatment of spurious divergences in YM3d}
\index{spurious divergences}
The general form for the divergent term in three dimensions is then
\begin{align}
 Z_\text{spur}(p^2,\Lambda)=\frac{C_\text{sub}}{p^2}=a_\text{sub}\frac{g^2\,N_c\,\Lambda}{p^2} + b_\text{sub}\frac{g^4\,N_c^2\,\ln{\Lambda}}{p^2}.
\end{align}
The coefficients $a_\text{sub}$ and $b_\text{sub}$ can be calculated perturbatively \cite{Huber:2016hns}.
A term proportional to $\log g$ \cite{Jackiw:1980kv} would require a quadratic cutoff dependence, which, however, was not found in the actual calculations.
However, once full vertices and propagators are employed numerically, this influences their values.
In the present truncation, differences of up to $5\,\%$ were observed.
Thus, it is best to fit the coefficients to the cutoff $\Lambda$.
In this way, also two-loop diagrams can be handled without having to calculate them analytically in the perturbative regime.
Furthermore, the contribution of the tadpole diagram is automatically absorbed in $C_\text{sub}$, as it can produce only powers of the cutoff or a logarithm thereof.
The perfect agreement of the fit with the analytic result can be illustrated by plotting $p^2 \partial Z_\Lambda^{-1}/\partial\Lambda^2$, where $Z_\Lambda^{-1}$ corresponds to the unrenormalized right-hand side of the \gls{dse}, see \fref{fig:YM3d_varCsub}.
The reason for the perfect agreement is that the fit function contains all possible dependencies on the cutoff $\Lambda$, namely $\ln \Lambda$ and $\Lambda$.
This is different in four dimensions where fitting is thus not practicable.
The subtraction coefficient $C_\text{sub}$ must indeed be known very precisely.
As an example, consider the solution when it is artificially modified as $C_\text{sub}\rightarrow 0.9999 C_\text{sub}$ as shown in \fref{fig:YM3d_varCsub}.
Any larger change has a correspondingly larger influence.

\begin{figure}[tb]
\centerline{
\includegraphics[width=0.47\textwidth]{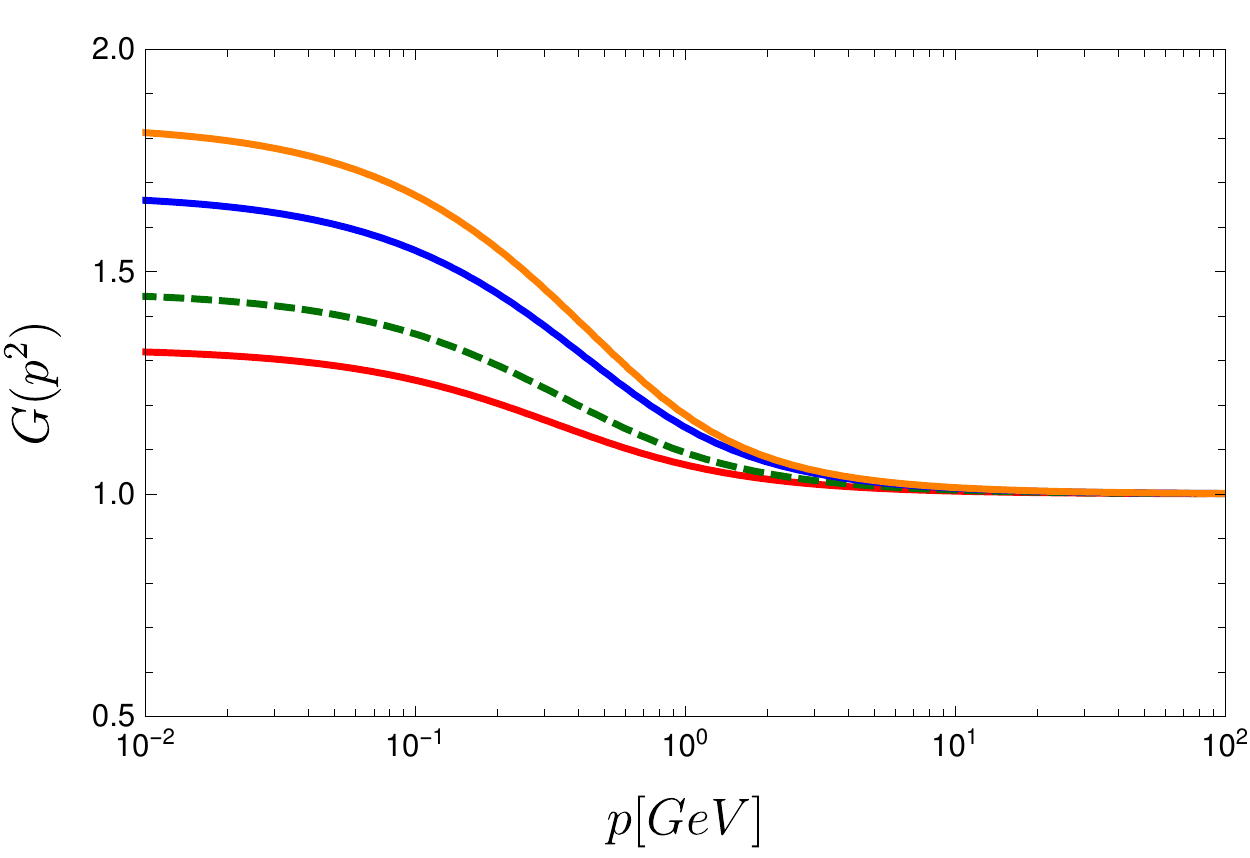}
\hfill
\includegraphics[width=0.47\textwidth]{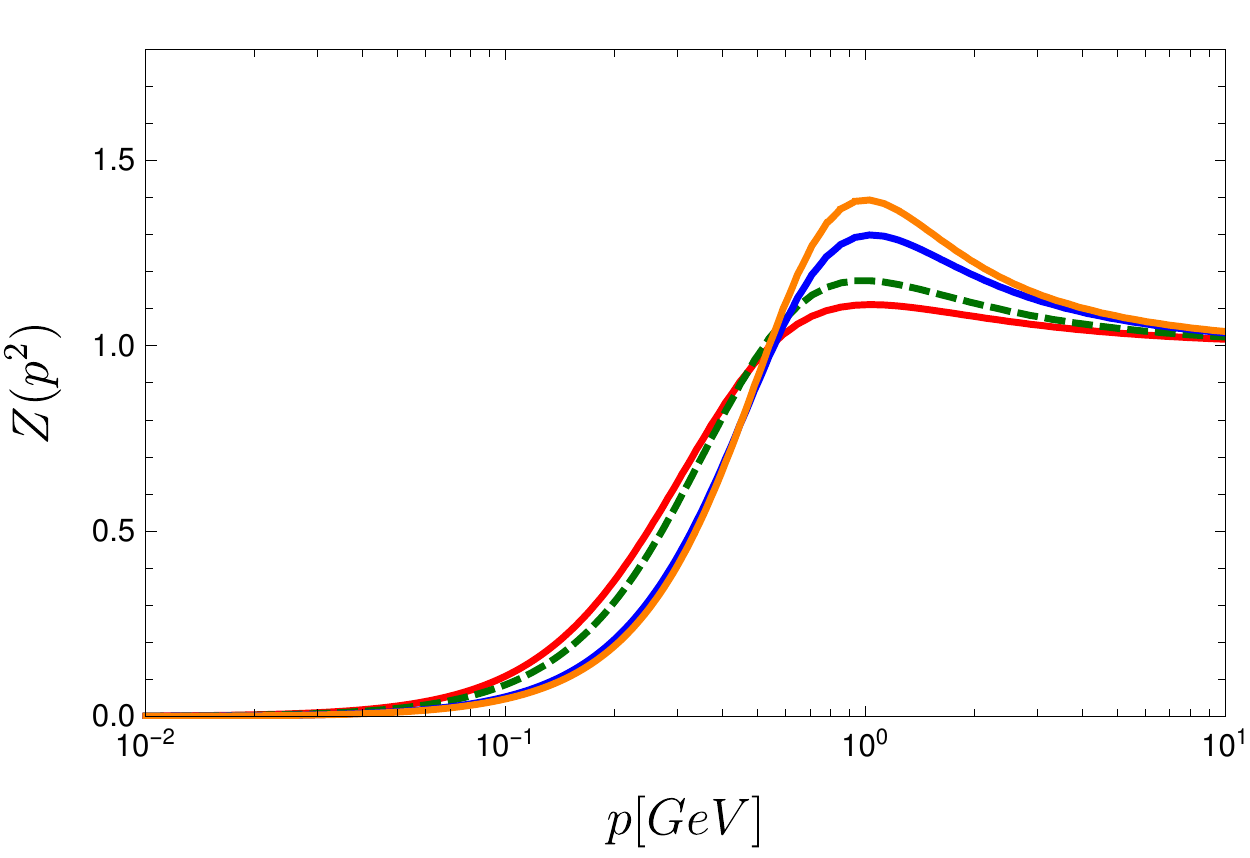}}
\caption{Ghost/Gluon dressing functions from the one-loop truncation using bare ghost-gluon and three-gluon vertices \cite{Huber:2016hns}.
Different solutions correspond to different values of the gluon propagator at zero momentum.}
\label{fig:YM3d_ghGlFamily}
\end{figure}

\myboxmargin{spurious divergences and family of solutions}
\index{decoupling solution}
The subtraction of spurious divergences contains a possible ambiguity, namely, a constant term can always be added to $C_\text{sub}$ without changing the asymptotic \gls{uv} behavior or the cutoff independence of the results.
Thus, technically the subtraction term can be used to generate a family of solutions.
In this case, instead of fitting $C_\text{sub}$, it is set such as to obtain a given value for the gluon propagator at zero momentum in a similar way as discussed in Sec.~\ref{sec:spurDivs}.
In \fref{fig:YM3d_ghGlFamily}, this possibility is shown for a simple test system that consists of the propagators and uses bare vertices \cite{Huber:2016hns}.
Clearly, different solutions arise akin to the family of decoupling solutions.
However, since the back reaction on the vertices is not contained in the system, the observed effects are most likely larger than they would be in a more elaborate truncation.
It should also be noted that the ghost dressing function is \textit{not} set to any value in the \gls{ir} and, hence, there is only one effective parameter for choosing different solutions.
In fact, it was found that doing so leads to the wrong \gls{uv} behavior of the form $a+c\,g^2/p$ with $a\neq 1$.
In four dimensions, the procedure of fixing the ghost dressing function via a boundary condition \cite{Fischer:2008uz} does not face this problem because of the anomalous running of the dressings.

As mentioned above, the coupling $g$ is dimensionful in three dimensions.
Thus, one could express all dimensionful quantities as multiples of powers of $g$.
A physical value is determined using lattice results where the scale is set via a string tension of $\sigma=(440\,\text{MeV})^2$.
To transfer this scale to the continuum results, the peak of the gluon dressing function is matched to that of lattice results.
However, since the heights of the dressing functions are not the same, one has some freedom to match the forms of the peaks.
From this an interval of $90$ to $125\,\%$ around the lattice maximum is used to set the continuum maximum.
In the plots where a band is shown, it represents this freedom in setting the scale.
In plots with single lines, the maxima are matched using a value of $1.025\,\text{GeV}$.

\myboxmargin{couplings in YM3d}
\myboxmargin{scale setting}
\myboxmargin{decoupling, on the lattice, comparison with a grain of salt}

\subsubsection{Results for the primitively divergent correlation functions}
\label{sec:YM3d_fullResults}

Before the details of testing the truncation dependence are discussed, I present the results for all primitively divergent correlation functions and compare them to lattice results in the minimal Landau gauge.
The results shown below were calculated with the largest truncation up to date: The propagator equations were fully included and the vertex equations were truncated to the perturbative one-loop level.
In other words: All correlations functions are dynamical except for the non-primitively divergent ones which are set to zero.
For the three- and four-gluon vertices only the dressings of the tree-level tensors were used.

\begin{figure}[tb]
 \includegraphics[width=0.48\textwidth]{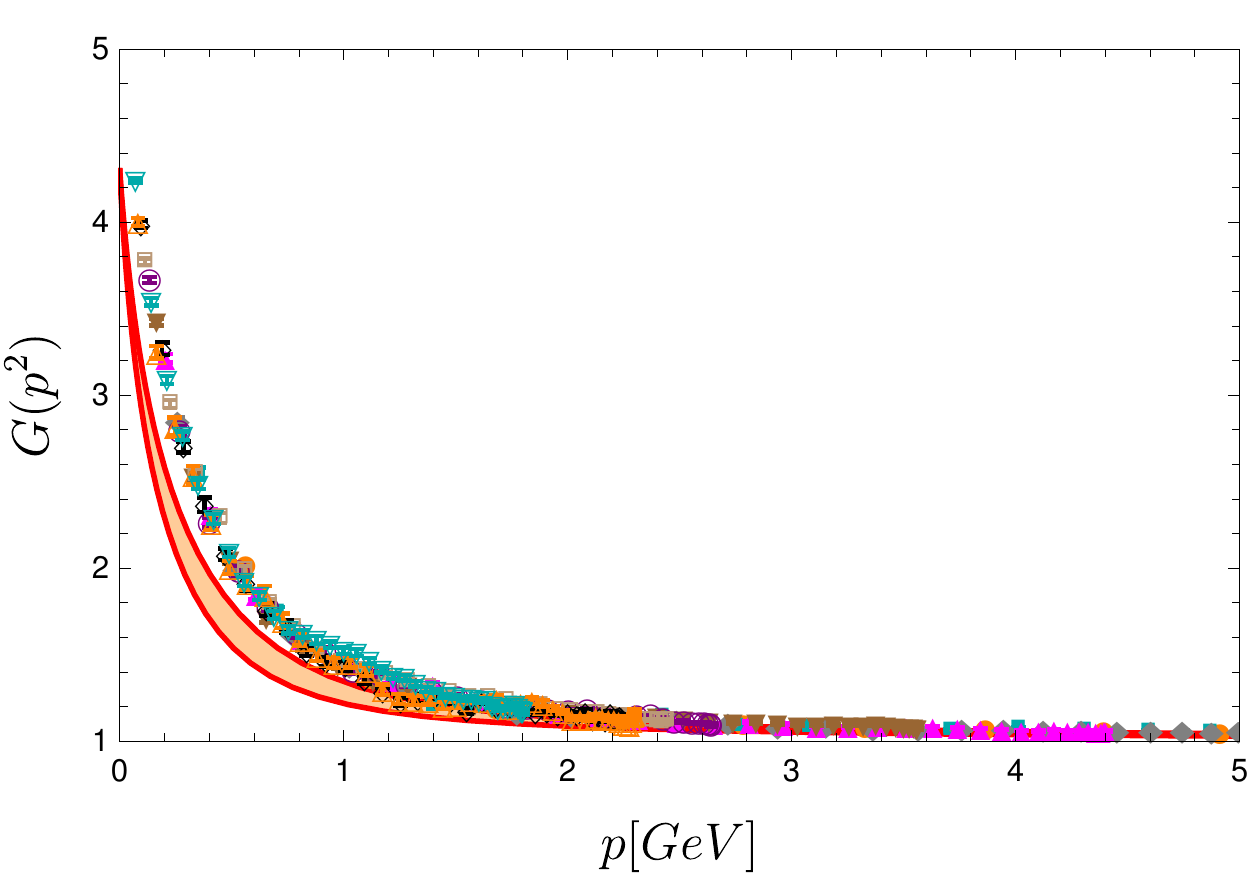}\hfill 
 \includegraphics[width=0.48\textwidth]{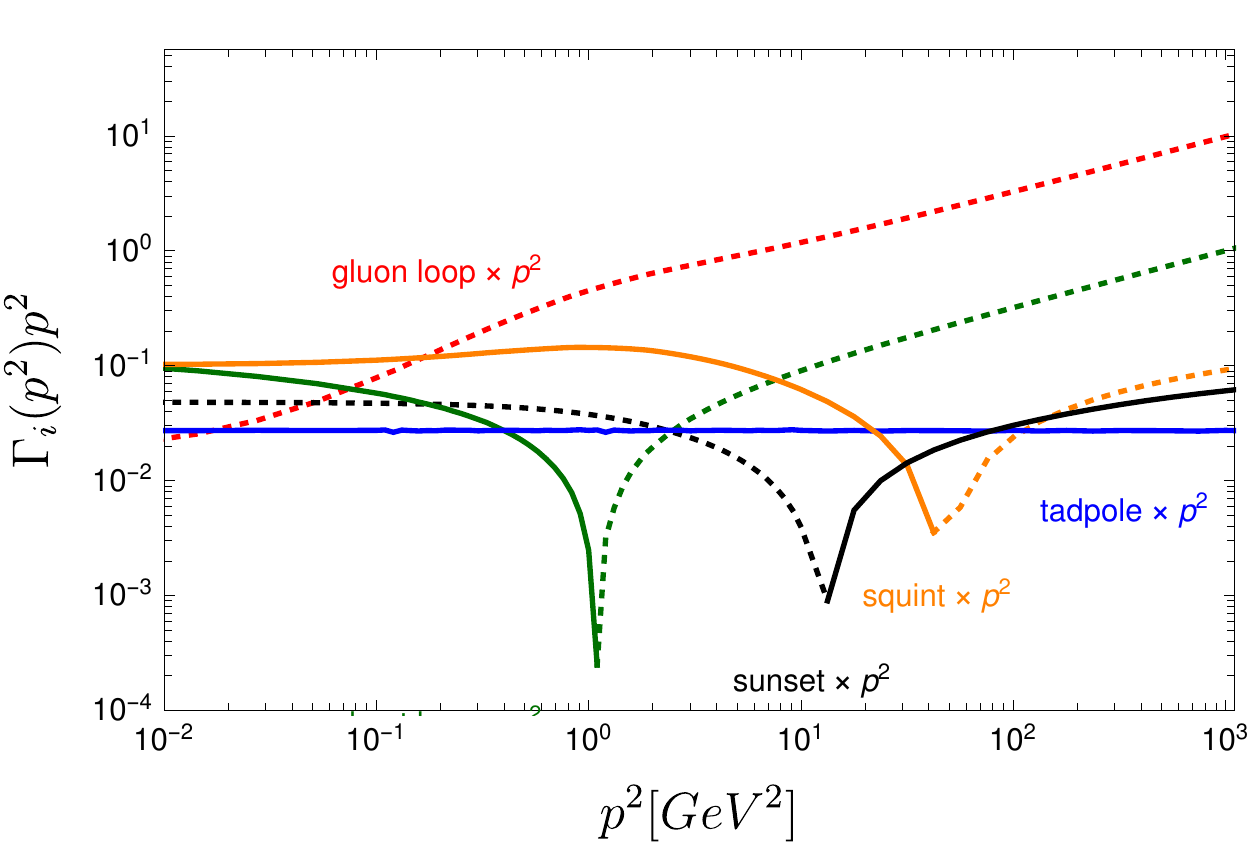}
 \caption{Left: Ghost dressing from the full system \cite{Huber:2016tvc} in comparison to lattice results \cite{Maas:2014xma}.
 Right: Contributions of individual diagrams to the gluon propagator DSE \cite{Huber:2016tvc}.
 Continuous/dotted lines denote positive/negative values.}
 \label{fig:YM3d_ghBand+glDiagrams}
\end{figure}

\myboxmargin{ghost propagator}
\index{ghost propagator}
The result for the ghost propagator is shown and compared to lattice results in \fref{fig:YM3d_ghBand+glDiagrams}.
Below $1\,\text{GeV}$, the \gls{dse} result is systematically below the lattice result.
Since the equation for the ghost propagator is complete, the source of this discrepancy must be caused by deviations in the other correlation functions.
Such deviations are indeed found, as discussed below.
The sensibility of the ghost dressing function on the gluon propagator was already visible in \fref{fig:YM3d_ghGlFamily}.
However, also the existence of a family of solutions is reflected in the ghost propagator.
Thus, the overall effect is most likely a mixture of both.

\myboxmargin{gluon propagator}
\index{gluon propagator}
\index{spurious divergences}
For the gluon propagator there is also a difference to lattice results.
It is largest in the midmomentum regime.
In particular, the peak in the dressing function is lower than in lattice results.
As for the ghost propagator, the existence of the family of solutions has most likely an effect here.
In particular, one can see in \fref{fig:YM3d_ghGlFamily}, that the height of the bump varies for different solutions.
In addition, while the equation for the gluon propagator is complete in terms of diagrams, two vertices are not taken with their full tensor structure.
Thus, those missing contributions may also have their share in the missing strength of the gluon dressing function.

\myboxmargin{gluon propagator: single diagrams}
\index{gluon propagator}
Calculating each single diagram from the obtained solution, one can also assess the quantitative importance of single diagrams.
Spurious divergences are subtracted in the same way as for the full equation, viz., by fitting the cutoff dependence.
Note that for other methods it might not be possible to calculate individual diagrams.
The contributions of all five diagrams are shown in \fref{fig:YM3d_ghBand+glDiagrams}.
The hierarchy in the \gls{uv} is as expected: The one-loop diagrams are dominant with the gluon loop being larger than the ghost loop as expected perturbatively.
In the low momentum regime, the ghost loop becomes important.
The sunset diagram is overall very small.
The second most important contribution in the midmomentum regime comes from the squint diagram.
As far as two-loop diagrams are concerned, it is interesting to see that the squint diagram clearly dominates over the sunset diagram.
However, one has to bear in mind that in this calculation only restricted tensor bases for the gluonic vertices were taken into account and it cannot be excluded that the full bases would alter the relative importance of the diagrams.

\begin{figure}[tb]
 \includegraphics[width=0.48\textwidth]{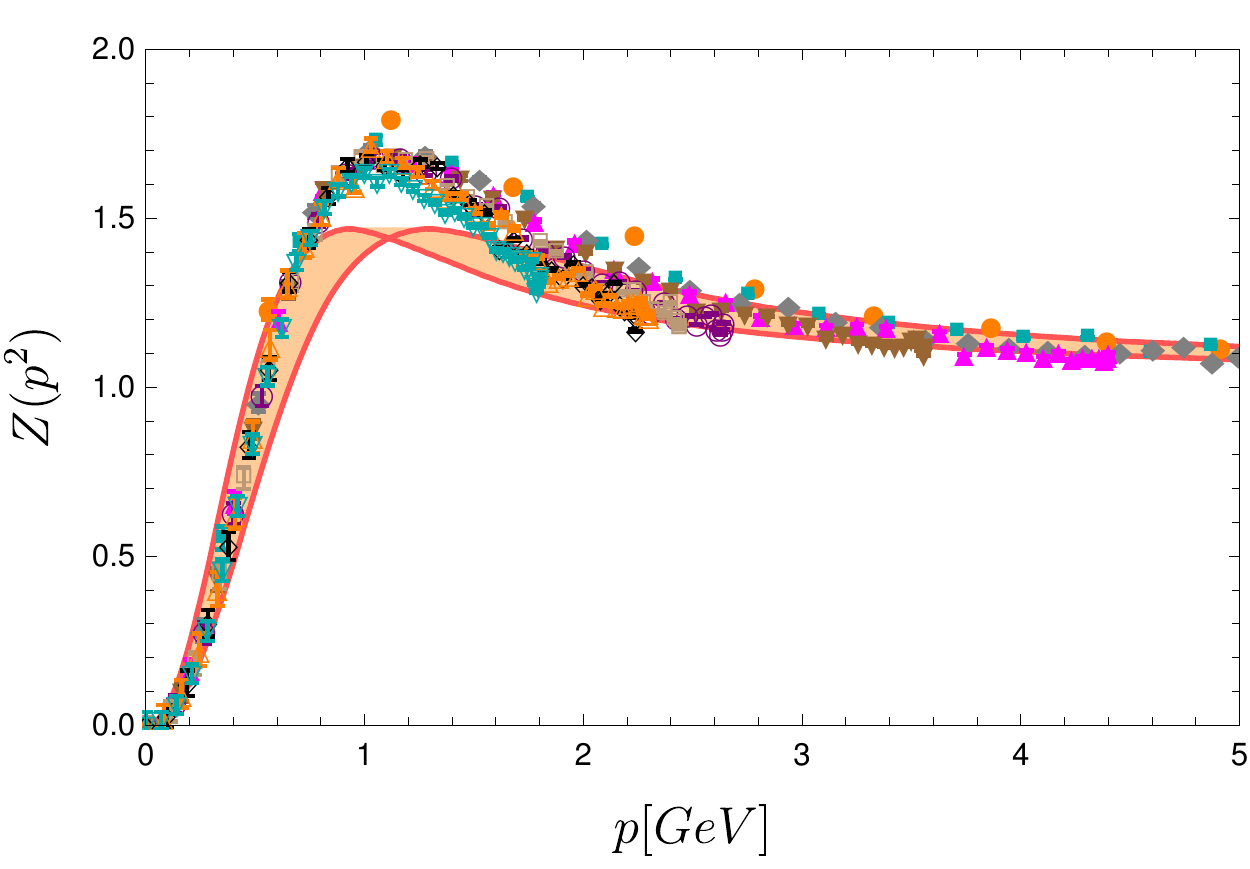}
 \hfill
 \includegraphics[width=0.48\textwidth]{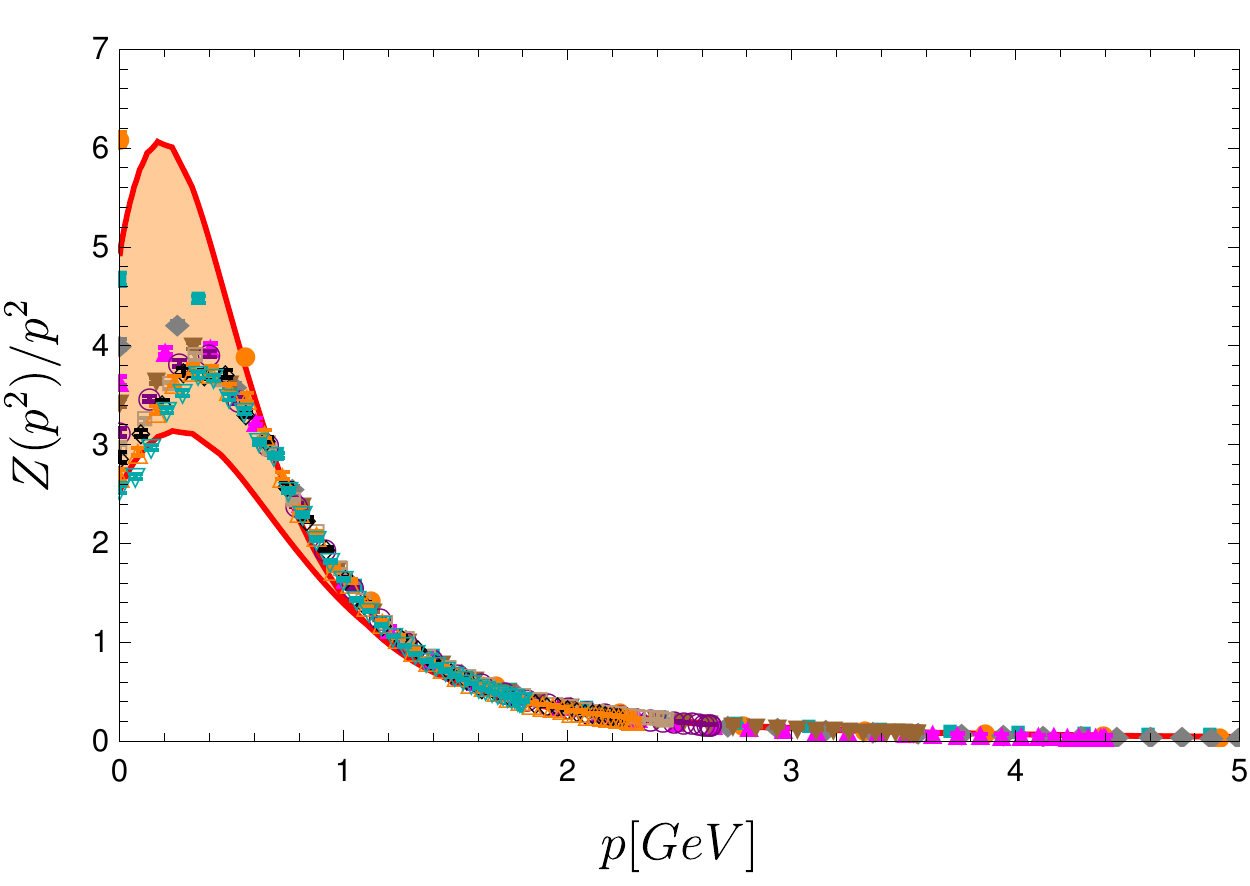}
 \caption{\label{fig:YM3d_glBand}Gluon dressing function (left) and gluon propagator (right) from the full system \cite{Huber:2016tvc} in comparison to lattice results \cite{Maas:2014xma}.
 Lattice momenta are along one axis.
 The shown lattice results correspond to lattice sizes between $N=68$ and $N=88$ and $\beta$ values between $3.48$ and $19.2$.
 The band is obtained by varying the maximum of the gluon dressing function between $922$ and $1282\,\text{MeV}$.}
\end{figure}

\begin{figure}[tb]
 \includegraphics[width=0.48\textwidth]{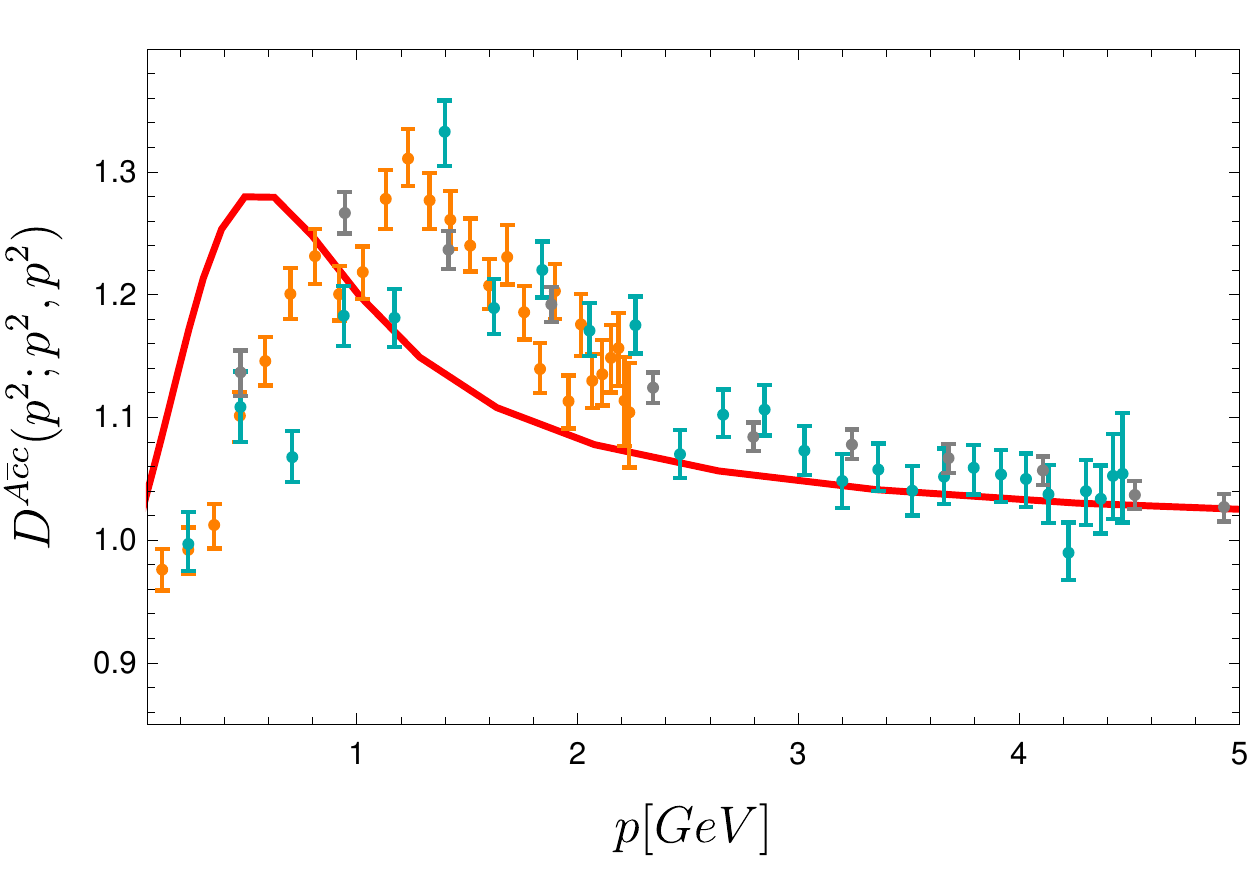}
 \hfill
 \includegraphics[width=0.48\textwidth]{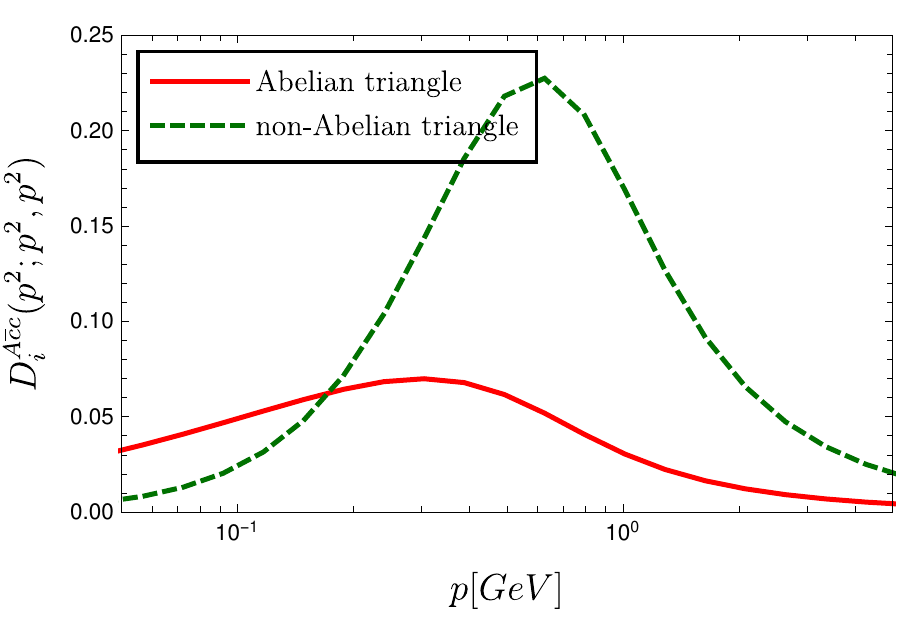}
 \caption{Ghost-gluon vertex dressing from the full system \cite{Huber:2016tvc} in comparison with lattice results \cite{Maas:2016ip} (left) and the contributions from single diagrams (right).
 The shown lattice results correspond to lattice size $N=60$ and $\beta=3.18$, $5.61$ and $10.5$.} 
 \label{fig:YM3d_ghg}
\end{figure}

\myboxmargin{ghost-gluon vertex}
\index{ghost-gluon vertex}
As discussed in Sec.~\ref{sec:Acbc}, the ghost-gluon vertex has two \glspl{dse}.
Here, the one with the bare vertex attached to the ghost leg is employed.
The obtained dressing function features some angle dependence.
In particular, the height of the bump around $1\,\text{GeV}$ varies.
The contributions of the two diagrams contained in the employed truncation are shown in \fref{fig:YM3d_ghg}.
As can be seen, the non-Abelian diagram is the more important one.
The comparison to lattice results \cite{Cucchieri:2008qm} shown in \fref{fig:YM3d_ghg} makes evident that the position of the bumps do not match.
This is also known from results for the gauge group $SU(2)$ in four dimensions, see Sec.~\ref{sec:res_ghg}.
A change in scale within the range discussed in Sec.~\ref{sec:YM3d_specifics} is insufficient to explain this.
The source of this discrepancy is currently not known.

\myboxmargin{three-gluon vertex}
\index{three-gluon vertex}
The three-gluon vertex has only a small angle dependence.
The structure of the dressing function is very simple, being close to one, the asymptotic value, at high momenta and only moving away below $1\,\text{GeV}$.
In the \gls{ir} it diverges linearly.
In general, the result is remarkably close to lattice results \cite{Cucchieri:2008qm}.
Seeing this structure, it is surprising to find that the individual diagrams are \emph{not} small.
They start to deviate from zero roughly below $10\,\text{GeV}$.
The smallness of deviation from the tree-level is thus the consequence of cancellations between the contributions of the individual diagrams.
Only the ghost triangle is not compensated and causes the divergence in the \gls{ir}.

\begin{figure}[tb]
 \includegraphics[width=0.48\textwidth]{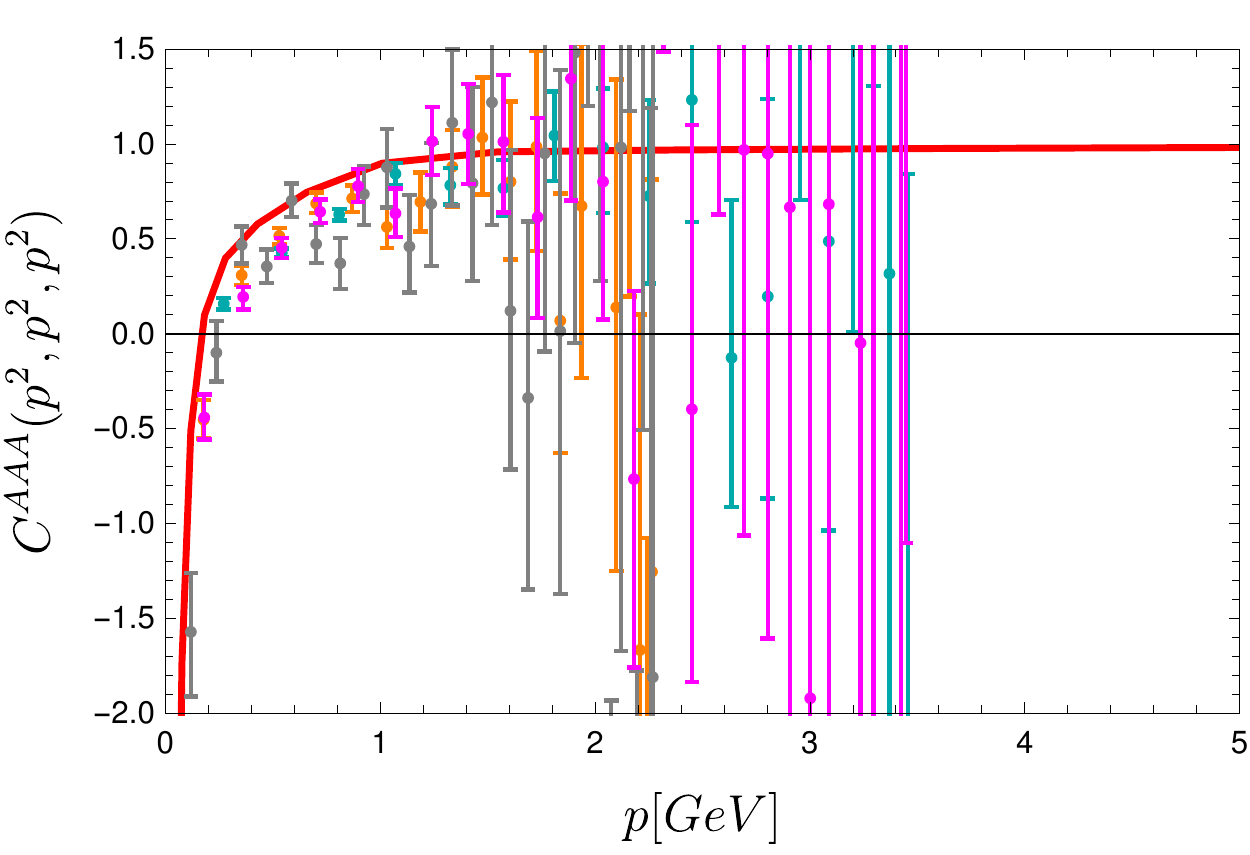}
  \hfill
 \includegraphics[width=0.48\textwidth]{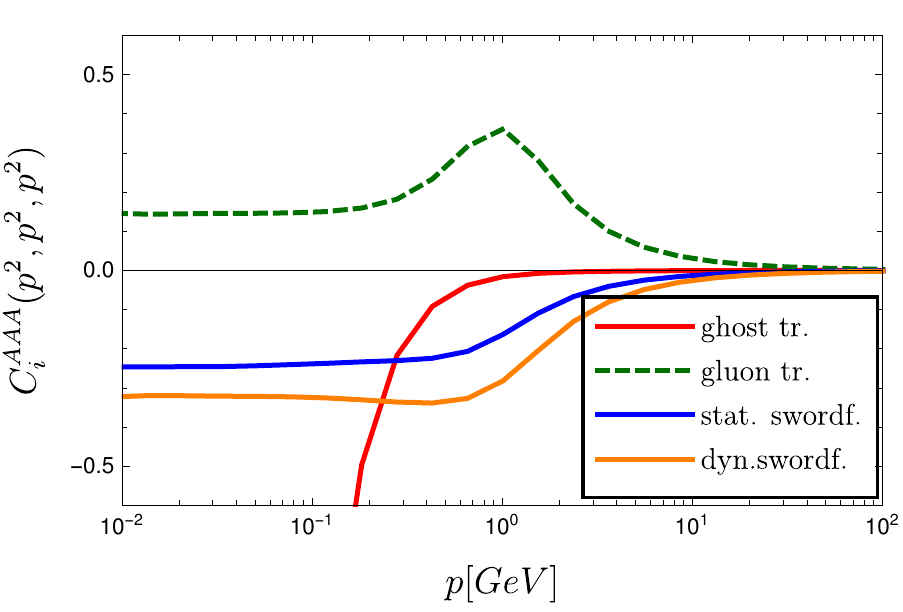}
 \caption{Three-gluon vertex dressing from the full system \cite{Huber:2016tvc} in comparison with lattice results \cite{Cucchieri:2008qm} (left) and the contributions from single diagrams (right).
 The shown lattice results correspond to lattice sizes $N=40$ and $60$ and $\beta=4.1$ and $6$.}
 \label{fig:YM3d_3g}
\end{figure}

\begin{figure}[tb]
 \includegraphics[width=0.48\textwidth]{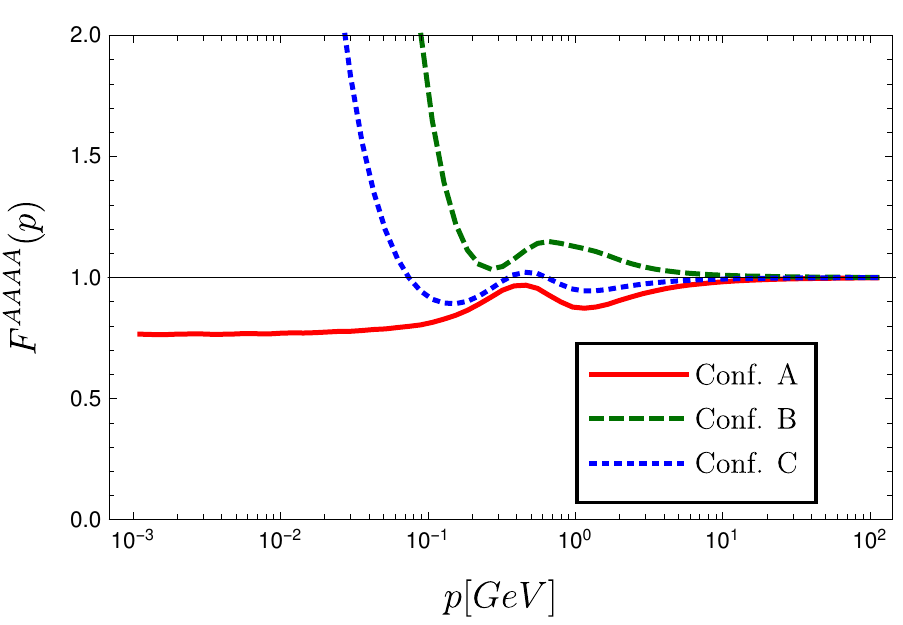}
 \hfill
 \includegraphics[width=0.48\textwidth]{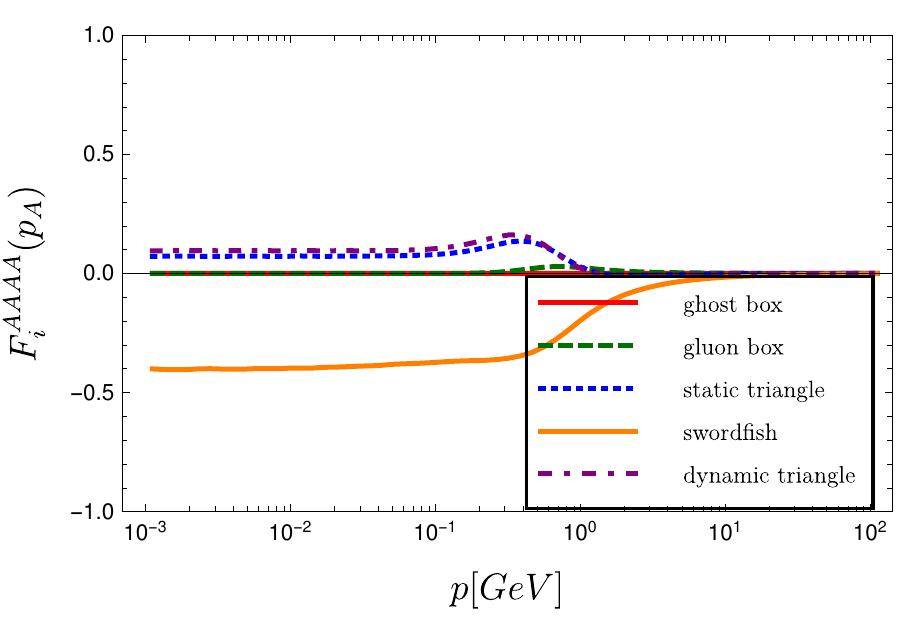}\\
 \vskip2mm
 \includegraphics[width=0.48\textwidth]{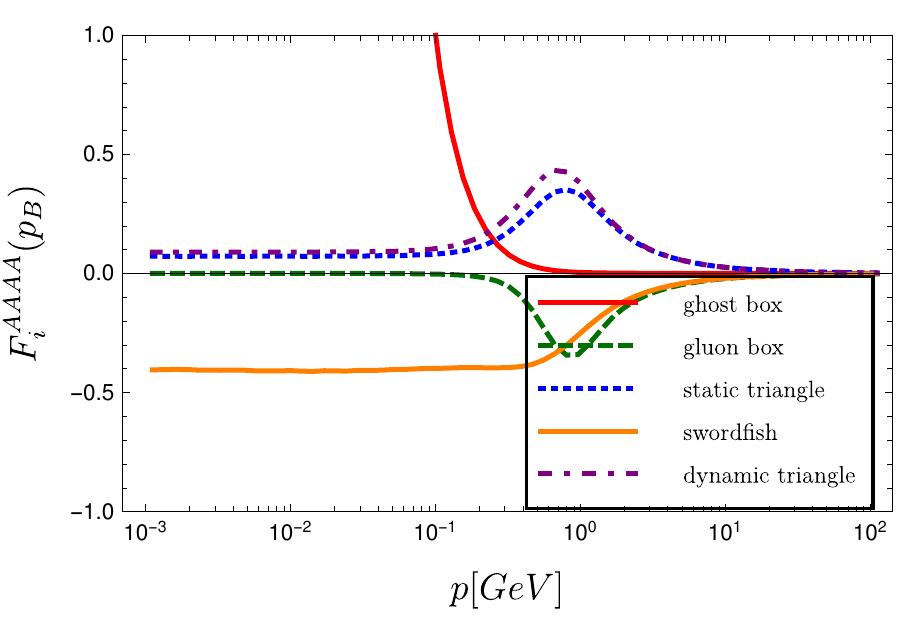}
 \hfill
 \includegraphics[width=0.48\textwidth]{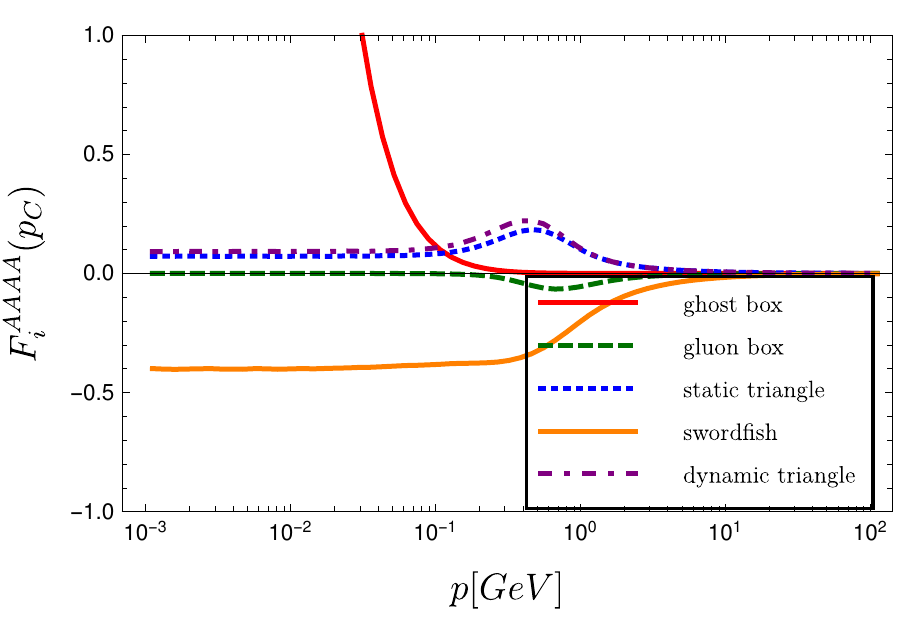}
 \caption{\label{fig:YM3d_fgV1}Tree-level dressing of the four-gluon vertex for different kinematic configurations (top left) and the individual contributions of single diagrams (top right and bottom) \cite{Huber:2016tvc}.}
\end{figure}

\begin{figure}[tb]
 \includegraphics[width=0.48\textwidth]{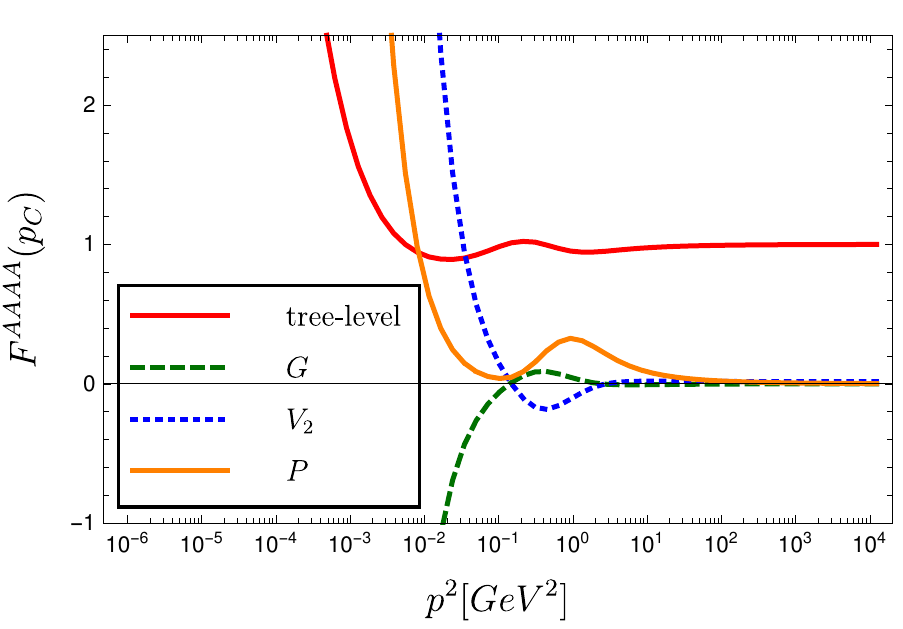}
 \hfill
 \includegraphics[width=0.48\textwidth]{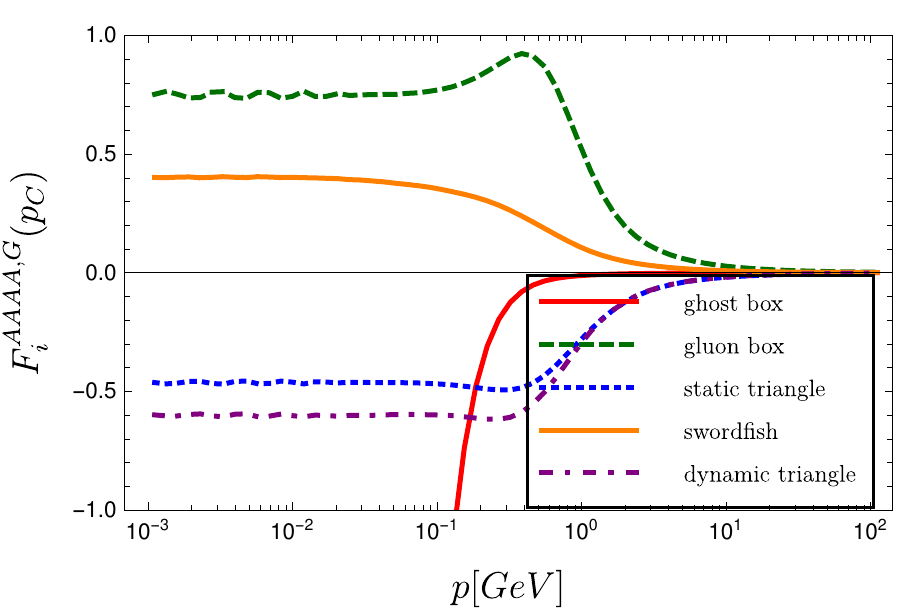}\\
 \vskip2mm
 \includegraphics[width=0.48\textwidth]{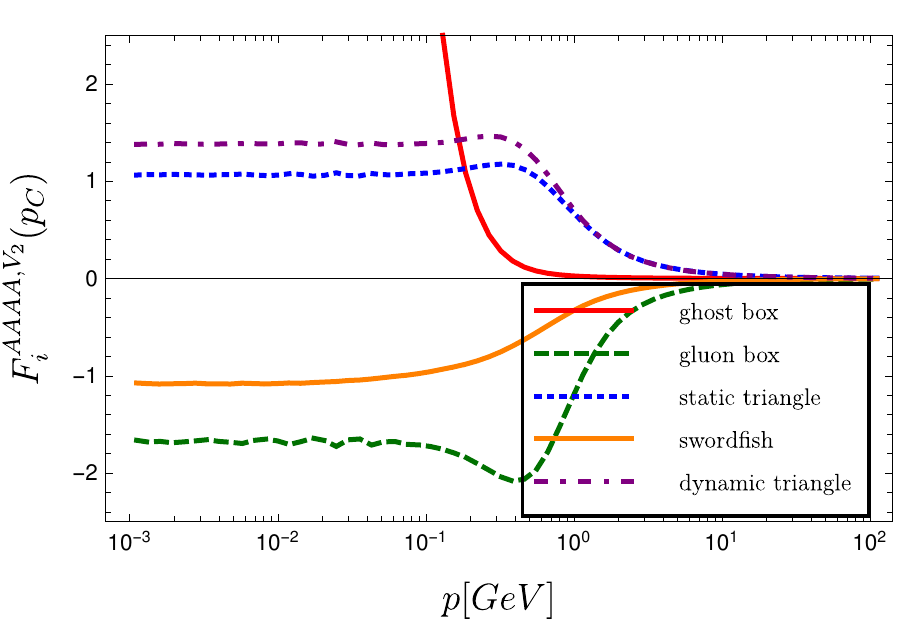}
 \hfill
 \includegraphics[width=0.48\textwidth]{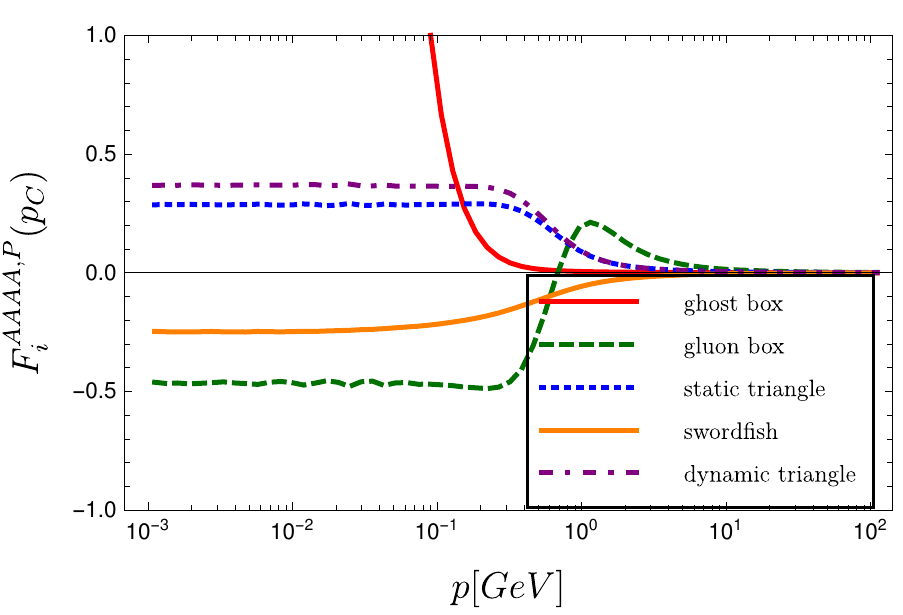}
 \caption{\label{fig:YM3d_fgOtheProjs}Various dressing functions of the four-gluon vertex (top left) and the individual contributions of single diagrams (top right and bottom) \cite{Huber:2016tvc}.}
\end{figure}

\myboxmargin{four-gluon vertex}
\index{four-gluon vertex}
The last correlation function to be discussed is the four-gluon vertex.
Similar to Sec.~\ref{sec:res_fg}, three specific configurations are chosen for illustration purposes.
The calculation itself was done for the full momentum dependence.
As can be seen in \fref{fig:YM3d_fgV1}, there is a similar pattern as for the three-gluon vertex although not as extreme.
Still, the deviation from the tree-level is small.
The dominant contribution at low momenta is again the ghost diagram.
The gluonic diagrams, on the other hand, partially cancel each other.
As can also be seen in \fref{fig:YM3d_fgV1}, the four-gluon vertex has a stronger angle dependence, at least in the variables chosen here, than the three-gluon vertex.
To get an idea about the possible influence of other dressing functions not taken into account, alternative projections were evaluated taken the previously obtained solution as input.
The results for three further projections including the corresponding contributions from single diagrams is shown in \fref{fig:YM3d_fgOtheProjs}.
The only sizable contributions appear around $1\,\GeV$, but compared to the tree-level dressing, which gets a constant contribution from the tree-level diagram, they are small.

\begin{figure}[tb]
 \includegraphics[width=0.47\textwidth]{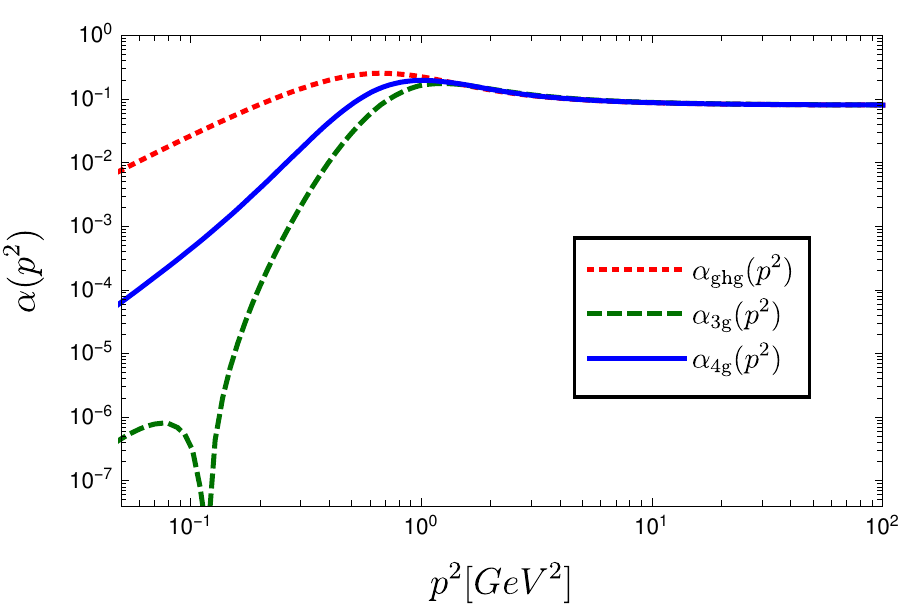}
 \hfill
 \includegraphics[width=0.45\textwidth]{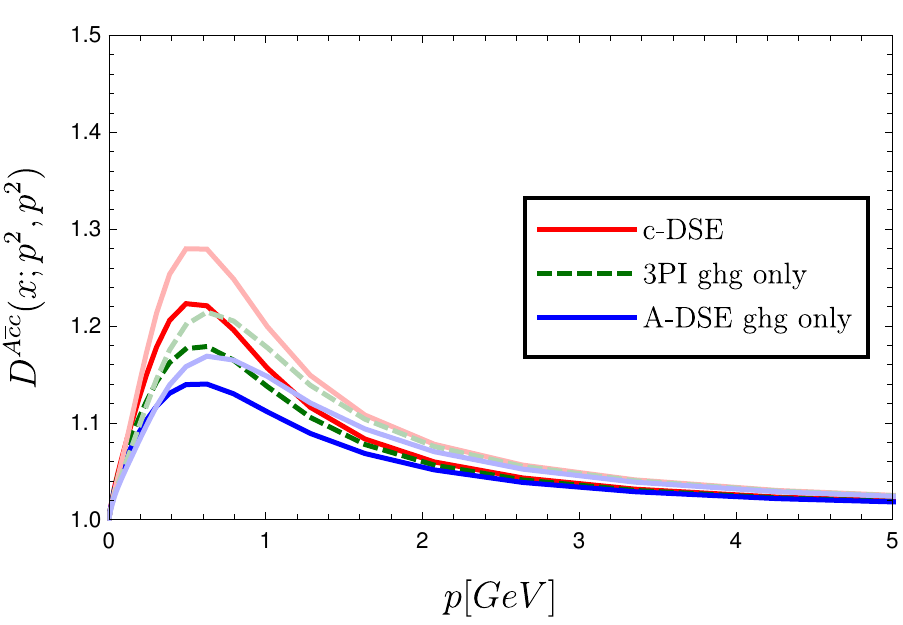}
 \caption{Left: The coupling constants from the ghost-gluon, three-gluon and four-gluon vertices \cite{Huber:2016hns}.
 Right: Ghost-gluon vertex dressing function calculated with fixed input from different equations \cite{Huber:2016tvc}.
 Dark/light lines correspond to $x=0$/$x=p^2$.}
\label{fig:YM3d_couplings+ghgCompEqs}
\end{figure}

\myboxmargin{couplings}
\index{running coupling}
Finally, the couplings as extracted from the different vertices, see \eref{eq:couplings}, are shown in \fref{fig:YM3d_couplings+ghgCompEqs}.
The couplings from the ghost-gluon, the three-gluon and the four-gluon vertices agree very well from the perturbative regime down to a few \GeV.

\subsubsection{Testing the truncation}
\label{sec:YM3d_testing}

\myboxmargin{What to be done: testing truncation in two ways}
While the truncation used in the previous section is already quite large, it is not clear whether it is already sufficient to obtain quantitatively reliable results.
However, since the truncation is large enough, it allows various modifications to test the influence of different parts.
For small truncations such tests are often not possible, because all the ingredients are crucial and tampering with them too much changes the solution drastically.
In this section, the truncation dependence is tested in two different ways.
First, the influence of choosing different equations is tested.
Second, the influence of the highest included correlation function, the four-gluon vertex, is assessed.

\myboxmargin{different equations for ghg: intro}
\index{3PI effective action}
\index{ghost-gluon vertex}
\index{$A$-DSE}
\index{$c$-DSE}
As explained in Sec.~\ref{sec:Acbc}, the ghost-gluon vertex has two different \glspl{dse}, one with the gluon leg attached to the bare vertex, the $A$-DSE, and one with the ghost leg attached to the bare vertex, the $c$-DSE.
The ghost-gluon vertex also fulfills an equation of motion derived from the \gls{3pi} effective action, see \fref{fig:ghg_DSE}.
Here, the results from all three equations are tested.
The two \glspl{dse} are truncated to the two triangle diagrams.
The difference between the three equations lies only in the number and position of dressed vertices.
One should keep in mind, though, that the truncations drop a different number and different types of diagrams:
For the $A$-DSE, nine diagrams are dropped (seven two-loop diagrams and two one-loop diagrams containing a quartic ghost vertex or a two-ghost-two-gluon vertex) and for the $c$-DSE only one diagram containing the two-ghost-two-gluon vertex is dropped.
The \gls{3pi} effective action is expanded up to three loops.
The resulting equation of motion contains the same diagrams as the \glspl{dse}, but all three vertices are dressed.
So while formally the truncation specified above is the same in all three cases, it has different effects on the equations.
The resulting difference can thus be interpreted as the systematic error of this truncation.

\begin{figure}[tb]
 \includegraphics[width=0.45\textwidth]{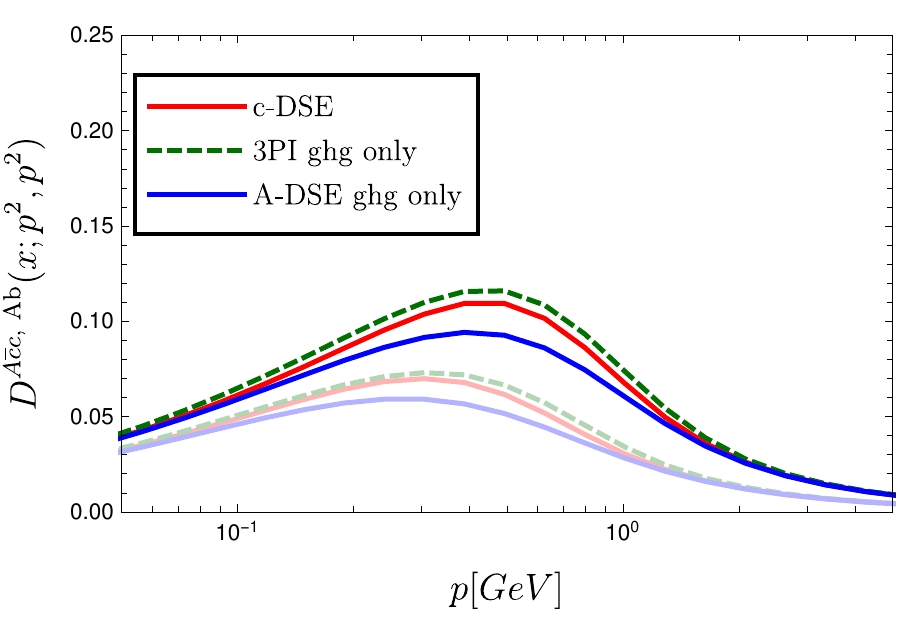}
 \hfill
 \includegraphics[width=0.45\textwidth]{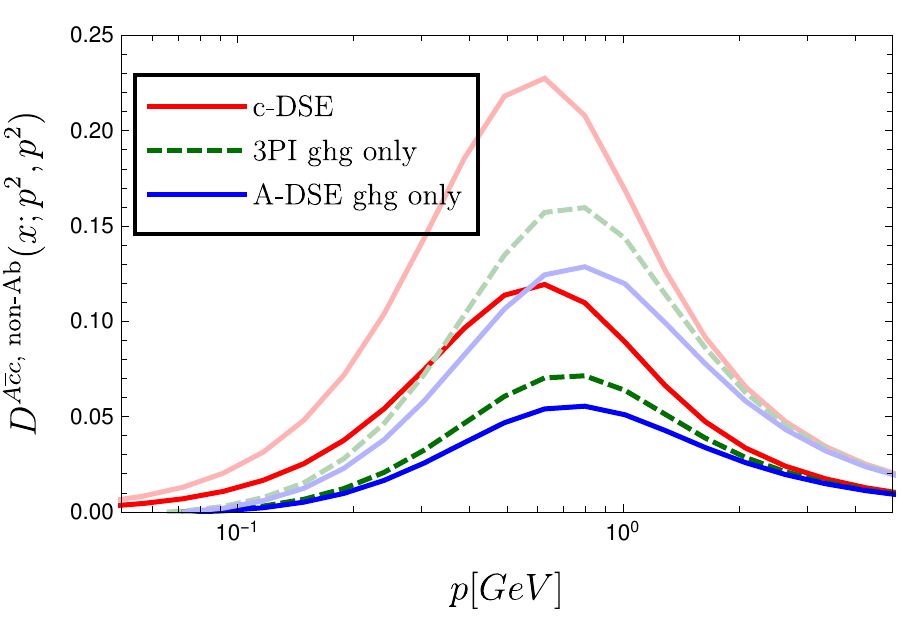}
 \caption{Contributions of the Abelian (left) and non-Abelian (right) diagrams of the ghost-gluon vertex DSE calculated with fixed input from different equations \cite{Huber:2016tvc}.
 Dark/light lines corresponds to $x=0$/$x=p^2$.} 
 \label{fig:YM3d_ghgCompEqs-diags}
\end{figure}

\myboxmargin{different equations for ghg: only ghg}
\index{3PI effective action}
\index{ghost-gluon vertex}
\index{$A$-DSE}
\index{$c$-DSE}
For this comparison fixed input from the full system discussed in Sec.~\ref{sec:YM3d_fullResults} is used to avoid a backcoupling of the truncation effects.
Fig.~\ref{fig:YM3d_couplings+ghgCompEqs} shows the results for two different kinematic configurations from the three equations.
The contributions of single diagrams are shown in \fref{fig:YM3d_ghgCompEqs-diags}.
Taking as a measure of the truncation error $e_\text{ghg}$ the maximum of the ratios of dressings, we obtain $e_\text{ghg}=13\,\%$.
The difference for the configurations shown is even lower and typically below $10\,\%$.

\begin{figure}[tb]
 \begin{center}
 \includegraphics[width=0.45\textwidth]{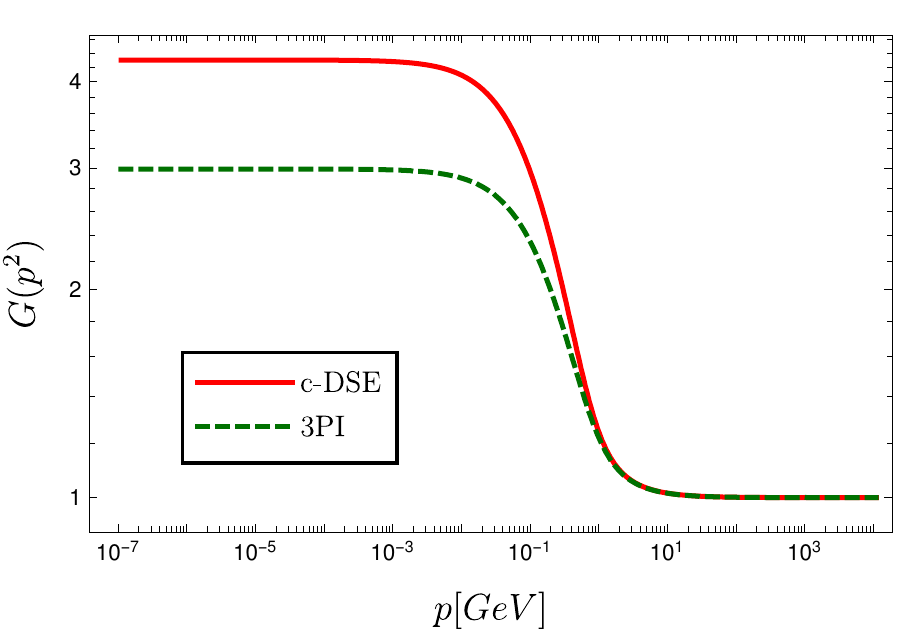}
 \caption{Comparison of results from the DSE and the 3PI systems for the ghost dressing function \cite{Huber:2016tvc}.} 
 \label{fig:YM3d_ghCompDSE3PI}
 \end{center}
\end{figure}

\begin{figure}[tb]
 \includegraphics[width=0.45\textwidth]{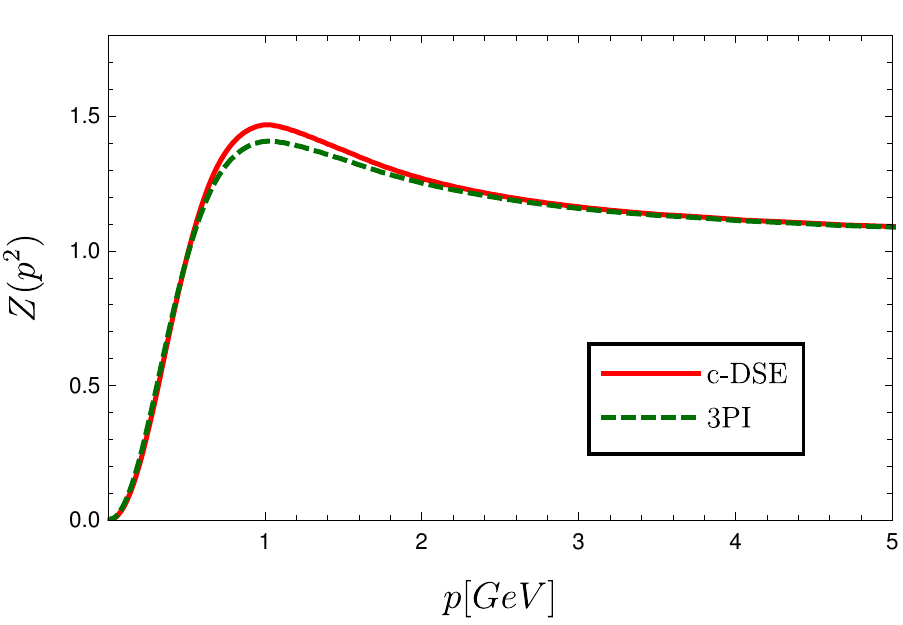}
 \hfill
 \includegraphics[width=0.45\textwidth]{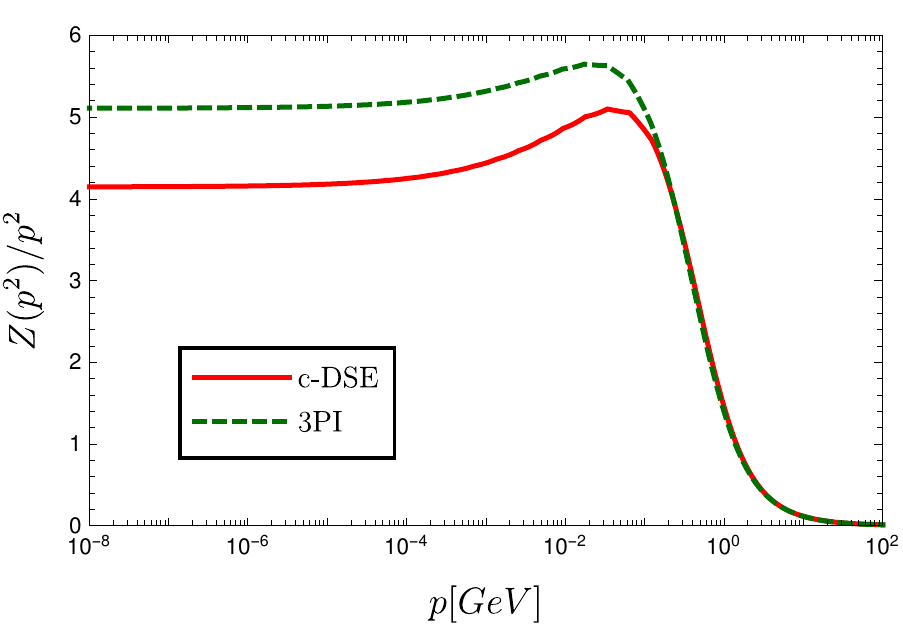}
 \caption{Comparison of results from the DSE and the 3PI systems for the gluon dressing function (left) and the gluon propagator \cite{Huber:2016tvc}.} 
 \label{fig:YM3d_glCompDSE3PI}
\end{figure}

\begin{figure}[tb]
 \includegraphics[width=0.45\textwidth]{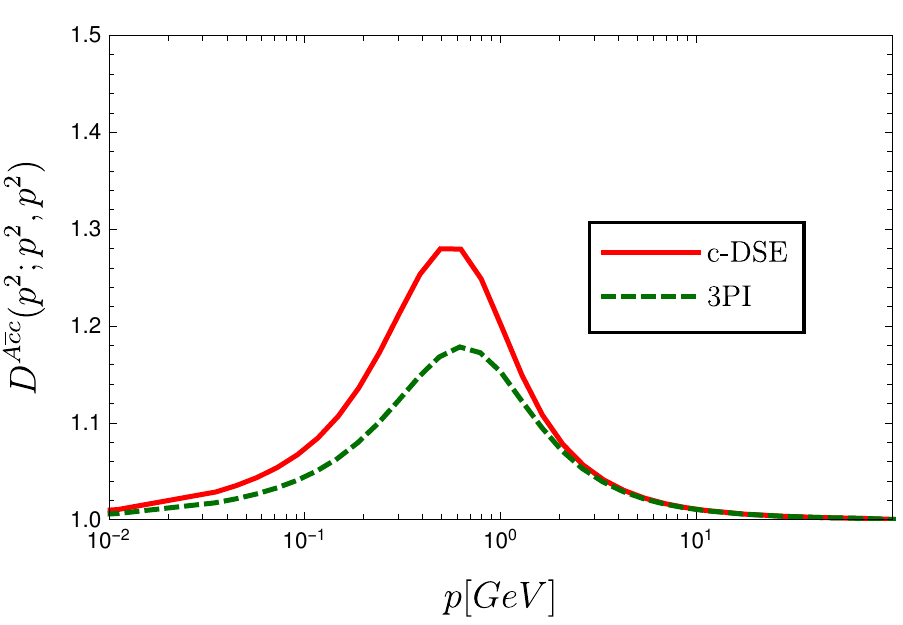}
 \hfill
 \includegraphics[width=0.45\textwidth]{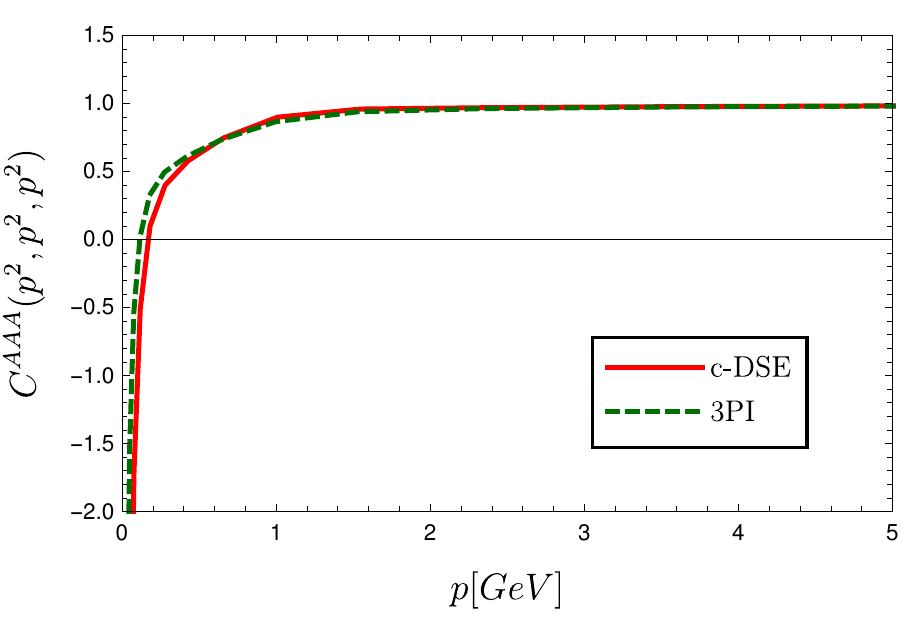}
 \caption{\label{fig:YM3d_ghgTgCompDSE3PI}Comparison of results from the DSE and the 3PI systems for the ghost-gluon vertex (left) and three-gluon vertex (right) dressing functions \cite{Huber:2016tvc}.} 
\end{figure}

\myboxmargin{different equations for ghg: full system}
\index{3PI effective action}
\index{ghost-gluon vertex}
\index{$A$-DSE}
As a second test, the full system was calculated with the equations of motion from \gls{1pi} and \gls{3pi} effective actions.
For the \gls{dse} system, the $c$-DSE was used.
The ghost dressing functions from the two systems differ only below $1\,\GeV$, see \fref{fig:YM3d_ghCompDSE3PI}.
For the gluon propagators the situation is similar and the main difference in the gluon dressing functions occurs at $1\,\GeV$, see \fref{fig:YM3d_glCompDSE3PI}.
Fig.~\ref{fig:YM3d_ghgTgCompDSE3PI} shows the differences for the ghost-gluon and three-gluon vertices.
For the former, it is similar to the ghost-gluon vertex-only calculation discussed above and depicted in \fref{fig:YM3d_couplings+ghgCompEqs}.
For the three-gluon vertex the main difference is a shift of the zero crossing for the \gls{3pi} system towards the \gls{ir}.
Overall the differences between the \gls{dse} and the \gls{3pi} setup are relatively small with the largest effects observed in the deep \gls{ir}.

\begin{figure}[tb]
 \includegraphics[width=0.48\textwidth]{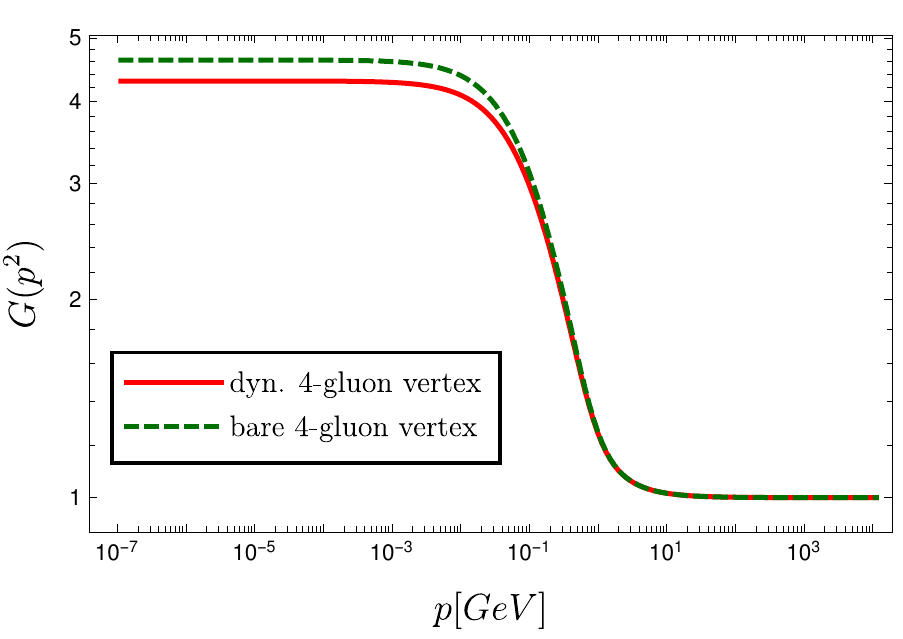}
 \hfill
 \includegraphics[width=0.48\textwidth]{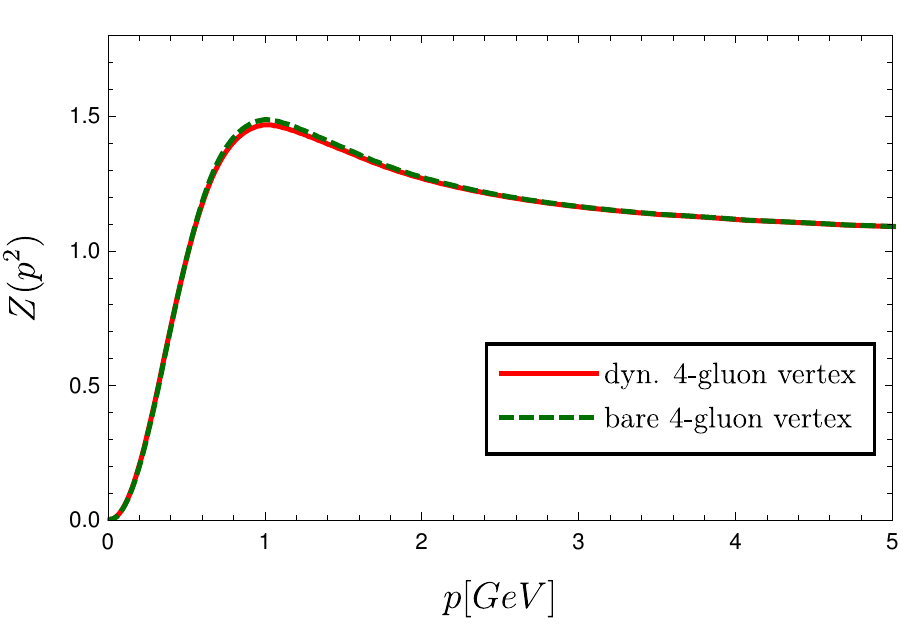}
 \caption{Ghost (left) and gluon (right) dressing functions from the full system with a bare (green, dashed line) and a dynamic four-gluon vertex (red, continuous line) \cite{Huber:2016tvc}.}
 \label{fig:YM3d_ghGlComp4g}
\end{figure}

\begin{figure}[tb]
 \includegraphics[width=0.48\textwidth]{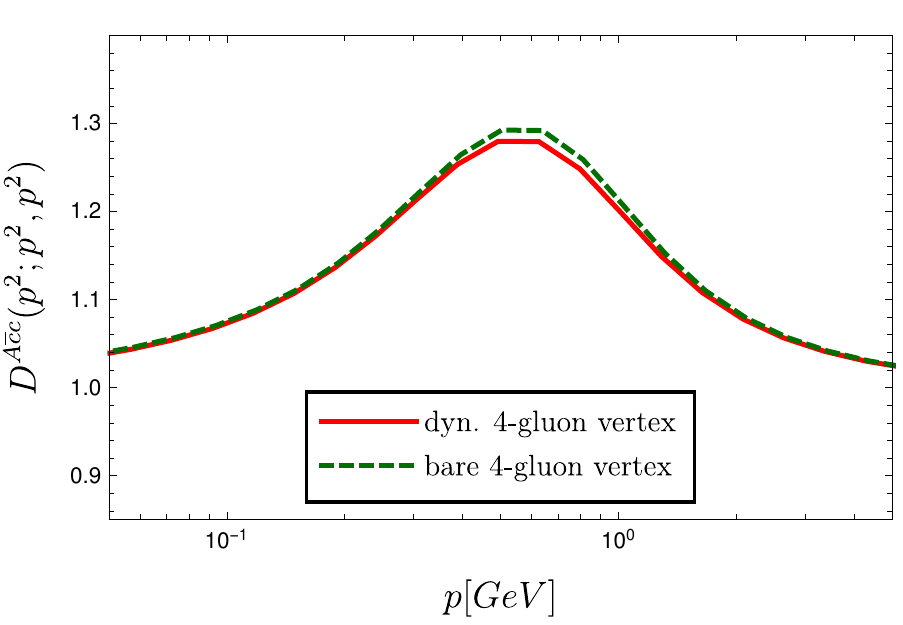}
 \hfill
 \includegraphics[width=0.48\textwidth]{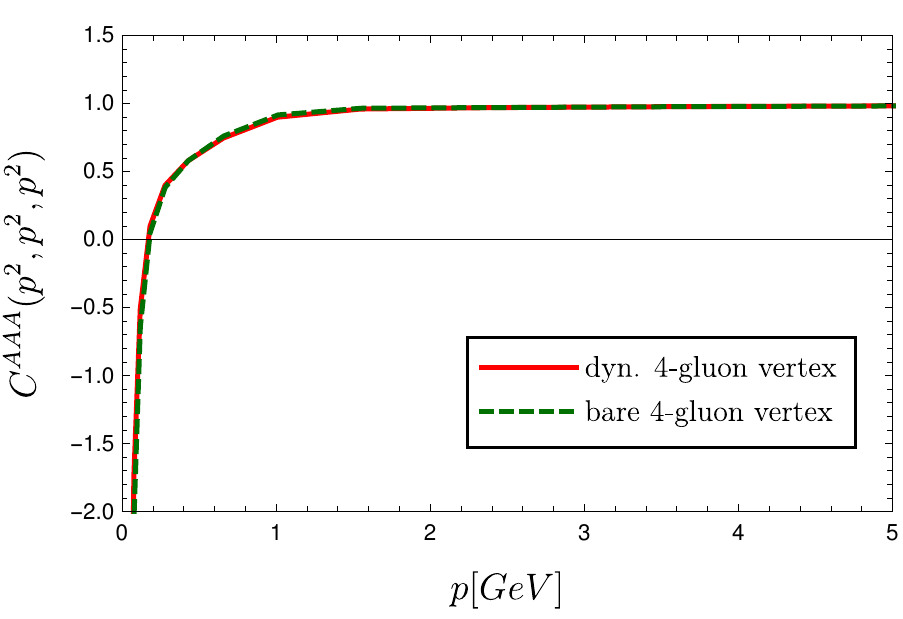}
 \caption{Ghost-gluon vertex (left) and three-gluon vertex (right) dressing from the full system with a bare (green, dashed line) and a dynamic four-gluon vertex (red, continuous line) \cite{Huber:2016tvc}.}
 \label{fig:YM3d_ghgTgComp4g}
\end{figure}

\myboxmargin{higher correlation functions: four-gluon vertex}
\index{four-gluon vertex}
Finally, the effect of the four-gluon vertex is discussed.
The corresponding results of this vertex, which show only a small deviation from the tree-level behavior, suggest that it has only a minor influence on lower correlation functions.
It appears in the gluon propagator and the three-gluon vertex equations and we already saw that the corresponding diagram in the former, the sunset diagram, has only a minor quantitative influence.
The ghost propagator and the ghost-gluon vertex are only influenced indirectly.
To quantify the effects, the system is solved with a bare four-gluon vertex and compared to the results from Sec.~\ref{sec:YM3d_fullResults}.
The comparison is shown in \fref{fig:YM3d_ghGlComp4g} and \fref{fig:YM3d_ghgTgComp4g}.
Clearly, the effect is very small indeed.
However, it should be noted that the four-gluon vertex plays a decisive role in the convergence of the three-gluon vertex \gls{dse} in four dimensions as mentioned in Sec.~\ref{sec:res_fg}.

\subsubsection{Discussion}
\label{sec:YM3d_disc}

Overall, the tests of the truncation performed here show that the changes are only of quantitative nature.
On the one hand, it is promising that finally a truncation was found that exhibits this degree of stability.
First of all, because simpler truncations are more sensitive to the input, and second, because the present truncation does not have the freedom to tune any model parameters.
On the other hand, the question is now why there is still a discrepancy to results from lattice calculations.
Several possible explanations exist.

\myboxmargin{discussion: meaningfulness of comparison with lattice}
\index{gauge fixing}
First, one must ask how meaningful a quantitative comparison between lattice and continuum results is at all, because the ways how the gauge is fixed in the two methods are different and no direct one-to-one correspondence is known.
The dependence on the gauge fixing algorithm is well established on the lattice for three \cite{Maas:2009se,Bornyakov:2011fn,Maas:2011se,Maas:2013vd,Bornyakov:2013ysa} and also four dimensions \cite{Cucchieri:1997dx,Bogolubsky:2005wf,Sternbeck:2006rd,Maas:2009se,Maas:2011se,Sternbeck:2012mf}.
Results exist \cite{Maas:2013vd,Maas:2017csm} where the difference between two ghost dressing functions is $50\,\%$ at $500\,\MeV$ and even more below.
As demonstrated in Sec.~\ref{sec:YM3d_specifics}, also with functional equations different solutions can be obtained.
However, at the moment it is not clear how close the present results should be to the lattice results using the minimal Landau gauge, because the gauge employed here is most likely a different gauge.

\myboxmargin{discussion: truncation sufficient?}
Of course, another source for the differences are contributions that are still missing in the continuum results.
There are two types of such contributions: Incomplete equations and incomplete tensor bases.
In the present truncation, neither applies to the propagators and the latter also does not apply to the ghost-gluon vertex.

\myboxmargin{discussion: ghg}
\index{ghost-gluon vertex}
For the ghost-gluon vertex different equations were tested in Sec.~\ref{sec:YM3d_testing}.
The found differences of the order of $10\,\%$ can clearly be attributed to truncation artifacts.
Thus, it will be necessary to include more diagrams and check how the results change.
In case of the $c$-DSE, this requires only the two-ghost-two-gluon vertex, which was discussed for four dimensions in Sec.~\ref{sec:res_AAcbc}, to make the equation complete, whereas the $A$-DSE contains two-loop diagrams as well.

\myboxmargin{discussion: 3g}
\index{three-gluon vertex}
In addition to missing diagrams, the calculation of the three-gluon vertex also lacks a full tensor basis.
In four dimensions, the three-gluon vertex \gls{dse} was solved with fixed input with a full tensor basis \cite{Eichmann:2014xya}.
It was found that the dressing of the tree-level dressing is the dominant contribution and the other three transverse dressings are quite small.
This should hold also in three dimensions and hence one would expect the effect of completing the basis to be correspondingly small.
However, one should not forget about the non-linearity of the equations which could make the suppressed dressings at least somewhat important.
On the other hand, the three-gluon vertex shows the best agreement with lattice results.
From this one would expect that the truncation works quite well for this particular quantity.

\myboxmargin{discussion: 4g}
\index{four-gluon vertex}
For the four-gluon vertex, the cancellations between diagrams do not work as well as for the three-gluon vertex.
In addition, there are many more dressing functions so that the sum of even small contributions could, if no cancellations appear, be sizable.
Thus, for the four-gluon vertex the choice of a proper basis could well be quite important.
An enlarged tensor basis could also change the importance of the sunset diagram if additional tensors are taken into account.

\myboxmargin{discussion: summary truncation}
In summary, there remain a few extensions of the present truncation the effects of which still have to be tested.
In the meantime, it is reassuring that with the \gls{frg} results with similar deviations from lattice results were found \cite{Cyrol:2017phd,Corell:2018yil}.
Taking into account also information from four dimensions, extensions of the three and four-gluon vertex bases currently seems to be a promising approach, especially since these vertices directly enter in the gluon propagator \gls{dse}.

\subsection{Correlation functions of two-dimensional Yang-Mills theory}
\label{sec:YM2d}

\index{lattice calculations}
\index{Yang-Mills theory!in two dimensions}
Yang-Mills theory in two dimensions is another testing ground for lattice and functional methods.
One of the reasons for a more detailed investigation with lattice methods was the hints found in three and four dimensions that the three-gluon vertex could become negative at low momenta \cite{Cucchieri:2006tf,Cucchieri:2008qm}.
A calculation in two dimensions can probe even deeper into the \gls{ir}.
Besides the confirmation of the zero crossing of the three-gluon vertex, it was also found that the correlation functions in two dimensions behave qualitatively different to three and four dimensions in the \gls{ir} regime \cite{Maas:2007uv}.
Within the available statistics, the scaling solution seemed to be realized for propagator and vertices \cite{Zwanziger:2001kw,Maas:2007uv,Huber:2007kc}.
Later studies, however, raised some questions, since the scaling laws seem not to respect the sum rule \cite{Pawlowski:2009iv,Maas:2009ph,Cucchieri:2011um,Cucchieri:2011ig,Maas:2015nva}.

\index{RGZ framework}
The absence of a decoupling solution sparked some additional interest in the study of two-dimensional Yang-Mills theory.
A study with the (refined) \gls{gz} framework found that the condensates responsible for making the gluon propagator finite in three and four dimensions do not exist in two dimensions due to \gls{ir} divergences \cite{Dudal:2008xd}.
Also in other approaches, \gls{ir} divergences were found to play a decisive role \cite{Tissier:2010ts,Tissier:2011ey,Weber:2011nw,Cucchieri:2012cb,Huber:2012zj,Zwanziger:2012xg}.

\begin{figure}[tb]
 \begin{center}
 \includegraphics[width=0.48\textwidth]{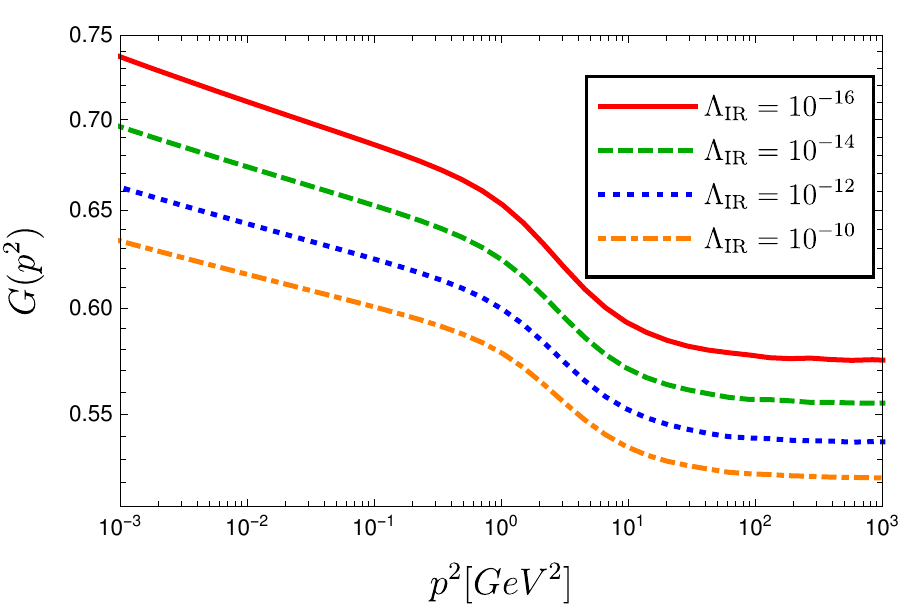}\hfill   
 \caption{Ghost dressing functions in two dimensions resulting from a decoupling type gluon propagator \cite{Huber:2012zj}.
 The various curves correspond to different IR cutoffs.}
 \label{fig:gh_only_gh_decoupling_2d}
 \end{center}
\end{figure}

\index{IR divergences}
\index{ghost propagator}
In \gls{dse} calculations the relevant divergences appear in the ghost propagator equation.
They need to be counteracted by a gluon propagator that vanishes at zero momentum \cite{Cucchieri:2012cb}.
This holds as long as the ghost-gluon vertex does not vanish itself, which, however, is not found in explicit calculations \cite{Huber:2012zj}.
Note, though, that these calculations of the vertex are done for the scaling case.
For the vertex to vanish at zero momentum in the decoupling case, the contribution of the tree-level would need to be canceled by the loop contributions.
Thus, a decoupling solution is unlikely to be realized in two dimensions.
Quantitatively, for the decoupling case the \gls{ir} divergences lead to an explicit dependence on the \gls{ir} cutoff as shown in \fref{fig:gh_only_gh_decoupling_2d}.

Another effect in two dimensions is the mixing of different momentum regimes.
In four dimensions, a clear separation is possible.
For example, the \gls{uv} behavior can be determined self-consistently without referring to low scales.
In three dimensions, the influence of different momentum regimes on each other is already noticeable \cite{Maas:2004se,Huber:2016tvc}, see also Sec.~\ref{sec:YM3d}.
However, a perturbative calculation can still be performed for high momenta.
In two dimensions, on the other hand, perturbation theory does not work anymore again due to arising \gls{ir} divergences.
They are cured by the nonperturbative behavior of the gluon propagator that vanishes at zero momentum.
Thus, the asymptotic high momentum behavior in two dimensions can only be determined qualitatively from dimensional arguments as
\begin{align}
 G(p^2)^{-1}&=1+c_\text{gh}\frac{g^2}{p^2},\nnnl
 Z(p^2)^{-1}&=1+c_\text{gl}\frac{g^2}{p^2}.
\end{align}
The coefficients $c_\text{gh}$ and $c_\text{gl}$ can be determined numerically in a nonperturbative calculation.

The strong entanglement of different momentum regimes makes it difficult to recover the correct \gls{uv} behavior in a truncated system, since any defects in other regimes have an impact on the \gls{uv} regime.
This problem appears also in three dimensions but is less severe there, see Sec.~\ref{sec:YM3d}.
However, as long as models are used for the vertices, once can use them to tune the \gls{uv} behavior of the dressing functions to approach $1$ in the \gls{uv}.

\begin{figure}[tb]
 \begin{center}
 \includegraphics[width=0.46\textwidth]{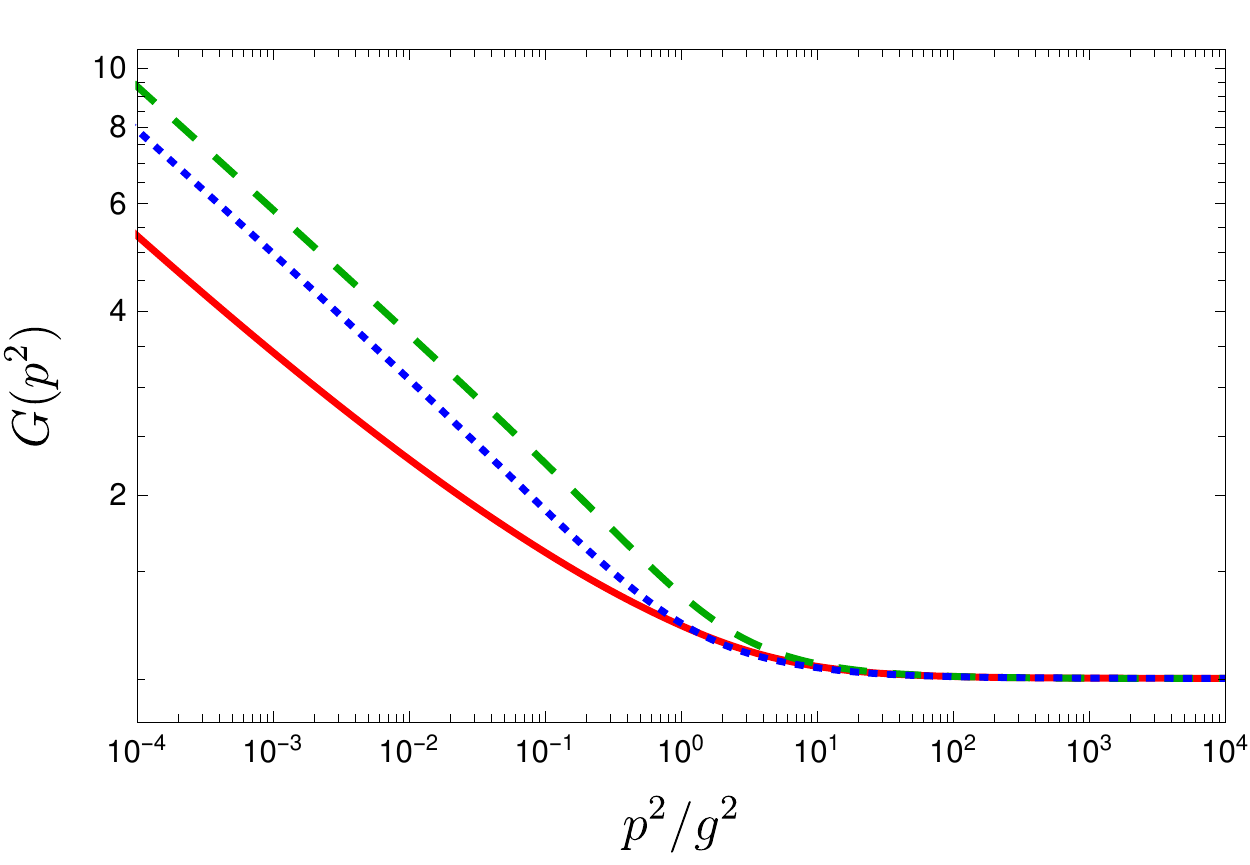}
 \hfill
 \includegraphics[width=0.46\textwidth]{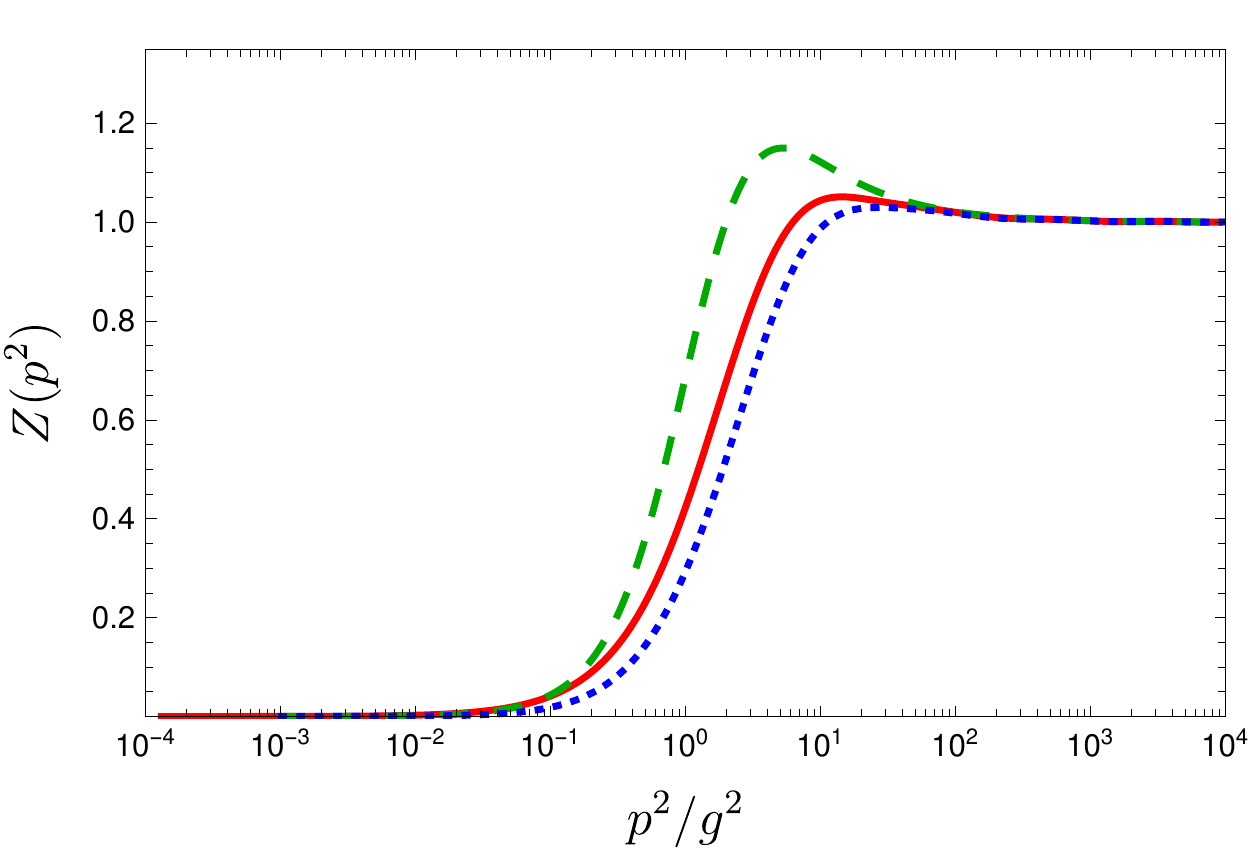}\\ 
 \includegraphics[width=0.46\textwidth]{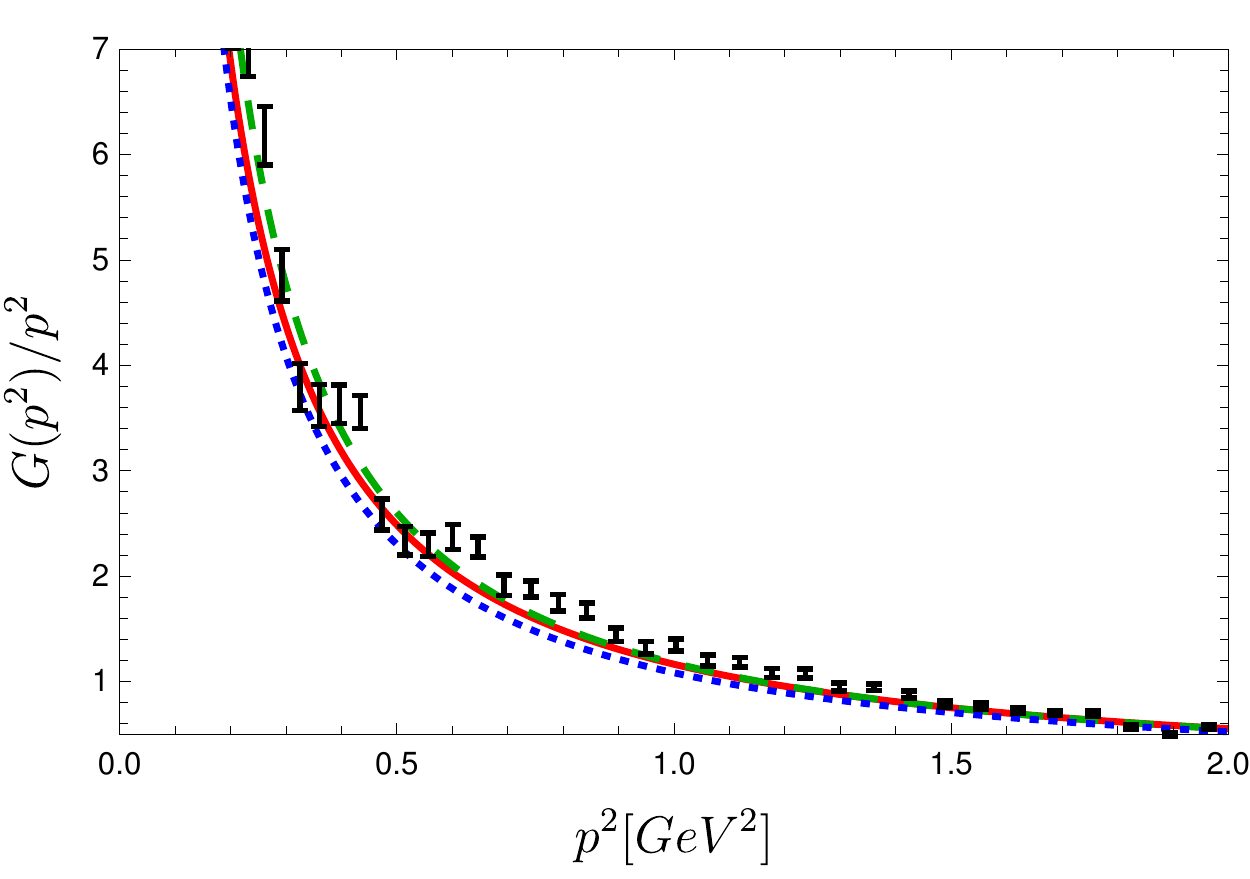}
 \hfill
 \includegraphics[width=0.46\textwidth]{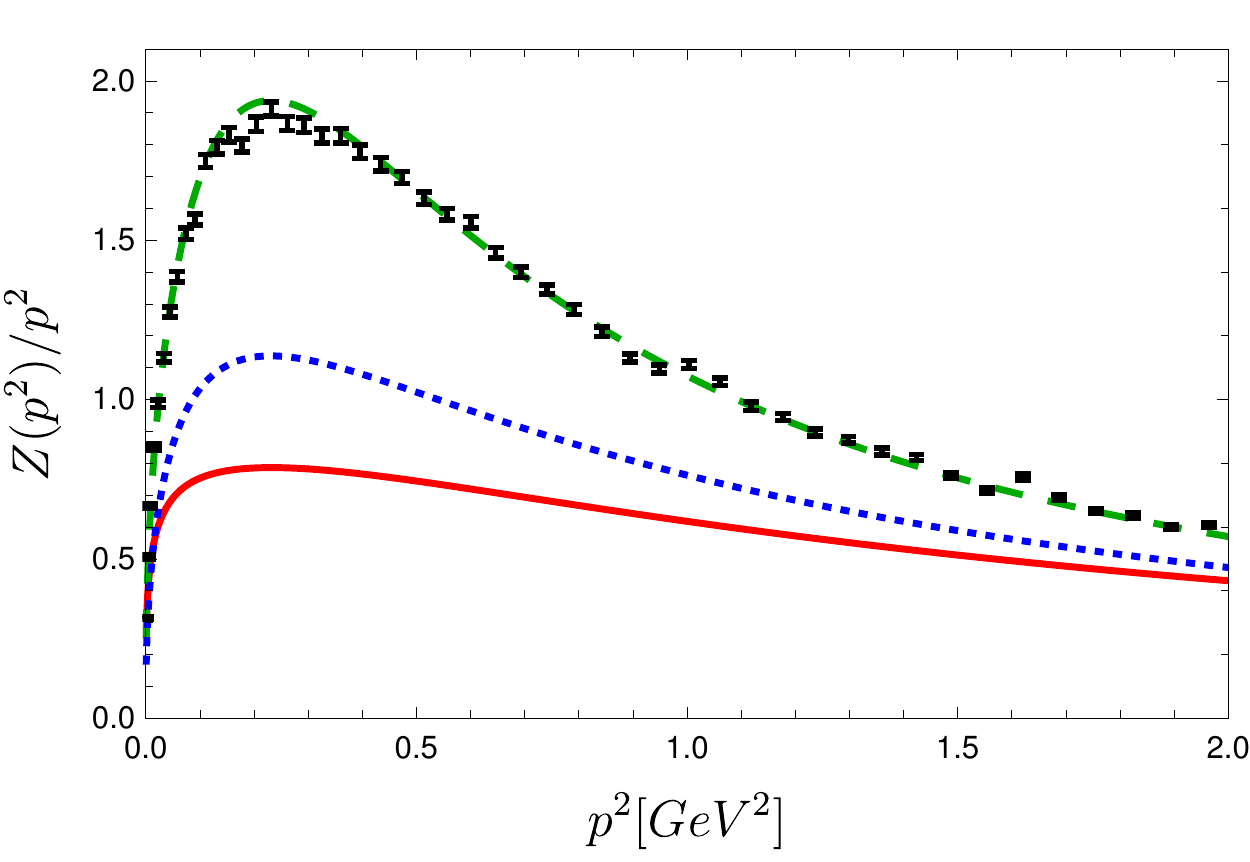}
 \caption{Ghost (left) and gluon (right) dressing functions (top) and propagators (bottom) obtained with different ghost-gluon vertices \cite{Huber:2012zj}:
  For the red/continuous lines a dynamic ghost-gluon vertex was used, for the green/dashed lines a bare ghost-gluon vertex, and for the blue/short-dashed lines the model from
  \eref{eq:ghg_ansatz_2d}.
  The three-gluon vertex model of \eref{eq:tg_ansatz_2d} was used with the parameters given in \tref{tab:YM2d_params}.
  The bottom row shows a comparison of the propagators for $SU(2)$ with lattice data (black points) from \cite{Maas:2011se}.
  The scale was set by matching the positions of the maxima in the gluon propagators to the lattice results.}
  \label{fig:YM2d_props}
 \end{center}
\end{figure}

Truncating the system of propagators to the one-loop diagrams without tadpole diagram \cite{Huber:2012zj}, only the ghost-gluon and three-gluon vertices are required.
In the following, three choices for the vertices will be tested.
Two use a model for the ghost-gluon vertex and one includes it dynamically.
In the former case, it is once taken as bare and once the following model is employed:
\begin{align}\label{eq:ghg_ansatz_2d}
 D^{A\bar{c}c}_{\text{mod}}(r,p,q)=1+\frac{1}{\Lambda^2+p^2+q^2+r^2}\left(f_\text{IR}+f_\text{IM}\frac{\Lambda^2(p^2+q^2)}{\Lambda^4+p^4+q^4}\right).
\end{align}
For the three-gluon vertex, two models will be tested.
The first one, which is constructed to mimic the known behavior of the vertex, is
\begin{align}\label{eq:tg_ansatz_2d}
 C^{AAA}_{\text{mod},1}(p,q,r)=h_\text{IR}(p^2+q^2+r^2)^{-3\ka-1}+\frac{p^2+q^2+r^2}{p^2+q^2+r^2+h_\text{IM}\Lambda^2}.
\end{align}
The correct UV behavior is ensured by the second term, whereas the first term implements the \gls{ir} divergence \cite{Huber:2012zj}.
Choosing $h_\text{IR}$ negative leads to a zero crossing.
The position of the zero crossing resulting from the employed parameters is higher than one would expect from lattice results.
In this way, the gluon loop can make up some missing strength coming from the neglected two-loop diagrams.
An alternative model, inspired by a model in three-dimensions \cite{Maas:2004se} that was created before the qualitative \gls{ir} behavior of the vertex was known, is
\begin{align}
\label{eq:tg_ansatz_2d_from3d}
 C^{AAA}_{\text{mod},2}(p,q,r)=\left(\frac{(G(p^2)G(q^2)G(r^2))^{-2-1/\kappa}}{Z(p^2)Z(q^2)Z(r^2)}\right)^\alpha.
\end{align}

The three combinations of vertices tested in the following are:
\begin{itemize}
 \item A bare ghost-gluon vertex and \eref{eq:tg_ansatz_2d_from3d} for the three-gluon vertex.
 \item Eq.~(\ref{eq:ghg_ansatz_2d}) for the ghost-gluon vertex and \eref{eq:tg_ansatz_2d} for the three-gluon vertex.
 \item A dynamically coupled ghost-gluon vertex and \eref{eq:tg_ansatz_2d} for the three-gluon vertex.
\end{itemize}
The parameter values for the models are listed in \tref{tab:YM2d_params}.

\begin{figure}[tb]
 \includegraphics[width=0.48\textwidth]{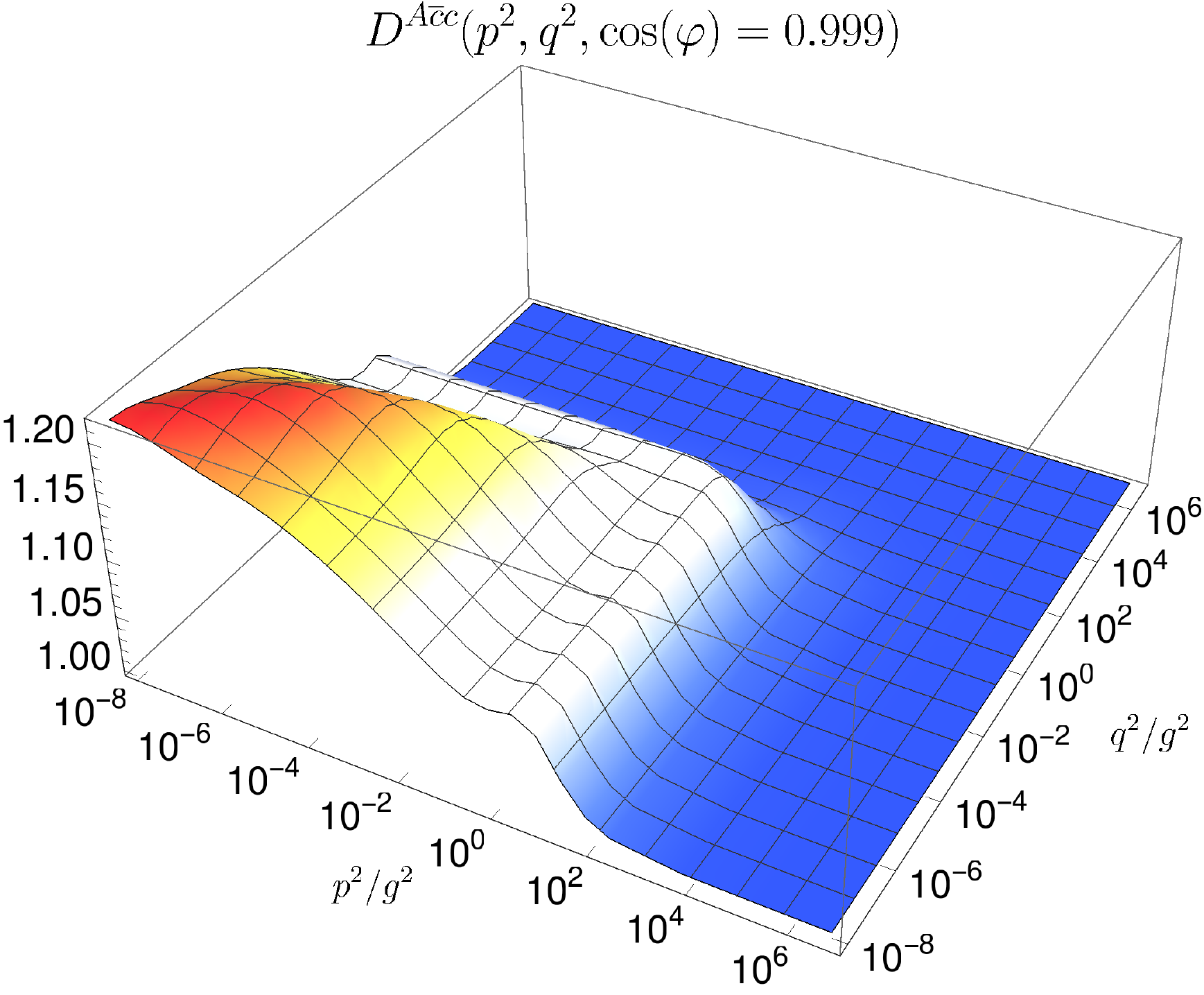}
  \hfill
 \includegraphics[width=0.48\textwidth]{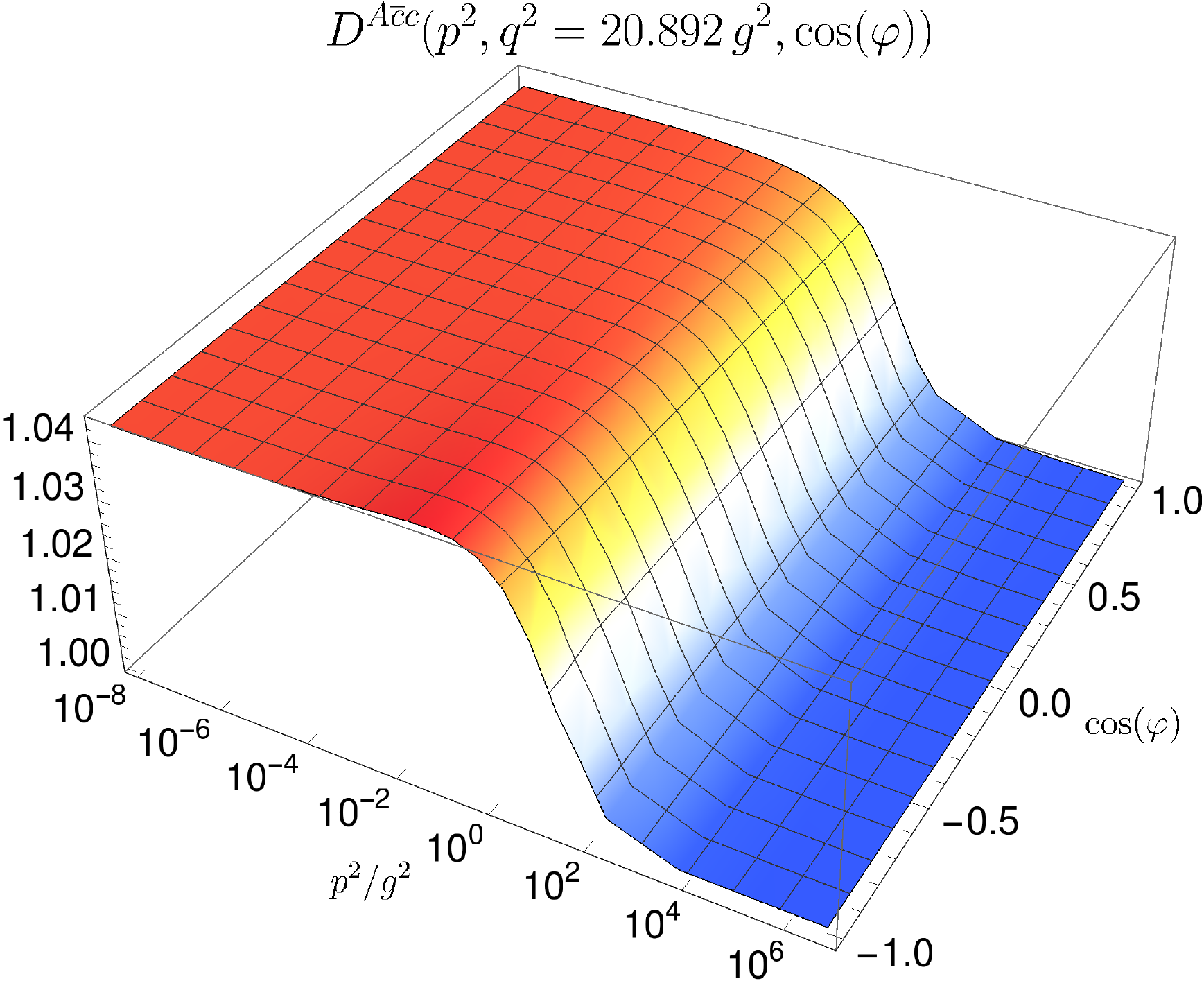}
 \caption{Dressing of the ghost-gluon vertex for various momentum configurations obtained from the coupled system of propagators and the vertex itself \cite{Huber:2012zj}.
 For the three-gluon vertex the ansatz of \eref{eq:tg_ansatz_2d} with the parameters given in \tref{tab:YM2d_params} was used.
 Fixed angle (left) and fixed ghost momentum $q^2$ (right) as indicated at the top of each plot.
 Note the different scales on the $z$-axes.
 }
 \label{fig:YM2d_ghg}
\end{figure}

Two-dimensional Yang-Mills theory is in principle superrenormalizable.
However, also here spurious divergences arise in the gluon propagator \gls{dse}, see Sec.~\ref{sec:spurDivs}.
They are logarithmic and appear in the following form:
\begin{align}
\frac{1}{Z(p^2)}=1+\Sigma(p^2)=1+\frac{g^2}{p^2}\Sigma'(p^2)+c\,g^2\frac{\ln\Lambda^2}{p^2}.
\end{align}
Here, $\Sigma(p^2)$ is logarithmically divergent and $\Sigma'(p^2)$ is finite.
The logarithmic divergence can be handled by a subtraction of the self-energy at $s$ multiplied by $s^2/p^2$:
\begin{align}\label{eq:Z_subtracted}
 \frac{1}{Z(p^2)}-\frac{1}{Z(s^2)}\frac{s^2}{p^2}=\Sigma(p^2)-\Sigma(s^2)\frac{s^2}{p^2}=\frac{g^2}{p^2}\Sigma'(p^2)-\frac{g^2}{p^2}\Sigma'(s^2),
\end{align}
Alternatively, other methods discussed in Sec.~\ref{sec:spurDivs} could be used.

\begin{table}[]
\begin{center} 
\begin{tabular}{|l||l|l||l|l||l|l|}
\hline
\textbf{Ghost-gluon vertex} & \multicolumn{2}{|l||}{Bare} & \multicolumn{2}{|l||}{Model \ref{eq:ghg_ansatz_2d}} & \multicolumn{2}{|l|}{Dynamic}\\
\hline
\textbf{Three-gluon vertex} & \multicolumn{2}{|l||}{Model \ref{eq:tg_ansatz_2d_from3d}} & \multicolumn{2}{|l||}{Model \ref{eq:tg_ansatz_2d}} & \multicolumn{2}{|l|}{Model \ref{eq:tg_ansatz_2d}}\\
\hline
\hline
\multicolumn{1}{l|}{} & $\alpha$ & $0.09$ & $\Lambda$ & $1\,\text{GeV}$ & $\Lambda$ & $1\,\text{GeV}$\\
\cline{2-7}
\multicolumn{3}{l||}{} &$f_\text{IR}$ & $2.14189$ & $h_\text{IR} $ & $-31.5/-14.75$\\
\cline{4-7}
\multicolumn{3}{l||}{} &$f_\text{IM}$ & $0.6$ & $h_\text{IM}$ & $9.88$\\
\cline{4-7}
\multicolumn{3}{l||}{} &$h_\text{IR} $ & $-109.616/-54.2$ & \multicolumn{2}{}{} \\
\cline{4-5}
\multicolumn{3}{l||}{} & $h_\text{IM}$ & $9.88$ & \multicolumn{2}{}{} \\
\cline{4-5}
\end{tabular}
\caption{Values employed for the parameters of the ghost-gluon and three-gluon vertex ans\"atze given in eqs.~(\ref{eq:ghg_ansatz_2d}), (\ref{eq:tg_ansatz_2d}) and (\ref{eq:tg_ansatz_2d_from3d}).
The second value for $h_\text{IR}$ is for $SU(2)$.}
\end{center}
\label{tab:YM2d_params}
\end{table}

\index{ghost propagator}
\index{gluon propagator}
\index{ghost-gluon vertex}
The results for the three setups are shown in \fref{fig:YM2d_props}.
Clearly, the models have a large impact on the midmomentum regime.
The ghost-gluon vertex result is shown in \fref{fig:YM2d_ghg}.
Due the large deviations between the functional and the lattice results in the propagators of the combined calculation, the agreement of the ghost-gluon vertex result with lattice results is not very good either \cite{Huber:2012zj}.
The employed level of truncation is rather basic, though, and it would be interesting to see how the results of all quantities would change in a more elaborate truncation as used in Sec.~\ref{sec:YM3d} for three dimensions.

\begin{figure}[tb]
 \begin{center}
 \includegraphics[width=0.47\textwidth]{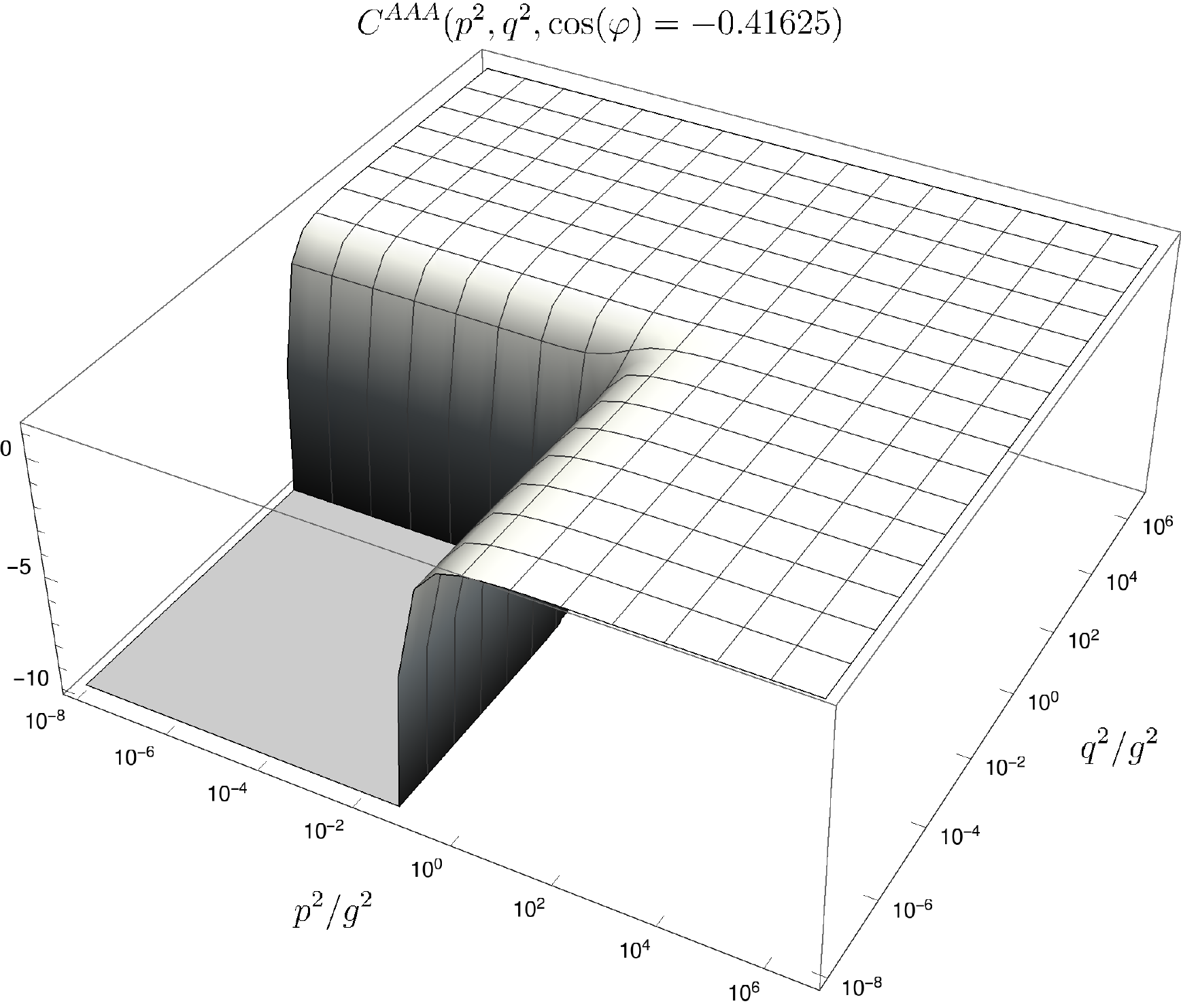}
 \hfill
 \includegraphics[width=0.47\textwidth]{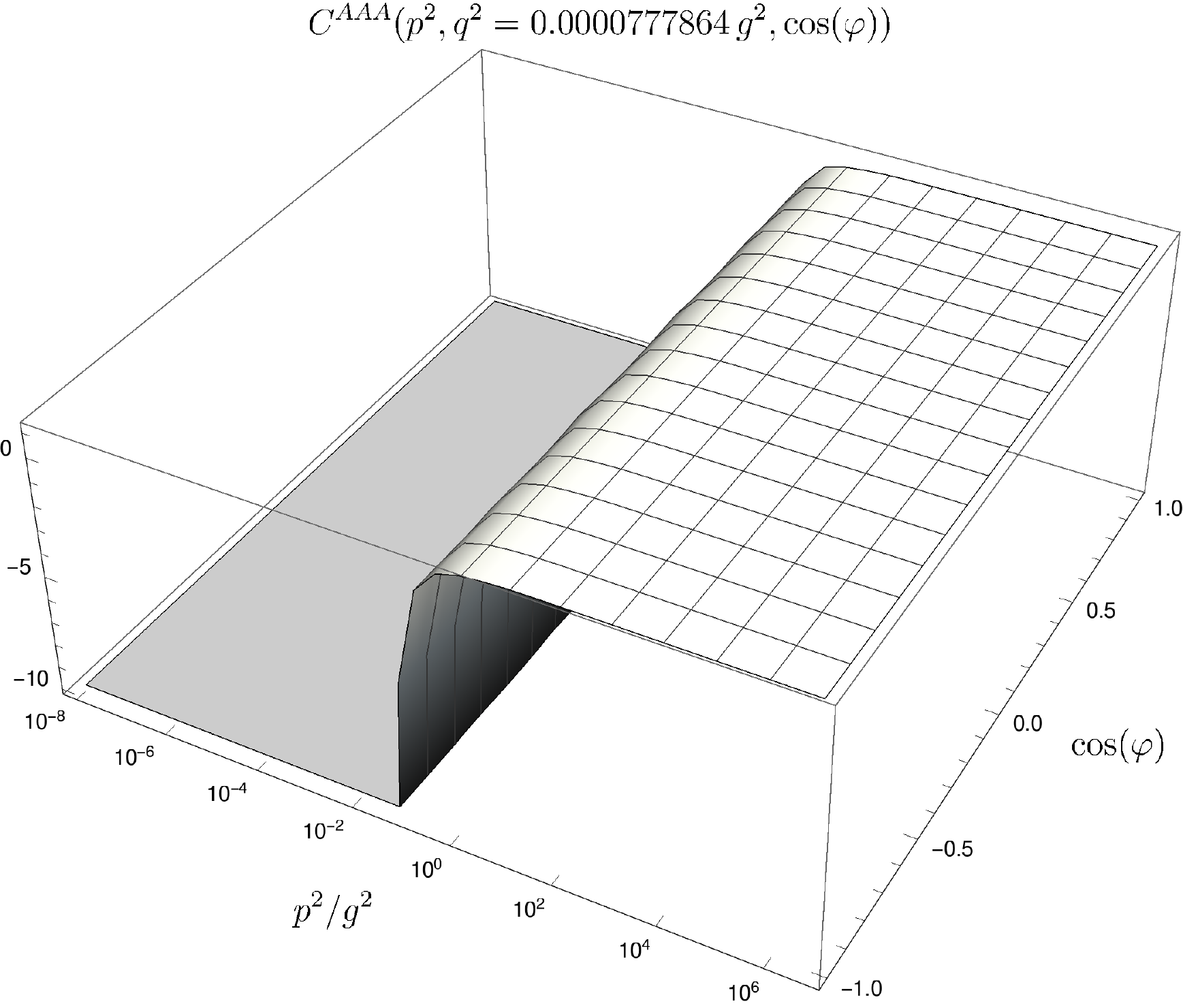}\\
 \vskip3mm
 \includegraphics[width=0.47\textwidth]{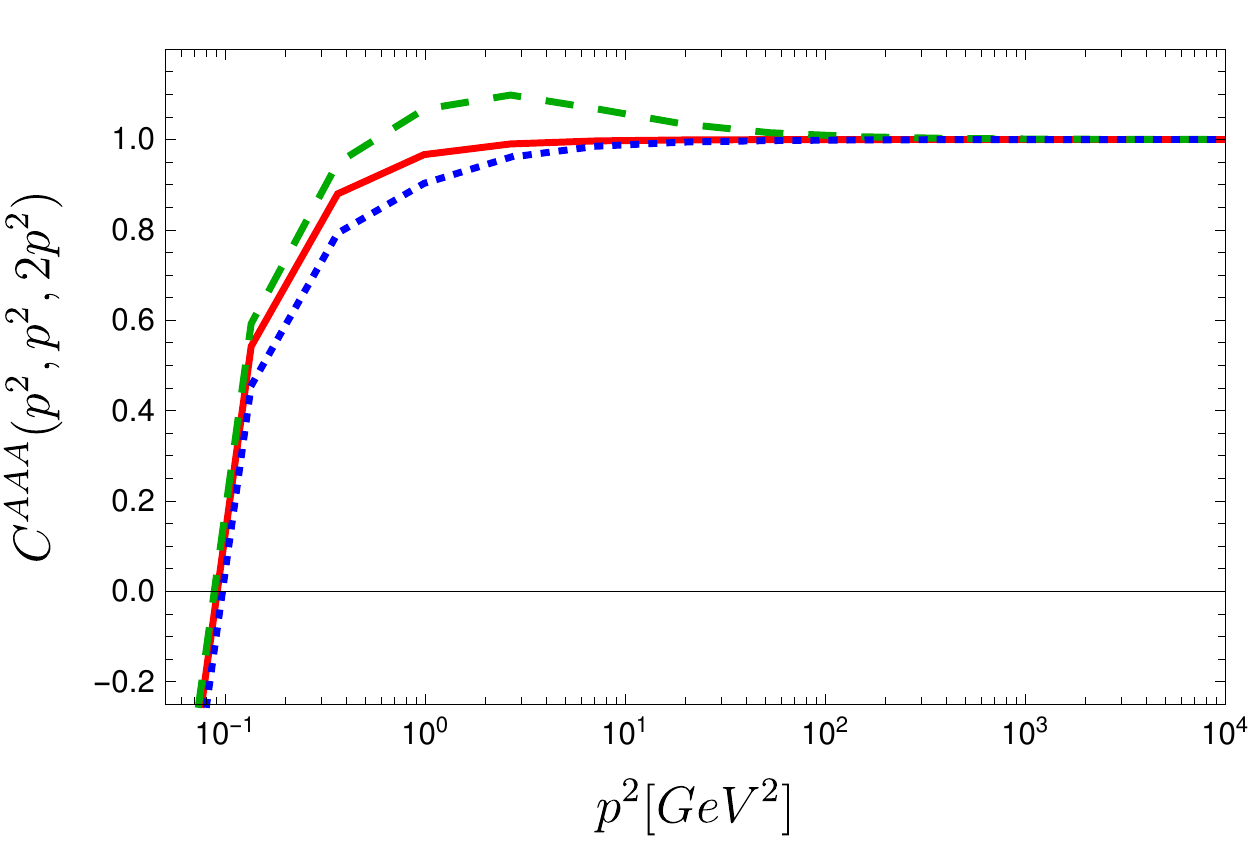}
 \hfill
 \includegraphics[width=0.47\textwidth]{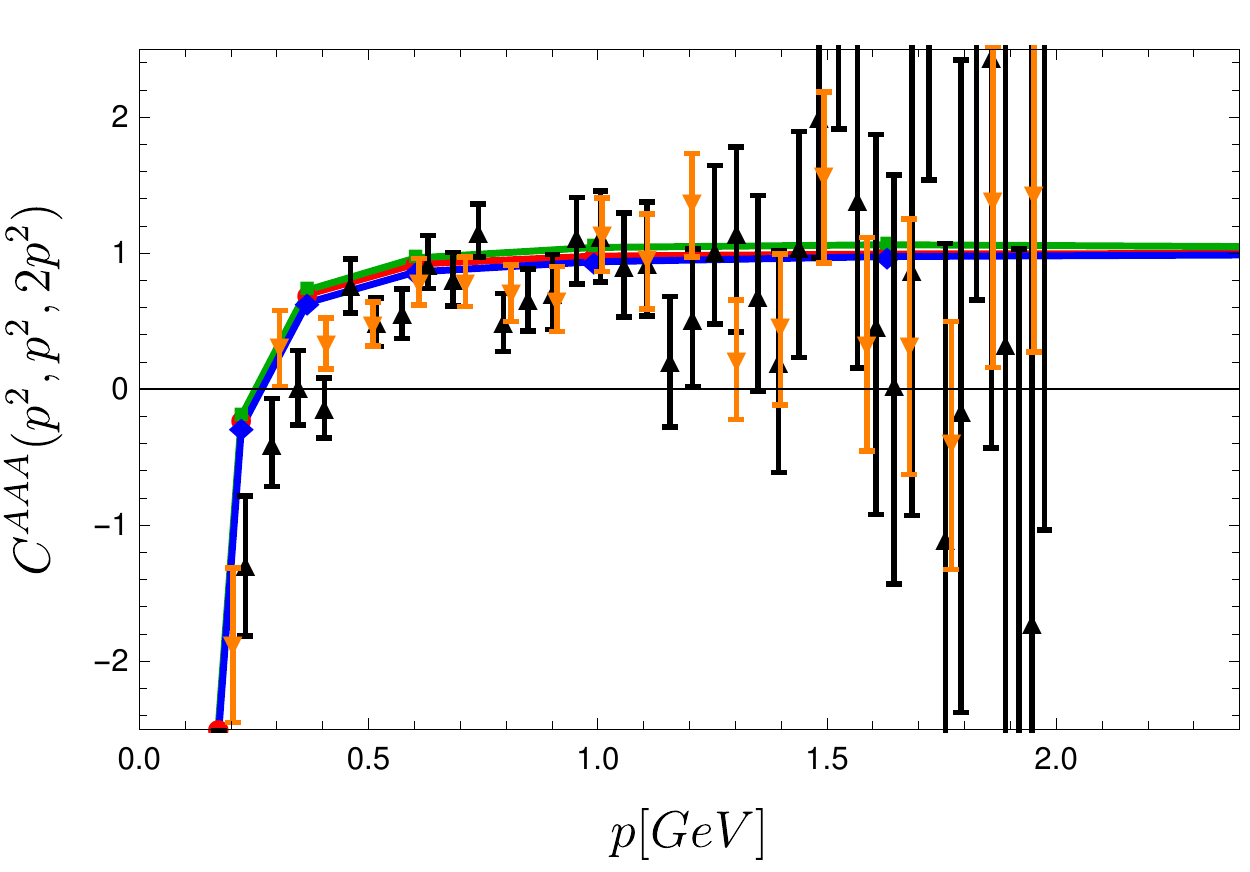}
  \caption{Tree-level dressing of the three-gluon vertex \cite{Huber:2012zj}.
  Top: Angle fixed at $\arccos(-0.41625)$ (left) and one momentum fixed at $\sqrt{0.00007786}\,g$ (right).
  Bottom: Three-gluon vertex in the symmetric configuration (left) and for two orthogonal momenta (right).
  The continuous/red line is with the ghost triangle only, the dashed/green one with both triangles and the dotted/blue line with all five diagrams.
  The comparison with lattice data is done for $SU(2)$ \cite{Maas:2007uv}.
  Black up-triangles are for $\beta=10$/$L=21\,\text{fm}^{-1}$ and orange down-triangles for $\beta=22.5$/$L=12\,\text{fm}^{-1}$.}
  \label{fig:YM2d_tg}
 \end{center}
\end{figure}

\index{three-gluon vertex}
For the three-gluon vertex, results are shown in \fref{fig:YM2d_tg}.
They were calculated from the solution of the propagators and the ghost-gluon vertex using a bare four-gluon vertex.
The employed truncation is the same one used in Sec.~\ref{sec:YM3d}, viz., all diagrams with two loops or non-primitively divergent vertices are discarded.
This provides the correct IR behavior and includes also the leading corrections to the tree-level behavior in the UV.
Fig.~\ref{fig:YM2d_tg} also shows the effect of different diagrams.
Clearly, the ghost triangle is leading in the \gls{ir}.
As in higher dimensions, a cancellation between the gluon triangle and the swordfish diagrams is observed: The gluon triangle introduces a bump in the dressing which is removed completely by the inclusion of the swordfish diagrams.

In summary, two-dimensional Yang-Mills theory, for which vertex \glspl{dse} were solved for the first time self-consistently, proved to be a useful testing ground for developing the necessary methods and get a first impression of the general problems like cancellations between diagrams in the gluonic vertices.
The results also hinted at the fact that, in the absence of resummation, the three-gluon vertex can be recovered in good agreement with lattice results, while a good description of the propagators is more difficult.

\section{Correlation functions in different gauges}
\label{chp:otherGauges}

\index{gauges}
The correlation functions of quarks and gluons are necessarily gauge dependent.
As explained in Sec.~\ref{sec:why_LG}, the Landau gauge has many advantages that make it very convenient to work with.
However, beyond that, also other gauges are investigated.
An obvious reason for using different gauges is to check the gauge dependence of gauge invariant quantities introduced by the truncation.
Another reason to investigate different gauges is that they might provide alternative perspectives at specific physical problems.
For example, the color-Coulomb potential, viz., the instantaneous part of the $00$-component of the gluon propagator in Coulomb gauge, is related to the potential of a quark--anti-quark pair \cite{Zwanziger:2002sh}.
This is known as 'no confinement without Coulomb confinement'.
Another example is the maximally Abelian gauge which allows the investigation of the Abelian dominance conjecture related to the dual superconductor picture of confinement.

In this section, results for various gauges beyond the Landau gauge are reviewed.
The first example are linear covariant gauges in Sec.~\ref{sec:linearCovGauges}.
The Landau gauge is actually just one specific gauge of this class.
In Sec.~\ref{sec:mag}, the maximally Abelian gauge is discussed and the Coulomb gauge in Sec.~\ref{sec:Coulomb}.
Beyond these examples, infinitely many more gauges exist, for example, the axial gauge, the planar gauge, and the light-cone gauge, see, e.g., Ref.~\cite{Leibbrandt:1987qv}, or exotic ones like the Palumbo gauge \cite{Palumbo:1985mj,Palumbo:1990kg}.
However, they are hardly used in functional equations and some gauges like the Laplacian gauge \cite{deForcrand:2000yr} are not accessible with functional methods at all.

\myboxmargin{Interpolating gauges}
\index{gauges!interpolating}
Before the discussion of results from the different specific gauges, an interesting possibility should be mentioned here, namely the interpolation between different gauges.
In the strict sense, also linear covariant gauges are such an interpolating gauge, but it is also possible to interpolate, for instance, between linear covariant and maximally Abelian gauges.
However, with the different levels of sophistication in studies of different gauges, calculations using interpolating gauges are currently rather limited.
Nevertheless, the prospect of connecting confinement scenarios that are more easily accessible in one or the other gauge makes them interesting for possible future calculations.

\mybox{Coulomb gauge and Landau gauge}
\index{Coulomb gauge}
\index{Landau gauge}
\index{$\lambda$ gauge}
A prime example of an interpolating gauge is the so-called $\lambda$ gauge which interpolates between the Landau and the Coulomb gauges via the gauge fixing condition
\begin{align}
 f_{\lambda}[A]=[\partial'_\mu A_\mu(x)]^2, \quad \partial'_\mu=\lambda \partial_0+\partial_i.
\end{align}
The case $\lambda=1$ corresponds to the Landau gauge, while the limit $\lambda \rightarrow 0$ corresponds to a Coulomb-like gauge.
For $\lambda=0$ an additional condition must be supplemented for a full gauge fixing condition.
Note that also the Faddeev-Popov operator depends now on $\lambda$:
\begin{align}
 M^{rs}	(x,y)=-\delta(x-y)\partial'_\mu D^{rs}.
\end{align}
The $\lambda$ gauge was studied on the lattice in Refs.~\cite{Maas:2006fk,Cucchieri:2007uj,Maas:2011ej}.

\mybox{list of interpolating gauges}
\index{maximally Abelian gauge}
\index{Curci-Ferrari gauge}
\index{Landau gauge}
\index{Coulomb gauge}
Besides the Coulomb-Landau gauges example many other realizations of interpolating gauges exist, for example, formulations for interpolation between the Landau gauge and the maximally Abelian gauge \cite{Dudal:2004rx,Huber:2009wh,Huber:2010ne}; linear covariant gauges, the Coulomb gauge and the maximally Abelian gauge \cite{Capri:2005zj,Capri:2006bj}; or linear covariant gauges, the maximally Abelian gauge and the Curci-Ferrari gauge \cite{Dudal:2005zr}.

\subsection{Linear covariant gauges}
\label{sec:linearCovGauges}

\mybox{
\begin{itemize}[label=\Square]
 \item GZ
 \item lattice
 \item DSEs
 \item Nielsen identities
\end{itemize}
}

\myboxmargin{general definition of linear covariant gauges}
\index{linear covariant gauges}
\index{Landau gauge}
The most natural choice of gauge beyond the Landau gauge are linear covariant gauges, also called $R_\xi$ gauges.
They have the same gauge fixing condition as the Landau gauge, $\partial A=0$, but it is not enforced strictly.
Rather, the gauge orbit is sampled around the Landau gauge with a Gaussian weight $e^{-\frac{1}{2\xi} (\partial A)^2}$ where the gauge fixing parameter $\xi$ determines the width:
\begin{align}
 \mathcal{L}_\text{gf}= \frac{1}{2\xi}(\partial_\mu A^r_\mu(x))^2-\int dy \,\bar{c}^r(x)\,M^{rs}(x,y)\,c^s(y).
\end{align}
The endpoint $\xi=0$ is the Landau gauge where the Gaussian distribution becomes a delta functional.
For $\xi>0$, the gauge field is not transverse.
Consequently, the Faddeev-Popov operator $M^{rs}(x,y)=M^{rs}(x)\delta(x-y)=-\partial_\mu D_\mu^{rs}\delta(x-y)$ is no longer Hermitian because of the term $-g f^{rst}(\partial A)$.
Thus, its eigenvalues are not real as in the Landau gauge and a Gribov region cannot be defined in the same manner.
However, there is a generalization of the Gribov-Zwanziger procedure \cite{Capri:2016aqq} that can be applied also to linear covariant gauges \cite{Capri:2015ixa}.

\myboxmargin{lattice definition of linear covariant gauges}
For lattice calculations, the gauge fixing process of linear covariant gauges poses an additional challenge, since it cannot be formulated in the standard way as a minimization problem \cite{Giusti:1996kf}.
A suitable implementation was suggested in Ref.~\cite{Cucchieri:2009kk}.

\myboxmargin{longitudinal correlations functions}
\index{Slavnov-Taylor identity}
\index{gluon propagator}
The relaxation of the gauge fixing condition entails that the gluon propagator in linear covariant gauges has a second dressing function.
However, it is fixed by gauge symmetry via a corresponding \gls{sti}.
Thus, the gluon propagator has the form
\begin{align}\label{eq:glP_linCov}
 D_{\mu\nu}(p)=P_{\mu\nu}(p)\frac{Z(p^2)}{p^2} + \xi \frac{p_\mu p_\nu}{p^4}.
\end{align}
Due to the longitudinal part of the propagator also the longitudinal parts of the vertices are relevant in functional equations.
In the Landau gauge, on the other hand, it is straightforward to see that the transverse part constitutes a closed set \cite{Fischer:2008uz}.

In Sec.~\ref{sec:linCov_props}, results for the propagators in linear covariant gauges are presented.
Details on how the longitudinal parts of the vertices are obtained are described in Sec.~\ref{sec:NI_WI}.
There, also the dependence of correlation functions on the gauge fixing parameter $\xi$ is discussed using Nielsen identities.

\subsubsection{Nielsen and Ward identities}
\label{sec:NI_WI}

In this section the Nielsen and Ward identities for linear covariant gauges are derived.
The former, encoding the gauge parameter dependence, can be used to calculate correlation functions for different values of the gauge fixing parameter from a solution for a fixed value.
The latter can be used to fix the longitudinal parts of correlation functions.

\index{Nielsen identity}
The fact that physical quantities must be independent of the gauge fixing parameter $\xi$ can be expressed via the \emph{Nielsen identity}.
One starts with the derivative of the effective action with respect to a gauge fixing parameter:
\begin{align}
 \frac{\partial\Gamma[\Phi]}{\partial \xi}&=-\frac{\partial W[J]}{\partial \xi}=- \frac{\partial}{\partial \xi}\ln\int D[\phi] e^{-S + \phi_i J_i}\nnnl
 &=\left\langle \frac{\partial S}{\partial \xi} \right\rangle =-\left\langle \frac{1}{2\xi^2} f[A]^2 \right\rangle.
\end{align}
The notation from Sec.~\ref{sec:der_dse} was used where $\Phi$ represents all classical \gls{qcd} fields and $J$ the corresponding sources.
For linear covariant gauges, we have $f[A]=\partial A$.
The Nielsen identity reads then
\begin{align}
\frac{\partial\Gamma[\Phi]}{\partial \xi}=\frac{1}{2\xi^2}\int_q q_\mu q_\nu \langle A^a_\mu(q) A^a_\nu(-q) \rangle=\frac{1}{2\xi^2}\int_q q_\mu q_\nu D^{aa,J}_{\mu\nu}.
\end{align}

Equations for correlation functions can be obtained by applying further derivatives with respect to fields.
For the ghost and gluon two-point functions, $\Gamma^{ab}(p)$ and $\Gamma_{\mu\nu}^{ab}(p)$, respectively, one obtains
\begin{align}
\partial_\xi \Gamma^{ab}(p)= & \int_q \Gamma_{\rho}^{A\bar c c,cda}(-q;p-q,p)\Gamma_{\rho'}^{A\bar c c,cbd}(q;p,p-q)\frac{q_\rho q_{\rho'}}{q^4}D(p-q)+\nnnl
 +&\frac1{2}\int_q \Gamma_{\rho\rho'}^{AA\bar cc,abcc}(q,-q,p,-p)\frac{q_\rho q_{\rho'}}{q^4},\\
\partial_\xi \Gamma_{\mu\nu}^{ab}(p)=& \int_q \Gamma_{\mu\rho\sigma}^{AAA,ade}(p,-q,-p+q)\Gamma_{AAA,\nu\rho'\sigma'}^{bde}(-p,q,p-q)D_{\rho\rho'}(p-q)\frac{q_\sigma q_{\sigma'}}{q^4}+\nnnl
 +&\frac1{2}\int_q \Gamma_{\mu\nu\rho\rho'}^{AAAA,abcc}(p,-p,q,-q)\frac{q_\rho q_{\rho'}}{q^4}.
\end{align}
Following a different route, the explicit equations for the ghost and gluon propagators were also derived in \cite{Aguilar:2015nqa}.

\myboxmargin{derivation Ward identities}
\index{Ward identities}
\emph{Ward identities} are derived from the invariance of the path integral under gauge transformations, see Refs.~\cite{Litim:1998qi,Freire:2000bq,Pawlowski:2005xe,Gies:2006wv} for details:
\begin{align}\label{eq:WI}
\frac1{Z}\int D\Phi \, \mathcal{G}^a e^{-S_{\text{YM}}-S_{\text{gf}}-S_{\text{gh}}-S_{\text{sources}}}=0,
\end{align}
where the Ward operator $\mathcal{G}^a$,
\begin{align}
\mathcal{G}^a=D_\mu^{ab}(x)\frac{\de}{\de A_\mu^b(x)}+g\,f^{abc}\left(c^c \frac{\de}{\de c^b}+\bar{c}^c \frac{\de}{\bar{c}^b} \right),
\end{align}
is the generator of infinitesimal gauge transformations.
Eq.~(\ref{eq:WI}) is the generating equation for the Ward identities of all Yang-Mills correlation functions.
By applying appropriate derivatives, the Ward identities for the ghost-gluon and three-gluon vertex follow.
Keeping only the (dressed) tree-level terms, as indicated by $\overset{\text{TL}}{=}$, they read
\begin{align}\label{eq:WI_ghg}
 i\,k_\mu\,\Gamma_\mu^{abc}(k;p,q)\overset{\text{TL}}{=}-g\,f^{abc}(q^2\,G^{-1}(q^2)-p^2\,G^{-1}(p^2)-k^2-p\cdot k)\\
\end{align}
for the ghost-gluon vertex and
\begin{align}\label{eq:WI_tg}
i\,p_\mu\,\Gamma_{\mu\nu\rho}^{abc}(p,q,r)\overset{\text{TL}}{=}g\,f^{abc}\left(\Gamma_{\nu\rho}^{T}(r)-\Gamma_{\nu\rho}^{T}(q)\right)
\end{align}
for the three-gluon vertex.
Note that the result for the three-gluon vertex does not assume any specific choice of basis tensors.
The inhomogeneous terms of the ghost-gluon vertex are a direct consequence of the Faddeev-Popov term.

\subsubsection{Propagators of linear covariant gauges from Dyson--Schwinger equations}
\label{sec:linCov_props}

\myboxmargin{special cases of linear covariant gauges}
\index{Yennie gauge}
\index{Feynman gauge}
Generally, linear covariant gauges have not been extensively investigated with the exception for their endpoint, the Landau gauge.
In the following, I will only refer to the case $\xi>0$, as the Landau gauge is discussed in detail in Sec.~\ref{chp:results_YM_Landau}.
Another special case of linear covariant gauges is the Feynman gauge, defined by $\xi=1$.
However, its special role is limited to perturbation theory because the gluon propagator reduces to the simple form $g_{\mu\nu}/p^2$ at tree-level.
This property does not extend to the non-perturbative regime, where the transverse part acquires a dressing.
The case $\xi=3$ is called the Yennie gauge.
Its special property is that the ghost one-loop self-energy becomes finite.
The same happens for the gluon for the value $\xi=13/3$.

\myboxmargin{nonperturbative behavior of linear covariant gauges}
In recent years, some progress has been made in nonperturbative investigations of linear covariant gauges:
In lattice techniques the obstacle of how to fix the gauge was overcome \cite{Cucchieri:2009kk,Bicudo:2015rma}, and also continuum methods have shed some light on the nonperturbative behavior of its correlation functions \cite{Aguilar:2015nqa,Huber:2015ria,Capri:2015pja,Capri:2015nzw,Capri:2016aif,Capri:2016gut,Aguilar:2016ock,DeMeerleer:2018txc,DeMeerleer:2019kmh}.
The picture that has emerged during the last few years by combining several methods is the following: The gluon propagator is affected only very little by varying the gauge fixing parameter.
In particular, it stays constant at zero momentum.
This was seen in lattice calculations \cite{Bicudo:2015rma}, Dyson--Schwinger calculations \cite{Huber:2015ria}, the \gls{pt-bfm} \cite{Aguilar:2016ock} and the refined Gribov-Zwanziger framework \cite{Capri:2015pja,Capri:2015nzw,Capri:2016aif,Capri:2016gut}.
The ghost dressing function, on the other hand, vanishes logarithmically for $\xi>0$, as seen with Nielsen identities \cite{Aguilar:2015nqa} and \glspl{dse} \cite{Aguilar:2015nqa,Huber:2015ria}.
Also in the refined Gribov-Zwanziger framework this holds \cite{Capri:2015nzw}.
First lattice results do not reach far enough into the \gls{ir} to say anything about the asymptotic behavior \cite{Silva:2018ta}.

\myboxmargin{longitudinal parts of correlation functions}
For $\xi>0$, calculations are aggravated compared to the Landau gauge by the explicit appearance of the longitudinal parts of correlation functions.
In the case of the gluon propagator, this can easily be handled, because the exact result is know from an \gls{sti}, see \eref{eq:glP_linCov}.
However, due to the necessary truncations, the calculation of the longitudinal part from its \gls{dse} or flow equation would show deviations from the exact result.
For higher correlation functions, the \glspl{sti} cannot be solved exactly.
Alternatively, one can use Ward identities to derive closed expressions.
The form of these equations is similar to \glspl{dse} and to render them useful they have to be truncated.
As the simplest truncation, the \gls{uv} leading terms can be kept, viz., only the tree-level terms are retained, see Sec.~\ref{sec:NI_WI}.

\myboxmargin{transverse parts of vertices}
\index{ghost-gluon vertex}
For an explicit calculation, also the transverse parts of the three-point functions need to be specified.
As a minimal requirement, the models should obey the correct anomalous running in the perturbative regime.
Since the ghost-gluon vertex is no longer finite for $\xi>0$, it also runs logarithmically.
This can be accommodated by the following model for the transverse part:
\begin{align}
 D^{A\bar c c,T}(k^2;p^2,q^2)=G(\bar{p}^2)^{\alpha_1^{\text{gg}}} Z(\bar{p}^2)^{\beta_1^{\text{gg}}}.
\end{align}
The exponents $\alpha_1^{\text{gg}}$ and $\beta_1^{\text{gg}}$ are determined by demanding the correct \gls{uv} behavior of the dressing.
A natural choice for the second condition would be \gls{ir} finiteness.
However, this cannot be achieved, since the ghost dressing function vanishes logarithmically.
So the exponents are determined assuming that the ghost dressing function is constant which should give a reasonable approximation.
The exponents can be found in \tref{tab:linCov_exponents_RGI}.
The full ghost-gluon vertex is then given by
\begin{align}\label{eq:ghg-tr-long}
&\Gamma_\mu^{abc}(k;p,q)=i\,g f^{abc}\left(D^{A\bar c c,T}(k^2;p^2,q^2)P_{\mu\nu}(k)p_\nu+D^{A\bar c c,L}(k^2;p^2,q^2)k_\mu\right)
\end{align}
with 
\begin{align}\label{eq:ghg_WI}
 D^{A\bar c c,L}(k^2;p^2,q^2)=\frac{q^2\,G^{-1}(q^2)-p^2\,G^{-1}(p^2)-k^2-p\cdot k}{k^2}.
\end{align}

\index{three-gluon vertex}
The three-gluon vertex consists of four transverse dressing functions.
In the Landau gauge, the dressing of the tree-level term is dominant \cite{Eichmann:2014xya}, so as a first approximation for the transverse part we will take into account the transversely projected tree-level tensor only.
For the corresponding dressing function we make a similar ansatz as for the ghost-gluon vertex:
\begin{align}
 D^{\text{3g}}(k^2,p^2,q^2)=G(\bar{p}^2)^{\alpha_1^{\text{3g}}} Z(\bar{p}^2)^{\beta_1^{\text{3g}}}.
\end{align}
The exponents are determined in the same manor and can be found in \tref{tab:linCov_exponents_RGI}.
Whenever the three-gluon vertex is longitudinally contracted,  \eref{eq:WI_tg} is used.

The anomalous dimensions of propagators and vertices are given in \tref{tab:linCov_anom_dims}.
The sum rules in linear covariant gauges are $\gamma+2\delta+2\gamma^\text{ghg}+1=0$ and $3\gamma+2\gamma^\text{3g}+1=0$.

\myboxmargin{IR behavior of ghost dressing function}
\index{ghost propagator}
Having specified these vertex models, the propagator \glspl{dse} can be investigated.
First, the ghost propagator \gls{dse} is analyzed from which the \gls{ir} behavior of the ghost propagator can directly be inferred.
After plugging in \eref{eq:ghg-tr-long}, it reads
\begin{align}
 G(x)^{-1}&=\widetilde{Z}_3+\frac{g^2\,N_c \widetilde{Z}_1}{8\pi^3} \int dy \,d\varphi \frac{\sin(\varphi)^2 }{z}G(z)\nnnl
 &\quad \times\left(-\sin(\varphi)^2  D^{A\bar c c,T}(y,z,x)Z(y)+\xi\frac{q \cos \varphi}{p}D^{A\bar c c,L}(y,z,x) \right)\\
 &=\widetilde{Z}_3+\frac{g^2\,N_c\,\widetilde{Z}_1}{8\pi^3} \int dy\,d\varphi \frac{\sin(\varphi)^2}{z}G(z)\nnnl
 &\quad\times \left(-\sin(\varphi)^2 D^{A\bar c c,T}(y,z,x)Z(y)+\xi\, \frac{p\, \cos \varphi}{q \,G(x)}+\xi\,\cos(\varphi)^2 \right),
\end{align}
where $x=p^2$, $y=q^2$ and $z=(p+q)^2$.
From the third term in the parentheses, the \gls{ir} behavior of the ghost dressing function can be determined \cite{Huber:2015ria}.
For low external momentum $p$ it becomes
\begin{align}
\xi\frac{g^2\,N_c\,\widetilde{Z}_1}{8\pi^3} \int dy\,d\varphi \frac{\sin(\varphi)^2}{y}\cos(\varphi)^2 G(y).
\end{align}
If $G(y)$ were \gls{ir} divergent or constant, this term would be \gls{ir} divergent what would require the ghost dressing function on the left-hand side to vanish, in contradiction to the original assumption.
Consequently, the ghost dressing function must vanish for low momenta.
As no terms exist that would allow for a power law divergence on the right-hand side, corresponding to a power law suppression of the ghost dressing function, it must vanish logarithmically.

\index{gluon propagator}
The gluon propagator \gls{dse} is truncated by retaining only the one-loop diagrams.
Furthermore, the tadpole is dropped.
The missing two-loop diagrams, however, would most likely contribute quantitatively.
Thus, this study gives only a qualitative picture.

\index{RG improvement}
The level of truncation of this system of equations requires the usual additional tweak of using an \gls{rg} improvement term discussed in Sec.~\ref{sec:resummation_oneLoop} to obtain the correct one-loop resummed anomalous running.
Here, in contrast to the Landau gauge, it is necessary also for diagrams with a ghost-gluon vertex, since the ghost-gluon vertex acquires an anomalous running for $\xi>0$.
Thus, also the ghost propagator \gls{dse} is affected.
In addition, in the gluon loop several terms are required due to different combinations of dressing functions introduced by the longitudinal terms.
The final equations read
\begin{align} 
\label{eq:gh-DSE-final-wRGI}
G^{-1}(x)&=\widetilde{Z}_3+\frac{g^2\,N_c}{8\pi^3} \int dy\,d\varphi \frac{\sin(\varphi)^2
}{z}\Bigg(-\sin(\varphi)^2 D^{A\bar c c,T}(y,z,x)Z(y)F(\alpha_1^{\text{gg}},\beta_1^{\text{gg}};\bar{p}^2)\nnnl
&+\xi F(\alpha_2^{\text{gg}},\beta_2^{\text{gg}};\bar{p}^2)\left( \frac{p\, \cos \varphi}{q \,G(x)}+\cos(\varphi)^2 \right)\Bigg)G(z),\\
\label{eq:gl-DSE-final-wRGI}
 Z^{-1}(p^2)&=Z_3+g^2\,N_c\,\int_q\,  G(y)G(z) K_{Z}^{gh}(x;y,z)D^{A\bar c c,T}(x;y,z)F(\alpha_1^{\text{gg}},\beta_1^{\text{gg}},\bar{p}^2)\nnnl
 &+g^2\,N_c\,\int_q \Big( Z(y)Z(z) K_{Z}^{gl}(x;y,z)D^{\text{3g}}(x,y,z)F(\alpha_1^{\text{3g}},\beta_1^{\text{3g}};\bar{p}^2) \nnnl
 &\quad+\xi \widetilde{K}_Z^{gl,\xi} +
 \xi^2 K_Z^{gl,\xi^2} Z(x)^{-1}\Big).
\end{align}
The detailed kernels are given in \ref{sec:app_kernels_linCov}.
The renormalization group improvement is encoded in the function $F(\alpha,\beta;\bar{p}^2)$:
\begin{align}
 F(\alpha,\beta;\bar{p}^2)=G(\bar{p}^2)^\alpha Z(\bar{p}^2)^\beta
\end{align}
with $\bar{p}^2=(x+y+z)/2$.
For different terms, different exponents $\alpha$ and $\beta$ are required which are detailed in \tref{tab:linCov_exponents_RGI}.
They are determined such as to reproduce a self-consistent \gls{uv} behavior of the equations while being constant in the \gls{ir} assuming a constant ghost dressing function.
The kernel $\widetilde{K}_Z^{gl,\xi}$ contains additional factors of $F(\alpha,\beta;\bar{p}^2)$.

\begin{table}[tb]
\begin{center}
 \begin{tabular}{|l|c|}
 \hline
 Correlation function & Anomalous dimension \\
 \hline \hline
  Ghost propagator & $\delta=-\frac{9-3\xi}{44} $\\
  \hline
  Gluon propagator & $\gamma=-\frac{13-3\xi}{22}$\\
  \hline
  Ghost-gluon vertex & $\gamma^\text{ghg}=-\frac{3\xi}{22}$\\
  \hline
  Three-gluon vertex & $\gamma^{\text{3g}}=\frac{17-9\xi}{44}$\\
  \hline
 \end{tabular}
\caption{The anomalous dimensions of the propagators and vertices in linear covariant gauges.}
\label{tab:linCov_anom_dims}
\end{center}
\end{table}

\begin{table}[tb]
\begin{center}
 \begin{tabular}{|l|c||l|c|}
 \hline
  $\alpha_1^{\text{3g}}$ & $-\frac{9 \xi -17}{3 (\xi -3)}$ & $\alpha_1^{\text{gg}}$ & $-\frac{2\xi}{\xi-3}$ \\
  \hline
  $\alpha_2^{\text{3g}}$ & $-\frac{4 (3 \xi -2)}{3 (\xi -3)}$  & $\alpha_2^{\text{gg}}$ & $-\frac{2 (3 \xi +13)}{3 (\xi -3)}$ \\
  \hline 
  $\alpha_3^{\text{3g}}$ & $-\frac{2 (\xi +3)}{\xi -3}$ & & \\
 \hline
 \end{tabular}
\caption{The exponents $\alpha$ for the RG improvement terms.
The $\beta$'s are all $0$ because the exponents were derived as if the ghost dressing function were IR constant.}
\label{tab:linCov_exponents_RGI}
\end{center}
\end{table}

\index{spurious divergences}
The spurious divergences in the gluon propagator \gls{dse} are subtracted analytically by subtracting the perturbatively calculated counterterms \cite{Huber:2014tva}, see also Sec.~\ref{sec:spurDivs}.
The corresponding coefficient $b$ of the subtraction term
\begin{align}\label{eq:C_sub_LC}
C_\mathrm{sub}&=
\LQ \,b\,\omega^{-1-\gamma}\sum_{n=0}^{\infty}\frac{\left(\ln\left(\LL/\LQ\right)\right)^{-\gamma+n}}{n!(-\gamma+n)}
\end{align}
is given by
\begin{align}
 b&=\frac{g^2\,N_c}{64\pi^2}\Big( G(s)^{2+2\alpha_1^{\text{gg}}} Z(s)^{2\beta_1^{\text{gg}}}- 6 G(s)^{2\alpha_1^{\text{3g}}} Z(s)^{2+2\beta_1^{\text{3g}}}
 - 3\xi G(s)^{\alpha_3^{\text{3g}}} Z(s)^{\beta_3^{\text{3g}}}
 \Big).
\end{align}

\begin{figure}[tb]
 \includegraphics[width=0.48\textwidth]{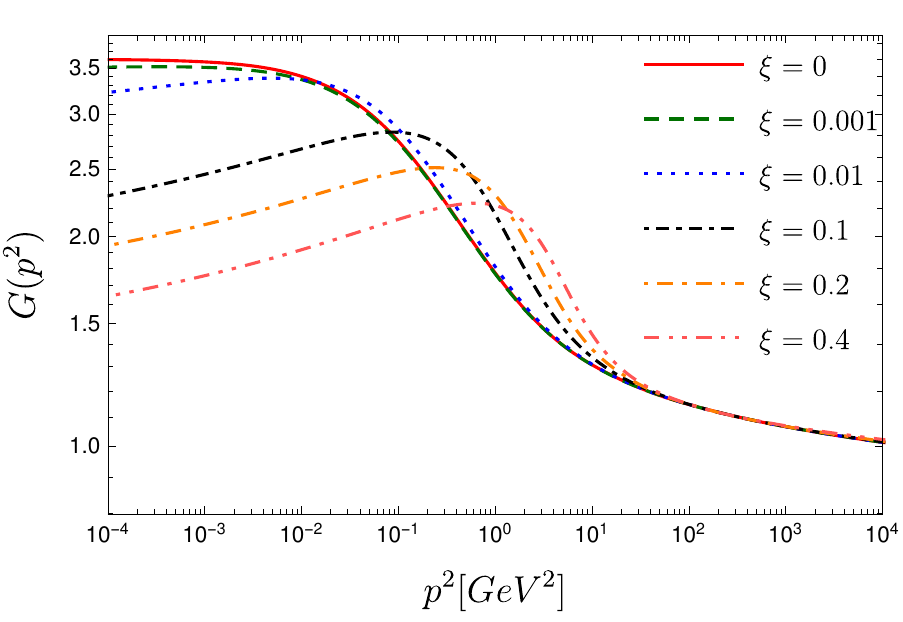}
 \hfill
 \includegraphics[width=0.48\textwidth]{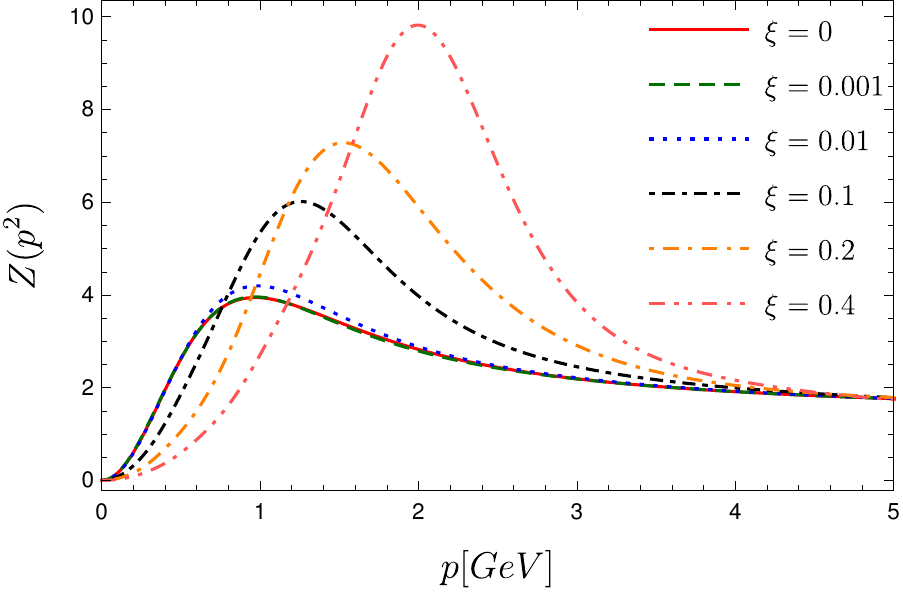}
 \caption{Ghost (left) and gluon (right) dressing functions for various values of $\xi$ \cite{Huber:2015ria}.}
 \label{fig:linCov_props}
\end{figure}

Solving the coupled system of ghost and gluon propagator \glspl{dse} yields the results depicted in \fref{fig:linCov_props}.
The ghost dressing function vanishes in the \gls{ir} as expected logarithmically, while the gluon propagator remains constant.
The bump in the gluon dressing function becomes larger for increasing $\xi$ and moves to higher momenta.
This shows that the employed truncation is not very suitable for higher values of the gauge fixing parameter, since the bump finally enters the regime which should be determined mainly by perturbation theory.
For low momenta, on the other hand, the results should suffer only from the same quantitative effects as the equivalent solution for the Landau gauge to which it connects smoothly, viz., the missing two-loop terms lead to missing strength in the mid-momentum regime of the gluon dressing function.

The results for the propagators in linear covariant gauges obtained with a simple truncation show that calculations with functional equations are feasible.
To push such investigations further, also vertices need to be calculated.
However, in that case the longitudinal parts could be problematic and it would be interesting to see how well the approximations of the Ward identities derived here work.
Other improvements, like the inclusion of two-loop terms, which would discard the necessity of \gls{rg} improvement terms, are expected to be straightforward.

\subsection{Maximally Abelian gauge}
\label{sec:mag}

\mybox{
\begin{itemize}
 \item dual superconductor picture of confinement
 \item Abelian dominance
\end{itemize}
}

\glsreset{mag}

\myboxmargin{maximally Abelian gauge}
\index{maximally Abelian gauge}
The \gls{mag} is an example of a non-linear covariant gauge.
It can be defined by a minimization procedure similar to the Landau gauge and is thus also amenable to lattice simulations in a standard way.
The physical motivation for this gauge is the role played by the \emph{diagonal part of the gauge field} in the dual superconductor picture of confinement.
This gauge breaks the global color symmetry and treats the diagonal and off-diagonal parts separately.
The plain \gls{mag} only gauge fixes the off-diagonal part and is thus an incomplete gauge fixing which has to be supplemented by an additional condition on the diagonal part.
The standard choice for this is the Landau gauge.

In this section, investigations of the \gls{ir} behavior of the \gls{mag} are presented.
In general, investigations in this gauge are by far not as advanced as in the Landau gauge.
The results presented here offer an explanation why this gauge constitutes a particular challenging gauge for functional equations.

\subsubsection{The dual superconductor picture of confinement and Abelian infrared dominance}

\myboxmargin{dual superconductor picture, monopoles and Abelian gauges}
\index{dual superconductor picture}
\index{Abelian gauges}
\index{Cartan subalgebra}
{\pretolerance=1000
The \emph{dual superconductor picture of confinement} explains confinement of chromoelectric charges by the QCD vacuum being a dual superconductor \cite{Mandelstam:1974pi,'tHooft:1975pu}.
In this scenario, chromoelectric charges are confined by chromomagnetic vortices which correspond to the flux tubes between quarks.
Such vortices can form when chromomagnetic monopoles condense.
The condensation of monopoles was observed later in lattice calculations \cite{Kronfeld:1987ri,Kronfeld:1987vd}.
Motivated by the dual superconductor picture, an order parameter based on the chromomagnetic charge was developed \cite{DiGiacomo:1993jy,DelDebbio:1995yf,DiGiacomo:1997sm,DiGiacomo:1999yas,Bonati:2011jv,DiGiacomo:2017blx}.
By choosing a certain class of gauges, the so-called \emph{Abelian gauges}, monopoles can be identified \cite{'tHooft:1981ht,Chernodub:1997ay}.
These gauges impose only a partial gauge fixing and leave the Abelian part of the gauge field unfixed.
For a theory with gauge group $SU(N)$, this is the subgroup $U(1)^{N-1}$.
There are infinitely many possibilities to realize an Abelian gauge.
One is the maximally Abelian gauge (MAG), in which the gauge field itself is made as Abelian as possible.
Thereby, Abelian refers to the part of the gauge field living in the Cartan subalgebra, which consists only of diagonal matrices and is thus also referred to as diagonal part.\footnote{The Cartan subalgebra is defined by the generators that commute with all other generators.
For $SU(2)$, this is $\sigma^3/2$ and for $SU(3)$ $\lambda^3/2$ and $\lambda^8/2$, where $\sigma^i$ are the Pauli matrices and $\lambda^i$ the Gell-Mann matrices.
For general $SU(N)$, the Cartan subalgebra consists of the matrices
\begin{align}
T^j=\frac{1}{2}\left(\frac{2}{j(j+1)}\right)^{1/2} \times diag(\underbrace{1,\ldots,1}_{j \text{ times}},-j,0,\ldots,0), \quad j=1,\ldots, N-1.
\end{align}}
The gauge field is maximally Abelian if the norm of the parts of the gauge field not in the Abelian subalgebra is minimized.

\myboxmargin{Abelian dominance}
\index{Abelian dominance}
\index{monopole dominance}
Based on the assumption that chromomagnetic monopoles cause confinement, Ezawa and Iwazaki conjectured that the Abelian parts of the gauge fields should be dominant at large distances, since classic magnetic monopoles live in the Cartan subalgebra \cite{Ezawa:1982bf}.
This conjecture is called \emph{Abelian \gls{ir} dominance}.
Indeed, for many quantities it is observed that they are only slightly affected by removing the off-diagonal part, a procedure called \emph{Abelian projection}.
For example, the string tension between a static pair of quarks has almost the same value if calculated from the diagonal part of gauge fields alone or from the complete fields, see, for example Refs.~\cite{Suzuki:1989gp,Polikarpov:1996wd}.
An extension of Abelian dominance is \emph{monopole dominance}.
In that case, only the monopoles are retained and still many observables remain close to the values from the full theory \cite{Stack:1994wm,Shiba:1994ab}.
Monopoles are also investigated as a mechanism to create a gauge invariant mass  in Yang-Mills theory \cite{Nishino:2018mwi}.
}

\subsubsection{Definition of the maximally Abelian gauge}

\index{maximally Abelian gauge}
As explained above, the \gls{mag} is defined by minimization of the norm of the off-diagonal part of the gauge field denoted by $B_\mu(x)$:
\begin{align}
 R_\text{MAG}=\frac1{2} \int dx \, B^a_\mu(x) B^a_\mu(x).
\end{align}
This leads to
\begin{align}\label{eq:MAGCondition}
D^{ab}_\mu B_\mu^b=0,
\end{align}
where $D^{ab}$ is the covariant derivative with respect to the diagonal field $A_\mu(x)$ only:
\begin{align}\label{eq:MAGCovDeriv}
 D_\mu^{ab}:=&\delta^{ab}\partial_\mu+g\,f^{abi} A_\mu^i.
\end{align}
The full gluon field is defined by
\begin{align}
 A_\mu=T^i A^i_\mu+T^a B^a_\mu.
\end{align}
Here, $T^i$ denotes the generators of the Cartan subalgebra and $T^a$ the other generators.
The identification of the two parts is often done in the literature via using indices $i,j,\ldots$ and $a,b,\ldots$, respectively.
In addition, the two different symbols $A$ and $B$ are chosen here for the gauge fields.
This splitting leads to an increase in the number of interaction vertices.
However, some vertices are not allowed because the corresponding structure constants vanish.
It can easily be inferred from the definition of the Cartan subalgebra,
\begin{align}\label{eq:structure-constants}
 [T^i,T^j]=0,
\end{align}
that the structure constants with two or more diagonal indices vanish, $f^{ija}=0$, $f^{ijk}=0$.
In addition, $f^{abc}$ is zero for $SU(2)$, because it only has two off-diagonal generators and the structure constants are fully anti-symmetric.

The Lagrangian density of \gls{qcd}, \eref{eq:LagrangianDensityQCD}, is written for the \gls{mag} explicitly in diagonal and off-diagonal components:
\begin{align}\label{eq:L-diag-offdiag}
 \mathcal{L}_\text{QCD}=\overline{\psi}(-\slashed{D}(A+B)+m )\psi+\frac{1}{4}F_{\mu\nu}^i F_{\mu\nu}^i+\frac{1}{4}F_{\mu\nu}^a F_{\mu\nu}^a.
\end{align}
The gauge fixing for the off-diagonal components can directly be inferred from \eref{eq:Lagrangian-gaugeFixing} using the gauge fixing condition \eref{eq:MAGCondition}.
However, an additional complication arises, because the action is not renormalizable: A quartic ghost interaction emerges that requires the introduction of a quartic ghost term in the bare action with a corresponding counterterm \cite{Min:1985bx,Fazio:2001rm}.
The prefactor of this term is determined by a Ward identity called the diagonal ghost equation \cite{Fazio:2001rm}.
It requires the prefactor to be equal to the gauge fixing parameter $\alpha$.
Since the limit $\alpha\rightarrow 0$ corresponds to the \gls{mag}, also the quartic ghost term vanishes.
However, this limit can only be performed at the end, since the vertices contain terms proportional to $1/\alpha$.
The complete gauge fixing term plus the additional quartic ghost interactions for the \gls{mag} reads then
\begin{align}\label{eq:L-MAG_gf_offdiag}
 \mathcal{L}_{\text{gf+R},\text{MAG}}&=\int dx\, \Big(\bar{c}^a {D}_\mu^{ab}D_\mu^{bc} c^c-g\,f^{bcd}\bar{c}^a {D}_\mu^{ab} B_\mu^c c^d - \nnnl
 &-g^2\, f^{abi}f^{cdi} B_\mu^b B_\mu^c \bar{c}^a c^d+\frac1{2\alpha}({D}_\mu^{ab} B_\mu^b)^2+\frac{\alpha}{8} g^2 f^{abc}f^{ade} \bar{c}^b c^c \bar{c}^d c^e-\nnnl
 &-\frac{1}{2} g\,f^{abc} ({D}_\mu^{ad} B_\mu^d) \bar{c}^b c^c +\frac1{4}g^2\alpha f^{abi}f^{cdi}\bar{c}^a\bar{c}^b c^c c^d+\alpha\frac1{8}g^2 f^{abc}f^{ade} \bar{c}^b \bar{c}^c c^d c^e\Big).
\end{align}

The diagonal part is still invariant under $U(N-1)$ gauge transformations.
A typical choice to fix it to a gauge is a Landau gauge-like condition, viz., $\partial A=0$.
However, in this case, the diagonal ghosts can be integrated out \cite{Capri:2010an,Capri:2005tj} similar to \gls{qed} and the gauge fixing part with gauge fixing parameter $\xi$ reads
\begin{align}\label{eq:L-MAG_gf_diag}
\mathcal{L}_{\text{gf},\text{diag}}&=\int dx\, \frac1{2\xi}(\partial_\mu A_\mu^i)^2 .
\end{align}
The total Lagrangian density is the sum of Eqs.~(\ref{eq:L-diag-offdiag}), (\ref{eq:L-MAG_gf_offdiag}) and (\ref{eq:L-MAG_gf_diag}).
For the gauge group $SU(2)$ some terms vanish due to $f^{abc}=0$.
Thus, often $SU(2)$ is considered instead of $SU(3)$.

Finally, let me remark that the \gls{mag} has a BRST symmetry based on the splitting of the \gls{brst} transformations from \eref{eq:BRST}:
\begin{align}
s\,B_\mu^a&=-(D_\mu^{ab} c^b-g\,f^{abc}B_\mu^b c^c-g\,f^{abi} B_\mu^b c^i),& 
s\,A_\mu^i&=-(\partial_\mu c^i-g\,f^{iab}B_\mu^a c^b),\\
s\,c^a&=-\mhalfo g\,f^{abc} c^b c^c-g\,f^{abi}c^b c^i,&
s\,c^i&=-\mhalfo g\,f^{iab}c^a c^b,\\
s\,\bar{c}^a&=i \, b^a,&
s\,\bar{c}^i&=i \, b^i,\\
s\,b^a&=0,&
s\,b^i&=0.
\end{align}
Consequently, gauge fixing can also be derived from BRST exact terms.
They are for the off-diagonal and diagonal terms:
\begin{align}
 \mathcal{L}_{\text{gf},\text{MAG}}&=s\bar{c}^a({D}_\mu^{ab} B_\mu^b-i \frac{\alpha}{2} b^a),\\
 \mathcal{L}_{\text{gf},\text{diag}}&=s\bar{c}^i(\partial_\mu A_\mu^i-i\frac \xi{2} b^i).
\end{align}
For full BRST invariance it is necessary that also the additional quartic ghost part can be written as a BRST exact term:
\begin{align}
\mathcal{L}_\text{R}&=  s (-\frac{1}{2} \lambda\, g\,f^{abi} \bar{c}^a \bar{c}^b c^i-\frac 1{4}\lambda'\, g\,f^{abc} \bar{c}^a \bar{c}^b c^c).
\end{align}
Here, the two possible parameters are made explicit.
Above it was argued that they both have to be equal to the gauge fixing parameter of the off-diagonal part, $\alpha$.

\subsubsection{Abelian dominance in the correlation functions of the maximally Abelian gauge}

\mybox{Correlation functions in the MAG: perturbatively, GZ, RGZ, lattice, DSEs}
The correlation functions of the \gls{mag} were studied in perturbation theory in different renormalization schemes \cite{Min:1985bx,Gracey:2005vu,Bell:2013xma,Bell:2015dbr}.
For example, the propagators are known up to three loops in the \gls{msbar} scheme \cite{Gracey:2005vu} and the three-point functions up to two-loops in the \gls{mom} scheme \cite{Bell:2015dbr}.
Since the \gls{mag} can be introduced in the same way as the Landau gauge by minimizing a Lorentz invariant gauge fixing functional, it is not only also directly accessible for lattice calculations \cite{Bornyakov:2003ee,Mendes:2006kc,Gongyo:2012jb,Gongyo:2013sha,Gongyo:2014lxa,Schrock:2015pna}, but also the Gribov-Zwanziger construction of an improved gauge fixing can be done in a similar way \cite{Capri:2005tj,Capri:2006cz,Capri:2008ak,Gongyo:2013rua}.
Finally, also functional methods were used to investigate the correlation functions of the \gls{mag} \cite{Huber:2009wh,Huber:2011fw,Alkofer:2011di,Mader:2013ru}.

\myboxmargin{scaling solution from functional equations}
\index{scaling solution}
With functional equations, the focus was on the study of the scaling solution.
For details on solutions of the scaling type see Sec.~\ref{sec:why_LG} below \eref{eq:powerLaws}.
The propagators in the \gls{mag} can be parametrized as
\begin{align}
 D_A^{ij}(p^2)&=\delta^{ij}\frac{c_A(p^2)}{p^2}
 \left(g_{\mu\nu}-\frac{p_\mu p_\nu}{p^2}\right),\\
 D_B^{ab}(p^2)&=\delta^{ab}\frac{c_B(p^2)}{p^2}
 \left(g_{\mu\nu}-(1-\alpha)\frac{p_\mu p_\nu}{p^2}\right),\\
 D_c^{ab}(p^2)&=-\delta^{ab}\frac{c_c(p^2)}{p^2} .
\end{align}
For the scaling solution, all dressing functions obey power laws:
\begin{align}
 c_A(p^2)&\overset{p^2\rightarrow 0}{=} d_A \cdot (p^2)^{\delta_A},&\quad
 c_B(p^2)&\overset{p^2\rightarrow 0}{=} d_B \cdot (p^2)^{\delta_B},&\quad
 c_c(p^2)&\overset{p^2\rightarrow 0}{=} d_c \cdot (p^2)^{\delta_c}.
\end{align}
With a scaling analysis, relations between the different exponents can be determined.
However, only after a general method was developed \cite{Huber:2009wh}, the behavior of the correlation functions could be unraveled.
The reason lies in the more complex structure of the functional equations compared to the Landau gauge which is not only a consequence of having three instead of two different fields (in the Yang-Mills case) but also of the more intricate possibilities of interactions.
The latter contain many four-point functions which are responsible for changing the qualitative behavior compared to the Landau gauge.

The scaling relation found is
\begin{align}
 \ka_\text{MAG}:=-\de_A=\de_B=\de_c\geq 0
\end{align}
and identifies certain two-loop diagrams as \gls{ir} leading \cite{Huber:2009wh}.
This complicates finding a suitable but simple truncation for the propagator equations, since besides the \gls{ir} dominant two-loop diagrams also all \gls{uv} leading diagrams are required even only for an exploratory but self-consistent solution.
What makes the found scaling solution interesting is that the diagonal gluon dressing function was found to be \gls{ir} divergent.
This is in marked contrast to the Landau gauge, where the ghost dressing function is \gls{ir} divergent for the scaling solution.
Another difference is that the other propagators in the \gls{mag} do not vanish in the \gls{ir}.
In fact, they can be \gls{ir} divergent but less than the tree-level propagator.
A numeric value for the exponent of the power laws found by taking into account only the sunset diagrams is $\kappa_\text{MAG}\approx 0.74$ \cite{Alkofer:2011di}.

\index{Abelian dominance}
With hindsight, the dominance of the diagonal gluon propagator is not so surprising.
First of all, the Gribov region looks different in the \gls{mag} and in the Landau gauge \cite{Capri:2008vk,Capri:2010an}.
It is bounded in the latter case in all directions, while the diagonal direction is unbounded for the \gls{mag}.
Furthermore, Abelian dominance in terms of correlation functions corresponds to dominance of the diagonal gluon propagator.
Finally, although lattice calculations find a decoupling type solution, these results are in agreement with what one would expect if in the \gls{mag} the relation between the scaling and decoupling solution is similar to the Landau gauge.
Interpreting the lattice results as massive propagators, it turns out that the diagonal gluon propagator has the lowest mass \cite{Gongyo:2012jb,Gongyo:2013sha}.
Thus, it is quantitatively \gls{ir} enhanced and it is conceivable that the scaling solution is the endpoint of a family of decoupling type solutions.

In summary, the study of the maximally Abelian gauge with functional methods is currently only exploratory.
The reason is the importance of two-loop diagrams.
Although this was found for a scaling type solution, it can be expected that the hierarchy of diagrams is similar for a decoupling solution.
In a quantitative calculation of the propagators in the \gls{mag}, the importance of the vertices is currently unclear.
A key point for the success of such calculations is probably a better understanding of them.
However, albeit cumbersome, even a brute force attempt seems possible that calculates propagators and vertices.

\subsection{Coulomb gauge}
\label{sec:Coulomb}

The Coulomb gauge could for a long time be considered the gauge investigated most often after the Landau gauge until in recent years linear covariant gauges received more attention.
The list of quantities calculated in Coulomb gauge includes the 't Hooft loop \cite{Reinhardt:2007wh}, the color dielectric function of the Yang-Mills vacuum \cite{Reinhardt:2008ek}, and the transition temperatures at vanishing density \cite{Reinhardt:2011hq,Heffner:2012sx, Reinhardt:2012qe,Reinhardt:2013iia,Reinhardt:2016pfe,Ebadati:2017ymp}.
Correlation functions were studied with continuum, e.g., \cite{Epple:2006hv,Schleifenbaum:2006bq,Campagnari:2010wc,Campagnari:2011bk,Huber:2012kd,Guimaraes:2015bra,Campagnari:2016chq} and lattice methods, e.g., \cite{Burgio:2008jr,Burgio:2016nad}.
An inconvenient aspect of the Coulomb gauge is the breaking of Lorentz covariance which makes its investigation technically more complex.
Furthermore, the question of the renormalizability of Coulomb gauge has not been answered yet.

In Sec.~\ref{sec:Hamilton} the Hamiltonian approach to Yang-Mills theory was described and functional identities leading to equations similar to \glspl{dse} were introduced.
The largest difference to \glspl{dse} is the replacement of purely gluonic bare vertices by variational kernels.
These kernels contain themselves already some nonperturbative information.
The equations for the ghost-gluon and three-gluon vertices are shown in Figs.~\ref{fig:dse_coulomb_3g} and \ref{fig:dse_coulomb_ghg}.
Only the lowest order diagrams are shown.

To solve these equations, the variational kernels are required in explicit form.
They can be calculated by minimization of the vacuum energy $\vev{H_\mathrm{YM}} = E[\omega,\gamma_3,\gamma_4] $ \cite{Campagnari:2010wc}.
The kernel for the three-gluon vertex is given by
\begin{align}
\gamma_{ijk}^{abc}(\vp,\vq,\vk) = \frac{2\,g\, T^{abc}_{ijk}(\vp,\vq,\vk)}{\Omega(\vp)+\Omega(\vq)+\Omega(\vk)}
\end{align}
and that for the four-gluon vertex by
\begin{align}\label{eq:gamma4}
\bigl[ \Omega(\vk_1) &+ \Omega(\vk_2) + \Omega(\vk_3) + \Omega(\vk_4) \bigr] \, \gamma^{abcd}_{ijkl}(\vk_1,\vk_2,\vk_3,\vk_4) =
2 \, g^2 \, T^{abcd}_{ijkl} \\
-\frac12 &\biggl\{
\gamma^{abe}_{ijm}(\vk_1,\vk_2,-\vk_1-\vk_2) \, P_{mn}(\vk_1+\vk_2) \, \gamma^{cde}_{kln}(\vk_3,\vk_4,\vk_1+\vk_2) \\
&{}\qquad + \gamma^{ace}_{ikm}(\vk_1,\vk_3,-\vk_1-\vk_3) \, P_{mn}(\vk_1+\vk_3) \, \gamma^{bde}_{jln}(\vk_2,\vk_4,\vk_1+\vk_3) \\
&{}\qquad\qquad + \gamma^{ade}_{ilm}(\vk_1,\vk_4,-\vk_1-\vk_4) \, P_{mn}(\vk_1+\vk_4) \gamma^{bce}_{jkn}(\vk_2,\vk_3,\vk_1+\vk_4)
\biggr\} \\
-2 g^2 & \biggl\{
f^{abe} f^{cde} \delta_{ij} \delta_{kl}
\bigl[\Omega(\vk_1) - \Omega(\vk_2)\bigr] F(\vk_1+\vk_2) \bigl[\Omega(\vk_3) - \Omega(\vk_4)\bigr] \\
&{}\qquad + f^{ace} f^{bde} \delta_{ik} \delta_{jl}
\bigl[\Omega(\vk_1) - \Omega(\vk_3)\bigr] F(\vk_1+\vk_3) \bigl[\Omega(\vk_2) - \Omega(\vk_4)\bigr] \\
&{}\qquad\qquad + f^{ade} f^{bce} \delta_{il} \delta_{jk}
\bigl[\Omega(\vk_1) - \Omega(\vk_4)\bigr] F(\vk_1+\vk_4) \bigl[\Omega(\vk_2) - \Omega(\vk_3)\bigr]
\biggr\}.
\end{align}
$\Omega(\vk)$ is the gluon energy defined by the static gluon propagator
\begin{align}
\vev{A_i^a(\vk) A_j^b(\vq)} = \delta^{ab} \, \frac{P_{ij}(\vk)}{2\Omega(\vk)} \, (2\pi)^4\delta(\vp+\vq),
\end{align}
where $P_{ij}(\vk)$ is the transverse projector
\begin{align}
P_{ij}(\vk) = \delta_{ij} - \frac{k_i k_j}{\vk^2}.
\end{align}
In \eref{eq:gamma4} the Coulomb interaction kernel $F(\vk)$ appears which is approximated by
\begin{align}\label{eq:F}
F(\vp) = \vp^2 G^2(\vp).
\end{align}
The quantities $T^{abc}_{ijk}$ and $T^{abcd}_{ijkl}$ correspond to the negative spatial parts of the three- and four-gluon vertices without coupling, see \eref{eq:bare_three-gluon_vertex} and \eref{eq:bare_four-gluon_vertex}, respectively:
\begin{align}
T^{abc}_{ijk}(\vp,\vq,\vk) = i \, f^{abc}\bigl[ \delta_{ij} (p-q)_k + \delta_{jk} (q-k)_i + \delta_{ki} (k-p)_j \bigr]
\end{align}
and
\begin{align}
T^{abcd}_{ijkl} =
f^{abe} f^{cde} (\delta_{ik} \, \delta_{jl} - \delta_{il} \, \delta_{jk})
+ f^{ace} f^{bde} (\delta_{ij} \, \delta_{kl} - \delta_{jk} \, \delta_{il})
+ f^{ade} f^{bce} (\delta_{ij} \, \delta_{kl} - \delta_{ik} \, \delta_{jl}).
\end{align}

The solutions for the propagators obtained in $SU(2)$ are fitted very well by the following expressions \cite{Epple:2006hv,Burgio:2008jr,Campagnari:2010wc}:
\begin{align}
 \Omega(\vk) &= \sqrt{\vk^2 + \frac{m_A^4}{\vk^2}}, \quad m_A^2=0.6\sigma_\text{C}\simeq(880\,\text{MeV})^2\\
 G(\vk) &= \frac{d(\vk^2)}{g \, \vk^2},\\
 d(x^2) &= a \sqrt{\frac{1}{x^2}+\frac{1}{\ln(x^2+c^2)}} \, , \qquad x^2\equiv\frac{\vp^2}{\sigma_\text{C}}, \qquad c\simeq 4, \qquad a\simeq5.
\end{align}
$\sigma_\text{C}$ is the Coulomb string tension.
The coupling is set in the following to $g=3.5$ what corresponds to a renormalization of $\mu=2.4\sqrt{\sigma_\text{C}}$.
With this input the equations for the three-point functions were solved.

\begin{figure}[tb]
 \begin{center}
 \includegraphics[width=0.48\textwidth]{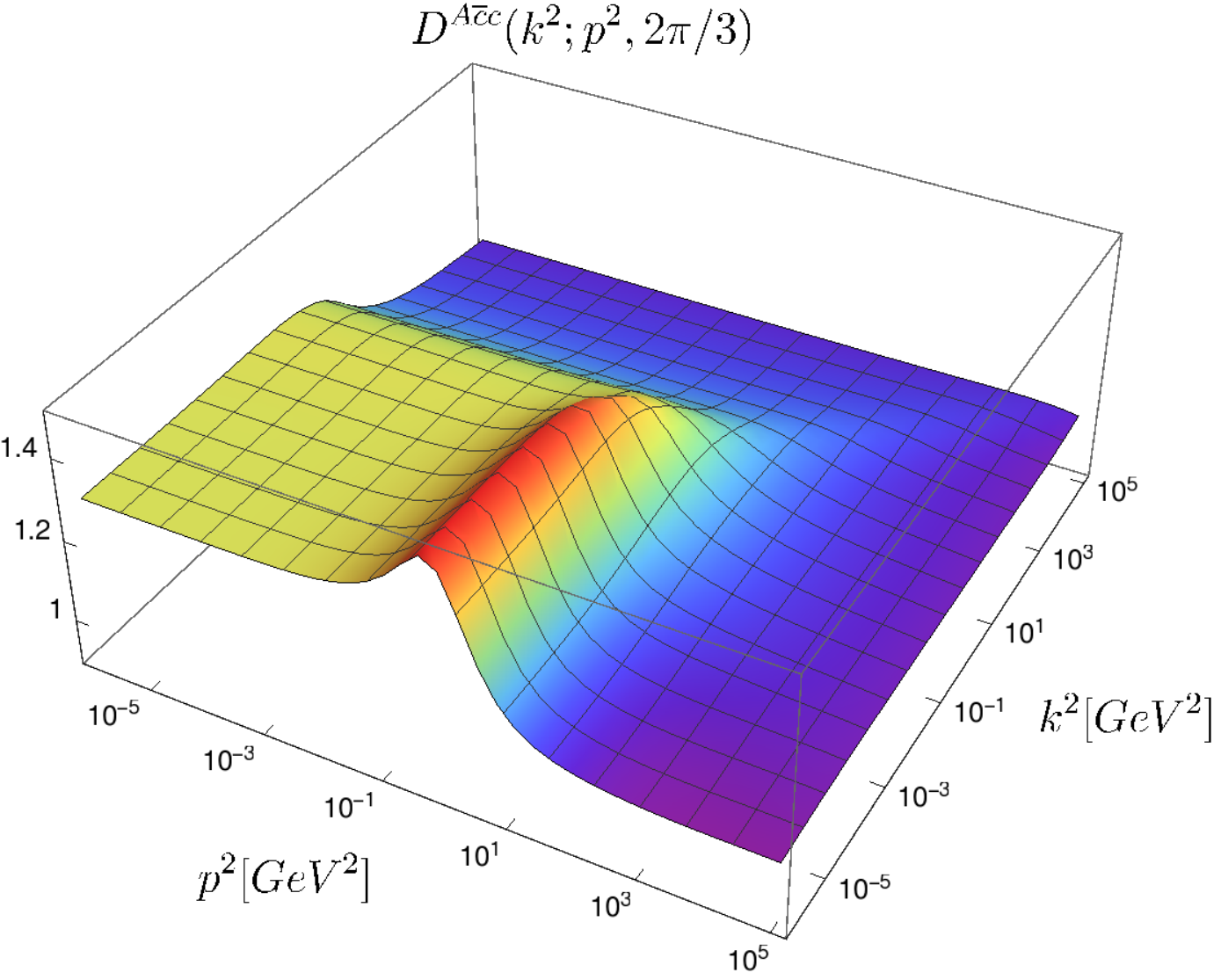} 
 \hfill
 \includegraphics[width=0.48\textwidth]{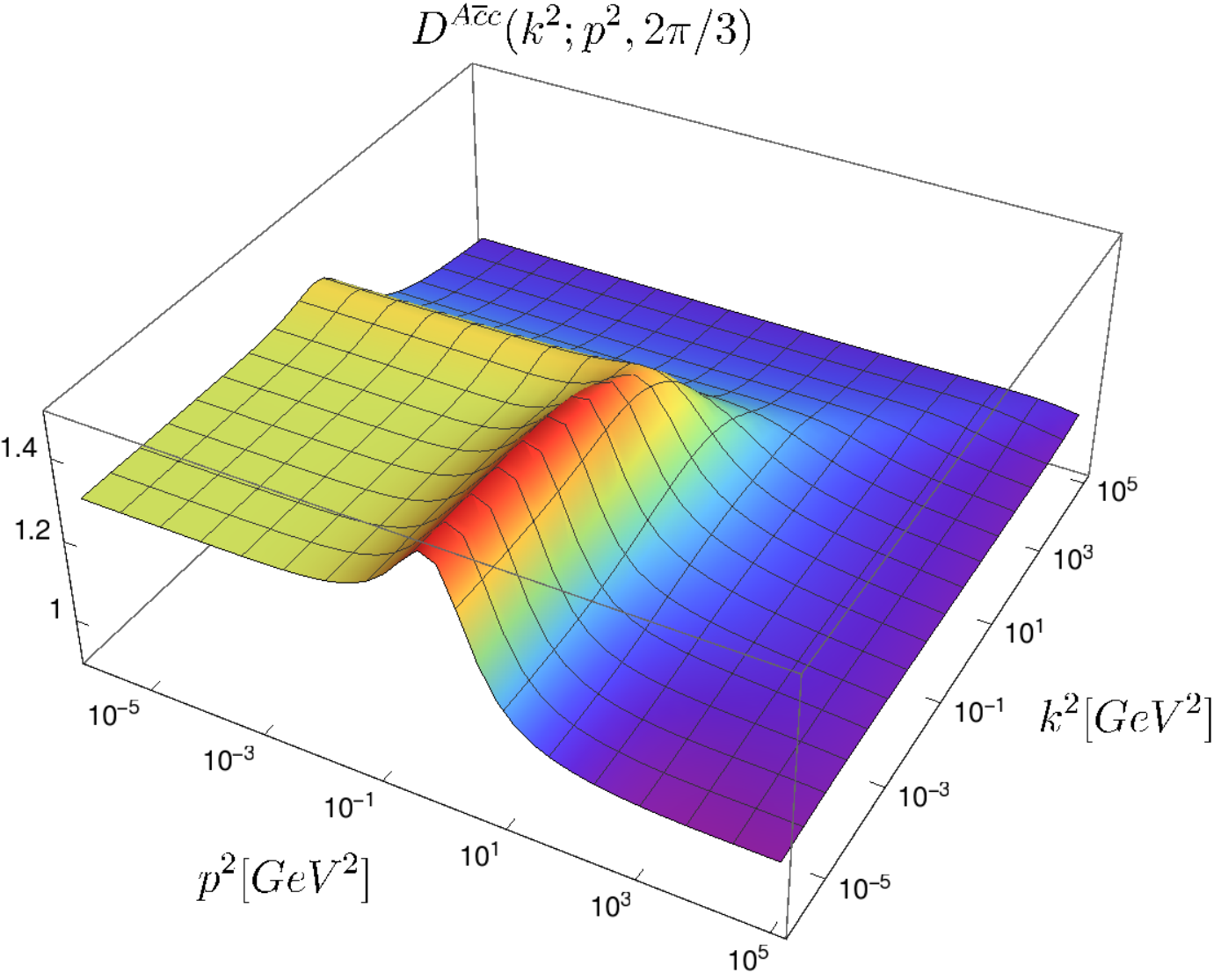}
 \caption{Dressing function of the ghost-gluon vertex \cite{Huber:2014isa}.
 The anti-ghost momentum is denoted by $p$, the gluon momentum by $k$.
 Left/Right: Ghost/Gluon legs attached to bare vertices, see top/bottom plot of \fref{fig:dse_coulomb_ghg}.}
 \label{fig:Coulomb_ghg_3d}
 \end{center}
\end{figure}

\begin{figure}[tb]
 \begin{center}
 \includegraphics[width=0.48\textwidth]{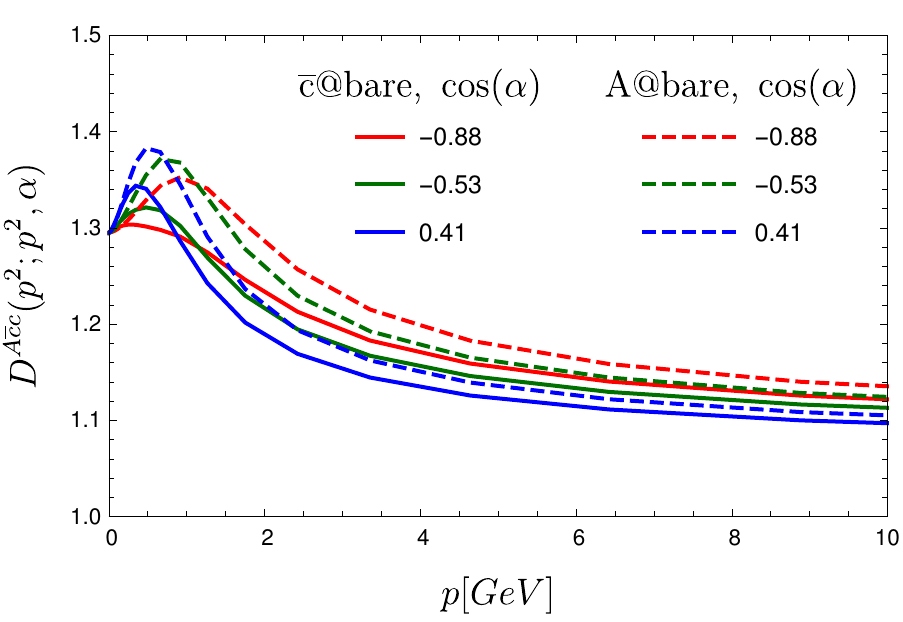}
 \hfill
 \includegraphics[width=0.48\textwidth]{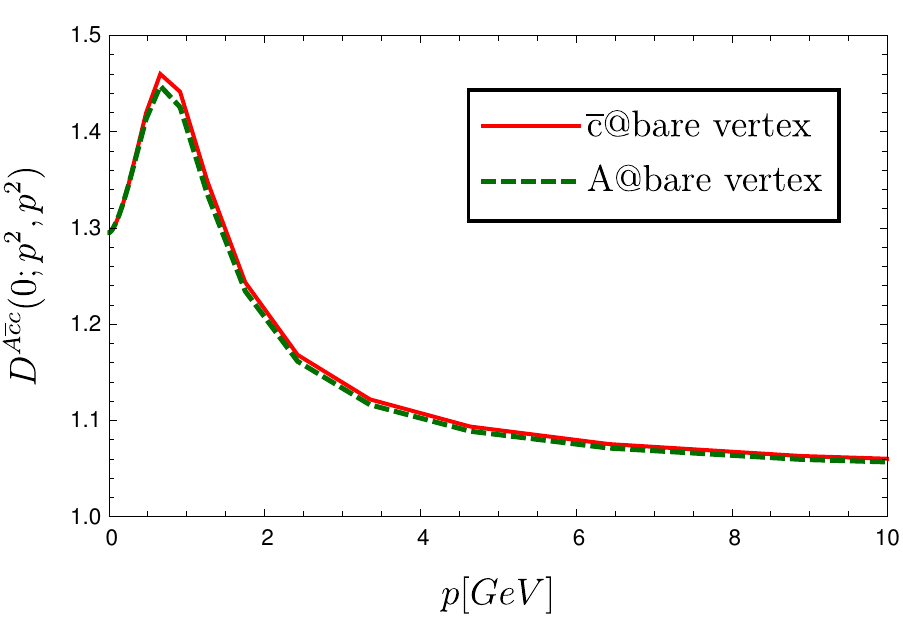}
 \caption{Comparison of the results from the two different ghost-gluon vertex equations \cite{Huber:2014isa}.
 Continuous/Dashed lines are from the versions with the ghost/gluon legs attached to the bare vertices.
 Left: Equal ghost and gluon momenta, different angles.
 Right: Zero gluon momentum.}
 \label{fig:Coulomb_ghg_comp_cbA}
 \end{center}
\end{figure}

The momentum dependence of the ghost-gluon vertex with the angle fixed at $2\pi/3$ is shown in \fref{fig:Coulomb_ghg_3d} for its two equations.
For equal momenta, the angle dependence is shown in more detail in the left plot of \fref{fig:Coulomb_ghg_comp_cbA}.
For vanishing gluon momenta, the results from the two equations basically agree as can be seen in the right plot of \fref{fig:Coulomb_ghg_comp_cbA}.
In general, differences between the two equations are very small, even smaller than the ones observed for a similar test in three-dimensional Yang-Mills theory, see Sec.~\ref{sec:YM3d_testing}.

Lattice data for the ghost-gluon vertex in Coulomb gauge is not available.
However, a qualitative comparison can be done with results form three-dimensional Yang-Mills theory \cite{Cucchieri:2008qm}.
A quantitative agreement can naturally not be achieved.
Not only do the lattice results contain too many uncertainties, but also some basic differences exist.
The foremost difference is expected in the \gls{uv}.
In Coulomb gauge, the ghost-gluon vertex possess an anomalous dimension while in three-dimensional Yang-Mills theory it approaches one rapidly.
At small momenta, the shown results approach a value above one, whereas the lattice results are compatible with one.
This is due to the lattice results supporting a solution of the decoupling type, whereas the results from the variational approach are of the scaling type.
This difference in the \gls{ir} behavior of the ghost-gluon vertex is known already from Landau gauge \cite{Huber:2012kd} and can be traced back to the fact that for the scaling type solution the integrals can have a nonzero contribution in the \gls{ir} limit, whereas for decoupling they vanish in the \gls{ir}.
Despite these differences, it is instructive to compare the results to three-dimensional Yang-Mills theory due to the similar structure of the equations.
And indeed the qualitative behavior of the ghost-gluon vertex is very similar in both cases.
Also, in agreement with an IR analysis in three dimensions \cite{Alkofer:2008dt}, no kinematic singularities are seen.

\begin{figure}[tb]
 \begin{center}
 \includegraphics[width=0.48\textwidth]{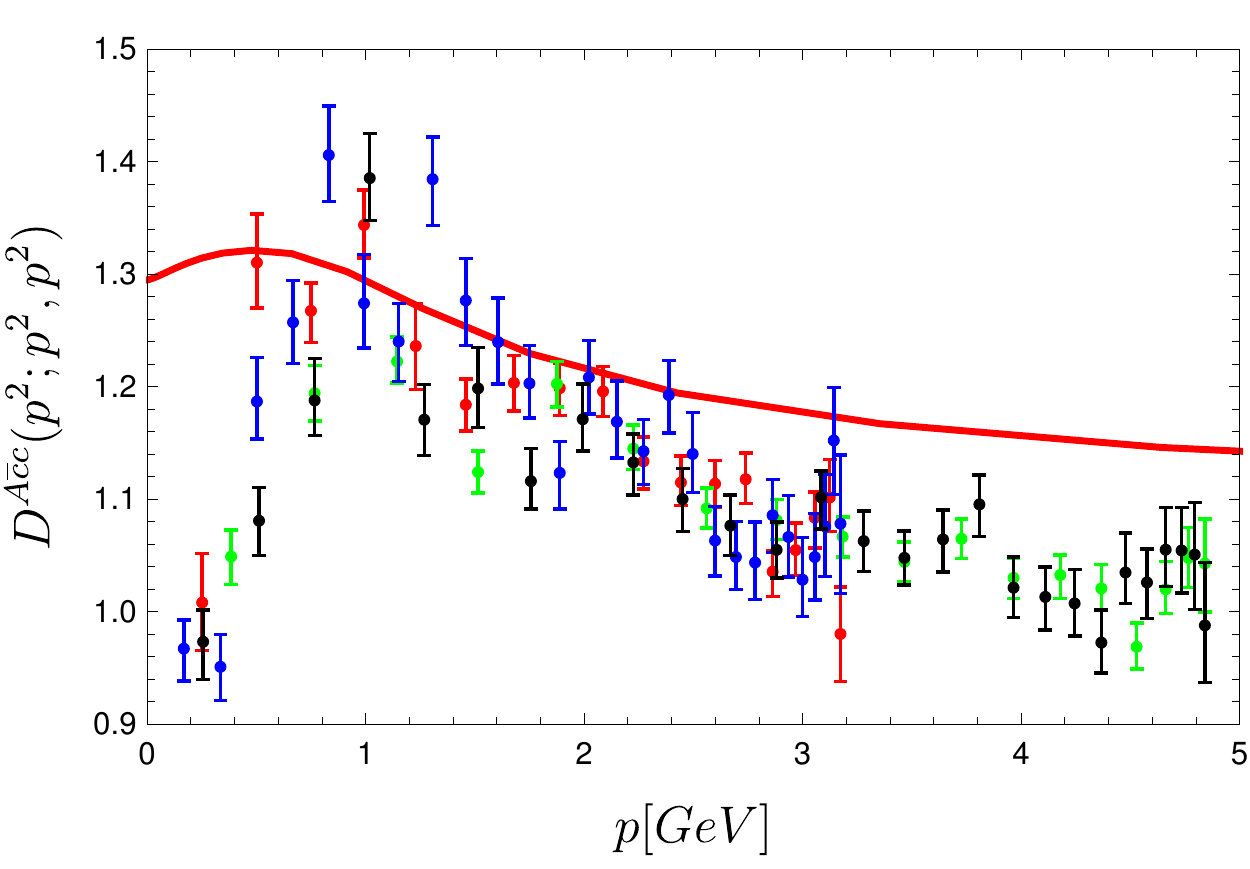}
 \hfill
 \includegraphics[width=0.48\textwidth]{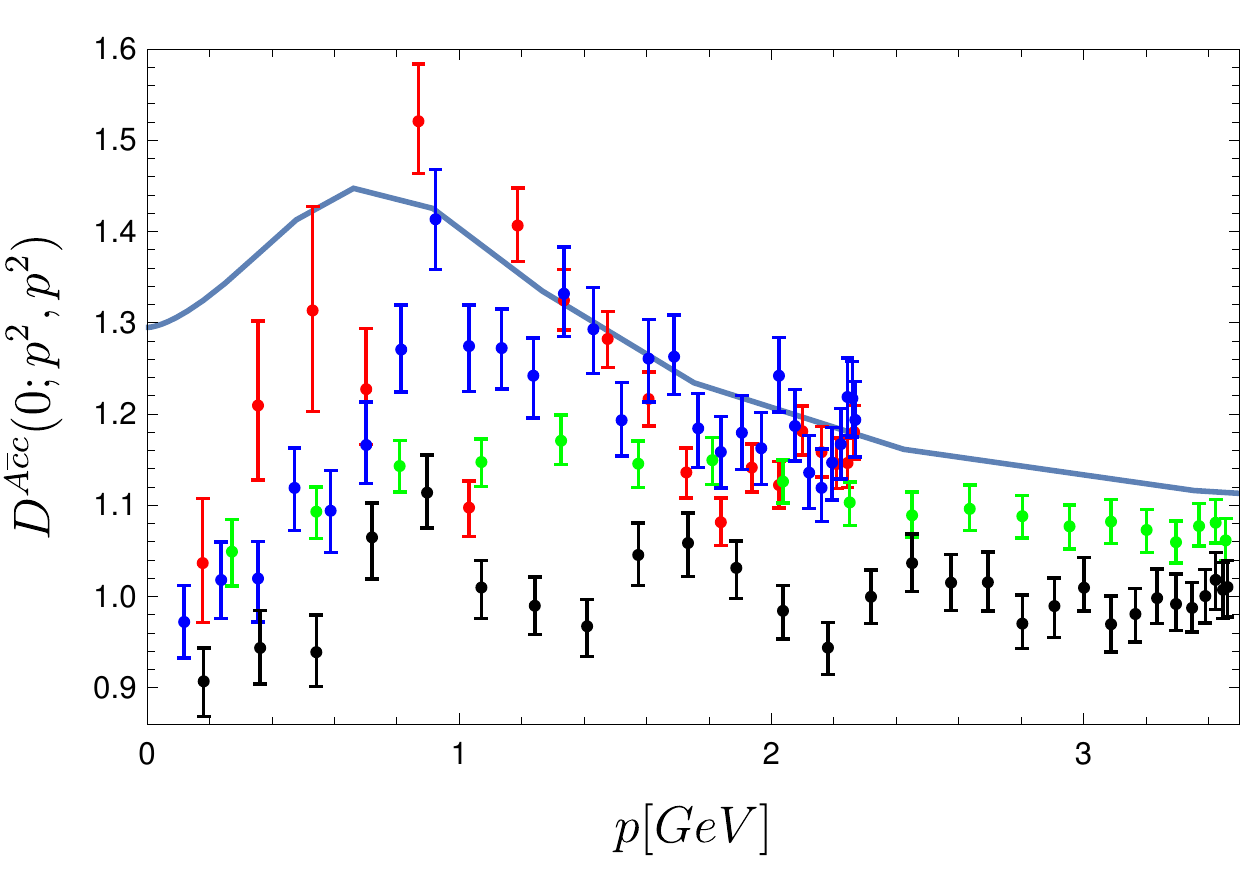}
 \caption{Comparison to lattice results \cite{Cucchieri:2008qm} at the symmetric point (left) and for vanishing gluon momentum (right) \cite{Huber:2014isa}.
 Different colors correspond to different lattice sizes $N\in \{40,60\}$ and values for $\beta \in \{4.2,6\}$; see Ref.~\cite{Cucchieri:2008qm} for details.}
 \label{fig:Coulomb_ghg_compL}
 \end{center}
\end{figure}

\begin{figure}[tb]
 \begin{center}
 \includegraphics[width=0.48\textwidth]{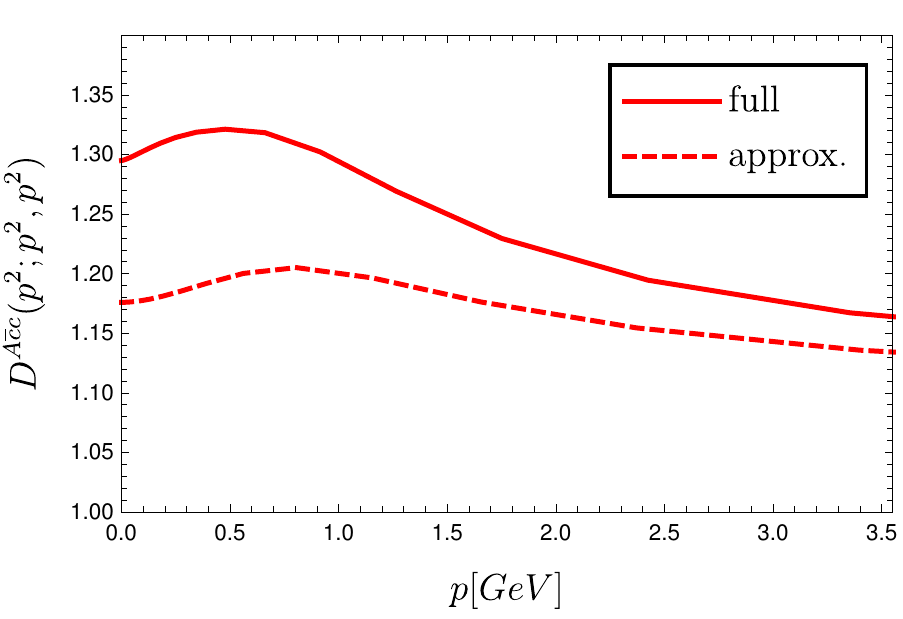} 
 \includegraphics[width=0.48\textwidth]{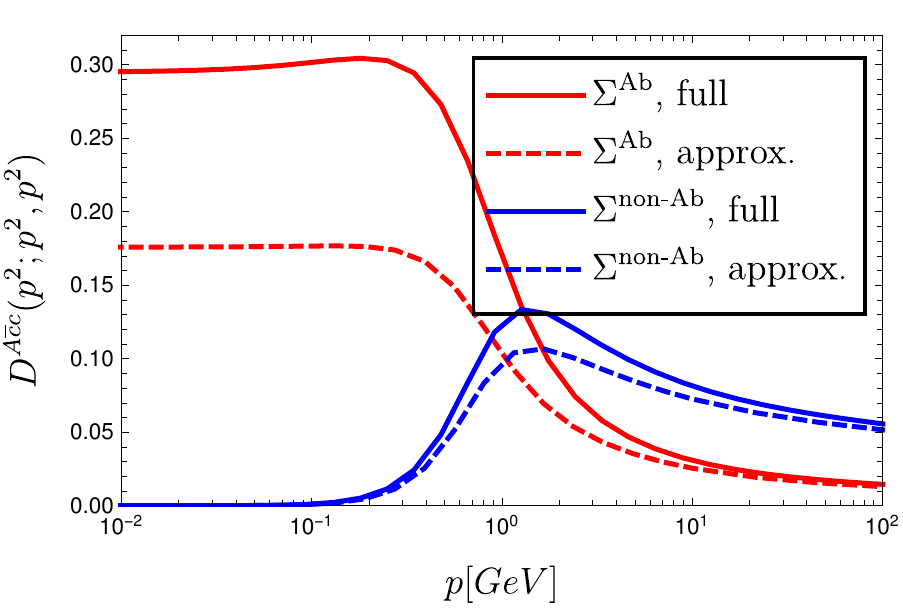}
 \caption{Comparison of the full non-perturbative calculation \cite{Huber:2014isa} with the semi-perturbative calculation of Ref.~\cite{Campagnari:2011bk}.
 Left: Dressing function of the ghost-gluon vertex.
 Right: Contributions from the Abelian and non-Abelian diagrams, $\Sigma^\text{Ab}$ and $\Sigma^\text{non-Ab}$, respectively.
 For this comparison the coupling constant $g$ was not factored out as in Fig.~2 of Ref.~\cite{Campagnari:2011bk} explaining the difference to that plot.}
\label{fig:Coulomb_ghg_comp_bareV}
 \end{center}
\end{figure}

In \fref{fig:Coulomb_ghg_comp_bareV} the effect of self-consistency is shown by comparing results for the $c$-DSE to a semiperturbative calculation \cite{Campagnari:2011bk}.
In that calculation, dressed propagators but bare ghost-gluon vertices were used.
Clearly, solving the equation self-consistently increases the strength of the vertex.
This is mostly due to the Abelian diagram.
The non-Abelian diagram has in the $c$-DSE only one ghost-gluon vertex so that backcoupling effects are not that large.

\begin{figure}[tb]
 \begin{center}
 \includegraphics[width=0.48\textwidth]{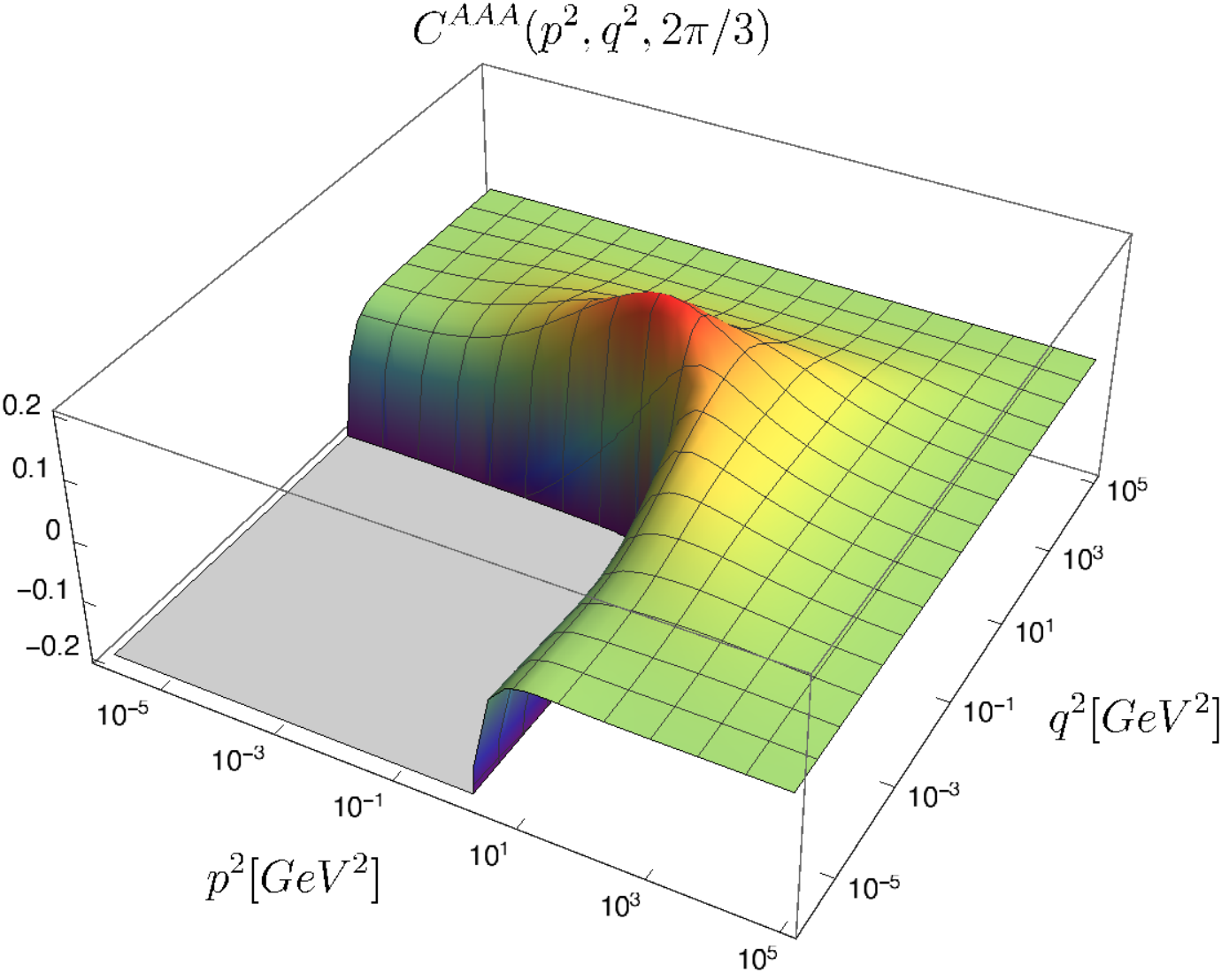}
 \includegraphics[width=0.48\textwidth]{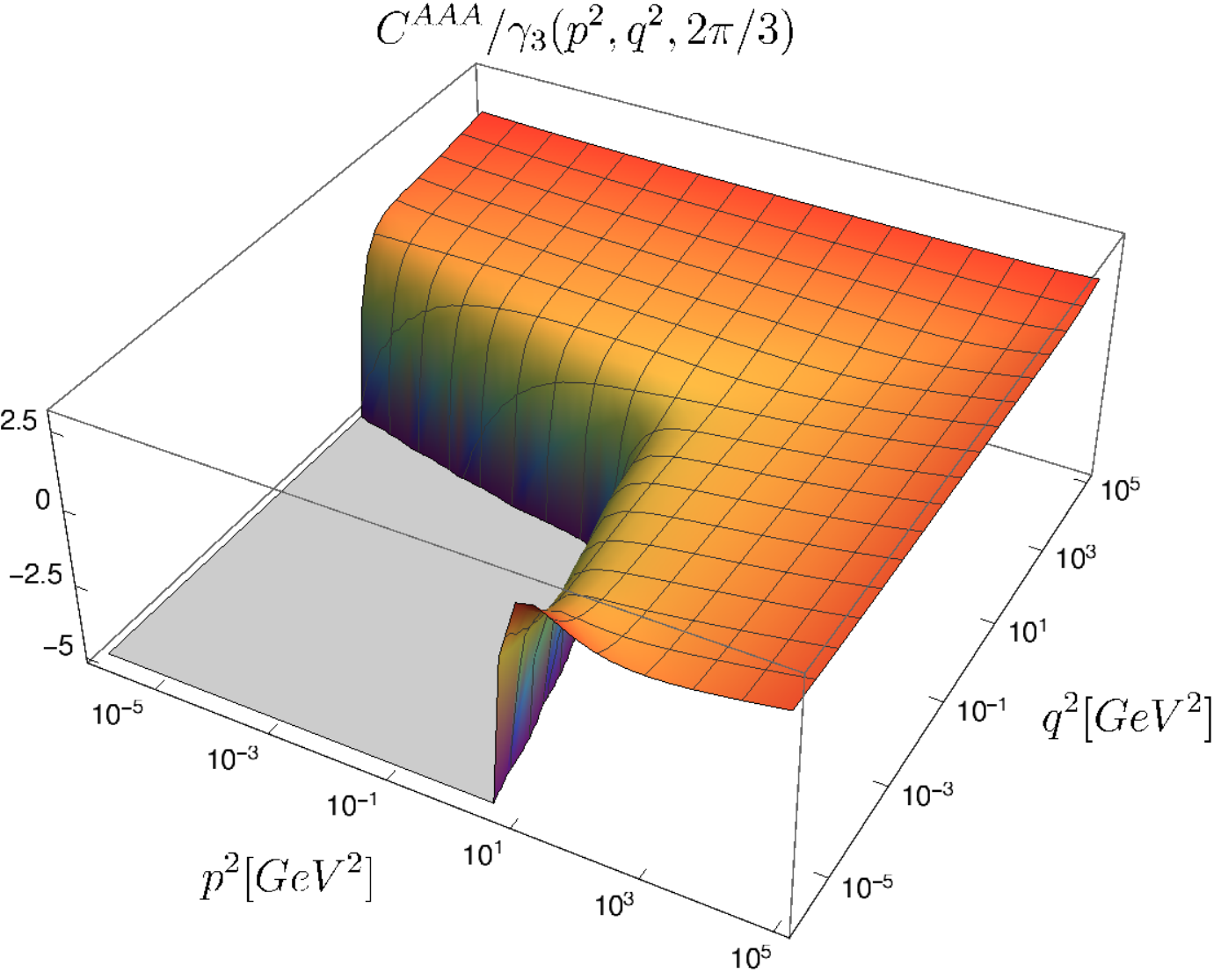}
 \caption{Momentum dependence of the three-gluon vertex with the angle fixed to $2\pi/3$ \cite{Huber:2014isa}.
 Left: Dressing of the three-gluon vertex.
 Right: The ratio of three-gluon vertex over the variational kernel.
 The deviation from Bose symmetry at the boundaries is a numerical artifact due to the smallness
 of $\gamma_3$ which enhances small numerical errors considerably.}
 \label{fig:Coulomb_tg_3d}
 \end{center}
\end{figure}

For the three-gluon vertex only the tree-level dressing is considered.
It is then convenient to consider the dressing function divided by the variational kernel, $C^{AAA}/\gamma^{AAA}$.
This quantity is more similar to the dressing function in other functional approaches, since the variational kernel plays here the role of the tree-level dressing.
Thus, in some figues both $C^{AAA}$ and $C^{AAA}/\gamma^{AAA}$ are shown.
The calculation was done using a bare ghost-gluon vertex.
It should also be noted that the appearance of the variational kernel instead of a bare vertex at the tree-level requires some additional steps in the renormalization procedure \cite{Huber:2014isa}.

\begin{figure}[tb]
 \begin{center}
 \includegraphics[width=0.48\textwidth]{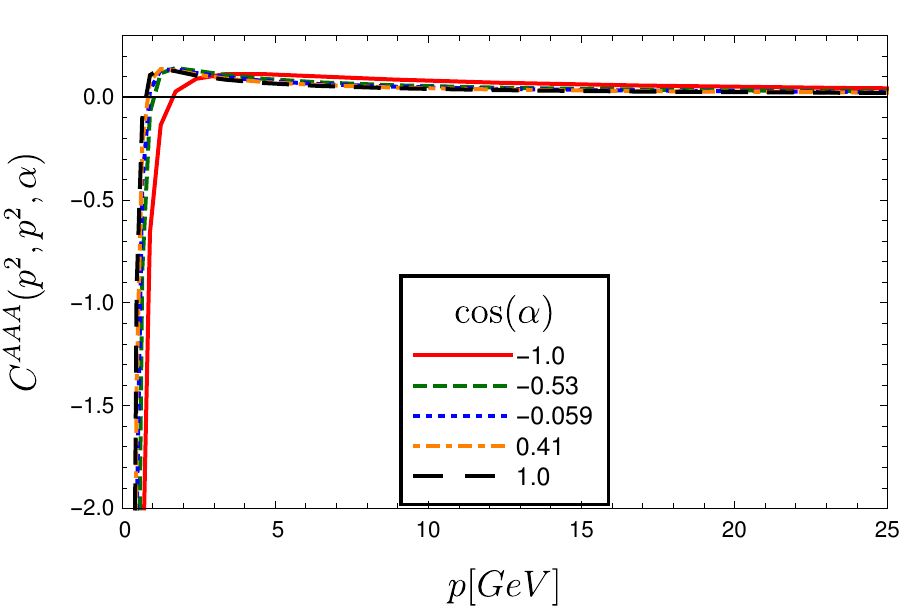}
 \includegraphics[width=0.48\textwidth]{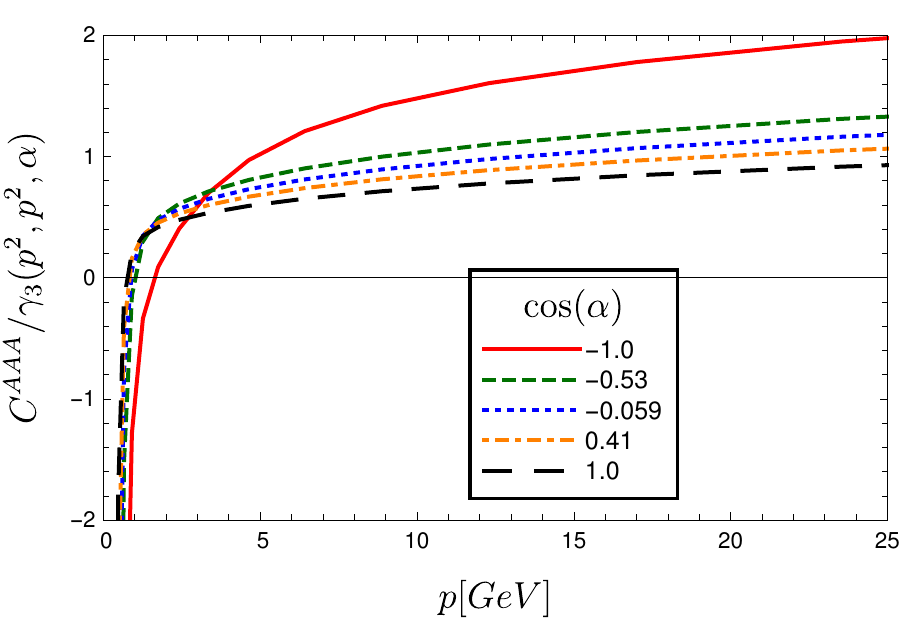}
 \caption{Angle dependence of the three-gluon vertex \cite{Huber:2014isa}.
 Two momenta are chosen with equal magnitude.
 Left: Dressing function of the three-gluon vertex.
 Right: Dressing function over variational kernel.}
 \label{fig:Coulomb_tg_equalMoms}
 \end{center}
\end{figure}

\begin{figure}[tb]
 \begin{center}
 \includegraphics[width=0.48\textwidth]{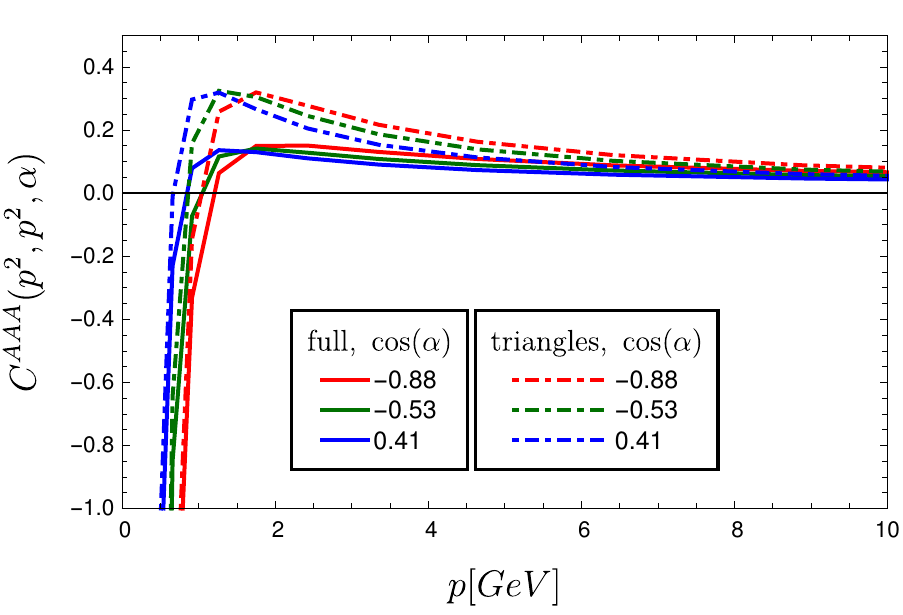}
 \includegraphics[width=0.48\textwidth]{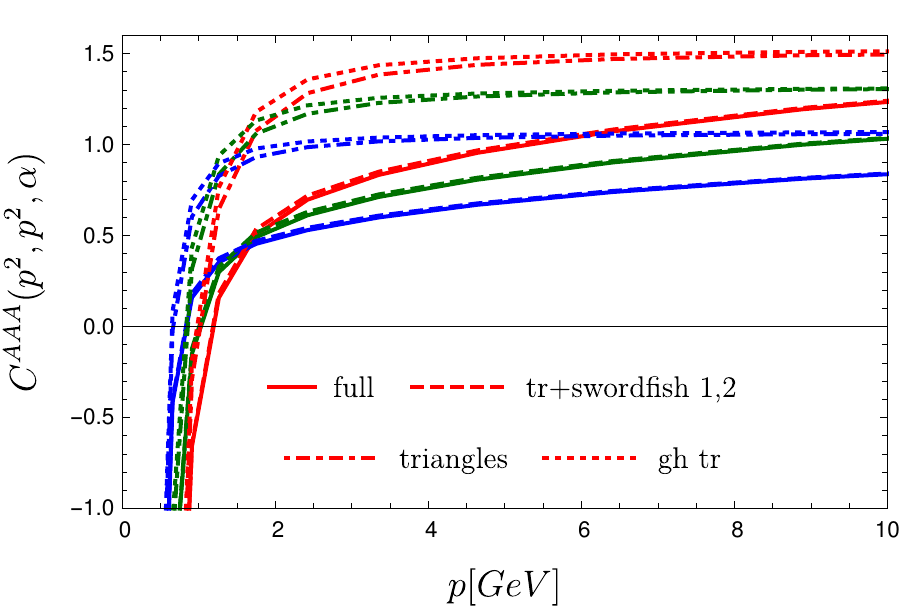}
 \caption{The three-gluon vertex from the full calculation (continuous line), from a simplified four-gluon kernel (dashed line), from a triangles-only calculation (dot-dashed line) and from a ghost-triangle-only calculation (dotted line) \cite{Huber:2014isa}.
 Left: Three-gluon vertex dressing function.
 Right: Ratio of the three-gluon vertex over the variational kernel.
 Different colors correspond to different angles as indicated in the left plot.}
 \label{fig:Coulomb_tg_comparisonApprox}
 \end{center}
\end{figure}

The kinematic dependence for the angle fixed to $2\pi/3$ is shown in \fref{fig:Coulomb_tg_3d}.
The bump in the midmomentum regime appears due to the variational kernel and is not present in $C^{AAA}/\gamma^{AAA}$.
The results are qualitatively indeed similar to calculations of the three-gluon vertex in three dimensions.
However, here a larger angle dependence is observed, as shown in \fref{fig:Coulomb_tg_equalMoms}.
In the \gls{ir}, the vertex becomes negative and diverges with a power law, since the solution corresponds to a scaling type solution.
The overall \gls{ir} exponent of the full three-gluon vertex is $-3$, in agreement with the results obtained in Refs.~\cite{Huber:2007kc,Campagnari:2010wc}.
As in Landau gauge, the \gls{ir} divergence is driven by the ghost triangle.
The gluon triangle has only a small impact, as can be seen in \fref{fig:Coulomb_tg_comparisonApprox}, where the results from solving only with different subsets of diagrams are shown.
From this figure, it is also evident that the swordfish diagrams do have a quantitative impact that should not be neglected.
Due to the variational kernel of the four-gluon vertex, $\gamma^{AAAA}$, one swordfish diagram also diverges as the ghost triangle with $p^{-5}$.
This is in contrast to the Landau gauge, where all swordfish diagrams are \gls{ir} suppressed compared to the ghost triangle.
However, the coefficient of this power law is much smaller than that of the ghost triangle.
It has only $8\,\%$ of the magnitude of the ghost triangle with the present truncation \cite{Huber:2014isa}.
In general, the corresponding swordfish diagram does not have a large impact on the solution.
Although its contribution is small, a reliable calculation of this diagram requires an increased precision compared to similar calculations in the Landau gauge \cite{Blum:2014gna,Huber:2012kd,Eichmann:2014xya}.

\begin{figure}[tb]
 \begin{center}
 \includegraphics[width=0.48\textwidth]{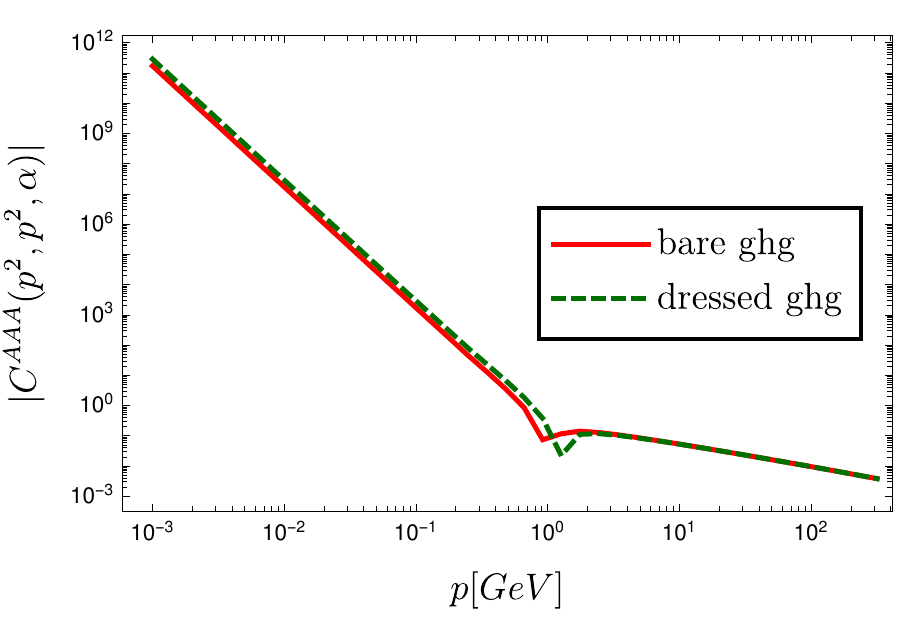}
 \hfill
 \includegraphics[width=0.48\textwidth]{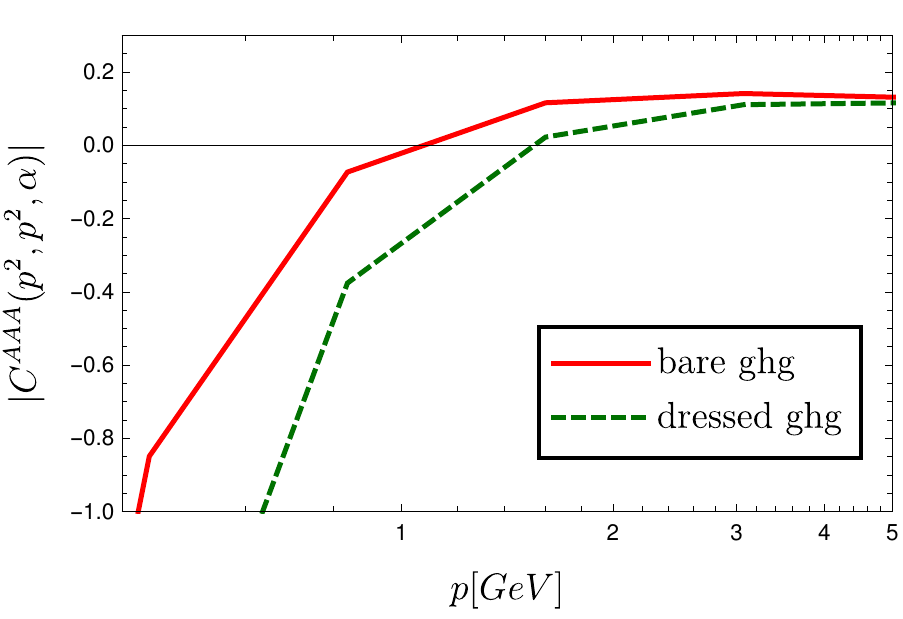}
 \caption{The three-gluon vertex dressing function at the symmetric point from the calculated ghost-gluon vertex (green, dashed line) compared to results calculated with a bare ghost-gluon vertex (red, continuous line) \cite{Huber:2014isa}.
 The right panel shows a zoom to the region around the zero crossing.}
 \label{fig:Coulomb_tg_compWithGhTrOnly}
 \end{center}
\end{figure}

With the ghost triangle being the dominant part, the question of the importance of the ghost-gluon vertex arises.
Fig.~\ref{fig:Coulomb_tg_compWithGhTrOnly} shows the ghost triangle calculated with a bare and with a dressed ghost-gluon vertex.
As can be expected, the ghost-gluon vertex shifts the zero crossing, but overall the effect is not very large and smaller than, e.g., the impact of the swordfish diagrams.

Results from the Hamiltonian approach in the Coulomb gauge including the ones presented here show that this gauge constitutes a reasonable alternative choice to the Landau gauge.
With corresponding effort, calculations using similar sophisticated truncations as in the Landau gauge should be possible.
Although the available results indicate a favorable answer, the question of the quantitative reliability of such calculations needs to be asked again.
In any case, the Hamiltonian approach provides an interesting alternative perspective on the nonperturbative regime and can also be applied to full \gls{qcd} \cite{Vastag:2015qjd,Campagnari:2016wlt,Campagnari:2018flz,Campagnari:2019zso}.

\section{Conclusions}
\label{chp:conclusions}

The investigation of the strong interaction has been an important topic in particle physics for the past decades, and we have not yet run out of problems to solve.
Despite its deceptively simple form, the Lagrangian density of \gls{qcd} still holds many secrets to be revealed.
Functional methods constitute a promising approach in this endeavor.
However, the fact that the exact equations have to be truncated is problematic as long as it is not known what the effect of a certain truncation is.
By pushing truncations further and enlarging the systems of equations to be solved, the use of models can be reduced or in some cases even be eliminated.
Such calculations demand a certain level of automatization for deriving and solving the equations which was made possible by tools that have been developed in recent years.
Some of them were reviewed in Sec.~\ref{sec:tools}.

The positive outcome of these efforts is increasing evidence that a self-contained and quantitative description of \gls{qcd} from functional methods is achievable.
To realize this, individual steps of understanding correlation functions and their mutual influence on each other must be taken and technical improvements must continue to be adopted.
Within the last few years, some correlation functions were calculated for the first time nonperturbatively and we are learning to separate the important from the irrelevant facts.
The emerging picture is promising, and although calculations are not getting easier, the potential benefit from establishing functional methods as a reliable nonperturbative method is motivating, for example, with regard to studies of the phase diagram of \gls{qcd}.

The way to improve the reliability of functional equations seems simple: enlarge the truncation.
However, it must be noted that this is not a plain brute force attack on the problem but involves also conceptual challenges.
In particular, when replacing models with dynamic quantities new problems can surface and results might initially even get worse before the full picture is understood.
The reason is simply that when models are used for unknown quantities, they provide a level of freedom that helps to overcome certain problems.
Some of these problems may not even appear as such and one has to deal with them only once a certain level of truncation is reached.
One example of the latter are the spurious divergences of the gluon propagator as discussed in detail in Sec.~\ref{sec:spurDivs}.
Various methods to deal with them are known, but the results differ quantitatively.
Only once we want quantitative results, care has to be taken how to subtract these divergences in order not to artificially influence the result.
Another aspect that becomes relevant when one wants to obtain a self-contained solution is perturbative resummation.
In this case, the problem lay dormant for a long time, because either no correct anomalous dimension was obtained, or the employed models guaranteed the self-consistency of the equation at the one-loop level.
How to correctly realize resummation in \glspl{dse} is treated in Sec.~\ref{sec:resummation}.

To understand quantum field theory, functional equations and correlation functions better, it is often useful to consider variations of the physical theory, for example, by changing the number of dimensions.
In the case of Yang-Mills theory, investigations of two and three dimensions proved to be helpful to scrutinize the structure of the solutions and the impact of different truncations as discussed in Sec.~\ref{sec:YM3d} and \ref{sec:YM2d}.
Since in these dimensions the theory is finite, no anomalous running occurs.
This simplifies the analysis.
A consistent treatment of the anomalous running poses a challenge in four dimensions and links back to Sec.~\ref{sec:resummation}.

As argued in Sec.~\ref{sec:why_LG}, the Landau gauge is the most widely adopted gauge for calculations with functional equations.
To complete the picture, it is also useful to look at other gauges.
Some examples were described in Sec.~\ref{chp:otherGauges}.
Among them, in particular linear covariant gauges are receiving more attention lately.
The reason is that their endpoint is the better understood Landau gauge and recently several problems related to the non-transverse gluon field have been addressed successfully.

For the future, the understanding of individual equations and insights into their interplay will be useful to unravel the nonperturbative regime of \gls{qcd}.
A particular challenge will be the study of its phase diagram.
This is a field where functional equations can exert their advantages of being a continuum method without any inherent technical problems akin to the complex action problem of lattice \gls{qcd}.
Of course, nonvanishing temperature and density introduce another two layers of complexity.
Having a good control over the vacuum results is a prerequisite to get on the right track to overcome also these problems.

\section*{Acknowledgments}

During the last years I enjoyed many useful discussions and had the pleasure to learn from and collaborate with many colleagues.
Among them I would like to thank in particular Reinhard Alkofer, Christian S. Fischer, Holger Gies, Jan M. Pawlowski, Lorenz von Smekal, and Andreas Wipf for their collaboration and their support.

For fruitful collaborations and helpful discussions I would like to thank Laurent Baulieu, Adrian L. Blum, Jens Braun, Davide Campagnari, Marcio A. L. Capri, Romain Contant, Anton K. Cyrol, David Dudal, Marcelo S. Guimaraes, Axel Maas, Mario Mitter, Hugo Reinhardt, H\`elios Sanchis Alepuz, Kai Schwenzer, Silvio P. Sorella, Nele Vandersickel, Richard Williams, Andreas Windisch, and Daniel Zwanziger.

In addition, discussions with Arlene Aguilar, Daniele Binosi, Gernot Eichmann, Leonard Fister, John Gracey, Markus Hopfer, Ernst-Michael Ilgenfritz, Christian Kellermann, Valentin Mader, Joannis Papavassiliou, Bernd-Jochen Schaefer, Andr\'e Sternbeck, Ralf-Arno Tripolt, Selym Villalba-Ch\'avez, Bj\"orn Wellegehausen, and Nicolas Wink were very helpful.

The most important thanks go to my family, to Birgit, Elisabeth and Katharina.
Although they are only mentioned in this one line, they also contributed to this work in their own unique ways.

\subsection*{Funding}

This research was supported by the Alexander von Humboldt foundation, the Helmholtz International Center for FAIR within the LOEWE program of the State of Hesse, the FWF (Austrian Science Fund) under Contract No. P 27380-N27 and the DFG (German Research Foundation) under Contract No. Fi970/11-1.

\subsection*{Technicalities}

Some plots were created with \textit{Jaxodraw} \cite{Binosi:2003yf}. The programs \textit{C++}, \textit{Mathematica} \cite{Wolfram:2004}, \textit{FORM} \cite{Vermaseren:2000nd,Kuipers:2012rf,Kuipers:2013pba,Ruijl:2017dtg}, \textit{DoFun} \cite{Alkofer:2008nt,Huber:2011qr,Huber:2019dkb} and \textit{CrasyDSE} \cite{Huber:2011xc} were used.

\appendix

\section{Kernels}
\label{sec:app_kernels}

For reference some \glspl{dse} kernels are collected here.
Two-loop kernels and vertex kernels are not given.
Except for the ghost-gluon vertex, the kernels of which can be found in Ref.~\cite{Huber:2012kd}, they are too long to be useful in written form.
To solve the corresponding \glspl{dse}, it is advised to create the required kernels with the help of some algebra program; see Sec.~\ref{sec:tools} for more information.

\subsection{Landau gauge}
\label{sec:app_kernels_LG}

\index{Landau gauge}
For generality, the bare ghost-gluon vertex from general covariant gauges of Refs.~\cite{Curci:1976ar,Baulieu:1981sb,ThierryMieg:1985yv,Alkofer:2003jr} is employed:
\begin{align}
 \Gamma^{A\bar{c}c,abc}_\mu(k;p,q)=i\,g\,f^{abc} \left( \eta \,p_\mu - \hat{\eta}\, q_\mu\right)
\end{align}
with $\eta+\hat\eta=1$.
For most applications one can set $\eta=1$.
For the dressed vertex the dressing $D^{A\bar{c}c,T}(k^2;p^2,q^2)$ is added.
For the gluon \gls{dse} the projector
\begin{align}
 P^{\zeta}_{\mu\nu}(p)=g_{\mu\nu}-\zeta\frac{p_\mu p_\nu}{p^2} 
\end{align}
is used.
$P^{\zeta=1}_{\mu\nu}$ corresponds to the transverse projector which is typically chosen.
This leads to the following expressions for the one-loop truncated propagator \glspl{dse}:
\begin{align}
 \frac{1}{G(p^2)}&=1+N_c\,g^2\,\int_q Z(q^2)G((p+q)^2) K_{G}(p,q)D^{A\bar{c}c}(q^2;(p+q)^2,p^2),\\
 \frac{1}{Z(p^2)}&=\widetilde{Z}_3+N_c\,g^2\,Z_4\int_q Z(q^2)K_{Z}^{tad,\zeta}(p,q)\nnnl
 &+N_c\,g^2\,\widetilde{Z}_1\int_q G(q^2)G((p+q)^2) K_{Z}^{gh,\zeta}(p,q)D^{A\bar{c}c,T}(p^2;q^2,(p+q)^2)\nnnl
 &+N_c\,g^2\,Z_1\int_q Z(q^2)Z((p+q)^2) K_{Z}^{gl,\zeta}(p,q)C^{AAA}(p^2,q^2,(p+q)^2).\label{eq:Z}
\end{align}
There is no dependence on $\eta$ in the ghost equation, because its ghost-gluon vertex is contracted with a transverse projector, and the ghost-gluon vertex dressing was put explicitly.
In the ghost loop of the gluon equation, the two structures of the bare vertex are contained in the kernel $K_{Z}^{gh,\zeta}(p,q)$.
The integral measure is defined as
\begin{align}
 \int_q=\int \frac{d^4q}{(2\pi)^4}.
\end{align}
The kernels are given by
\begin{align}
 K_{G}(p,q)&=\frac{ \left(x^2+(y-z)^2-2 x (y+z)\right)}{4 x y^2 z},\\
 K_{Z}^{gh,\zeta}(p,q)&=\frac{x^2 
\left(\zeta - 2 - 4\, \eta\hat\eta \, (\zeta-1)  \right)
+2 x (y+z)-\zeta  (y-z)^2}{12 x^2 y z}, \label{eq:KZgh}\\
 K_{Z}^{gl,\zeta}(p,q)&=\frac{z^2 \zeta }{24 x^2 y^2}+\frac{z (5 x-x \zeta +4 y \zeta )}{12 x^2 y^2}+\frac{x^2 (-19+\zeta )+2 x y (-17+\zeta )-18 y^2 \zeta }{24 x^2 y^2}\nnnl
 &+\frac{(x-y)^2 \left(x^2+10 x y+y^2 \zeta \right)}{24 x^2 y^2 z^2}+\frac{4 x^3+x y^2 (-17+\zeta )+4 y^3 \zeta -x^2 y (15+\zeta )}{12 x^2 y^2 z},\\
 K_{Z}^{tad,\zeta}(p,q)&=-\frac{\zeta  x^2-2 x ((18-5 \zeta ) y+\zeta  z)+\zeta  (y-z)^2}{12 x^2 y^2},
\end{align}
where $x=p^2$, $y=q^2$ and $z=(p+q)^2$.

\subsection[Linear covariant gauges]{Linear covariant gauges}
\label{sec:app_kernels_linCov}

\index{linear covariant gauges}
The kernels for the gluon propagator \gls{dse} in linear covariant gauges, \eref{eq:gl-DSE-final-wRGI}, are
\begin{align}
K_{Z}^{gh}(x;y,z)&=-\frac{  \left(x^2-2 x (y+z)+(y-z)^2\right) }{12 x^2 y z},\\
K_{Z}^{gl}(x;y,z)&=\frac{ \left(x^2-2 x (y+z)+(y-z)^2\right) \left(x^2+10 x (y+z)+y^2+10 y z+z^2\right)}{24 x^2 y^2 z^2},\\
\widetilde{K}_Z^{gl,\xi}(x;y,z)&=\frac{  \left(x^3 (y+z)+x^2 \left(9 y^2-4 y z+9 z^2\right)\right)}{24 x^2 y^2 z^2} F(\alpha_3^{\text{3g}}, \beta_3^{\text{3g}}; \bar{p}^2)\nnnl
&+\frac{  \left(x \left(-9 y^3+y^2 z+y z^2-9 z^3\right)-(y-z)^2 \left(y^2+z^2\right)\right)}{24 x^2 y^2 z^2} F(\alpha_3^{\text{3g}}, \beta_3^{\text{3g}}; \bar{p}^2)\nnnl
&-\frac{Z(z) (x-z) \left(x^2-2 x (y-5 z)+(y-z)^2\right)}{24 x y^2 z^2Z(x)}Z(x)F(\alpha_2^{\text{3g}}, \beta_2^{\text{3g}}; \bar{p}^2)\nnnl
 &-\frac{  Z(y) (x-y) \left(x^2+2 x (5 y-z)+(y-z)^2\right)}{24 x y^2 z^2 Z(x)}Z(x)F(\alpha_2^{\text{3g}}, \beta_2^{\text{3g}}; \bar{p}^2),\\
 K_Z^{gl,\xi^2}(x;y,z)&=\frac{ \left(x^2-2 x (y+z)+(y-z)^2\right)}{24 y^2 z^2}.
\end{align}

\section{Calculation of two-loop diagrams with a hard momentum cutoff}
\label{sec:app_twoLoop}

\begin{figure}[tb]
 \centering
 \includegraphics[width=0.4\textwidth]{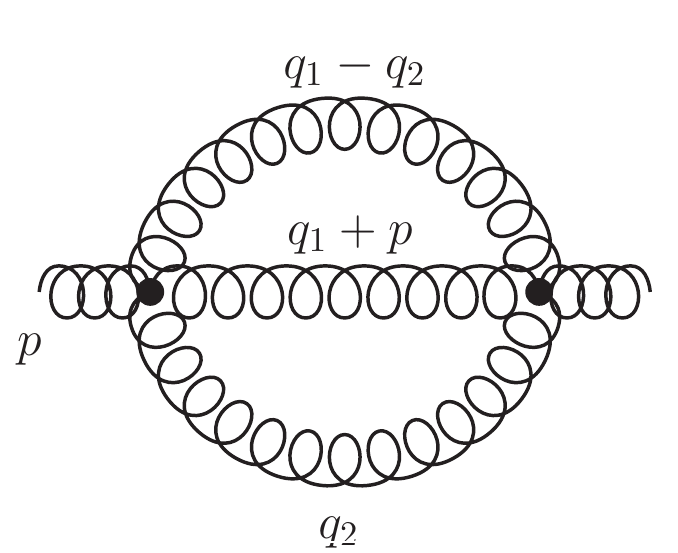}
 \hfill
 \includegraphics[width=0.4\textwidth]{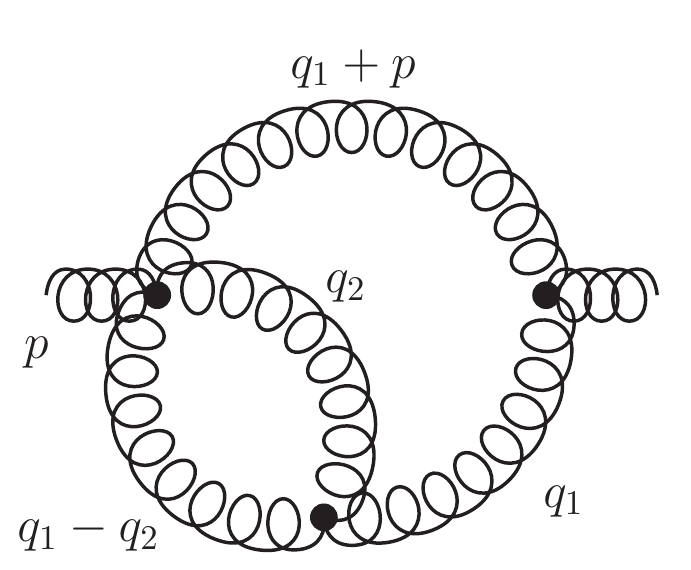}\\
 \caption{The sunset (left) and squint (right) diagrams and the routing of momenta.}
 \label{fig:app_twoLoop}
\end{figure}

\index{sunset diagram}
\index{squint diagram}
\index{two-loop diagrams}
\index{cutoff}
The calculation of a two-point two-loop integral of the type sunset or squint using a hard \gls{uv} cutoff is described here.
The generic integral reads
\begin{align}\label{eq:sunset_integral}
 I(p^2) = \intdq{q_1}\intdq{q_2} \frac{f(p,q_1,q_2)}{(q_1^2)^{n_1} (q_2^2)^{n_2} [(q_1-q_2)^2]^{n_3}[(q_1+p)^2]^{n_4}}.
\end{align}
As can be seen in \fref{fig:app_twoLoop}, the squint and sunset diagrams have similar integral structures except for the factor $(q_1^2)^{n_1}$ in the denominator.
In actual perturbative calculations, the exponents $n_i$ are only one or two.

The external momentum $p$ is chosen to define the $4$-direction, the loop momentum $q_2$ the $3$-$4$-plane and the loop momentum $q_1$ the $2$-$3$-$4$ volume:
\begin{align}
 p=P\begin{pmatrix}0\\0\\0\\1\end{pmatrix},
 \quad q_1=Q_1\fourvec{0}{\sin \theta_{11}\sin\theta_{12}}{\cos \theta_{11} \sin \theta_{12}}{\cos \theta_{12}},
 \quad q_2=Q_2\fourvec{0}{0}{ \sin \theta_{22}}{\cos \theta_{22}}.
\end{align}
Two angles of $q_2$ and one angle of $q_1$ can be integrated trivially as they do not appear in any scalar products.
The integral measure reduces then to
\begin{align}
 \intdq{q_1}\intdq{q_2} \rightarrow \int  \frac{dy_1 dy_2 d\theta_{11}d\theta_{12}d\theta_{22} \sin \theta_{11}\sin^2 \theta_{12}  \sin^2 \theta_{22}}{2(2\pi)^6}.
\end{align}
The integrals over the moduli of $q_1$ and $q_2$ have been rewritten to integrals over $y_1=q_1^2$ and $y_2=q_2^2$.

In the one-loop case, the angle integrals have to be considered separately for $q<p$ and $q>p$.
The same applies here, only that now there are four different regions:
\begin{itemize}
 \item[(I)] $p<q_1<q_2$
 \item[(II)] $q_1<p$, $q_1<q_2$
 \item[(III)] $p<q_1$, $q_2<q_1$
 \item[(IV)] $q_1<p$, $q_2<q_1$
\end{itemize}
For illustration, the angle integrals for $f(p,q_1,q_2)=1$, $n_1=n_2=0$ and $n_3=n_4=1$ are calculated in the following for $q_2<q_1$.

Since for the angle integrals only the ratio of $q_1^2$ and $q_2^2$ is relevant, we introduce $a^2=q_2^2/q_1^2<1$:
\begin{align}
 J_1(p^2,q_1^2,q_2^2) &= \int_0^\pi\int_0^\pi\int_0^\pi  \frac{d\theta_{11}d\theta_{12}d\theta_{22} \sin \theta_{11} \sin \theta_{12}^2 \sin \theta_{22}^2}{(q_1-q_2)^2(q_1+p)^2}=\nnnl
 &=\int_0^\pi\int_0^\pi\int_0^\pi  \frac{d\theta_{11}d\theta_{12}d\theta_{22} \sin \theta_{11} \sin \theta_{12}^2 \sin \theta_{22}^2 a^4}{q_2^2(a^2 p^2+q_2^2+2\,a\,p\,q_2\,\cos \theta_{12})}\nnnl
 &\quad\times\frac{1}{1+a^2-2\,a\,\cos \theta_{12}\cos\theta_{22}-2\,a\,\cos\theta_{11}\sin \theta_{12}\sin\theta_{22}}.
\end{align}
First, the integral over $\theta_{11}$ is performed.
It should be noted that this reduces the power of $\sin \theta_{22}$ by one:
\begin{align}\label{eq:tl_J1}
 J_1(p^2,q_1^2,q_2^2) &=\int_0^\pi\int_0^\pi\int_0^\pi  \frac{d\theta_{12}d\theta_{22} \sin  \theta_{12}^2 \sin \theta_{22}\, a^3\ln\left( \frac{1+a^2-2\,a\,\cos(\theta_{12}+\theta_{22})}{1+a^2-2\,a\,\cos(\theta_{12}-\theta_{22})} \right)}{2q_2^2(a^2 p^2+q_2^2+2\,a\,p\,q_2\,\cos \theta_{12})}\nnnl
 &=\int_0^\pi\int_0^\pi  \frac{d\theta_{12} \sin  \theta_{12}^2 \, a^3\,q_2^2}{2q_2^2(a^2 p^2+q_2^2+2\,a\,p\,q_2\,\cos \theta_{12})}K_1(a,\theta_{12}).
\end{align}
$K_1(a,\theta_{12})$ is the integral over $\theta_{22}$.
To calculate it, the logarithm is split into a sum and the variable transformations $\theta_{22}\rightarrow u-\theta_{12}$ and $\theta_{22}\rightarrow u+\theta_{12}$ are employed:
\begin{align}
 K_1(a,\theta_{12})&=\int_{\theta_{12}}^{\pi+\theta_{12}}du \sin(u-\theta_{12})\ln (1+a^2-2\,a\,\cos u)\nnnl
 &\quad +\int_{-\theta_{12}}^{\pi-\theta_{12}}du \sin(u+\theta_{12})\ln (1+a^2-2\,a\,\cos u).
\end{align}
Using the symmetry properties of the integrals, this expression can be rewritten to
\begin{align}
 K_1(a,\theta_{12})&=-2\int_0^\pi du \cos u \sin \theta_{12}\ln (1+a^2-2\,a\,\cos u)=\frac{2\pi \sin\theta_{12}}{a}
\end{align}
and \eref{eq:tl_J1} becomes
\begin{align}
 J_1(p^2,q_1^2,q_2^2) &=\pi \int_0^\pi \frac{d\theta_{12} \sin  \theta_{12}^2  }{q_2^2(p^2+q_1^2+2\,a\,p\,q_1\,\cos \theta_{12})},
\end{align}
where $a^2=q_2^2/q_1^2$ was plugged in.
The final result is
\begin{align}
 J_1(p^2,q_1^2,q_2^2)=\frac{\pi^2}{2q_2^2}\left(\frac{1}{q_1^2}\theta(q_1^2-p^2) + \frac{1}{p^2}\theta(p^2-q_1^2)\right).
\end{align}
$\theta$ is the unit step function.
The case $q_1<q_2$ can be done in the same way, only that in this case one sets $a^2=q_1^2/q_2^2$.

For the complete calculation of two-loop diagrams, many different $f(p,q_1,q_2)$ are required.
To keep calculations manageable, the calculation is automated.
To this end, all required integrals are extracted from the expressions and precalculated analytically with \textit{Mathematica}.
Unfortunately, the correctness of the results is not guaranteed and different versions of \textit{Mathematica} give different results.
The trick to have correct results is quite simple, though.
Every integral is checked numerically.
If a wrong result is detected, the integral is recalculated after performing some simple rewritings like $\cos^2\theta_{12} \rightarrow 1-\sin^2\theta_{12}$.
The final list of results is converted into a list of simple replacement rules.

A tricky part is the required transformation of variables, as many different combinations appear.
However, one can deal with that algorithmically and bring all expressions into the desired form before applying the rules.
As a final test, all analytically derived results can be tested numerically at random points.

\bibliographystyle{utphys_mod}
\bibliography{literature_thesis}

\printindex

\end{document}